\documentclass[preprint,12pt]{elsarticle}
\pdfoutput=1 % allows using 'latex mcreview.tex' (arXiv style)

\usepackage[utf8]{inputenc}
\usepackage{amsmath}
\usepackage{xspace}
\usepackage{subfigure}

\usepackage{amssymb}
\biboptions{square,sort&compress}

\def\useparts{1}

\usepackage{xspace}
\usepackage{xcolor}

\usepackage{setspace}
\usepackage{threeparttable}
\usepackage{hyperref}
\usepackage{relsize}
\usepackage{ifthen}

\newcommand{\EqRef}[1]{Eq.~\eqref{#1}\xspace}

\newcommand{\SecRef}[1]{Section~\ref{#1}\xspace}
\newcommand{\SecsRef}[1]{Sections~\ref{#1}\xspace}
\newcommand{\AppRef}[1]{\ref{#1}\xspace}
\newcommand{\FigRef}[2][]{Fig.~\ref{#2}#1\xspace}
\newcommand{\FigsRef}[1]{Figs.~\ref{#1}\xspace}
\newcommand{\TabRef}[1]{Tab.~\ref{#1}\xspace}

\DeclareRobustCommand{\plusplus}{\raisebox{0.2ex}{\smaller++}}

\def\jimmy{\textsc{Jimmy}\xspace}

\def\herwig{\textsc{Herwig}\xspace}

\def\herwigpp{Herwig\plusplus\xspace}
\newcommand{\Herwig}{\herwig}
\newcommand{\Herwigpp}{\herwigpp}

\def\ariadne{\textsc{Ariadne}\xspace}
\def\Ariadne{\ariadne}

\def\thepeg{\textsc{ThePEG}\xspace}

\def\pythia{\textsc{Pythia}\xspace}
\def\Pythia{\pythia}
\def\pythiasix{\pythia~6\xspace}
\def\pythiaseven{\pythia~7\xspace}
\def\pythiaeight{\pythia~8\xspace}

\def\jetset{\textsc{Jetset}\xspace}
\def\professor{\textsc{Professor}\xspace}
\def\rivet{Rivet\xspace}
\def\agile{AGILe\xspace}

\newcommand{\Sherpa}{\textsc{Sherpa}\xspace}
\def\sherpa{\Sherpa}

\def\alpgen{AlpGen\xspace}
\def\Alpgen{\alpgen}

\def\hepdata{HepData\xspace}

\def\rivet{Rivet\xspace}
\def\professor{Professor\xspace}

\def\fastjet{FastJet\xspace}
\def\hepmc{HepMC\xspace}
\def\agile{AGILe\xspace}
\def\hztool{HZTool\xspace}

\def\evtgen{EvtGen\xspace}
\def\tauola{TAUOLA\xspace}%sc?
\def\cteq#1{\ensuremath{\text{CTEQ#1}}\xspace}

\newcommand{\Helac}{HELAC\xspace}
\newcommand{\Phegas}{PHEGAS\xspace}
\newcommand{\Madgraph}{MadGraph\xspace}
\newcommand{\Madevent}{MadEvent\xspace}
\newcommand{\Whizard}{Whizard\xspace}
\newcommand{\OmegaCode}{O'Mega\xspace}
\newcommand{\Vincia}{\textsc{Vincia}\xspace}
\newcommand{\Comix}{\textsc{Comix}\xspace}
\newcommand{\Apacic}{\textsc{Apacic\plusplus}\xspace}
\newcommand{\Amegic}{\textsc{Amegic\plusplus}\xspace}
\newcommand{\Ahadic}{\textsc{Ahadic\plusplus}\xspace}

\newcommand{\Hadrons}{\textsc{Hadrons\plusplus}\xspace}
\newcommand{\Photons}{\textsc{Photons\plusplus}\xspace}

\newcommand{\FeynRules}{FeynRules\xspace}

\def\mcnet{MCnet\xspace}

\def\lep{LEP\xspace}

\def\lhc{LHC\xspace}

\def\mcnet{MCnet\xspace}

\def\Cpp{C\plusplus\xspace}
\def\cpp{\Cpp}

\def\fortran{Fortran\xspace}

\newcommand{\MCatNLO}{MC@NLO\xspace}
\newcommand{\POWHEG}{\textsc{Powheg}\xspace}
\newcommand{\MENLOPS}{\textsc{MEnloPS}\xspace}
\newcommand{\Vegas}{VEGAS\xspace}
\newcommand{\ALPHA}{ALPHA\xspace}

\newcommand{\abr}[1]{\langle #1\rangle}

\newcommand{\done}{{\rm d}}
\newcommand{\dthree}{{\rm d}^3}

\newcommand{\oforder}[1]{\ensuremath{\mathcal{O}(#1)}\xspace}

\newcommand{\mb}[1]{\mathbf{#1}}
\newcommand{\mr}[1]{\mathrm{#1}}
\newcommand{\drm}{\mr{d}}

\newcommand{\bea}{\begin{eqnarray}}
\newcommand{\eea}{\end{eqnarray}}
\newcommand{\beq}{\begin{equation}}
\newcommand{\eeq}{\end{equation}}
\newcommand{\bi}{\begin{itemize}}
\newcommand{\ei}{\end{itemize}}
\newcommand{\nnb}{\nonumber}

\newcommand{\pt}{\ensuremath{p_\perp}\xspace}
\newcommand{\pT}{\pt}
\newcommand{\pPerp}[1]{\ensuremath{p_{\perp #1}}\xspace}
\newcommand{\kt}{\ensuremath{k_\perp}\xspace}
\newcommand{\kT}{\kt}

\newcommand{\chisq}{\ensuremath{\chi^2}\xspace}
\newcommand{\pp}{\ensuremath{pp}\xspace}
\newcommand{\ppbar}{\ensuremath{p\bar{p}}\xspace}
\newcommand{\epluseminus}{\ensuremath{e^+ e^-}\xspace}
\newcommand{\ee}{\epluseminus}
\newcommand{\ep}{\ensuremath{e^\pm p}\xspace}
\newcommand{\alphaS}{\ensuremath{\alpha_\text{s}}\xspace}

\newcommand{\LambdaQCD}{\ensuremath{\Lambda_\text{QCD}}\xspace}
\newcommand{\pdf}[2][]{\ensuremath{f_{#2}^{#1}}}
\newcommand{\xpdf}[2][]{\ensuremath{x\!f_{#2}^{#1}}}
\newcommand{\Nc}{\ensuremath{N_c}}

\ifnum\useparts > 0
\newcommand{\mcintrosection}[1]{\section{#1}}
\newcommand{\mcintrosubsection}[1]{\subsection{#1}}
\newcommand{\mcpart}[1]{\part{#1}}
\newcommand{\mcsection}[1]{\section{#1}}
\newcommand{\mcsubsection}[1]{\subsection{#1}}
\newcommand{\mcsubsubsection}[1]{\subsubsection{#1}}
\newcommand{\PartRef}[1]{Part~\ref{#1}\xspace}
\else
\newcommand{\mcintrosection}[1]{\section{#1}}
\newcommand{\mcintrosubsection}[1]{\subsection{#1}}
\newcommand{\mcpart}[1]{\section{#1}}
\newcommand{\mcsection}[1]{\subsection{#1}}
\newcommand{\mcsubsection}[1]{\subsubsection{#1}}
\newcommand{\mcsubsubsection}[1]{\paragraph{#1}}
\newcommand{\PartRef}[1]{Chapter~\ref{#1}\xspace}
\fi

\newcommand{\eg}{e.g.~}
\newcommand{\ie}{i.e.~}
\newcommand{\cf}{cf.~}

\newcommand{\gensectionintro}{Introduction}
\newcommand{\gensectionhard}{Hard processes}
\newcommand{\gensectionshower}{Parton showering}
\newcommand{\gensectionMPI}{Multiple parton interactions and beam remnants}
\newcommand{\gensectionhadronize}{Hadronization}
\newcommand{\gensectiondecay}{Hadron decays and QED radiation}
\newcommand{\gensectionoutlook}{Summary}

\newcommand{\llBlob}[2]{\put(#1,#2){\circle{13}}\put(#1,#2){\circle{11}}\put(#1,#2){\circle{9}}\put(#1,#2){\circle{7}}\put(#1,#2){\circle{5}}\put(#1,#2){\circle{3}}\put(#1,#2){\circle{1}}}
\newcommand{\llmrm}[1]{{\mathrm{#1}}}
\newcommand{\lld}{\ensuremath{\done}}

\newcommand{\llxsec}{\ensuremath{\sigma}\xspace}
\newcommand{\llPS}{\ensuremath{\Phi}\xspace}
\newcommand{\llPScut}[2]{\ensuremath{\llPS_{#1(_>#2)}}\xspace}
\newcommand{\llPSmax}[2]{\ensuremath{\llPS_{#1(_<#2)}}\xspace}

\newcommand{\llW}{\ensuremath{W}\xspace}

\newcommand{\llord}{\ensuremath{q^2}\xspace}
\newcommand{\llordms}{\ensuremath{Q^2_{\llmrm{MS}}}\xspace}
\newcommand{\llordmax}{\ensuremath{Q^2}\xspace}
\newcommand{\llordcut}{\ensuremath{Q^2_0}\xspace}
\newcommand{\llaux}{\ensuremath{z}\xspace}
\newcommand{\llsup}[1]{\ensuremath{^{\llmrm{#1}}}\xspace}
\newcommand{\llsub}[1]{\ensuremath{_{\llmrm{#1}}}\xspace}

\newcommand{\llsud}[1]{\ensuremath{\Delta}}
\newcommand{\llsudb}[1]{\ensuremath{\Bar{\Delta}}}
\newcommand{\llsplitP}{\ensuremath{P}\xspace}

\newcommand{\lleetoj}{\ensuremath{e^+e^-\to\llmrm{jets}}\xspace}

\newcommand{\lletal}{et~al.\xspace}
\newcommand{\llee}{\ensuremath{e^+e^-}\xspace}
\newcommand{\llep}{\ensuremath{ep}\xspace}
\newcommand{\llD}{\ensuremath{D}\xspace}
\newcommand{\llipt}{\ensuremath{p_\perp}\xspace}
\newcommand{\llipti}[1]{\ensuremath{p_{\perp #1}}\xspace}
\newcommand{\llirap}{\ensuremath{y}\xspace}
\newcommand{\llsdip}{\ensuremath{S\llsub{dip}}\xspace}
\newcommand{\llqrk}{\ensuremath{q}\xspace}
\newcommand{\llg}{\ensuremath{g}\xspace}
\newcommand{\llqbar}{\ensuremath{\bar{q}}\xspace}
\newcommand{\llqqbar}{\ensuremath{q\bar{q}}\xspace}
\newcommand{\llDIPSY}{\protect\scalebox{0.8}{DIPSY}\xspace}

\journal{Physics Reports}

\begin{document}
%\linenumbers

\begin{frontmatter}



\title{General-purpose event generators for LHC physics}


\author[Edinburgh]{Andy~Buckley}
\author[UCL]{Jonathan~Butterworth}
\author[Karlsruhe]{Stefan~Gieseke}
\author[Durham]{David~Grellscheid}
\author[SLAC]{Stefan~H{\"o}che}
\author[Durham]{Hendrik~Hoeth}
\author[Durham]{Frank~Krauss}
\author[Lund,CERN]{Leif~Lönnblad}
\author[UCL]{Emily~Nurse}
\author[Durham]{Peter~Richardson}
\author[Heidelberg]{Steffen~Schumann}
\author[Manchester]{Michael~H.~Seymour}
\author[Lund]{Torbjörn~Sjöstrand}
\author[CERN]{Peter~Skands}
\author[Cambridge]{Bryan~Webber}

\address[Edinburgh]{PPE Group, School of Physics \& Astronomy, University of Edinburgh, EH25 9PN, UK}
\address[UCL]{Department of Physics \& Astronomy, University College London, WC1E 6BT, UK}
\address[Karlsruhe]{Institute for Theoretical Physics, Karlsruhe Institute of Technology, D-76128 Karlsruhe}
\address[Durham]{Institute for Particle Physics Phenomenology, Durham University, DH1 3LE, UK}
\address[SLAC]{SLAC National Accelerator Laboratory, Menlo Park, CA 94025, USA}
\address[Lund]{Department of Astronomy and Theoretical Physics, Lund University, Sweden}
\address[CERN]{PH Department, TH Unit, CERN, CH-1211 Geneva 23, Switzerland}
\address[Heidelberg]{Institute for Theoretical Physics, University of Heidelberg, 69120 Heidelberg, Germany}
\address[Manchester]{School of Physics and Astronomy, University of Manchester, M13 9PL, UK}
\address[Cambridge]{Cavendish Laboratory, J.J.~Thomson Avenue, Cambridge CB3 0HE, UK}

\begin{abstract}
\hyphenation{perturb-ative}
  We review the physics basis, main features and use of general-purpose
  Monte Carlo event generators for the simulation of
  proton-proton collisions at the Large Hadron Collider.
  Topics included are:
  the generation of hard-scattering matrix elements for
  processes of interest, at both leading and next-to-leading QCD
  perturbative order; their matching to approximate treatments of
  higher orders based on the showering approximation;
  the parton and dipole shower formulations; parton distribution
  functions for event generators; non-perturbative aspects such as
  soft QCD collisions, the underlying event and diffractive processes;
  the string and cluster models for hadron formation; the treatment of
  hadron and tau decays; the inclusion of QED radiation and
  beyond-Standard-Model processes. We describe the principal features of
  the \ariadne, \herwigpp, \pythiaeight and \Sherpa generators, together with
  the \rivet and \professor validation and tuning tools, and discuss the
  physics philosophy behind the proper use of these generators and
  tools.  This
  review is aimed at phenomenologists wishing to understand better how
  parton-level predictions are translated into hadron-level events as
  well as experimentalists wanting a deeper insight into the tools
  available for signal and background simulation at the LHC.
  \begin{minipage}{1.0\linewidth}
    \vspace*{-42cm}%
    \begin{minipage}{0.325\linewidth}
      \flushright 
      CERN-PH-TH-2010-298\\
      Cavendish-HEP-10/21\\
      MAN/HEP/2010/23\\
      SLAC-PUB-14333\\
      HD-THEP-10-24
    \end{minipage}
    \begin{minipage}{0.25\linewidth}
      \flushright 
      KA-TP-40-2010\\
      DCPT/10/202\\
      IPPP/10/101\\
      LU TP 10-28\\
      MCnet-11-01
    \end{minipage}
  \end{minipage}
\end{abstract}


\end{frontmatter}

% \linenumbers

\tableofcontents
\newpage

\mcintrosection{General introduction}
\label{sec:general-introduction}
Understanding the final states of high energy particle
collisions such as those at the Large Hadron Collider (LHC) is an extremely challenging
theoretical problem.  Typically hundreds of particles are produced,
and in most processes of interest their momenta range over
many orders of magnitude.  All the particle species of the Standard
Model (SM), and maybe some beyond, are involved.  The relevant matrix
elements are too laborious to compute beyond the first few orders of
perturbation theory, and in the case of QCD processes they involve the
intrinsically non-perturbative and unsolved problem of confinement.
Once these matrix elements have been computed within some
approximation scheme, there remains the problem of dealing with their
many divergences and/or near-divergences.  Finally they must be
integrated over a final-state phase space of huge and variable
dimension in order to obtain predictions of experimental observables.

Over the past thirty years an armoury of techniques has been developed
to tackle these seemingly intractable problems.  The crucial tool
of factorization allows us to separate the treatment of many processes
of interest into different regimes, according to the scales of
momentum transfer involved.  At the highest scales, the constituent
partons of the incoming beams interact to produce a relatively small
number of energetic outgoing partons, leptons or gauge bosons.  The matrix
elements of these hard subprocesses are perturbatively computable.
At the very lowest scales, of the order of 1~GeV, incoming partons are
confined in the beams and outgoing partons interact non-perturbatively
to form the observed final-state hadrons.  These soft processes cannot
yet be calculated from first principles but have to be modelled.
The hard and soft regimes are distinct but connected
by an evolutionary process that can be calculated in principle from
perturbative QCD.  One consequence of this scale evolution is the
production of many additional partons in the form of initial- and
final-state parton showers, which eventually participate in the
low-scale process of hadron formation.

All three regimes of this highly successful picture of hard collisions
are eminently suited to computer simulation using Monte Carlo
techniques.  The large and variable dimension of the phase space,
$3n-4$ dimensions\footnote{Three components of momentum per produced
  particle,  minus four constraints of overall energy-momentum
  conservation.}
plus flavour and spin labels for an $n$-particle final state,  makes Monte
Carlo the integration method of choice: its accuracy improves inversely as the
square root of the number of integration points, irrespective of the dimension.
The evolution of scales that leads to parton showering is a
Markov process that can be simulated efficiently with Monte Carlo
techniques, and the available hadronization models are
formulated as Monte Carlo processes from the outset. 
Furthermore the factorized nature of the
problem means that the treatment of each regime can be improved
systematically as more precise perturbative calculations or more
sophisticated hadronization models become available.   

Putting all
these elements together, one has a Monte Carlo event generator capable
of simulating a wide range of the most interesting processes that are
expected at the LHC, which can be used for several distinct purposes in
particle physics experiments.  Event generators are usually required to extract a
signal of new physics from the background of SM
processes. Comparisons of their predictions to the data can be used to
perform measurements of SM parameters. They also provide realistic
input for the design of new experiments, or for new selection or
reconstruction procedures within an existing experiment. 

Historically, the development of event generators 
began shortly after the discovery of the partonic structure of
hadrons and of QCD as the theory of strong interactions.\footnote{For an early
review, see~\cite{Webber:1986mc}.}  Some important features of hard
  processes, such as deep inelastic scattering and hadroproduction of jets and lepton
  pairs, could be understood simply in terms of parton interactions.
  To describe final states in more detail, at first
simple models were used to fragment the primary partons
directly into hadrons, but this could not account for the transverse
broadening of jets and lepton pair distributions with increasing
hardness of the interaction.  It was soon appreciated that the primary
partons, being coloured, would emit gluons in the same way that
scattered charged particles emit photons, and that these gluons,
unlike photons, could themselves radiate, leading to a parton cascade
or shower that might account for the broadening.  It was then evident
 that hadron formation would occur
naturally as the endpoint of parton showering, when the typical scale
of momentum transfers is low and the corresponding value of the QCD
running coupling is large.  However, this very fact renders the
hadronization process non-perturbative, so hadronization models,
inspired by QCD but not so far derivable from it, were developed with
tunable parameters to describe the hadron-level properties of final states.

Although most of the signal processes of interest at the LHC fall into
the category of hard interactions that can be treated by the above
methods, the vast majority of collisions are soft, leading to
diffractive scattering or multiparticle production with low transverse
momenta.  These soft processes also need to be simulated
but, as in the case of hadronization, their non-perturbative nature
means that we have to resort to models with tunable parameters to
describe the data.  A related phenomenon is the component of the final
state in hard interactions that is not associated with the primary hard process
-- the so-called ``underlying event''.  There is convincing evidence
that this is due to secondary interactions of the other constituent
partons of the colliding hadrons.  The hard tails of these
interactions are described by perturbative QCD, but again the soft
component has to be modelled.  The same multiple-parton interaction
model can serve for the simulation of soft collisions,
provided there is no conflict between the parameter values needed to
describe the two phenomena. 

The main purpose of this review is to provide a survey of how all
the above components are implemented in the general-purpose event
generators that are currently available for the simulation of LHC
proton-proton collisions.  The authors are members of
MCnet,\footnote{http://www.montecarlonet.org/} a European Union funded
Marie Curie Research Training Network dedicated to developing the next
  generation of Monte Carlo event generators and providing training of
  its user base; the review seeks to contribute to those objectives.

Our discussion is aimed at phenomenologists wishing to understand better
the simulation of hadron-level events as well as experimentalists
wanting a deeper insight into the tools available for signal and
background simulation at the LHC.  We have tried to start at a level
that does not assume expertise beyond graduate particle physics
courses.  However, some sections dealing with current developments,
such as the matching of matrix elements and parton showers, are
necessarily more technical.  In those cases the treatment is less
pedagogical but we provide references to further discussion and proofs. 
Each section ends with a set of bullet points summarizing the main
points. In many cases we illustrate points by reference to plots of event
generator output, and compare with experimental data where available.

We begin in \PartRef{sec:revi-phys-behind} with a more detailed
discussion of the physics involved in event generators, starting with
an overview in \SecRef{sec:event-structure} of the structure of an event
and the steps by which it is generated.  We then describe the hard
subprocess in \SecRef{sec:subprocesses} before going on to the
parton showers in \SecRef{sec:parton-showers}. The precision of these
perturbative components of the simulation has been improved in recent
years by various schemes to include higher-order QCD corrections
without double counting, which we review in
\SecRef{sec:me-nlo-matching}.

Next we turn to the non-perturbative aspects of event generation,
starting in \SecRef{sec:pdfs-event-gener} with the parton distribution
functions of the incoming hadrons, which are used not only to compute
the hard subprocess cross section but also for the generation of
initial-state parton showers.  We go on to discuss the modelling of
soft collisions, the underlying event and diffraction in
\SecRef{sec:minim-bias-underly}, and then in \SecRef{sec:hadronization}
we describe the principal hadronization models used in present-day
event generators.

It is well established that a large fraction of produced particles
come from the decays of unstable hadronic resonances, and therefore the
accurate simulation of these decays, together with electroweak decays
that occur before particles have exited a typical beampipe or detector, is an essential
part of event generation, reviewed in \SecRef{sec:hadron-decays}.  Next we
describe the available techniques for simulating QED
radiation. \PartRef{sec:revi-phys-behind} ends with a discussion of
the simulation of physics beyond the Standard Model.

\PartRef{sec:spec-revi-main} contains brief reviews of the individual
event generators that were developed as part of the MCnet Network,
referring back to \PartRef{sec:revi-phys-behind} for the physics
involved and the modelling options that are implemented.  Then
in \PartRef{sec:comp-gener-with} we discuss issues involved in the use
of event generators, their validation and tuning, and the tools that
have been developed for these purposes. In particular, guidelines for
making experimental measurements that are optimally useful for Monte
Carlo validation and tuning are given.

\PartRef{sec:comp-gener-with} ends with some illustrative plots of
results from the MCnet event generators for a wide range of processes.  It
should be emphasised that these results are only ``snapshots'' of the
current state of the generators, which have not yet been thoroughly
tuned for use at the LHC.  For up-to-date comparisons with LHC data
one should consult the repository of plots at {\tt mcplots.cern.ch}.

A number of Appendices deal with important technical points in more
detail.  \AppRef{sec:mc-methods} gives a brief survey of the basic
Monte Carlo methods employed in event generators, while
\AppRef{sec:app_mcs} discusses methods for evaluating hard subprocess
matrix elements and phase space integration.
A particularly important Standard Model parameter is the top quark mass, and
we devote \AppRef{sec:top-quark-masses} to the meaning of this quantity as determined
by tuning the corresponding event generator parameters.

  As space is limited, and the emphasis of MCnet has been on
  general-purpose event generation for proton colliders, some topics
  relevant to the LHC programme, notably  heavy ion collisions, are
  not included.  We also do not cover specialized generators for
  specific processes, or programs that operate only at parton level
  and do not generate complete hadron-level final states. In
  most cases the latter can be interfaced to the MCnet generators
  through standard file formats,  as outlined in
  \AppRef{sec:MEinterfaces},  although care must be taken to avoid
  double counting, as discussed in \SecRef{sec:me-nlo-matching}.

For reference and to avoid repetition, we have collected in
\TabRef{tab:abbrev} the common abbreviations used throughout the review.

\begin{table}
\begin{center}
\begin{tabular}{|l|l|}
\hline
BSM & Beyond Standard Model \\ 
DIS & Deep inelastic (lepton) scattering \\ 
FSR & Final-state (QCD) radiation \\  
ISR & Initial-state (QCD) radiation \\
LL & Leading logarithm(ic) \\
LO & Leading order \\ 
MC & Monte Carlo \\ 
ME & Matrix element \\
MPI & Multiple parton interations \\
NLL & Next-to-leading logarithm(ic) \\
NLO & Next-to-leading order \\
PDF & Parton distribution function \\
PS & Parton shower \\
SM & Standard Model \\ 
UE & Underlying event \\
\hline
\end{tabular}
\caption{ Abbreviations used in this review.\label{tab:abbrev}}
\end{center}
\end{table}

%\input{introduction/mc-truth}

% Local Variables: 
% mode: LaTeX
% TeX-master: "../mcreview"
% End: 

\pagebreak
\mcpart{Review of physics behind MC event generators}
\label{sec:revi-phys-behind}

% Local Variables: 
% mode: LaTeX
% TeX-master: "../mcreview"
% End: 

\mcsection{Structure of an event}
\label{sec:event-structure}
We start this part of the review with a brief overview of the steps by
which event generators build up the structure of a hadron-hadron
collision involving a hard process of interest -- that is, a process
in which heavy objects are created or a large momentum transfer
occurs. As outlined already in \SecRef{sec:general-introduction},
there are several
basic phases of the process that need to be simulated: a primary hard
subprocess, parton showers associated with the incoming and outgoing
coloured participants in the subprocess, non-perturbative interactions
that convert the showers into outgoing hadrons and connect them to the
incoming beam hadrons, secondary interactions that give rise to the
underlying event, and the decays of unstable particles that do not
escape from the detector. There are corresponding steps in the event generation.

Of course, not all these
steps are relevant in all processes.  In particular, the majority of
events that make up the total hadron-hadron cross section are of
soft QCD type and rely more on phenomenological models.  At the
other extreme the simulation of new-physics events such as
supersymmetric particle production and decay, and the SM backgrounds to
them, rely on essentially all of the components.

We also briefly introduce two issues that affect all areas of the
simulation: the jet structure of the final state and a widely used
approximation to QCD~-- the large-$\Nc$ limit.

In most applications of event generators, one is interested in events of
a particular type.  Rather than simulating typical events and waiting
for one of them to be of the required type, which can be as rare as
1~in~$10^{15}$ in some applications, the simulation is built around the
hard subprocess.  The user selects hard subprocesses of given types and
partonic events are generated according to their matrix elements
and phase space, as described in \SecRef{sec:subprocesses} and in more
detail in \AppRef{sec:app_mcs}.  These are typically of LO for the given
process selected (which could be relatively high order in the QCD
coupling, for example for $Z$+4~partons) and calculated with the
tree-level matrix elements.  There has however been important progress
in including loop corrections into hard process generation, as described
in \SecRef{sec:me-nlo-matching}.

Since the particles entering the hard subprocess, and some of those
leaving it, are typically
QCD partons, they can radiate gluons.  These gluons can radiate
others, and also produce
quark--antiquark pairs, generating showers of outgoing partons.
This process is simulated with a step-wise Markov
chain, choosing probabilistically to add one more parton to the final
state at a time, called a parton shower algorithm, described in
\SecRef{sec:parton-showers}.  It is formulated as an evolution in some
momentum-transfer-like variable downwards from a scale defined by the
hard process, and as both a forwards evolution of the outgoing partons
and a backwards evolution of the incoming partons progressively towards
the incoming hadrons.

The incoming hadrons are complex bound states of strongly-interacting
partons and it is possible that, in a given hadron-hadron collision,
more than one pair of partons may interact with each other.  These
multiple interactions go on to produce additional partons throughout the
event, which may contribute to any observable, in addition to those from
the hard process and associated parton showers that we are primarily
interested in.  We therefore describe this part of the event structure
as the underlying event.  As described in
\SecRef{sec:minim-bias-underly}, it can also be formulated as a downward
evolution in a momentum-transfer-like variable.

As the event is evolved downwards in momentum scales it ultimately
reaches the region, at scales of order 1~GeV, in which QCD becomes
strongly interacting and perturbation theory breaks down.  Therefore at
this scale the perturbative evolution must be terminated and replaced by
a non-perturbative hadronization model that describes the confinement of the system of
coloured partons into colourless hadrons. A key feature of these
models, described in \SecRef{sec:hadronization},  is that individual
partons do not hadronize independently, but rather colour-connected
systems of partons hadronize collectively.  These models are not
derived directly from QCD and consequently have more free parameters
than the preceding components.  However, to a good approximation
they are universal~-- the hadronization of a given coloured system is
independent of how that system was produced, so that once tuned on one
data set the models are predictive for new collision types or energies.

Finally, many of the hadrons that are produced during hadronization are
unstable resonances.  Sophisticated models are used to simulate their
decay to the lighter hadrons that are long-lived enough to be considered
stable on the time-scales of particle physics detectors,
\SecRef{sec:hadron-decays}.  Since many of the particles involved with
all stages of the simulation are charged, QED radiation effects can also
be inserted into the event chain at various stages,
\SecRef{sec:qed-radiation}.

\mcsubsection{Jets and jet algorithms}

The final states of many subprocesses of interest include hard partons.
Radiation from the incoming partons is a source of additional partons in
the final state.  The parton shower evolution is dominated by the
emission of additional partons that are either collinear with the
outgoing partons or are soft.  The final state of the parton shower
therefore predominantly has a structure in which most of the energy is
carried by localized collinear bundles of partons, called jets.  The
hadronization mechanism is such that this jet structure is preserved and
it is experimentally observed that the final state of
high-momentum-transfer hadronic events is
dominated by jets of hadrons.  The distributions of the total momentum
of hadrons in jets are approximately described by perturbative
calculations of partons with the same total momentum.

Although jets are a prominant feature of hadronic events, they are not
fundamental objects that are defined by the theory.  In order to
classify the jet final state of a collision, define which hadrons belong
to which jet and reconstruct their total momentum, we need a precise
algorithmic jet definition, or jet algorithm.  There has been much
progress on the properties that such algorithms must satisfy in order to
be convenient theoretically and experimentally.  We are not able to
review this work here (for a recent thorough review, see
\cite{Salam:2009jx} for example), but we mention one important property
that we require of a jet algorithm.  One of the applications we will use
them for is the matching of perturbative calculations at different
orders and with different jet structures and in order for this to be
well-defined we must use an algorithm for which jet cross sections can
be calculated on the parton level to arbitrarily high order of
perturbation theory.  This is only true of jet algorithms that are
\emph{collinear and infrared safe}.  That is, for
any partonic configuration, replacing any parton with a collinear set of
partons with the same total momentum, or adding any number of infinitely
soft partons in any directions, should produce the identical
result.  One can show that, provided this property is satisfied, jet
cross sections are finite at any perturbative order and have
non-perturbative corrections that are suppressed by powers of the jet
momenta, so that at high momentum transfers the jet structure of the
hadronic final state of
a collision is very well described by a parton-level calculation.

\mcsubsection{The large-$\Nc$ limit}
\label{sec:large-nc-limit}

It is of course well established that QCD is an SU(3) gauge theory.
Nevertheless it is frequently useful to consider the generalization to
a theory with $\Nc$ colours, SU(\Nc). 
  We will see that various aspects of event simulation
simplify in the limit of large \Nc.  For any \Nc, one can combine a
fundamental colour with a fundamental anticolour to produce an adjoint
colour and a colour singlet, $\Nc\otimes\bar{\Nc}=(\Nc^2-1)\oplus1$.
Conversely, we can think of the colour of a gluon as being that of a
quark and an antiquark, up to corrections from the fact that the gluon
does not have a singlet component.  One can decompose the colour
structure of each of the Feynman rules, and hence of any Feynman diagram,
into a set of delta-functions between external fundamental colours.  We
call this the colour flow of the diagram.  In the limit of large \Nc,
only diagrams whose colour flow is planar, \ie for which the fundamental
colour connections can be drawn in a single plane, contribute.  Each
colour connection that needs to come out of the plane results in a
suppression of $1/\Nc^2$.  This connection between the topology of a
diagram and its colour flow is an extremely powerful organizing
principle, which we will see comes into several different aspects of
event modelling.  One should bear in mind that whenever we use the
large-$\Nc$ limit, corrections to it are expected to be suppressed by at
least $1/\Nc^2\sim10$\% and in practice, because of the connection with
the topology, are often further dynamically suppressed.

% Local Variables:
% mode: LaTeX
% TeX-master: "../mcreview"
% End:

\mcsection{Hard subprocesses}
\label{sec:subprocesses}
Many LHC processes of interest involve large momentum transfers, for
example to produce heavy particles or jets with high transverse momenta.
Thus the simulation of subprocesses with large invariant momentum
transfer is at the core of any simulation of collider events in
contemporary experiments through Monte Carlo event generators. As QCD
quanta are asymptotically free, such reactions can be described by
perturbation theory, thus making it possible to compute many features
of the subprocess in question by, for example, using Feynman diagrams.

%%%%%%%%%%%%%%%%%%%%%%%%%%%%%%%%%%%%%%%%%%%%%%%%%%%%%%%%%%%%%%%%%%%%%%
\mcsubsection{Factorization formula for QCD cross sections}
%%%%%%%%%%%%%%%%%%%%%%%%%%%%%%%%%%%%%%%%%%%%%%%%%%%%%%%%%%%%%%%%%%%%%%
Cross sections for a scattering subprocess $ab\to n$ at hadron colliders 
can be computed in collinear factorization through~\cite{Ellis:1991qj}
\begin{eqnarray}
\label{Eq::Master_For_XSec}
\sigma%_{h_1h_2\to n} 
&=&
\sum\limits_{a,b}\,\int\limits_{0}^{1}\done x_a\done x_b\,\int\,
\pdf[h_1]{a}(x_a,\mu_F)\pdf[h_2]{b}(x_b,\mu_F)\,
\done\hat\sigma_{ab\to n}(\mu_F,\mu_R)\\
&=&
\sum\limits_{a,b}\,\int\limits_{0}^{1}\done x_a\done x_b\,\int\done\Phi_n\,
\pdf[h_1]{a}(x_a,\mu_F)\pdf[h_2]{b}(x_b,\mu_F)\nonumber\\
&&\qquad\times\frac{1}{2\hat s}|{\cal M}_{ab\to n}|^2(\Phi_n;\mu_F,\mu_R)\,,
\nonumber
\end{eqnarray}
where
\begin{itemize}
\item $\pdf[h]{a}(x,\mu)$ are the parton distribution functions
      (PDFs), which depend on the light-cone momentum fraction $x$ of parton $a$ 
      with respect to its parent hadron $h$, and on the factorization 
      scale\footnote{One could imagine to have two factorization scales,
      one for each hadron.  This may be relevant for certain processes such as 
      the fusion of electroweak bosons into a Higgs boson, where, at leading 
      order, the two hadrons do not interact through the exchange of colour.}
      $\mu_F$;
\item $\hat\sigma_{ab\to n}$ denotes the parton-level cross section for the 
      production of the final state $n$ through the initial partons $a$ and $b$.
      It depends on the momenta given by the final-state phase space $\Phi_n$, 
      on the factorization scale and on the renormalization scale $\mu_R$.
      The fully differential parton-level cross section is given by the product 
      of the corresponding matrix element squared, averaged over initial-state 
      spin and colour degrees of freedom, $|{\cal M}_{ab\to n}|^2$, and the 
      parton flux $1/(2\hat s) = 1/(2x_ax_bs)$, where $s$ is the hadronic 
      centre-of-mass energy squared.
\item The matrix element squared $|{\cal M}_{ab\to n}|^2(\Phi_n;\mu_F,\mu_R)$ can
      be evaluated in different ways.  In \AppRef{Sec:TL_ME} we discuss
      some of the technology used for tree-level matrix elements.  Here it 
      should suffice to say that the matrix element can be written as a sum
      over Feynman diagrams, 
      \begin{equation}
      {\cal M}_{ab\to n} = \sum\limits_{i}\,{\cal F}^{(i)}_{ab\to n}\,. 
      \end{equation}
      However, any summation over quantum numbers can be moved outside the 
      square, allowing one to sum over helicity and colour orderings
      such that
      \begin{equation}
      |{\cal M}_{ab\to n}|^2(\Phi_n;\mu_F,\mu_R) = 
      \sum\limits_{h_i;\,c_j}\,|{\cal M}^{\{ij\}}_{ab\to n}|^2
      (\Phi_n,\{h_i\},\{c_j\};\mu_F,\mu_R)\,. 
      \end{equation}
      In the computation of cross sections, this allows one to Monte
      Carlo sample not only over the phase space, but also over the
      helicities and colour configurations.  Picking one of the latter
      in fact defines the starting conditions for the subsequent
      parton showering, as discussed in more detail in
      \SecRef{sec:parton-showers}.
\item $\done\Phi_n$ denotes the differential phase space element over the $n$ 
      final-state particles,
      \begin{equation}
      \done\Phi_n = \prod\limits_{i=1}^n\frac{{\rm d}^3p_i}{(2\pi)^32E_i}
      \cdot(2\pi)^4\delta^{(4)}(p_a+p_b-\sum\limits_{i=1}^n p_i)\,,
      \end{equation}
      where $p_a$ and $p_b$ are the initial-state momenta.  For
      hadronic collisions, they are given by $x_aP_a$ and $x_bP_b$,
      where the Bjorken variables, $x_a$ and $x_b$, are also
      integrated over, and $P_a$ and $P_b$ are the fixed hadron
      momenta.
\end{itemize}

This equation holds to all orders in perturbation theory.   However,
when the subprocess cross section is computed beyond leading order
there are subtleties, which will be discussed later, and therefore for
the moment we consider only the use of leading-order (LO) subprocess
matrix elements.

It should be noted that the integration over the phase space may contain
cuts, for two reasons. First of all there are cuts reflecting the
geometry and acceptance of detectors, which are relevant for
the comparison with measured cross sections and other
related quantities.  On top of that there are other cuts, which, although
their details may be dominated by similar considerations, reflect
a physical necessity.  These are, for instance, cuts on the transverse
momentum of particles produced in $t$-channel processes, which exhibit
the analogue of the Coulomb singularity in classical electron scattering
and are related to internal particles going on their mass shell.  In a
similar way, especially for QCD processes, the notion of jets defined
by suitable algorithms (see \SecRef{sec:event-structure}) shields the 
calculation of the cross section of a process from unphysical soft and/or
collinear divergences.  
At leading order, these correspond simply to a set of cuts on parton momenta, 
preventing them from becoming soft or collinear.
%%%%%%%%%%%%%%%%%%%%%%%%%%%%%%%%%%%%%%%%%%%%%%%%%%%%%%%%%%%%%%%%%%%%%%
\mcsubsection{Leading-order matrix-element generators}
%%%%%%%%%%%%%%%%%%%%%%%%%%%%%%%%%%%%%%%%%%%%%%%%%%%%%%%%%%%%%%%%%%%%%%
All multi-purpose event generators provide a comprehensive list of LO 
matrix elements and the corresponding phase-space parameterizations for 
$2\to 1$, $2\to 2$ and some $2\to 3$ production channels in the 
framework of the Standard Model and some of its new physics extensions. 
For higher-multiplicity final states they employ dedicated matrix-element
and phase-space generators, such as \Alpgen~\cite{Mangano:2002ea}, 
\Amegic~\cite{Krauss:2001iv}, \Comix~\cite{Gleisberg:2008fv}, 
\Helac/\Phegas~\cite{Kanaki:2000ey,Papadopoulos:2000tt}, 
\Madgraph/\Madevent~\cite{Stelzer:1994ta,Maltoni:2002qb} and
\Whizard/\OmegaCode~\cite{Moretti:2001zz,Kilian:2007gr}, which are 
either interfaced (see \AppRef{sec:MEinterfaces}) or built-in 
as for the case of \Sherpa.  These codes specialize in the efficient 
generation and evaluation of tree-level matrix elements for multi-particle 
processes, see \AppRef{Sec:TL_ME} and \ref{Sec:PS_ME}.  

In doing so they have to overcome a number of obstacles. First of all, the 
number of Feynman diagrams used to construct the matrix elements increases
roughly factorially with the number of final-state particles.  This
typically renders textbook methods based on the squaring of amplitudes through
completeness relations inappropriate for final-state multiplicities of four 
or larger.  Processes with multiplicities larger than six are even more
cumbersome to compute and usually accessible through recursive relations only.
Secondly, the phase space of final-state particles in such reactions
necessitates the construction of dedicated integration algorithms, based on
the multi-channel method.  This, and other integration techniques, will be
discussed in more detail in \AppRef{sec:mc-methods} and \ref{Sec:PS_ME}.

%%%%%%%%%%%%%%%%%%%%%%%%%%%%%%%%%%%%%%%%%%%%%%%%%%%%%%%%%%%%%%%%%%%%%%
\mcsubsection{Choices for renormalization and factorization scales}
%%%%%%%%%%%%%%%%%%%%%%%%%%%%%%%%%%%%%%%%%%%%%%%%%%%%%%%%%%%%%%%%%%%%%%
The cross section defined by \EqRef{Eq::Master_For_XSec} is fully
specified only for a given PDF set and a certain choice for the
unphysical factorization and renormalization scales. There exists no
first principle defining what are the \emph{correct} $\mu_F$ and
$\mu_R$. However, our knowledge of the logarithmic structure of QCD
for different classes of hard scattering processes limits the range
of reasonable values. This knowledge is used as a guide when setting
the default choices in the various generators. Considering the class
of $2\to 1$ and $2\to 2$ processes, typically one hard scale $Q^2$ is
identified such that $\mu_F=\mu_R=Q^2$. Examples thereof are the
production of an $s$-channel resonance of mass $M$, where $Q^2=M^2$ or
the production of a pair of massless particles with transverse
momentum $p_T$, where typically $Q^2=p_T^2$. In general-purpose event
generators the hard scale $Q^2$ has the further meaning of a starting
scale for subsequent initial- and final-state parton showers\footnote{
  The precise phase-space limits of course depend on the relation
  between the generator's shower evolution scale and $Q^2$.}.
Accordingly when choosing $\mu_F$ and $\mu_R$ for processes with
final-state multiplicity larger than two, care has to be taken not to
introduce any double counting between the matrix-element calculation
and the parton-shower simulation, see \SecRef{sec:me-nlo-matching}.

%%%%%%%%%%%%%%%%%%%%%%%%%%%%%%%%%%%%%%%%%%%%%%%%%%%%%%%%%%%%%%%%%%%%%%
\mcsubsection{Choices for PDFs}
%%%%%%%%%%%%%%%%%%%%%%%%%%%%%%%%%%%%%%%%%%%%%%%%%%%%%%%%%%%%%%%%%%%%%%
Regarding the PDF, one is in principle free to choose any parameterization
that matches the formal accuracy of the cross section calculation, see 
\EqRef{Eq::Master_For_XSec}. All generators provide access to commonly
used PDF sets via the LHAPDF interface~\cite{Whalley:2005nh}. However, each generator
uses a default PDF set and the predictions of certain tunes of parton shower, 
hadronization and underlying event model parameters might be altered when 
changing the default PDF set, see \SecRef{sec:rivet-professor}. For a 
detailed discussion on PDF issues in Monte Carlo event generators see 
\SecRef{sec:pdfs-event-gener}.

%%%%%%%%%%%%%%%%%%%%%%%%%%%%%%%%%%%%%%%%%%%%%%%%%%%%%%%%%%%%%%%%%%%%%%
\mcsubsection{Anatomy of NLO cross section calculations}
\label{sec:subprocesses:NLOcross_sections}
%%%%%%%%%%%%%%%%%%%%%%%%%%%%%%%%%%%%%%%%%%%%%%%%%%%%%%%%%%%%%%%%%%%%%%
Most of the current multi-purpose event generators currently employ
leading-order (LO) matrix elements to drive the simulation. This means
that the results are only reliable for the shape of distributions,
while the absolute normalization is often badly described, due to
large higher-order corrections. One therefore often introduces a
so-called $K$-factor when comparing results from event generators with
experimental data. This factor is normally just that, a single factor
multiplying the LO cross section, typically obtained by the ratio of
the total NLO cross section to the LO one for the relevant
process. However, in this report we use the concept in a broader
sense, where the $K$-factor can depend on the underlying kinematics of
the LO process.

However, in striving for a higher accuracy and a better control
of theoretical uncertainties, some processes have been made accessible
at next-to-leading order accuracy and have been included in the
complete simulation chain, properly matched to the subsequent parton
showers.  This motivates the introduction of some formalism here,
which will be used in \SecRef{sec:me-nlo-matching}, where NLO event
generation will be discussed in some detail.

A cross section calculated at NLO accuracy is composed of three parts,
the LO or Born-level part, and two corrections, the virtual and the 
real-emission one.  Schematically,
\begin{equation}
\done\sigma^{\rm NLO} =
\done\tilde\Phi_n\left[{\cal B}(\tilde\Phi_n) + \alphaS{\cal V}(\tilde\Phi_n)\right] +
\done\tilde\Phi_{n+1}\alphaS{\cal R}(\tilde\Phi_{n+1})\,,
\end{equation}
where the tildes over the phase space elements $\done\tilde\Phi_{n}$ denote integrals over
the $n$-particle final state {\em and} the Bjorken variables, and
include the incoming partonic flux, and where the terms ${\cal B}$,
${\cal V}$, and ${\cal R}$ denote the Born, virtual and real emission
parts.  They in turn include the PDFs, and the summation over flavours
is implicit.

An obstacle in calculating these parts is the occurrence of
ultraviolet and infrared divergences.  The former are treated in a
straightforward manner, by firstly regularizing them, usually in
dimensional regularization, before the theory is renormalized.  The
infrared divergences, on the other hand, are a bit more cumbersome to
deal with.  This is due to the fact that they show up both in the
virtual contributions, which lead to the same $n$-particle final
state, and in the real corrections, leading to an $n+1$-particle final
state.  According to the Bloch-Nordsieck~\cite{Bloch:1937pw} and
Kinoshita-Lee-Nauenberg theorems~\cite{Kinoshita:1962ur,Lee:1964is},
for sensible, \ie infrared-safe, observables these divergences must
mutually cancel.  This presents some difficulty, since they are
related to phase spaces of different dimensionality.  In order to cure
the problem several strategies have been devised, which broadly fall
into two categories: phase-space slicing methods, pioneered in
\cite{Giele:1991vf,Giele:1993dj}, and infrared subtraction
algorithms~\cite{Catani:1996vz,Catani:2002hc,
  Kosower:1997zr,Kosower:2003bh,GehrmannDeRidder:2005cm,Daleo:2006xa,
  Frixione:1995ms,Frixione:1997np}. Current NLO calculations usually
use the latter.  They are based on the observation that the soft and
collinear divergences in the real-emission correction $\cal R$ exhibit
a universal structure.  This structure can be described by the
convolution of (finite) Born-level matrix elements, ${\cal B}$, with
suitably chosen, universal splitting kernels, ${\cal S}$, which in
turn encode the divergent structure.  Therefore, the ``subtracted
real-emission term'' $[{\cal R}-{\cal B}\otimes{\cal S}]$ is infrared finite and can
be integrated over the full phase space $\Phi_{n+1}$ of the
real-emission correction in four space-time dimensions.  The
subtraction terms ${\cal B}\otimes{\cal S}$
are added back in and combined with the virtual
term, ${\cal V}$, after they have been integrated over the radiative
phase space.  This integration is typically achieved in $D$
dimensions, such that the divergences emerge as poles in $4-D$.
Taking everything together, the parton-level cross section at NLO
accuracy reads, schematically,
\begin{eqnarray}
\label{Eq::NLO_XSec_Subtracted}
\sigma^{\rm NLO} =
\int\limits_n\done\tilde\Phi_n^{(4)}\,{\cal B} &+&
\alphaS\int\limits_{n+1}\done\tilde\Phi_{n+1}^{(4)}
        \left[\vphantom{\int\limits_{n}}
        {\cal R}-{\cal B}\otimes{\cal S}\right]\nonumber\\
&+&\alphaS\int\limits_n\done\tilde\Phi_n^{(D)}\,
        \left[\tilde{\cal V}+
              {\cal B}\otimes
              \int\limits_1\done\Phi_1^{(D)}{\cal S}\right]\,,
\end{eqnarray}
where the dimensions of the phase space elements and the number of
final-state particles have been made explicit and where collinear counter-terms
have been absorbed into the modified virtual contribution, $\tilde{\cal V}$.

It is worth noting that the task of evaluating the above equation can be
compartmentalized in a straightforward way.  A natural division is between
specialized codes, so-called one-loop providers (OLPs), that provide the 
virtual part, ${\cal V}$, and generic tree-level matrix element generators
which will take care of the rest, including phase space integration.
For details see \AppRef{Sec:Inter_ME}.  In the long run this will allow
for an automated inclusion of NLO accuracy into the multi-purpose
event generators; first steps in this direction have been made
in \cite{Binoth:2010xt,Alioli:2010xd,Hoeche:2010pf}.

In order to go to even higher accuracy, \ie to the NNLO level,
the above equation would become even more cumbersome, with more
contributions to trace.  This, however, will most likely remain far beyond 
the anticipated accuracy reach of the multi-purpose event generators for a 
long time to come.

%%%%%%%%%%%%%%%%%%%%%%%%%%%%%%%%%%%%%%%%%%%%%%%%%%%%%%%%%%%%%%%%%%%%%%
\mcsubsection{Summary}
%%%%%%%%%%%%%%%%%%%%%%%%%%%%%%%%%%%%%%%%%%%%%%%%%%%%%%%%%%%%%%%%%%%%%%
\begin{itemize}
\item The factorization formula in \EqRef{Eq::Master_For_XSec}
  is employed to calculate cross sections at hadron colliders.  The
  necessary ingredients are the parton-level matrix element, the
  parton distribution functions and the integration over the
  corresponding phase space.
\item At leading order, \ie for tree-level processes, there is a plethora
      of fully automated tools, constructing and evaluating the matrix elements
      with different methods.  They typically do not rely on textbook methods
      but on the helicity method or recursion relations.  
\item Due to the complexity of the processes, the phase space integration
      is a complicated task, which is usually performed using Monte
      Carlo sampling methods, which extend to include also treatment of
      the sum over polarizations and, more recently, even colours.  
\item The choice of the renormalization and factorization scales is not
      fixed by first principles, but rather by experience.  Combining the
      matrix elements with the subsequent parton shower defines, to some 
      extent, which choices are consistent and therefore ``allowed''.
\item Higher-order calculations, \ie including loop effects, are not yet
      fully automated.  They consist of more than just one matrix element
      with a fixed number of final-state particles, but they include 
      terms with extra particles in loops and/or legs.  These extra emissions
      introduce infrared divergences, which must cancel between the
      various terms.  This also makes the combination with the parton shower
      more cumbersome.  
\end{itemize}

% Local Variables: 
% mode: LaTeX
% TeX-master: "../mcreview"
% End: 

\mcsection{Parton showers}
\label{sec:parton-showers}
The previous section discussed the generation of a hard process
according to lowest-order matrix elements.  These describe the momenta of
the outgoing jets well, but to give an exclusive picture of the process,
including the internal structure of the jets and the distributions of
accompanying particles, any fixed order is not sufficient.  The effect
of all higher orders can be simulated through a parton shower algorithm,
which is typically formulated as an evolution in momentum transfer down
from the high scales associated with the hard process to the low scales,
of order 1~GeV, associated with confinement of the partons it describes
into hadrons.

In this section, we describe the physics behind parton showering.  Much
of our language will be based on the conventional approach in which a
parton shower simulates a succession of emissions from the incoming and
outgoing partons.  Towards the end of the section, however, we will
describe a slightly different formulation based on a succession of
emissions from the coloured dipoles formed by pairs of these
partons.  As we will discuss there, at the level of detail of our
presentation, the two approaches are almost equivalent and most of this
section applies equally well to dipole-based showers.

\mcsubsection{Introduction: QED bremsstrahlung in scattering processes}

We are familiar with the fact that in classical electrodynamics charges
radiate when scattered (see for example \cite{jackson1999},~chapter 15).
Calculating a scattering process in perturbative QED, one finds that the
radiation pattern of photons at the first order agrees with this
classical calculation (an important fact proved in Low's
theorem~\cite{Low:1958sn}).  One also encounters loop diagrams, which
correct the non-emission process such that the sum of emission and
non-emission probabilities is unity.  At successively higher
orders, soft photons are effectively emitted independently.
The spectrum extends down to arbitrarily low
frequencies, so that the total number of photons emitted is ill-defined,
but the number of observable
photons above a given energy is finite.  The probability of no
observable photons is also finite, and exponentially suppressed for
small energy cutoffs (known as Sudakov suppression~\cite{Sudakov:1954sw}).

One important property of this QED bremsstrahlung is the fact that
emission from different particles involved in the same scattering event
is coherent.  One manifestation of this is that when a high energy
photon produces an $e^+e^-$ pair in the field of a nucleus and, due to
the high boost factor, the pair are extremely close to each other in
direction, they do not ionize subsequent atoms they pass near because,
while they are closer together than the atomic size, the atoms only see
their total charge, which is zero, and not their individual charges.
Only once their separation has reached the atomic size do they start to
ionize.  In effect, the charged particles only behave independently with
respect to observers in a forward cone of opening angle given by their
separation and at larger angles they behave as a coherent pair.  This is
observed in bubble-chamber photographs as a single line of very weak
ionization that becomes stronger and eventually separates into two lines
and is known as the Chudakov effect~\cite{Chudakov:55ce}.  We will see
that there is a corresponding effect in QCD.

Having recalled these basic features of QED bremsstrahlung, we will
calculate the equivalent processes in QCD and see many analogous
features, as well as crucial differences arising from the non-abelian
nature of QCD and the resulting strong interactions at low energy.

\mcsubsection{Collinear final state evolution}

Although the utility of parton showers comes from the fact that they are
universal (process-independent) building blocks, we find it instructive
to motivate their main features by considering a specific process,
$e^+e^-$ annihilation to jets.  The leading-order cross section is given
by the electroweak process $e^+e^-\to q\bar{q}$ and is finite.  We
define its total cross section to be~$\sigma_{q\bar q}$.

We are more interested in the next-order process, $e^+e^-\to q\bar{q}g$,
which we hope to formulate as the production of a $q\bar{q}$ pair,
accompanied by the emission of a gluon by that pair.  Parameterizing the
three-parton phase space with $\theta$, the opening angle between the
quark and the gluon, and $z$, the energy fraction of the gluon, we
obtain
\begin{equation}
  \label{sigmaeeqqg2}
  \frac{\done\sigma_{q\bar qg}}{\done\cos\theta\,\done z} \approx
  \sigma_{q\bar q} \, C_F \, \frac{\alphaS}{2\pi} \,
  \frac2{\sin^2\theta}\,\frac{1+(1-z)^2}z\,,
\end{equation}
where $C_F=\frac{\Nc^2-1}{2\Nc}$ is a colour factor that can be thought as
the colour-charge squared of a quark.  In \EqRef{sigmaeeqqg2} we see
that the differential cross section diverges at the edges of phase
space.  To illustrate this, we have approximated the full expression
(which can be found in \cite{Ellis:1991qj} for example) by neglecting
non-divergent terms.
Recalling that the bremsstrahlung distribution was also divergent in
QED, but that this did not matter for a physical description of the
final state of observable photons, this divergence may not be a problem.
But we will certainly want to understand its physical origin since, as
we approach the divergences, the emission distribution will be large and
these will be the regions that dominate the emission pattern.

In \EqRef{sigmaeeqqg2}, we also see the structure we were hoping for: the
cross section for $q\bar{q}g$ is proportional to that for $q\bar{q}$ and
therefore we may interpret the rest of the expression as the probability
for gluon emission, differential in the kinematics of the gluon.

The integrand of \EqRef{sigmaeeqqg2} can diverge in three
ways: $\theta\to0$, corresponding to the gluon being collinear to the
quark; $\theta\to\pi$, corresponding to the gluon being back-to-back
with the quark, \ie collinear with the antiquark; and $z\to0$,
the gluon energy going to zero for any value of the opening angle.  Each
of the first two divergences can be traced to a propagator in one of the
two Feynman diagrams going on-shell.  However, it should be emphasized
that \EqRef{sigmaeeqqg2} contains the sum of the two diagrams and
properly includes their interference.  The third divergence comes from
the propagators in both diagrams going on-shell simultaneously and much
more clearly involves the interference of the two diagrams.  We return
to discuss the soft region in \SecRef{parton-showers:soft-gluons} and
for now focus on the collinear regions.

We can separate the angular distribution into two components, each of
which is divergent in only one of the two collinear regions,
\begin{equation}
  \frac2{\sin^2\theta} = \frac1{1-\cos\theta} + \frac1{1+\cos\theta}
  \approx \frac1{1-\cos\theta} + \frac1{1-\cos\bar\theta},
\end{equation}
where $\bar\theta$ is the angle between $g$ and $\bar{q}$ and the
approximation is as good as the one in \EqRef{sigmaeeqqg2}.  The
distribution can therefore be written as the sum of two separate
distributions, describing the emission of a gluon close to the
directions of the quark or the antiquark.  Since the distributions are
summed, they are effectively independent.  We emphasize again though,
that they are derived from the proper sum of amplitudes for diagrams in
which the gluon is attached to either emitter; it is just convenient to
separate them into pieces that can be treated independently.

We can therefore write the emission distribution as
\begin{equation}
  \label{firstcollinear}
  \done\sigma_{q\bar qg} \approx \sigma_{q\bar q} \sum_{\mathrm{partons}}
  C_F \, \frac{\alphaS}{2\pi} \, \frac{\done\theta^2}{\theta^2}
  \, \done z\frac{1+(1-z)^2}z,
\end{equation}
where now $\theta$ is the opening angle between the gluon and the parton
that emitted it.  This is starting to look like something that can be
implemented and iterated in a Monte Carlo algorithm, with an independent
emission distribution for each parton.  Before generalizing it, we point
out one mathematically-trivial property of this equation, which will
turn out to be important for the physical properties of our parton
shower algorithm.  In writing down \EqRef{firstcollinear}, we have
focused on the small-$\theta$ region, which gives the collinear
divergence.  However, we would have obtained a mathematically-identical
expression if we had chosen to parameterize the phase space in terms of
any other variable proportional to $\theta^2$, for example the
virtuality of the off-shell quark propagator,
$q^2=z(1-z)\,\theta^2\,E^2$, where $E$
is its energy, or the gluon's transverse momentum with respect to the
parent quark's direction,
$\kt^2=z^2(1-z)^2\,\theta^2\,E^2$, since
\begin{equation}
  \label{eq:evolchoices}
  \frac{\done\theta^2}{\theta^2} = \frac{\done q^2}{q^2} =
  \frac{\done\kt^2}{\kt^2}.
\end{equation}
Any of these forms would give identical results in the collinear limit,
but different extrapolations away from it, \ie different finite terms
accompanying the divergence.

While it is not obvious from our derivation, the structure of
\EqRef{firstcollinear} is completely general.  For any hard process that
produces partons of any flavour $i$, the cross section for a hard
configuration that has cross section $\sigma_0$ to be accompanied by a
parton $j$ with momentum fraction $z$ is given by
\begin{equation}
  \label{collinear}
  \done\sigma \approx \sigma_0 \sum_{\mathrm{partons},i}
  \frac{\alphaS}{2\pi} \, \frac{\done\theta^2}{\theta^2}
  \, \done z\,P_{ji}(z,\phi) \done\phi,
\end{equation}
with $P_{ji}(z,\phi)$ a set of universal, but flavour-dependent (and,
through $\phi$, the azimuth of $j$ around the axis
defined by $i$, spin-dependent) functions.  The spin-dependence can be
found in, for example, Ref.~\cite{Ellis:1991qj}~-- we give the
spin-averaged functions:
\begin{equation}
  \label{DGLAP}
  \begin{array}{rcl@{\hspace*{2.5em}}rcl}
    P_{qq}(z) &=& C_F\,\frac{1+z^2}{1-z}, &
    P_{gq}(z) &=& C_F\,\frac{1+(1-z)^2}z, \\
    P_{gg}(z) &=& C_A\,\frac{z^4+1+(1-z)^4}{z(1-z)}, &
    P_{qg}(z) &=& T_R(z^2+(1-z)^2),
  \end{array}
\end{equation}
where $C_F$ was already defined above, $C_A=\Nc$ is a colour factor that
can be thought as the colour-charge squared of a gluon, and $T_R$
is a colour factor that is fixed only by convention, $T_R=\frac12$ (a
different value of $T_R$ would be compensated by a different definition
of $\alphaS$).  $P_{qq}$, $P_{gq}$, $P_{gg}$ and $P_{qg}$ correspond to the
splittings $q\to qg$, $q\to gq$, $g\to gg$ and $g\to q\bar{q}$
respectively\footnote{A fifth splitting function $P_{\bar{q}g}$
  corresponding to $g\to\bar{q}q$  is equal to $P_{qg}$ by symmetry.}.
In the collinear limit, in which these results are valid, they are
independent of the precise definition of $z$~-- it could be the energy
fraction, light-cone momentum fraction, or anything similar, of parton
$j$ with respect to parton~$i$.
We now have the basic building block to write an iterative algorithm:
since \EqRef{collinear} is a completely general expression for any hard
process to be accompanied by a collinear splitting, we can iterate it,
using it on the hard process to generate one collinear splitting and
then treating the final state of that splitting as a new hard process,
generating an even more collinear splitting from it, and~so~on.

However, we are not quite ready to do so yet, because we have not yet
learnt how to deal with the divergence.  We have seen where it comes
from and that it is universal, but not how to tame it to produce a
well-defined probability distribution.  This comes when we ask what we
mean by a final-state parton.  The point is that any physical
measurement cannot distinguish an exactly collinear pair of partons from
a single parton with the same total momentum and other quantum numbers.
The infinitely high probability is associated with a transition that has
no physical effect.  As in our discussion of QED, to produce
physically-meaningful distributions, we should introduce a resolution
criterion, saying that we will only generate the distributions of
resolvable partons.  A particular convenient choice, although by no
means the only one possible, is the transverse momentum: to say that two
partons are resolvable if their relative transverse momentum is above
some cutoff~$Q_0$.  This cuts off both the soft and collinear
divergences, and gives a total resolvable-emission probability that is
finite.  To calculate the non-resolvable-emission probability, one must
integrate the emission distribution below the cutoff and add it to the
loop-correction to the hard process.  The result is finite, but there is
an easier way to obtain it: unitarity tells us that the total
probability of \emph{something\/} happening, either emission or
non-emission, is unity, and therefore, knowing the emission probability,
we can calculate the non-emission probability as one minus it.  (This
unitarity argument is exact in the case of soft or collinear emission,
but in general hard non-collinear loops contribute a finite correction,
which can be absorbed into the normalization of the total cross section,
restoring unitarity.)
It is sometimes said that parton shower algorithms do not include loop
corrections, but if this were so the non-emission probability would be
ill-defined.  It is better to say that they construct the loop
corrections by unitarity arguments from the tree corrections.

We are almost ready to construct the probability distribution for one
emission from a hard process, the basic building block that we will
iterate to produce a parton shower.  To do this, we have to realize that
the distribution we have been calculating so far is the inclusive
emission distribution of all gluon emissions: their total energy is the
total energy carried away by all gluons emitted, given by the classical
result.  To calculate instead the distributions of exclusive multi-gluon
events, it is convenient to separate out the distributions of individual
gluons, for example by introducing an ordering variable.  Let us take as
an illustrative
example, the virtuality of the internal line, $q^2$.  The distribution
we have been calculating is the total probability for all branchings of
a parton of type $i$ between $q^2$ and $q^2+\done q^2$,
\begin{equation}
  \done\mathcal{P}_i=\frac{\alphaS}{2\pi}\,\frac{\done q^2}{q^2}
  \int_{Q_0^2/q^2}^{1-Q_0^2/q^2}\done z\,P_{ji}(z),
\end{equation}
where the limits on $z$ come from the requirement that the partons be
resolvable, and their precise form depends on the definition of the
resolution criterion and of~$z$.
In order to construct the probability distribution of the
first branching, \ie the one that yields the largest contribution to the
virtuality of the internal line, we need to calculate the probability
that there are no branchings giving virtualities greater than a given
$q^2$ value, given that it has a maximum possible virtuality of $Q^2$.
We define this function to be $\Delta_i(Q^2,q^2)$.  It is given by a
differential equation,
\begin{equation}
  \label{q^2distribution}
  \frac{\done\Delta_i(Q^2,q^2)}{\done q^2} = \Delta_i(Q^2,q^2)\,
  \frac{\done\mathcal{P}_i}{\done q^2},
\end{equation}
corresponding to the fact that, when changing $q^2$ by a small amount,
the probability $\Delta_i$ can only change by the branching probability
$\done\mathcal{P}_i$ if there are no branchings above $q^2$, which
has probability $\Delta_i$.  It is easy to check that this equation has
the solution
\begin{equation}
  \label{Sudakov}
  \Delta_i(Q^2,q^2) = \exp\left\{-\int_{q^2}^{Q^2}\frac{\done k^2}{k^2}\,
  \frac{\alphaS}{2\pi} \int_{Q_0^2/k^2}^{1-Q_0^2/k^2}\done z\,P_{ji}(z)
  \right\}\,.
\end{equation}
This formula has a close analogy with the well-known radioactivity decay
formula: if the rate of decay of nuclei is $\lambda$ per unit time, then
the probability that a given nucleus has not decayed by time $T$ is
given by $\exp\left\{-\int_0^T\done t\,\lambda\right\}$.  Or, in words,
the probability of non-branching over some region is given by $e$ to
the minus the total inclusive branching probability over that region.

A particular case of this non-branching probability is
$\Delta_i(Q^2,Q_0^2)$, the total probability to produce \emph{no\/}
resolvable branchings.  This is the Sudakov form factor we encountered in
our discussion of QED, given by
\begin{eqnarray}
  \label{exponentiation}
  \Delta_i(Q^2,Q_0^2) &=& \exp\Biggl\{-\int_{Q_0^2}^{Q^2}\frac{\done k^2}{k^2}\,
  \frac{\alphaS}{2\pi} \int_{Q_0^2/k^2}^{1-Q_0^2/k^2}\done z\,P_{ji}(z)
  \Biggr\} \\
  &\sim& \exp\Biggl\{-C_F\,\frac{\alphaS}{2\pi}\log^2\frac{Q^2}{Q_0^2}
  \Biggr\}\,,
\end{eqnarray}
for a quark, a probability that falls faster than any inverse power of
$Q^2$.

Finally, we have the building block we need to iteratively attach
additional partons to a hard process one at a time.  Since $\Delta_i$
describes the probability to have no branching above $q^2$, its
derivative, the right-hand-side of \EqRef{q^2distribution}, is the
probability distribution for the first branching.  Having produced such
a branching, the same procedure has to be applied to each of the child
partons, with their $q^2$ values required to be smaller than the one we
generated for this splitting, to prevent double-counting.  Evolution
continues until no more resolvable branchings are produced above $Q_0^2$.
The only missing ingredient now is the starting condition: the value of
$Q^2$ for the parton line that initiated the shower, which we return to
in \SecRef{parton-shower:initial-conditions}.

The Monte Carlo implementation of \EqRef{q^2distribution} is remarkably
straightforward in principle: a random number $\rho$ is chosen between
0~and~1 and the equation $\Delta_i(Q^2,q^2)=\rho$ is solved for $q^2$.  If
the solution is above $Q_0^2$, a resolvable branching is generated at scale
$q^2$ and otherwise there is no resolvable branching and evolution
terminates.  For a resolvable branching a $z$ value is chosen according to
$P_{ji}(z)$.  Such a shower algorithm implements numerically the all-order
summation inherent in the exponentiation of \EqRef{exponentiation}.
Since this correctly sums the terms with the greatest number of logs of
$Q_0^2$ at each order of $\alphaS$ it is called a leading collinear
logarithmic parton shower algorithm.

However, there are considerable ambiguities in constructing such an
algorithm.  We already mentioned that an identical form would be given
by any other choice of evolution scale proportional to $\theta^2$, we
simply chose $q^2$ as an illustrative example.  We also defined $z$ to
be the energy fraction of the emitted parton, but in fact in the exactly
collinear limit in which \EqRef{collinear} is valid, choosing the
longitudinal momentum fraction, the light-cone momentum fraction, or
anything else similar, would give identical results, but different
extrapolations away from that limit.  Finally, since the hard process
matrix element deals with on-shell partons, and the parton shower
process has generated a virtuality for the parton line, energy-momentum
must be shuffled between partons in some way to be conserved, but the
collinear approximation does not specify how this should be done. All
of these are formally allowable choices, with the same leading
collinear logarithmic accuracy, but they differ in the amount of
subleading terms they introduce.  In the case of the evolution scale,
we will see in \SecRef{parton-showers:soft-gluons} that a study of the
soft limit of QCD matrix elements gives us an indication of the best
choice.

Before turning to the soft limit, we discuss one important source of
higher-order corrections, namely running coupling effects.  A certain
tower of higher-order diagrams, including those with loops inserted into
an emitted gluon, can be summed to all orders and absorbed by the
simple replacement of $\alphaS$ by $\alphaS(\kt)$, the running
coupling evaluated at the scale of the transverse momentum of the
emitted gluon~\cite{Amati:1980ch}.  This can be easily absorbed into the
algorithm above,
but has a couple of important consequences.  Firstly, parton
multiplication becomes much faster: as $q^2$ decreases, $\alphaS$
becomes larger and it becomes easier to emit further gluons until at
small enough scales the emission probability becomes of order~1 and
phase space fills with soft gluons.  Secondly, since one has to avoid
the region for which $\alphaS$ becomes of order~1, $Q_0$ has to be
considerably above $\LambdaQCD$, and actually becomes a physical
parameter affecting observable distributions at the end of the parton
shower, rather than a purely technical cutoff parameter that can be
taken as small as one likes, as it is without running coupling effects.
These facts mean that in the parlance used in analytical resummation,
the parton shower is not a purely perturbative description but induces
power corrections $\sim(Q_0/Q)^p$, contributing to the non-perturbative
structure of the final state. Here $p\ge1$ is a constant that may depend on
the parton shower algorithm used and the observable calculated; usually
$p=1$.

The ingredients described in this section are sufficient to construct a
final-state collinear parton shower algorithm.  However, recall that in
\mbox{$e^+e^-\to q\bar{q}g$} (\EqRef{sigmaeeqqg2}) we found that the
matrix elements were enhanced in both the
collinear and soft limits.  In order to give a complete description of
all dominant regions of the emission distribution, we should consider
soft emission in as much detail.

\mcsubsection{Soft gluon emission}
\label{parton-showers:soft-gluons}
In studying the matrix elements for $e^+e^-$ annihilation to
$q\bar{q}g$, we discovered that they were divergent as the gluon energy
goes to zero, in any direction of emission, as well as in the collinear
limit.  One may also show that this soft divergence is a general feature
of QCD amplitudes and also that it can be written in a universal factorized
form.  However, the big difference relative to the collinear case is
that the factorization is valid at the amplitude level: the
amplitude is given by the product of the amplitude to produce the system
of hard partons, times a universal factor describing the emission of the
additional gluon.  The cross section is calculated by summing all
Feynman diagrams and squaring and in practice many diagrams contribute
at a similar level, so that interference terms between diagrams are
unavoidable.  This tells us that soft gluons should be considered to be
emitted by the scattering process as a whole, rather than any given
parton, and appears to spoil the picture of independent evolution of
each parton.

\begin{figure}
  \centerline{%
    \includegraphics[width=5cm]{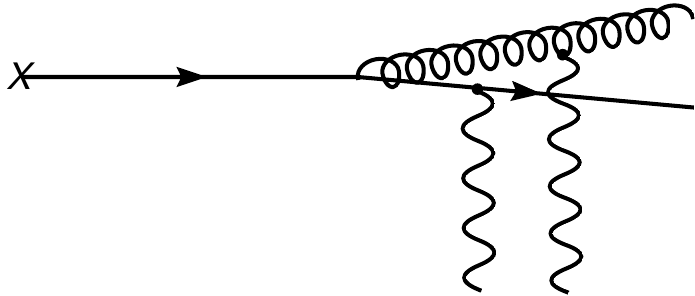}
    \hfill
    \includegraphics[width=5cm]{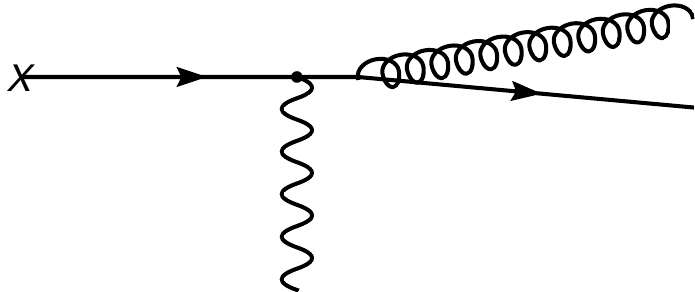}}
  \caption{Illustration of QCD coherence. The emission of a soft
    wide-angle gluon receives contributions from Feynman diagrams in
    which it is attached to any of the external partons (left). The
    coherent sum of these diagrams is equal to the emission from a
    single parton with the total momentum and colour of the
    partons. That is, as if it were emitted before the smaller-angle
    harder gluon (right).}
  \label{fig:softgluoncoherence}
\end{figure}

Consider, as a concrete example, the configuration shown in
\FigRef{fig:softgluoncoherence}.  A quark has been produced in a hard
process and has gone on to emit a reasonably hard, but reasonably
collinear, gluon, and we wish to calculate the probability that this
event is accompanied by a soft wide-angle gluon.  The soft factorization
theorem tells us that the amplitude for this process should be
calculated as the sum of amplitudes for the gluon to be attached to each
of the external partons, as indicated by the two placements of the gluon
on the left-hand-side of the figure.  The two resulting amplitudes are
of exactly the same order and have a non-trivial phase structure, so
that interference between them seems absolutely crucial.  It appears
impossible to reconcile this with the picture of independent collinear
evolution discussed in the previous section.

However, the coherence that we discussed in the context of QED
brems\-strahlung comes into play here and shows us that we \emph{can\/}
formulate soft emission within a parton shower approach.  Explicitly
calculating the amplitudes described above, one can show that in the
region shown in the figure, in which the softer gluon is at a larger
angle than the harder one, the interference is largely destructive,
reducing the emission distribution from the level it would be if
the two partons emitted independently to a term proportional to $C_F$.
Specifically, the result is identical to the one that would be obtained
from a configuration in which the collinear quark/gluon pair is replaced
by a single \emph{on-shell\/} quark with the same total longitudinal
momentum.  That is, we can think of the wide-angle emission as being
\emph{as if\/} it occurred before the more collinear one, summarized
pictorially in the right-hand side of \FigRef{fig:softgluoncoherence}.
However, it should be emphasized that this picture is the summary of the
proper interference between quantum mechanical amplitudes and does not
represent a Feynman diagram in which the gluon is emitted by the
internal line.  This should remind us of the Chudakov effect in QED:
there, wide angle emission from the $e^+e^-$ pair was absent, because
their total charge is zero.  Here, since the gluon is itself coloured,
the emission pattern is more complicated, but the result is the same:
the soft wide angle gluon sees the \emph{total\/} colour charge of the
system of partons to which it is attached.

On the other hand, calculating the case in which the angle between the
soft gluon and one of the other partons is much smaller than that
between them, one finds that the two contributions have very different
sizes and the cross section can be described as the sum of independent
emissions from the two partons.  These considerations can easily be
generalized to systems of more than two emitting partons.  They can be
summarized in a remarkably simple result: soft gluon effects can be
correctly taken into account by a collinear parton shower algorithm,
provided that, out of the choice of all possible evolution scales, one
uses the opening angle.  This is the central result that leads to
angular-ordered, or coherence-improved, parton showers, such as is
implemented in \herwig.  As a result, the first emission in the shower
is often not the hardest and it often happens that several soft
wide-angle gluons are emitted before the hardest gluon in the shower, a
fact that will lead to some complications when we try to match with
matrix element calculations in \SecRef{sec:me-nlo-matching}.

In \SecRef{sec:dipoles}, an alternative implementation of colour
coherence is discussed, based on colour dipoles between pairs of
emitters.  It is explained there that, at the level of detail discussed
here, the \kt-ordered dipole showers and angular-ordered parton showers
are effectively equivalent.

\mcsubsection{Initial state evolution \label{sec:isr}}
So far we have discussed the evolution of the partons produced in a hard
process.  According to the analogy with QED, we equally expect partons
to radiate on their way in to a scattering or annihilation process,
giving rise to an initial-state parton shower.  In principle, the
generation of initial-state showers can be set up in an extremely
similar way to the final-state showers already discussed, with an
incoming parton evolving through a series of $1\to2$ splittings to a
shower of partons, one of which is ultimately involved in the hard
process, with the rest being emitted as accompanying radiation.
An added complication in the initial-state case is that whole showers,
and branches off the sides of showers, can develop that do not
ultimately participate in any hard scatter.  These should be collapsed
back into the proton remnant in the same way that fluctuations that
develop in a freely-moving proton that does not have any interaction
collapse back into it.

In practice, simulating initial-state showers in this way is extremely
inefficient, because the majority
of partons have low energy and virtuality and if the showers were
produced with the same frequency as in nature, it would be extremely
rare to produce exactly the right kinematics to produce a hard process
of interest, such as Higgs production.  Moreover, as we shall discuss in
\SecRef{parton-shower:initial-conditions}, the properties of the parton
showers, both initial- and final-state, are correlated with the hard
process, and it would be difficult to build this in to such an
initial-state shower.  Instead,
event generators actually start by selecting the hard process and then
using the parton shower evolution to dress it with additional radiation.
That is, our basic building block is a \emph{backwards\/} step: one
generates the probability distribution for a parton with given momentum
fraction and value of evolution scale to have come from one at a higher
momentum fraction and lower scale.  This is iterated
until the evolution scale reaches the infrared cutoff, whereupon a
non-perturbative model of the remnant left behind by the extraction of a
parton from the incoming hadron is invoked.

The evolution of the PDFs with momentum
scale is given by the DGLAP
equations~\cite{Gribov:1972ri,Dokshitzer:1977sg,Altarelli:1977zs}.  They
can be viewed as
describing the flow of information about the PDFs to a given point in the
$(x,Q^2)$ plane from a boundary condition, usually a fixed line at some
input scale $Q^2=Q_0^2$, requiring information about all higher values
of $x$.  As pointed out in \cite{Sjostrand:1985xi} and further developed in
\cite{Marchesini:1987cf}, one can use the solution to the DGLAP
equations to guide the
backward evolution just described.  That is, for a parton of a given
flavour at a given $x$ and $q^2$ value, we can calculate the conditional
distribution that it came from a parton of the same or another flavour
at a higher $x$ and lower $q^2$ value.  Moreover, at leading order, this
distribution is positive definite and we can formulate it as a
probabilistic Markov chain.  The final result, which we do not derive
here (see the original references, \cite{Sjostrand:1985xi,Marchesini:1987cf}), is that the Sudakov form
factor, $\Delta_i(Q^2,q^2)$
(\EqRef{Sudakov}), which gives the probability that a final-state parton
does not produce any radiation at scales between $q^2$ and $Q^2$, is
replaced in the initial-state case\footnote{Note the interchange of the
  $ij$ indices on $P_{ij}$ relative to the final-state case, which comes
  about because this is a backward evolution.} by a non-emission
probability $\Delta_i(Q^2,q^2;x)$,
\begin{equation}
  \label{ISRsudakov}
  \!
  \Delta_i(Q^2,q^2;x) = \exp\left\{-\int_{q^2}^{Q^2}\frac{\done k^2}{k^2}\,
  \frac{\alphaS}{2\pi} \int_{Q_0^2/k^2}^{1-Q_0^2/k^2}\done z\,P_{ij}(z)
  \, \frac{x/z\,f_j(x/z,k^2)}{x\,f_i(x,k^2)}
  \right\}.
  \!\!
\end{equation}
The inclusive emission probability, which gets exponentiated to give the
non-emission probability, contains an extra factor of the ratio of
parton distribution functions at the `new', higher, value of $x$ that
the parton may evolve back to and its `current' value.
Thus, if our parton is in a region
in which the PDF decreases rapidly with increasing $x$, its non-emission
probability will be close to one, \ie its emission probability will be
small, and it is more likely that the parton came straight out of the
hadron at the infrared cutoff scale, rather than having been produced by
evolution of a higher-$x$ parton.  In the same way, if an emission is
generated, its $z$ value is not generated according to $P_{ij}(z)$, but
rather includes an extra factor of $x/z\,f_j(x/z,k^2)$.  In this way, one
can show that the algorithm is guaranteed to follow the same evolution
as the input parton distributions, provided the limit $Q_0\to0$ is
taken.  In reality, it is modified somewhat by infrared (\ie finite
$Q_0$) effects.  In addition, the emitted partons go on to produce
final-state parton showers of their own.

The arguments concerning the coherence of radiation from different
emitters and the scale of the running coupling apply equally well to
initial-state showers.  Indeed, one can show that along the
initial-state line, the opening angle of each emission relative to the
fixed direction of the incoming hadron is the correct ordering
variable \cite{Marchesini:1987cf} in a parton shower algorithm
and that the showers produced by the emitted
partons should have opening angles limited also by this angle.  Again
the appropriate scale for the running coupling is the transverse
momentum of emitted gluons.  Ref.~\cite{Catani:1990rr} showed that,
if these conditions are met, then the shower algorithm, even with LO
splitting functions, is correct to NLO accuracy in the limited phase
space region $x\to1$.
%% This accuracy allows one to understand the effective
%% scheme in which the parton shower is defined, relate it to the
%% $\overline{\mathrm{MS}}$ scheme and, in principle, use it to extract,
%% from measured scaling violations at large $x$, a value of
%% $\alphaS^{\overline{\mathrm{MS}}}(M_Z)$.  This relation is exploited in
%% \Herwigpp\ to make $\alphaS^{\overline{\mathrm{MS}}}(M_Z)$ the input
%% parameter specified by the user, but the fact that the relation is only
%% valid in a limited region of phase space makes it questionable how
%% useful in practice this relation is.  The value is still therefore
%% considered a freely adjustable parameter, not necessarily fixed to the
%% world average.

In the discussion of final-state parton showers, we emphasized that they
were derived from the full amplitudes of the theory, including
interference between different amplitudes, and describing a particular
gluon emission as being from a particular parton is a convenient
language to use, but is not truly what happens at the fundamental
level.  Since we properly include colour coherence effects, we do
account for interference between amplitudes.  The same is true for the
separation into final-state and initial-state emission: the separation
is arbitrary and only the sum of the two is physically meaningful and
reproduces the underlying quantum mechanical amplitude.  In the dipole
approach to parton showering, gluons are emitted by the colour dipole
that stretches between a colour--anticolour pair.  This picture works
also for scattered partons, where an incoming colour line behaves
effectively like an outgoing anticolour line and a scattered quark in
DIS, for example, radiates coherently with a radiation pattern that
peaks in the incoming and outgoing directions, without the need for an
explicit separation into initial- and final-state, as we shall discuss a
little more in \SecRef{sec:dipoles}.

Although not the main focus of this review, we note that for scattering
processes involving partons with very small momentum fractions (\ie at
given hard process kinematics, for very high energy incoming hadrons)
logarithms of the momentum fraction at each splitting can be large and a
different resummation is needed
(BFKL~\cite{Balitsky:1978ic,Kuraev:1977fs} or CCFM~\cite{Catani:1989sg}).
Such a resummation
can also be formulated in a probabilistic way, either as a dipole
cascade (see \SecRef{sec:ariadne}) or as a parton shower, as in the
SmallX~\cite{Marchesini:1992jw} and Cascade~\cite{Jung:2000hk} programs.
While quantifying how small a momentum fraction is needed before
including these effects becomes essential has proved elusive, it seems
very likely that a variety of hard processes at the LHC with momentum
fractions below $10^{-4\mbox{\scriptsize~or~}5}$ will be significantly
affected by them.

\mcsubsection{Connecting parton showers to the hard process}
\label{parton-shower:initial-conditions}
We have discussed the evolution of partons on their way in and out of a
scattering process, but not the starting conditions for the showers,
\ie the maximum values of $Q^2$.  Here again coherence plays a crucial
role.  In this section we discuss the simplest case of $2\to2$
scattering, but the general case is intimately linked with the question
of matrix element matching, which we discuss in detail in
\SecRef{sec:me-nlo-matching}.

A first consideration involves avoidance of double counting.  A QCD $2\to2$
scattering accompanied by an emission from one of the external legs that
is much harder than the hard scale gives the hard process a strong
recoil that boosts one or both of its outgoing partons to a significantly
higher transverse momentum.  The outcome is a configuration that is
indistinguishable from one that arises from a harder hard process
accompanied by a softer emission from one its external legs.  The fact
that one configuration can arise in two ways is a double counting and
should be resolved by only allowing one of them.  Since the parton
shower is built on the soft and collinear approximations, the
distribution of soft emission in hard scatters is more accurate than
that of hard emission in soft scatters and it is the former that should
be used.  This can be enforced by setting the upper limit of the parton
shower evolution to the scale of the hard scattering.  In the case of
processes for which the lowest order is purely electroweak, for example
gauge boson production, there is no analogous process with which hard
emission would be double-counted, but nevertheless event generators
typically also in this case limit emission to be below the hard scale,
where the parton shower approximations are most reliable, and instead
populate the region of phase space corresponding to harder emission
using matrix element corrections, as discussed
in \SecRef{sec:me-nlo-matching}.

\begin{figure}[t]
  \centerline{\raisebox{2cm}{(a)}\!\!%
    \raisebox{0pt}{\includegraphics[width=4cm]{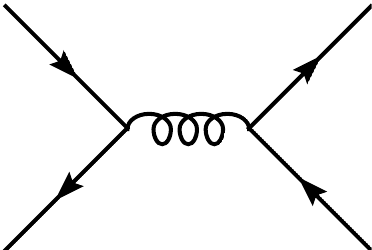}}
    \hfill\raisebox{2cm}{(b)}\!\!
    \raisebox{-5pt}{\includegraphics[width=4cm]{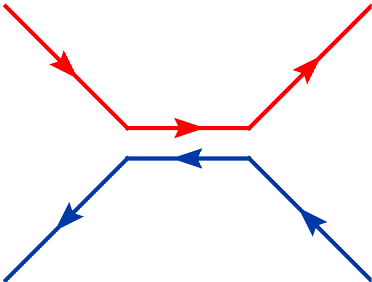}}
    \hfill\raisebox{2cm}{(c)}\!\!
    \raisebox{30pt}{\includegraphics[width=4cm]{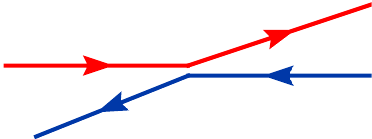}}}
  \caption{Illustration of colour coherence effect in hard scattering
    processes. In the quark--antiquark annihilation and production
    process (a), the quark's flavour is annihilated, but its colour
    flows onto the outgoing quark (b), such that in the
    centre-of-mass system, the colours are only scattered through small
    angles (c).}
  \label{fig:hardscattercoherence}
\end{figure}

The second consideration arises due to colour coherence.  In the limit
of a large number of colours, $\Nc\to\infty$, the colour structure of a
gluon can be considered to be a fundamental colour--anticolour pair, as
discussed in \SecRef{sec:large-nc-limit}.
In a scattering process such as $q\bar{q}\to q'\bar{q}'$, illustrated in
\FigRef{fig:hardscattercoherence}, although the flavour of the quark and
antiquark is annihilated, their colour lines flow onto the $s$-channel
gluon and hence onto the outgoing (anti)quarks, as illustrated in
\FigRef[b]{fig:hardscattercoherence}.
\begin{figure}[t]
  \centerline{\raisebox{2cm}{(a)}\!\!%
    \raisebox{0pt}{\includegraphics[width=2.7cm]{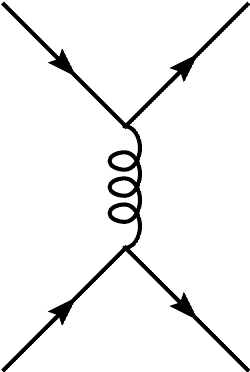}}
    \hfill\raisebox{2cm}{(b)}\!\!
    \raisebox{-5pt}{\includegraphics[width=3.0cm]{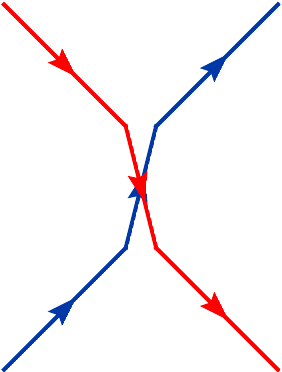}}
    \hfill\raisebox{2cm}{(c)}\!\!
    \raisebox{30pt}{\includegraphics[width=4cm]{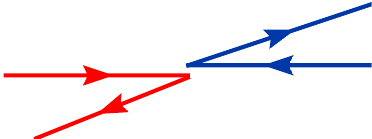}}}
  \caption{As \FigRef{fig:hardscattercoherence} but for a quark
    scattering process.}
  \label{fig:hardscattercoherence2}
\end{figure}
The outgoing quark--antiquark pair has a distribution over all polar
angles, but the colour coherence effect is best illustrated by an event
in which the outgoing quark's direction is only a small angle away from
the incoming quark's (see \FigRef[c]{fig:hardscattercoherence}).  Although
the quark lines are annihilated and created, the colour lines are
scattered through small angles.  Detailed analysis supports the
intuition that one might derive from analogy with the Chudakov effect in
QED, that the radiation pattern from such an event is effectively that
of two independent colour lines, each of which is scattered through only
a small angle.  Such a small-angle scattering does not radiate at large
angles, only into a forward cone of opening angle given by the
scattering angle.  This event would not, therefore, radiate significantly at
central rapidities.

This radiation pattern should be contrasted with the one from a quark
scattering event with identical kinematics (\ie $qq'\to qq'$ at a small
angle, \FigRef{fig:hardscattercoherence2}).  In this process the
scattering takes place via a $t$-channel
gluon so that the colour of the incoming $q$ quark is carried through
the gluon on to the outgoing $q'$ quark and vice versa.  Therefore,
although the quarks have been scattered through a small angle, their
colour charges have been scattered through a large angle, almost
$180^\circ$, and they radiate throughout almost the entire event.  This
is actually the norm for small angle scattering, since it is dominated
by $t$-channel gluon exchange, but in the general case it is essential
that the colour connection of each parton is identified so that the
coherence of the different emitters can be incorporated.

The algorithm to set the starting scale of the shower from each parton
that implements this coherence
can therefore be stated as follows: trace the colour line of the parton
through the hard process to find the parton to which it is
colour-connected, the ``colour partner''.  Start the shower from each
parton with a maximum allowed opening angle given by the angle to the
colour partner.

\begin{figure}[tbp]
  \centering
  \includegraphics[scale=0.7]{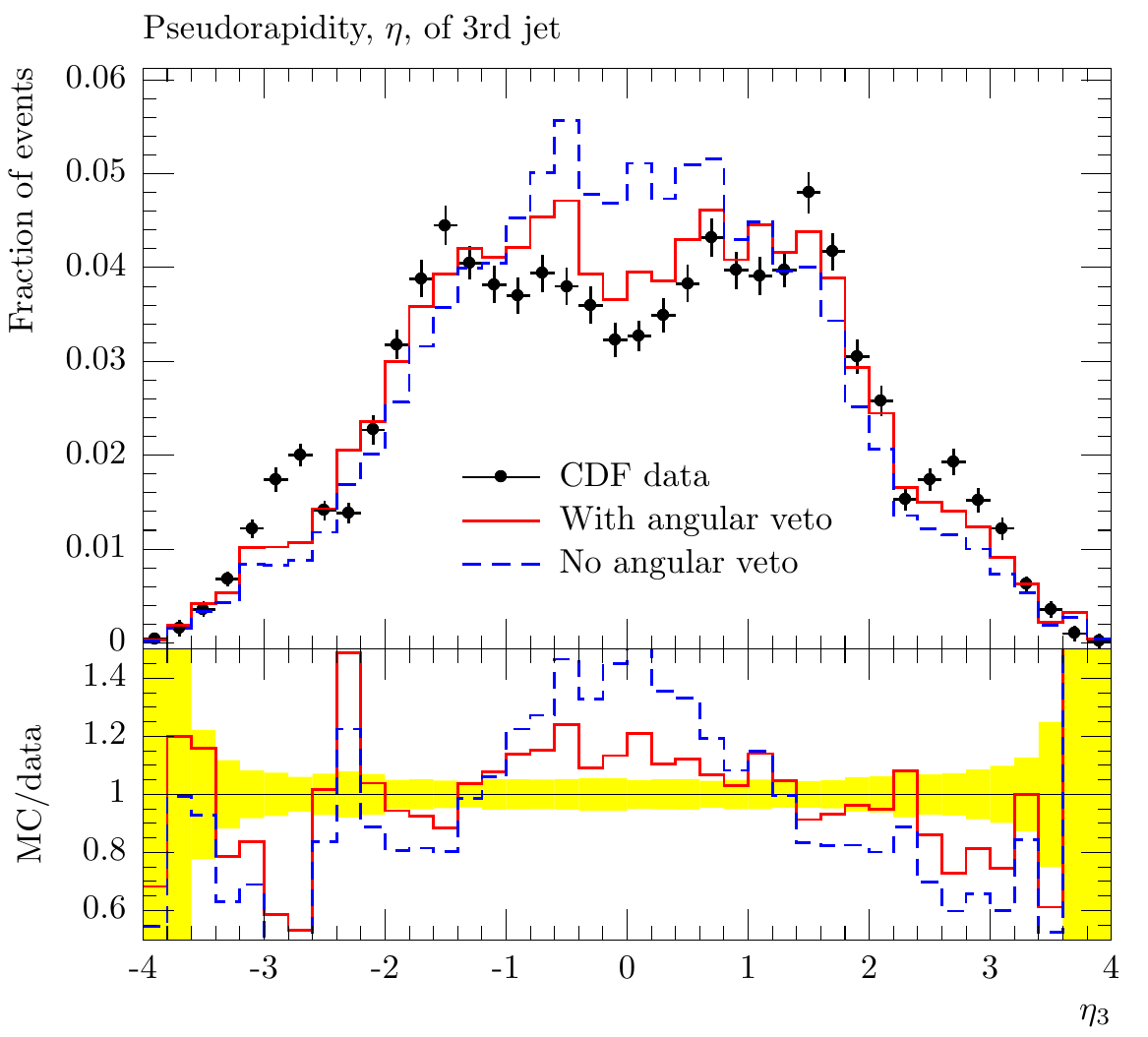}
  \caption{CDF's evidence for colour coherence in $p\bar{p}$ collisions at
    $\sqrt{s} = 1.8$~TeV. The pseudorapidity of the 3rd jet is plotted,
    uncorrected for detector effects, with the \pythiasix Monte Carlo generator
    for comparison with angular vetoing turned on and off in the parton
    showers.}
  \label{fig:CDFcoherence}
\end{figure}

A detailed analysis of three-jet events by the CDF
collaboration~\cite{Abe:1994nj}, showed that this colour coherence
effect is absolutely crucial to fit the data. They studied events with
three jets, the hardest of which had transverse energy\footnote{The transverse
  energy of a particle or jet is defined as $E_T=E\sin\theta$, where
  $E$ is its energy and $\theta$ is the polar angle of its direction of
  motion with respect to the beam axis.} $E_T>110$~GeV and the softest had
$E_T>10$~GeV, so that the sample was dominated by configurations with a
pair of roughly balancing high transverse energy jets and a relatively
soft third jet.  Distributions in the direction of this third jet were shown
to be particularly sensitive to the colour coherence effects in the
initial conditions of the shower.  Despite the fact that this analysis
was not corrected for detector effects, it is so important as a testing
ground for event generators that it has been implemented into \rivet
(see \SecRef{sec:rivet}) with
approximate detector corrections applied to the Monte Carlo events.  As an
example, we show in \FigRef{fig:CDFcoherence} the distribution of
$\eta_3$, the pseudorapidity of the third hardest jet.
Since colour coherence is so inherent to modern generators, to illustrate this
effect we have had to return to \pythiasix with its old virtuality-ordered
parton shower (\pythiasix now preferentially uses a dipole shower ordered in
transverse momentum and \pythiaeight only includes this version, see
\SecRef{sec:pythia8}).  Results are shown with angular ordering of the first
initial-state emission switched off and on.  We see
that the dip in the data at central $\eta_3$ is not reproduced by
the Monte Carlo simulation unless angular emission ordering is imposed to account for
coherence effects.  The comparison of these data with the latest versions of the
main generators discussed in this review are shown in
\SecRef{sec:physics-areas-where}.

The general case of arbitrary $2\to2$ scattering is slightly more
complicated, because a given hard process may have more than one colour
flow.  For example quark--gluon scattering $qg\to qg$, has three Feynman
diagrams, with an $s$-channel or $u$-channel quark, or a $t$-channel
gluon.  One can show that there are two independent colour flows and,
for example, the colour flow of the $t$-channel diagram can be written
as the difference between the other two.  The amplitude associated with
each diagram is gauge-dependent, but the contribution of the sum of all
diagrams to the amplitude for a given colour flow is gauge-invariant and
is therefore physically meaningful.  Finally, one finds that the
interference between the amplitudes for different colour flows is always
suppressed by factors of the number of colours, $\Nc$, and therefore
that in the large-$\Nc$ limit (see \SecRef{sec:large-nc-limit}),
different colour flows can be considered
as independent physical processes.  One can therefore construct, for a
given hard process configuration, probabilities for it to be in each of
the colour flow states in the large-$\Nc$ limit.  Having chosen a given
colour flow, one can trace the colour lines through the hard process to
find the colour partner of each parton.  As shown in
\cite{Ellis:1986bv}, this procedure results in an emission
distribution that is correct up to terms that are suppressed not only by
at least $1/\Nc^2$, but are also dynamically suppressed, having no
collinear enhancement.

In the large-$\Nc$ limit, a gluon has two colour connections and in the
parton shower emits with colour factor $C_A$.  In the dipole approach of
\SecRef{sec:dipoles}, each line emits with a colour factor $C_A/2$, but
typically in coherent parton showers a different approach has been used:
One chooses one of the two colour partners with equal probabilities and
generates a parton shower with colour factor $C_A$ limited by the
opening angle to the chosen partner.  This procedure gives the correct
inclusive distribution of emission, but as shown in \cite{Schofield}
it produces too much event-to-event fluctuation,
the wrong rate for any number of exclusive emissions and, in particular,
a too high rate of non-emission.  The correct procedure is the more
dipole-like one in which each colour line emits with factor $C_A/2$ into
a cone limited by its colour partner.

We
make a final comment concerning the Lorentz invariance of the whole
procedure.  The coherence-improved parton shower described above was
formulated in terms of an evolution in opening angle, which is
manifestly not Lorentz invariant, but in practice it is usually
implemented as an evolution in the energy of the emitter times the
opening angle, $\tilde{q}\sim E\theta$.  In the soft and collinear
limits in which it is valid, this evolution \emph{is\/} Lorentz
invariant.  However, its starting condition is not.  In fact, one can
show, \cite{Ellis:1986bv}, that the initial conditions of two
colour-connected partons $i$ and $j$ are given by
\begin{equation}
  \label{eq:psinitialconditions}
  \tilde{q}_{i,\max}\,\tilde{q}_{j,\max} = p_i\cdot p_j.
\end{equation}
That is, depending on the choice of Lorentz frame in which the parton
showering is performed, the maximum value of evolution variable for
parton $i$, $\tilde{q}_{i,\max}$, and for parton $j$,
$\tilde{q}_{j,\max}$, may take any values, but they must be related by
\EqRef{eq:psinitialconditions}.  Although each shower is separately
frame-dependent, provided
colour-connected pairs of partons are developed in the same frame, the
shower of the whole event is Lorentz-invariant.  The same considerations
apply also to the separation between initial- and final-state showers.
One can think of the corresponding boosts as being like a gauge
transformation: changing the gauge moves radiation between different
legs but the sum of emission from all legs is gauge invariant.  This
reflects the fact we have mentioned several times that we use the
language of a classical branching process, because it is convenient, but
it is derived from the underlying quantum structure of the gauge theory.

\mcsubsection{Quark mass effects}
The parton showers we have described so far are for massless partons.
In this section we shall consider how they are modified by parton mass
effects.  Although we discuss the explicit case of quarks, our comments
are equally applicable to any massive coloured particle, for example
squarks and gluinos in supersymmetry.  We shall mainly consider the case
of scattering or production processes with momentum transfers
significantly larger than the quark mass since, as we shall see,
radiation is suppressed in the threshold region, where the invariant
mass of the coherent system of which the quark is a part is not much
larger than its mass.  We therefore work in a
Lorentz frame in which the quark's energy is large relative to its mass,
\ie its velocity is close to~$c$.  We defer a more technical
discussion of precisely how the quark mass is defined, which is
ambiguous beyond the leading order of perturbation theory, until
\AppRef{sec:top-quark-masses}.

In the soft limit, the universality of the amplitude to emit a gluon is
unaffected by the parton mass.  One can therefore derive a general
formula for the angular distribution of soft gluons emitted by a dipole
consisting of one or two massive (anti)quarks.  It has the property that
at large angles it is identical to the distribution produced by a
massless quark of the same total momentum.  This again accords with our
picture of colour coherence~-- a soft wide angle gluon is not able to
resolve the details of the colour line that emits it, all it sees is a
colour charge moving in a given direction.  The details of whether that
colour charge is carried by one parton or a bundle of collinear partons
and whether those partons are massless or massive do not affect it.  On
the other hand, the distribution is modified at small angles.  Defining
the ratio $m/E\equiv\theta_0$, which we assume to be small, we end up
with a distribution that is identical to that from a massless quark, but
with the replacement
\begin{equation}
  \label{eq:deadcone}
  \frac{\done\theta^2}{\theta^2} \rightarrow
  \frac{\theta^2\,\done\theta^2}{(\theta^2+\theta_0^2)^2}.
\end{equation}
We see that emission at large angles, $\theta\gg\theta_0$ is indeed
unaffected, but that it is suppressed at small angles, falling to zero
in the exactly forward direction.  In older parton shower algorithms,
such as \Herwig\ \cite{Marchesini:1989yk}, this was used as the basis for the
`dead cone approximation', in which the showering of massive partons was
performed in an identical way to massless ones, but with a cutoff on
opening angle at~$\theta_0$.  With an appropriate choice of frame in
which this evolution is performed, it can be shown to produce
approximately the right amount of radiation, but it is clearly too
brutal an approximation, producing too much radiation at angles a little
above $\theta_0$ and none at all below it.

While \EqRef{eq:deadcone} is derived in the high energy limit,
$\theta_0\ll1$, it shows that near threshold, where $\theta_0\sim1$,
emission is suppressed at all angles.  Indeed, the more general
expression from which it is derived shows that at large angles, the
emission probability is proportional to the velocity-squared of the
emitting partons and hence that it goes to zero at threshold.

Ref.~\cite{Norrbin:2000uu} considered the matrix elements for gluon
emission in the decay of heavy objects to lighter partons with various
colour, spin and parity quantum number assignments.  It was found that the
radiation pattern for finite gluon energy, while always suppressed in
the forward direction, is significantly dependent on all of these
parameters and in most cases not going exactly to zero even in the
collinear limit.  These are implemented in \Pythia as process-dependent
mass corrections.

The authors of Ref.~\cite{Catani:2002hc} derived a generalization of the DGLAP splitting
function to the massive case, which they called the quasi-collinear
limit.  This is defined as the limit of $p_t^2\sim m^2\ll Q^2$, where
$Q^2$ is the hard scale.  They found
\begin{equation}
  \hspace*{-1cm}
  \begin{array}{rcl@{\hspace*{2.5em}}rcl}
    P_{QQ}(z) &=& C_F\left[\frac{1+z^2}{1-z\phantom{^2}}-\frac{m^2}{p_Q\cdot p_g}\right], &
    P_{gQ}(z) &=& T_R\left[z^2+(1-z)^2+\frac{2m^2}{(p_Q+p_{\bar{Q}})^2}\right].
  \end{array}
  \hspace*{-1cm}
\end{equation}
These are such that the subsequent limit $m\to0$ at fixed \pt\ smoothly
recovers the massless behaviour.  The limit $\pt\to0$ at fixed $m$
corresponds to the soft limit, if at fixed opening angle, and reproduces
\EqRef{eq:deadcone} or the collinear limit, if at fixed energy, and
recovers the result of Ref.~\cite{Norrbin:2000uu} for the generalization of
the dead-cone suppression to finite energy.  The quasi-collinear
splitting functions are used in \Herwigpp and \Sherpa, while the
matrix-element correction method of \cite{Norrbin:2000uu} is used in
\pythiaeight.

\mcsubsection{The dipole approach to parton showering}
\label{sec:dipoles}
In discussing the parton shower approach to simulating radiation from
the partons involved in a scattering process, we mentioned several times
the alternative formulation in terms of emission from sets of colour
dipoles.  This approach was first used in the \Ariadne\ program
\cite{Gustafson:1987rq}, described in more detail in
\SecRef{sec:ariadne}, and in fact is used by the majority of recent new
implementations, including \pythiaeight \cite{Sjostrand:2004ef}, see
\SecRef{sec:pythia8}, and \sherpa, with a choice of two different dipole
shower implementations, \cite{Winter:2007ye,Schumann:2007mg}, see
\SecRef{sec:sherpa}, as well as several standalone dipole cascade programs
\cite{Giele:2007di,Nagy:2007ty,Dinsdale:2007mf}.  For most
purposes it can be considered equivalent to the coherence-improved
parton showers discussed already, but it does have some advantages,
which we briefly describe in this section.

The basic observation is that, as discussed in
\SecRef{parton-shower:initial-conditions} and in more detail in
\SecRef{sec:cluster-model}, in the large-$\Nc$ limit the
colour structure of an arbitrarily-complicated system of partons can be
decomposed as a colour flow, \ie a set of colour lines each starting on an
incoming quark, outgoing antiquark or gluon, connecting it with an
outgoing quark, incoming antiquark or other gluon, and that in the
soft-gluon and large-$\Nc$ limits, each of these lines emits
independently.  Whether the configuration was produced by a
matrix-element calculation or by the parton shower itself, one can
calculate, again in the large-$\Nc$ limit, the probabilities of
different colour flows and hence choose in a given event a particular
colour flow.  One can quantify the validity of the dipole approximation
as being the limit in which the transverse momentum of the emitted
gluon, relative to the axes defined by the colour line from which it is
emitted, is much smaller than any scales involved in the production of
that colour line.  It is therefore natural to use transverse momentum as
the ordering variable for dipole showers.

The procedure is then to start from the hard process, decompose its
colour structure as described in
\SecRef{parton-shower:initial-conditions} and choose one colour flow.
This gives a unique initial condition for the subsequent dipole
evolution.  Each colour line connecting a pair of partons effectively
forms a colour--anticolour dipole and the emission from each dipole is
generated independently.  Although in the soft limit each dipole emits
independently with a classical radiation pattern, emission with a finite
transverse momentum results in a recoil.  Since a gluon carries the
colour lines of two dipoles, any recoil it experiences may affect the
subsequent evolution of the neighbouring dipoles.  Therefore the event
is evolved globally, with the highest transverse momentum emission from
any dipole being generated first, complete with its recoil, and its
transverse momentum giving the upper limit for the subsequent evolution
of the ensemble of dipoles.  Although this shower is formulated in the
soft limit, it was shown by \cite{Gustafson:1987rq} that
collinear effects can be incorporated in the large-$\Nc$ limit by a
simple modification to the rapidity distribution of the emitted gluon
(which is flat in the soft limit).  Recent implementations have gone
further by using the dipole splitting functions and kinematics defined
in the dipole subtraction method \cite{Catani:1996vz} to partition each dipole
into `monopole' pieces related to each of the two emitters of the dipole
(with the other taking the role of spectator) each with its own colour
factor, so that the collinear limit is exact in $\Nc$ and, like the
coherence-improved parton shower, neglected terms are both $\Nc^2$ and
dynamically suppressed.

Some doubt was cast on the validity of the dipole shower method in
Ref.~\cite{Dokshitzer:2008ia}, but it may be that the problem found
there is an artefact of the toy model
used \cite{Nagy:2009re,Skands:2009tb} and is not shared by full
implementations.  To our
understanding, transverse-momentum-ordered dipole showers with collinear
improvement are accurate to leading-collinear-logarithmic order and
hence are formally as accurate as angular-ordered parton showers.

In fact, despite this formal equivalence, the dipole shower has some
practical advantages.  Firstly, the fact that it is transverse momentum
ordered means that the hardest emission is generated first, making
matrix element corrections significantly more straightforward to
implement and obviating the need for truncated showers (see
\SecRef{sec:me-nlo-matching}) from the internal lines.  Secondly, the
fact that emission is a $2\to3$ process, rather than $1\to2$, means that
energy-momentum can be explicitly conserved, with all external partons
on mass shell, at each step of the shower.  In the parton shower approach,
the initially on-shell parton develops some virtuality and momentum
conservation must be violated at intermediate steps of generation.  Only
once all partons have developed their showers is a small amount of
momentum shuffled between partons to restore its conservation.  Finally,
the problem mentioned in \SecRef{parton-shower:initial-conditions}, of
how to generate emission from the two colour lines of a gluon is
obviated, since they simply radiate independently in the dipole
approach.
The kinematic variables used in dipole showers are Lorentz invariant and
they naturally combine the radiation from colour-connected parton pairs,
hence the issues discussed at the end
of \SecRef{parton-shower:initial-conditions} are automatically satisfied
by dipole showers.

We finally mention the extension of the dipole method to initial-state
radiation.  In the original implementation
in \Ariadne~\cite{Lonnblad:1992tz}, \SecRef{sec:ariadne}, there was no
explicit initial-state radiation.  The outgoing hadron remnant acted as
an emitter at one end of a dipole like any other parton except that,
being an extended object, radiation in its collinear direction was
suppressed.
%% This approximately mimicked the effect in the backward
%% evolution algorithm of valence quark distributions, which suppress
%% radiation from high-$x$ partons, but there is no explicit mechanism to
%% match the distribution with an input DGLAP-evolved PDF set.  Following
%% emission from a remnant dipole, the remnant itself does not take any
%% recoil, but rather, it emits a low-momentum gluon that takes its recoil,
%% which was shown to mimic some of the features of CCFM
%% evolution~\cite{Lonnblad:1992tz}.
More recent dipole showers have formulated the evolution of dipoles
containing an initial-state emitter in a more backward-evolution-like
way, guided by the PDF set.
%% However, a potential feature of this approach is the
%% fact that, in the evolution of Drell--Yan events for example, the
%% description of low transverse momentum events can be at odds with the known
%% asymptotic behaviour of the QCD cross section~\cite{Parisi:1979se}.  The
%% starting configuration is an initial--initial dipole, so that the
%% produced $\gamma^*/W/Z$ boson absorbs the recoil from the hardest
%% emission.  This emission splits the dipole into two initial--final
%% dipoles and the simplest implementation of the subsequent evolution
%% shares the recoil between the
%% partons without affecting the boson's momentum.
%% This is in contrast to the QCD behaviour, in which the full
%% Feynman-diagrammatic approach shows that resummation of multiple
%% emissions is essential to describe the low-$p_t$ region and the Sudakov
%% exponentiation implied by including only the hardest gluon is not
%% sufficient.  As discussed in
%% Ref.~\cite{Nagy:2009vg}, it is possible in principle to define
%% dipole-like algorithms that solve this problem by allowing the whole
%% event to absorb the recoil from each emission and \pythia and \sherpa
%% have both implemented solutions to this problem.

\mcsubsection{Summary}

\begin{itemize}
\item Scattered, annihilated and created partons radiate gluons.
\item Since gluons themselves are coloured, this radiation gives rise to
  further gluon radiation and parton multiplication.
\item These showers can be simulated as an evolution in some
  appropriately chosen scale down from the scale of the hard process
  towards an infrared scale at which non-perturbative confinement
  effects set in.
\item This evolution effectively sums to all orders terms enhanced by
  logarithms of the hard and soft scales, and can be formulated as
  coherence-im\-proved parton showers or transverse-momentum-ordered
  dipole showers.
\item Modern algorithms are extremely sophisticated implementations of
  all-order perturbative QCD.  Although they use the probabilistic
  language of time-ordered sequential emission, they are derived from
  the full quantum structure of the underlying gauge theory, QCD.
%Describing emission as
%being from a given parton or dipole is convenient, but is not physically
%meaningful, only the sum of emission from all emitters in the event is.
\end{itemize}

Nevertheless, parton shower algorithms are certainly not the whole story
in describing the exclusive structure of an event.  Firstly, they are
built on soft and collinear approximations to the full cross sections,
while many of the observables we are interested in are explicitly
sensitive to hard wide-angle emission and multi-jet final states, which
can only be described accurately with the help of higher-order matrix
elements, as discussed in \SecRef{sec:me-nlo-matching}.  And secondly,
they cannot be extended arbitrarily far into the infrared region where
QCD becomes strongly-interacting and must be cut off at some scale
beyond which non-perturbative models for hadronization and the
distributions of partons in incoming hadrons (\ie PDFs and primordial
$k_\perp$) must be invoked, as described in
\SecsRef{sec:minim-bias-underly} and \ref{sec:hadronization}.

% Local Variables:
% mode: LaTeX
% TeX-master: "../mcreview"
% End:

\mcsection{ME and NLO matching and merging}
\label{sec:me-nlo-matching}

\mcsubsection{Introduction}
\label{sec:matching:introduction}

In the previous sections we have described how to simulate
partonic final states with matrix elements and with parton showers,
and it should be clear to the reader that these approaches have
different merits and shortcomings. While fixed-order matrix elements
are excellent when simulating well separated, hard partons, they have
problems when trying to describe collinear and soft partons, due to the
occurrence of large logarithms. 
Also, obtaining the correct matrix element becomes very cumbersome when
we have more than a handful of partons. With parton showers it is the
other way around; hard, wide-angle emissions are poorly approximated,
while soft and collinear parton emissions are well described even for
very many partons. Clearly it would be desirable to combine the matrix
element and parton shower approaches to get a good description of any
partonic state. In particular, we note that a good description of soft
and collinear multi-parton states is necessary for hadronization
models such as string and cluster fragmentation (see
\SecRef{sec:hadronization}) to work properly.

To combine fixed-order matrix elements with parton showers is,
however, not a trivial task. The na\"ive procedure of simply adding a
parton shower to an event generated with a matrix element generator
does not work. One problem is related to the fact that tree-level
matrix elements are \textit{inclusive}, in that they give the
probability of having \textit{at least} $n$ partons in a state
calculated exactly to lowest order in \alphaS, while the
corresponding state generated by a parton shower is
\textit{exclusive}, given by the probability that there are
\textit{exactly} $n$ partons calculated approximately to all orders in
\alphaS. Another problem is that care must be taken not to double
count some regions of phase space or, conversely, to undercount other
regions.

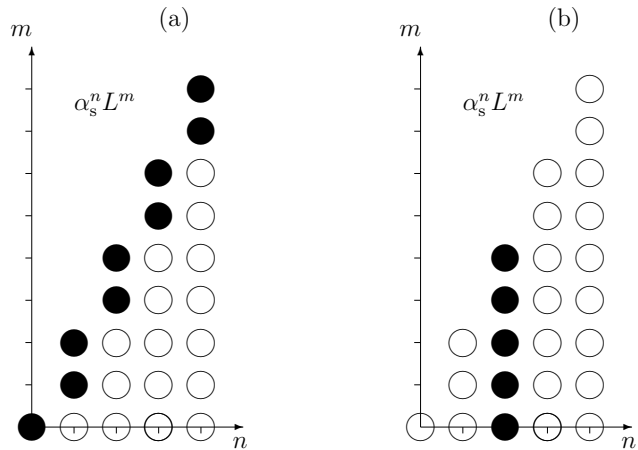
\begin{figure}
\begin{center}
\hfil\scalebox{0.8}{\mbox{\begin{picture}(120,200)
  \put(70,200){(a)}
  \put(10,10){\vector(0,1){180}}
  \put(0,195){$m$}
  \put(10,30){\line(-1,0){3}}
  \put(10,50){\line(-1,0){3}}
  \put(10,70){\line(-1,0){3}}
  \put(10,90){\line(-1,0){3}}
  \put(10,110){\line(-1,0){3}}
  \put(10,130){\line(-1,0){3}}
  \put(10,150){\line(-1,0){3}}
  \put(10,170){\line(-1,0){3}}
  \put(10,10){\vector(1,0){100}}
  \put(105,0){$n$}
  \put(30,10){\line(0,-1){3}}
  \put(50,10){\line(0,-1){3}}
  \put(70,10){\line(0,-1){3}}
  \put(90,10){\line(0,-1){3}}
  \put(30,160){$\alphaS^n L^m$}
  \put(10,10){\circle*{13}}
  \put(30,10){\circle{13}}
  \put(30,30){\circle*{13}}
  \put(30,50){\circle*{13}}
  \put(50,10){\circle{13}}
  \put(50,30){\circle{13}}
  \put(50,50){\circle{13}}
  \put(50,70){\circle*{13}}
  \put(50,90){\circle*{13}}
  \put(70,10){\circle{13}}
  \put(70,30){\circle{13}}
  \put(70,50){\circle{13}}
  \put(70,70){\circle{13}}
  \put(70,90){\circle{13}}
  \put(70,110){\circle*{13}}
  \put(70,130){\circle*{13}}
  \put(70,10){\circle{13}}
  \put(90,10){\circle{13}}
  \put(90,30){\circle{13}}
  \put(90,50){\circle{13}}
  \put(90,70){\circle{13}}
  \put(90,90){\circle{13}}
  \put(90,110){\circle{13}}
  \put(90,130){\circle{13}}
  \put(90,150){\circle*{13}}
  \put(90,170){\circle*{13}}
\end{picture}
% Local Variables: 
% mode: LaTeX
% TeX-master: "../mcreview"
% End: 
}}\hfil\scalebox{0.8}{\mbox{\begin{picture}(120,200)
  \put(70,200){(b)}
  \put(10,10){\vector(0,1){180}}
  \put(0,195){$m$}
  \put(10,30){\line(-1,0){3}}
  \put(10,50){\line(-1,0){3}}
  \put(10,70){\line(-1,0){3}}
  \put(10,90){\line(-1,0){3}}
  \put(10,110){\line(-1,0){3}}
  \put(10,130){\line(-1,0){3}}
  \put(10,150){\line(-1,0){3}}
  \put(10,170){\line(-1,0){3}}
  \put(10,10){\vector(1,0){100}}
  \put(105,0){$n$}
  \put(30,10){\line(0,-1){3}}
  \put(50,10){\line(0,-1){3}}
  \put(70,10){\line(0,-1){3}}
  \put(90,10){\line(0,-1){3}}
  \put(30,160){$\alphaS^n L^m$}
  \put(10,10){\circle{13}}
  \put(30,10){\circle{13}}
  \put(30,30){\circle{13}}
  \put(30,50){\circle{13}}
  \put(50,10){\circle*{13}}
  \put(50,30){\circle*{13}}
  \put(50,50){\circle*{13}}
  \put(50,70){\circle*{13}}
  \put(50,90){\circle*{13}}
  \put(70,10){\circle{13}}
  \put(70,30){\circle{13}}
  \put(70,50){\circle{13}}
  \put(70,70){\circle{13}}
  \put(70,90){\circle{13}}
  \put(70,110){\circle{13}}
  \put(70,130){\circle{13}}
  \put(70,10){\circle{13}}
  \put(90,10){\circle{13}}
  \put(90,30){\circle{13}}
  \put(90,50){\circle{13}}
  \put(90,70){\circle{13}}
  \put(90,90){\circle{13}}
  \put(90,110){\circle{13}}
  \put(90,130){\circle{13}}
  \put(90,150){\circle{13}}
  \put(90,170){\circle{13}}
\end{picture}
% Local Variables: 
% mode: LaTeX
% TeX-master: "../mcreview"
% End: 
}}\hfil
  \caption{Pictorial view of the terms in the \alphaS-expansion that
  enter into a jet cross section in \lleetoj. For each order in \alphaS,
  there are a number of large logarithms of the form
  $L^m=\log({Q\llsub{cm}/Q\llsub{jet}})^m$ (vertical axis). For
  $\alphaS^n$ the largest such logarithmic term is proportional to $L^{2n}$.
  For \eg a 4-jet observable we want to correctly include all
  coefficients from $\alphaS^2$ and onwards. In (a) we see the terms
  that would be correctly included in a NLL parton shower (filled
  blobs), while in (b) we see the terms correctly included in a
  tree-level matrix element.\label{fig:matching-ordersPS}}
\end{center}
\end{figure}
This problem can be understood on a more pictorial level in
\FigRef{fig:matching-ordersPS}.  There we have, for the process
\lleetoj, depicted the orders in the coupling constant $\alphaS$ on
the horizontal axis vs.\ the number of potentially occurring large
logarithms of the type $L^m=\log(Q\llsub{cm}/Q\llsub{jet})^m$ on the
vertical axis.  Here, $Q\llsub{cm}$ is an energy scale of the order of
the invariant mass of the produced system and $Q\llsub{jet}$ is
related to the resolution scale of a given jet algorithm (see
\SecRef{sec:event-structure}). The Born process is of order
$\alphaS^0$ and typically associated with the production of two
jets, the quark and antiquark in $e^+e^-\to q\bar q$.  Clearly, for each
additional emission, another factor of $\alphaS$ is necessary, such
that four jet production is of order $\alphaS^2$ and so on.  Also,
each emission can be related to at most two large logarithms,
associated with the soft and collinear divergences, see the previous
section on parton showers.

Now, the parton shower takes into account the exact leading and maybe
even next-to-leading logarithms, \ie\ it correctly takes into
account all real emissions and virtual corrections at all orders of
the type $\alphaS^nL^{2n}$ and $\alphaS^nL^{2n-1}$, while lower
powers of $L$ are treated approximately or
completely omitted. The leading $\alphaS^nL^{2n}$ term is easily
obtained by an \alphaS-expansion of the Sudakov form factor in
\EqRef{Sudakov}, while the next-to-leading term is obtained from the
hard collinear emission and from coherent treatment of soft emissions
in \SecRef{parton-showers:soft-gluons}. The treatment of these
logarithms will not impact on the total hadronic cross section, which
is still given by the Born-level value, due to the probabilistic
structure of the parton shower as discussed in
\SecRef{sec:parton-showers}.

On the other hand, differential distributions and observables
sensitive to the pattern of additional QCD radiation will be defined
by these logarithms.  Stated in other words: the parton shower will
not change the norm, but it will describe the shape of
radiation-sensitive distributions.

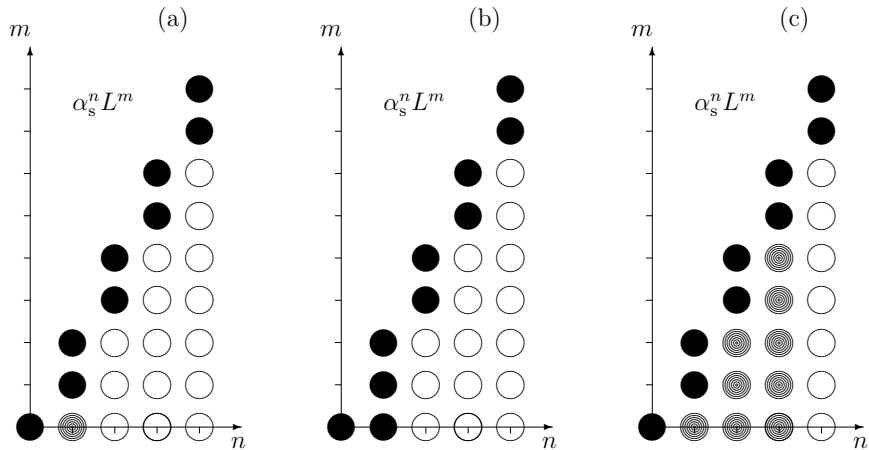
\begin{figure}
\begin{center}
\hfil\scalebox{0.8}{\mbox{\begin{picture}(120,200)
  \put(70,200){(a)}
  \put(10,10){\vector(0,1){180}}
  \put(0,195){$m$}
  \put(10,30){\line(-1,0){3}}
  \put(10,50){\line(-1,0){3}}
  \put(10,70){\line(-1,0){3}}
  \put(10,90){\line(-1,0){3}}
  \put(10,110){\line(-1,0){3}}
  \put(10,130){\line(-1,0){3}}
  \put(10,150){\line(-1,0){3}}
  \put(10,170){\line(-1,0){3}}
  \put(10,10){\vector(1,0){100}}
  \put(105,0){$n$}
  \put(30,10){\line(0,-1){3}}
  \put(50,10){\line(0,-1){3}}
  \put(70,10){\line(0,-1){3}}
  \put(90,10){\line(0,-1){3}}
  \put(30,160){$\alphaS^n L^m$}
  \put(10,10){\circle*{13}}
  \llBlob{30}{10}
  \put(30,30){\circle*{13}}
  \put(30,50){\circle*{13}}
  \put(50,10){\circle{13}}
  \put(50,30){\circle{13}}
  \put(50,50){\circle{13}}
  \put(50,70){\circle*{13}}
  \put(50,90){\circle*{13}}
  \put(70,10){\circle{13}}
  \put(70,30){\circle{13}}
  \put(70,50){\circle{13}}
  \put(70,70){\circle{13}}
  \put(70,90){\circle{13}}
  \put(70,110){\circle*{13}}
  \put(70,130){\circle*{13}}
  \put(70,10){\circle{13}}
  \put(90,10){\circle{13}}
  \put(90,30){\circle{13}}
  \put(90,50){\circle{13}}
  \put(90,70){\circle{13}}
  \put(90,90){\circle{13}}
  \put(90,110){\circle{13}}
  \put(90,130){\circle{13}}
  \put(90,150){\circle*{13}}
  \put(90,170){\circle*{13}}
\end{picture}
% Local Variables: 
% mode: LaTeX
% TeX-master: "../mcreview"
% End: 
}}\hfil\scalebox{0.8}{\mbox{\begin{picture}(120,200)
  \put(70,200){(b)}
  \put(10,10){\vector(0,1){180}}
  \put(0,195){$m$}
  \put(10,30){\line(-1,0){3}}
  \put(10,50){\line(-1,0){3}}
  \put(10,70){\line(-1,0){3}}
  \put(10,90){\line(-1,0){3}}
  \put(10,110){\line(-1,0){3}}
  \put(10,130){\line(-1,0){3}}
  \put(10,150){\line(-1,0){3}}
  \put(10,170){\line(-1,0){3}}
  \put(10,10){\vector(1,0){100}}
  \put(105,0){$n$}
  \put(30,10){\line(0,-1){3}}
  \put(50,10){\line(0,-1){3}}
  \put(70,10){\line(0,-1){3}}
  \put(90,10){\line(0,-1){3}}
  \put(30,160){$\alphaS^n L^m$}
  \put(10,10){\circle*{13}}
  \put(30,10){\circle*{13}}
  \put(30,30){\circle*{13}}
  \put(30,50){\circle*{13}}
  \put(50,10){\circle{13}}
  \put(50,30){\circle{13}}
  \put(50,50){\circle{13}}
  \put(50,70){\circle*{13}}
  \put(50,90){\circle*{13}}
  \put(70,10){\circle{13}}
  \put(70,30){\circle{13}}
  \put(70,50){\circle{13}}
  \put(70,70){\circle{13}}
  \put(70,90){\circle{13}}
  \put(70,110){\circle*{13}}
  \put(70,130){\circle*{13}}
  \put(70,10){\circle{13}}
  \put(90,10){\circle{13}}
  \put(90,30){\circle{13}}
  \put(90,50){\circle{13}}
  \put(90,70){\circle{13}}
  \put(90,90){\circle{13}}
  \put(90,110){\circle{13}}
  \put(90,130){\circle{13}}
  \put(90,150){\circle*{13}}
  \put(90,170){\circle*{13}}
\end{picture}
% Local Variables: 
% mode: LaTeX
% TeX-master: "../mcreview"
% End: 
}}\hfil\scalebox{0.8}{\mbox{\begin{picture}(120,200)
  \put(70,200){(c)}
  \put(10,10){\vector(0,1){180}}
  \put(0,195){$m$}
  \put(10,30){\line(-1,0){3}}
  \put(10,50){\line(-1,0){3}}
  \put(10,70){\line(-1,0){3}}
  \put(10,90){\line(-1,0){3}}
  \put(10,110){\line(-1,0){3}}
  \put(10,130){\line(-1,0){3}}
  \put(10,150){\line(-1,0){3}}
  \put(10,170){\line(-1,0){3}}
  \put(10,10){\vector(1,0){100}}
  \put(105,0){$n$}
  \put(30,10){\line(0,-1){3}}
  \put(50,10){\line(0,-1){3}}
  \put(70,10){\line(0,-1){3}}
  \put(90,10){\line(0,-1){3}}
  \put(30,160){$\alphaS^n L^m$}
  \put(10,10){\circle*{13}}
  \llBlob{30}{10}
  \put(30,30){\circle*{13}}
  \put(30,50){\circle*{13}}
  \llBlob{50}{10}
  \llBlob{50}{30}
  \llBlob{50}{50}
  \put(50,70){\circle*{13}}
  \put(50,90){\circle*{13}}
  \llBlob{70}{10}
  \llBlob{70}{30}
  \llBlob{70}{50}
  \llBlob{70}{70}
  \llBlob{70}{90}
  \put(70,110){\circle*{13}}
  \put(70,130){\circle*{13}}
  \put(70,10){\circle{13}}
  \put(90,10){\circle{13}}
  \put(90,30){\circle{13}}
  \put(90,50){\circle{13}}
  \put(90,70){\circle{13}}
  \put(90,90){\circle{13}}
  \put(90,110){\circle{13}}
  \put(90,130){\circle{13}}
  \put(90,150){\circle*{13}}
  \put(90,170){\circle*{13}}
\end{picture}
% Local Variables: 
% mode: LaTeX
% TeX-master: "../mcreview"
% End: 
}}\hfil
\caption{Pictorial view analogous to
  \FigRef{fig:matching-ordersPS}. (a) The terms included in tree-level
  matching of the first emission. Note that the $\alphaS^1L^0$ blob
  is only half filled, as it is correctly taken into account only in
  the real-emission contributions, not in the virtual ones, which
  means that the shapes of distributions will be correct but not their
  normalization. (b) The terms included in NLO matching of the first
  emission. (c) The terms included in tree-level CKKW-like merging up
  to 5-jet, where the half-filled blobs are only correctly taken into
  account for real-emission contributions above the merging
  scale.\label{fig:matching-ordersNLO}}
\end{center}
\end{figure}

Taken together, coherent parton showers will correctly include all
filled blobs in \FigRef[a]{fig:matching-ordersPS} (equivalent to the
terms $\alphaS^nL^{2n}$ and $\alphaS^nL^{2n-1})$.  The natural
question thus arises of how to include more terms into the picture in
an exact way.  To this end, a number of different procedures have been
devised in the past two decades, and will be discussed is some detail
in subsequent subsections:
\begin{itemize}
\item \underline{Tree-level matching:}\\
  The first procedure for matching matrix elements with parton showers
  was invented decades ago by Bengtsson and
  Sjöstrand\cite{Bengtsson:1986hr}.  Similar techniques were later
  used also in
  \cite{Gustafson:1987rq,Seymour:1994we,Seymour:1994df,Miu:1998ju,Lonnblad:1995ex}
  and concentrated on correcting the first or hardest parton shower
  emission (\SecRef{sec:matching-first-ps}).  In our pictorial
  language, this amounts to including one extra blob as in
  \FigRef[a]{fig:matching-ordersNLO}, of order $\alphaS^1 L^0$, but
  \emph{only on the shape}, while the norm for the inclusive process (its
  cross section) was still given at the LO Born-level.
\item \underline{NLO matching:}\\
  In order to include also the NLO correction to the total cross
  section of the inclusive sample, one could naively apply a constant
  $K$-factor, as discussed in
  \SecRef{sec:subprocesses:NLOcross_sections}.  This, however, will
  not necessarily be good enough for precision studies, since these
  higher-order corrections may also influence the kinematics of the
  Born-level configuration.  Therefore, in the past decade much effort
  has been put into correcting parton showers also with exact
  next-to-leading-order (NLO) matrix elements, allowing to take into
  account the full effect of the $\alphaS^1 L^0$-term (see
  \FigRef[b]{fig:matching-ordersNLO}).  Here the
  \MCatNLO\cite{Frixione:2002ik,Frixione:2006gn}
  (\SecRef{sec:matching-mcatnlo}) and \POWHEG\cite{Nason:2004rx}
  (\SecRef{sec:powheg}) procedures have been very successful for the
  correction of the first parton shower emission and for including the
  effect on the cross section.  These ideas have been implemented in
  various programs, and we refer to the later parts of this section
  and to the descriptions of the individual event generators in
  \PartRef{sec:spec-revi-main} for the corresponding references.
\item \underline{Multi-jet merging at LO:}\\
  Another active area of developments in the last decade has been the
  treatment of multi-jet topologies, starting with the merging
  algorithm by Catani, Kuhn, Krauss and Webber (CKKW)
  \cite{Catani:2001cc} and a similar procedure developed in parallel
  by L\"onnblad (later coined CKKW-L) \cite{Lonnblad:2001iq}. In these procedures
  also the second and higher emissions in the parton shower were
  corrected to the corresponding tree-level matrix element, but at the
  price of introducing a technical merging scale above which the
  corrections are made (\SecRef{sec:matching-at-tree}). So, the aim of
  these procedures is to simulate each jet multiplicity (with jets
  above the cut) with the corresponding tree-level matrix element,
  dressed with the parton shower. Pictorially it corresponds to taking
  into account all $\alphaS^nL^{2n}$ and $\alphaS^nL^{2n-1}$ blobs
  as in a normal PS, but also the impact \emph{on the shape} of all
  the other blobs up to $\alphaS^n$ for the $n$ first real emissions
  above the cut, as indicated in \FigRef[c]{fig:matching-ordersNLO}.

  The CKKW and CKKW-L approaches can be shown to maintain the
  logarithmic accuracy of the parton shower, at least in $e^+e^-$
  annihilations, without any double-counting of contributions.  In
  addition, more pragmatic versions of the CKKW(-L) merging have
  been introduced, with less focus on the formal accuracy; the
  MLM approach by Mangano \cite{Mangano:2010xx} (see also
  \cite{Alwall:2007fs}) and the Pseudo-Shower algorithm by
  Mrenna and Richardson\cite{Mrenna:2003if}.  By now, the field has
  matured quite a lot, and in some recent publications the emphasis
  shifted to a more careful discussion of formal accuracy and its
  preservation, as in \cite{Hoeche:2009rj,Hamilton:2009ne}.
\item \underline{Multi-jet merging at NLO:}\\
  For higher parton multiplicities, there have also been some
  suggestions for NLO corrections (\SecRef{sec:nlo-merging}). The
  \MENLOPS procedure \cite{Hamilton:2010wh} simply combines \POWHEG
  with CKKW so that the Born-level differential cross section becomes
  correct to NLO.  The NL$^3$ procedure \cite{Lavesson:2008ah} and the
  procedure being implemented in the \Vincia
  program\cite{Giele:2007di,Skands:2010xx} go further and try to
  also correct the higher jet cross sections to NLO, but so far no
  easily accessible implementation exists.
\end{itemize}

Clearly there are several different strategies for combining matrix
elements and parton showers. As indicated above, it is useful to
distinguish two groups of approaches.
% for combining matrix elements and parton showers.
With \textit{matching} we refer to those approaches
in which high-order corrections to an inclusive process are
integrated with the parton shower.  The other strategy involves a
\textit{merging} scale, usually defined in terms of a jet resolution
scale, where any parton produced above that scale is generated with a
corresponding higher-order matrix element and, conversely, any parton
produced below is generated by the shower.

\mcsubsection{Correcting the first emission}
\label{sec:matching-first}

We start out by looking at the first emission in the parton shower. In
a transverse-momentum-ordered parton shower, this is the hardest
emission and will determine the main structure of the final
state. Hence, it is important to get this right also far away from the
soft and collinear regions where the parton shower is a good
approximation.

\mcsubsubsection{The NLO cross section}

To guide us we refer to the typical inclusive NLO cross section for a
process with a given Born-level state, and we rewrite the schematic
formula in \EqRef{Eq::NLO_XSec_Subtracted} from
\SecRef{sec:subprocesses:NLOcross_sections} in a more explicit form,
\bea
\lld\llxsec\llsup{NLO} &=&
\lld\llPS_{0}\left[B(\llPS_{0})+\alphaS V_1(\llPS_{0})
  + \alphaS\int\lld\llPS_{1|0} S_1(\llPS_{1})\right]\nnb\\
&+& \lld\llPS_{1}
\left[\alphaS R_1(\llPS_{1})
  - \alphaS S_1(\llPS_{1})\vphantom{\int}\right]\,.
\label{eq:matching:NLO1}
\eea
Here we identify\footnote{Note that implicitly these terms also
  contain symmetry factors and parton luminosities.} the Born-level
and the real-emission phase space $\llPS_{0}$ and $\llPS_{1}$, together
with the corresponding tree-level matrix elements $B(\llPS_{0})$ and
$\alphaS R_1(\llPS_{1})$. We also have the virtual or loop
contribution $\alphaS V_1(\llPS_{0})$ and the subtraction term
$\alphaS S_1(\llPS_{1})$ which, when integrated over the
one-particle phase-space element, $\llPS_{1|0}$, renders the first
bracket finite, and also regularizes the real-emission term, making
everything finite.

Typically, using the universality of soft and collinear divergencies,
we can write the subtraction term in a factorized form as
\begin{equation}
  \label{eq:matching:factorized-subtaction}
  S_1(\llPS_{1}) = B(\llPS_0)\otimes\tilde{S}(\llPS_{1|0}),
\end{equation}
where $\tilde{S}(\llPS_{1|0})$ are the universal subtraction kernels
with analytically known integrals.  At this point it is useful to
split the real-emission correction into one part containing all
singularities, $R_1\llsup{s}$, and one non-singular part,
$R_1\llsup{ns}$:
\begin{equation}
  \label{eq:matching:factorized-subtaction-ns}
R_1(\llPS_1) = R_1\llsup{s}(\llPS_1) + R_1\llsup{ns}(\llPS_1)\,.
\end{equation}
The splitting is quite arbitrary, as long as $R_1\llsup{s}$ contains
all singularities, but will later be used for illustrating differences
in some of the matching algorithms.

We can now write the inclusive NLO cross section
\bea
\lld\llxsec\llsup{NLO} &=&
\lld\llPS_{0}\left[B(\llPS_{0})+\alphaS V_1(\llPS_{0})
  + \alphaS\int\lld\llPS_{1|0} S_1(\llPS_{1})\right]\nnb\\
&+& \lld\llPS_{1}
\alphaS\left[R_1\llsup{s}(\llPS_{1})
  - S_1(\llPS_{1})\vphantom{\int}\right]
+ \lld\llPS_{1}\alphaS R_1\llsup{ns}(\llPS_{1})\,,
\label{eq:matching:NLO2}
\eea
where, again, all individual terms in the sum are finite. We can now
absorb the second bracket into the first and we can define the
\emph{NLO-weighted Born contribution}, $\bar{B}$, by integrating out
the singular terms,
\bea
\bar{B}(\llPS_{0}) &=&
B(\llPS_{0})+\alphaS V_1(\llPS_{0})
  + \alphaS\int\lld\llPS_{1|0} S_1(\llPS_{1})\nnb\\
&+& \alphaS\int\lld\llPS_{1|0} \left[R_1\llsup{s}(\llPS_{1})
  - S_1(\llPS_{1})\vphantom{\int}\right].
\label{eq:matching:NLOBorn}
\eea
Here we stress again that the $S_1$ can be written as a convolution of
a Born term and a universal subtraction term, and in
\begin{equation}
  \label{eq:matching:factorized-subtaction-ps}
  \int\lld\llPS_{1|0}S_1(\llPS_{1}) =
  B(\llPS_0)\otimes\int\lld\llPS_{1|0}\tilde{S}(\llPS_{1|0}),
\end{equation}
the integral over $\tilde{S}$ can be calculated analytically using
dimensional regularization, allowing us to add it to the virtual part
and thus explicitly cancel the divergences there. The integral over
$R_1\llsup{s} - S_1$, in contrast, must typically be evaluated
numerically through Monte Carlo integration.

We now arrive at a form of the inclusive NLO cross section,
\begin{equation}
  \label{eq:matching:NLO3}
  \lld\llxsec\llsup{NLO}=
  \lld\llPS_0\bar{B}(\llPS_{0})+\lld\llPS_1\alphaS R_1\llsup{ns}(\llPS_1),
\end{equation}
which will serve as a starting point for our discussion about
higher-order corrections to the parton shower approach.

\mcsubsubsection{The first emission in a parton shower}
\label{sec:matching-first-ps}

We now look at the inclusive cross section as given by the first
emission in a parton shower:
\begin{equation}
  \label{eq:matching:PSfirst}
  \lld\llxsec\llsup{PS}=
  \lld\llPS_0B(\llPS_0)\left[\llsud{0}(\llordmax,\llordcut)+
    \int_{\llordcut}\frac{\lld\llord_1}{\llord_1}
    \int\!\!\lld\llaux_1\,\frac{\alphaS}{2\pi}\llsplitP(\llaux_1)
    \llsud{0}(\llordmax,\llord_1)\right],
\end{equation}
where we have denoted the ordering variable \llord (with $\llordmax$
giving the starting scale, and $\llordcut$ the cutoff scale for the
shower). We note again that, performing the integral, the bracket is
unity, reflecting the unitary nature of the parton shower.

In principle higher-order corrections will have two effects in a
parton shower. They alter
\begin{enumerate}
\item the shape of distributions related to the first, hardest emission;
\item the norm --- the total cross section --- of the produced sample;
\end{enumerate}
and there are methods that focus on one or the other, or both.
%It is thus not surprising that they will be introduced in various ways.

The first modification, which was also historically the first matching
procedure \cite{Bengtsson:1986hr}, is simply to replace the splitting
function above with the singular part of the real emission matrix
element,
\begin{equation}
  \label{eq:matching:first}
  \frac{\lld\llord_1}{\llord_1}
  \lld\llaux_1\frac{\alphaS}{2\pi}\llsplitP(\llaux_1)\rightarrow
  \lld\llPS_{1|0}\frac{R_1\llsup{s}(\llPS_1)}{B(\llPS_0)}
\end{equation}
Two things are worth noting here:
\begin{itemize}
\item In the soft and collinear limits of the real-emission matrix
  elements, the effect of the extra emission factorizes into universal
  terms, which exhibit {\em exactly the same singularity structure as
    the splitting kernels employed in parton shower Monte Carlos}.
  These singularities, through the cut in the emission phase space,
  given by $\llordcut$, lead to large logarithms, which in turn are
  resummed by the parton shower.  The same large logarithms are, of
  course, then also encoded in $R_1\llsup{s}/B$, and thus the
  logarithmic structure of the parton shower is preserved.
\item In this transition, the one-particle phase space element of the
  parton shower is replaced by $\lld\llPS_{1|0}$, the phase space
  element that, starting from a Born configuration, produces a
  real-emission phase space configuration.  Clearly, if the
  one-particle emission phase space is not completely covered by the
  parton shower, the replacement above on its own will not be
  sufficient, since parts of the true available phase space will be
  left out.  In this case, a hard matrix element correction,
  essentially through the non-singular term $R_1\llsup{ns}$, is
  mandatory.  This effect of not covering the full phase space may
  happen for two reasons:
  \begin{enumerate}
  \item the parton shower does not completely cover the emission phase
    space.  This is true, \eg, for angular-ordered parton showers;
  \item the Born configuration exhibits zeros, due to polarization
    effects or similar, that are not present after the first emission.
    This happens, \eg, for the zero in the lepton pseudo-rapidity distribution from
    hadronic $W$-boson production, as discussed in
    \cite{Alioli:2008gx}.
%for the radiation zero in hadronic di-boson production
%    \cite{do-we-have-a-reference-for-radiation-zeros-in-W-production}.
  \end{enumerate}
\end{itemize}
Note that the replacement above can fairly easily be carried over to
the Sudakov form factor
\begin{equation}
  \label{eq:matching:sudcorr}
  \llsudb{0}(\llordmax,\llord)=
  \exp\left[-\int\lld\llPScut{1|0}{\llord}\alphaS
    \frac{R_1\llsup{s}(\llPS_1)}{B(\llPS_0)}\right],
\end{equation}
using the so-called veto algorithm (see \AppRef{mcmethods:veto}).

We now get the inclusive cross section
\begin{eqnarray}
  \label{eq:matching:PScorrfirst}
  \lld\llxsec\llsup{PScorr}&=&
  \lld\llPS_0B(\llPS_0)\left[\llsudb{0}(\llordmax,\llordcut)+
    \int\lld\llPScut{1|0}{\llordcut}\alphaS
    \frac{R_1\llsup{s}(\llPS_1)}{B(\llPS_0)}
    \llsudb{0}(\llordmax,\llord_1)\right]\nnb\\
  &+&\lld\llPS_1\alphaS  R_1\llsup{ns}(\llPS_1)\,,
%  \left[R_1(\llPS_1)-R_1\llsup{s}(\llPS_1)\vphantom{\int}\right],
\end{eqnarray}
and by undoing the integral over the real emission we get the first
emission of the parton cascade, properly weighted by the Sudakov form
factor to give the higher-order $\alphaS^nL^{2n}$ and
$\alphaS^nL^{2n-1}$ blobs in \FigRef[a]{fig:matching-ordersPS}, from
which we can now continue with the subsequent emissions in the parton
shower to get fully exclusive partonic final states.

The next logical step is to also achieve ${\cal O}(\alphaS)$
accuracy at the cross section level.  There are two ways to do this,
which go under the names of \POWHEG and \MCatNLO, respectively.

\mcsubsubsection{\POWHEG}\label{sec:powheg}

The \POWHEG approach, effectively, is an advanced matrix element
reweighting procedure, where the Born-level term in front of the first
square bracket in \EqRef{eq:matching:PScorrfirst} is replaced by the
NLO weighted Born-level term.  Furthermore, the whole first-order
real-emission term $R_1$ is used for $R_1\llsup{s}$, so that
$R_1\llsup{ns}=0$. Thus
\begin{equation}
  \label{eq:matching:powheg}
  \lld\llxsec\llsup{POWHEG} =
  \lld\llPS_0\bar{B}(\llPS_0)\left[\llsudb{0}(\llordmax,\llordcut)+
    \int\lld\llPScut{1|0}{\llordcut}
    \alphaS\frac{R_1(\llPS_1)}{B(\llPS_0)}
    \llsudb{0}(\llordmax,\llord_1)\right],
%  &+&\lld\llPS_1\alphaS
%  \left[R_1(\llPS_1)-R_1\llsup{s}(\llPS_1)\vphantom{\int}\right],
\end{equation}
and parton showering will give rise to similar emissions as the first
term in \EqRef{eq:matching:PScorrfirst} but with a global NLO-reweighting,
$\bar{B}/B$, acting as a local $K$-factor.

\paragraph{Truncated and vetoed parton showers}
\label{sec:matching:truncated}

Here we digress a bit to consider the problems arising if a parton
shower is not ordered in hardness or \kt. In the \POWHEG approach,
the first emission is supposed to be the hardest one, and if the parton
shower is not ordered in hardness one cannot simply add it to the
states generated by \POWHEG, as that would not ensure that the
subsequent emissions would be less hard than the first one.

The simplest solution is to start the shower at its maximum possible
ordering scale, but veto any emission that is harder than the first
\POWHEG one. However, as pointed out in \cite{Nason:2004rx}, this
means that the colour structure and kinematics of the parton shower
would be altered, as the emission with the highest ordering scale
would be emitted from the $+1$-parton state rather than from the Born
state, as it would have been in the normal shower.

The solution is to first reconstruct the parton shower variables
$(\llord_1,\llaux_1)$ for the emission given by \POWHEG. Then the
shower is started from the corresponding Born-state with the maximum
ordering variable $\llordmax$ and is allowed to evolve down to
$\llord_1$, vetoing any emission harder than the first emission. Then
the $(\llord_1,\llaux_1)$ emission is inserted, and the shower can
continue evolving, still vetoing any emission harder than the first
emission. This procedure is called a truncated, vetoed
shower\cite{Nason:2004rx}.

\mcsubsubsection{\MCatNLO}\label{sec:matching-mcatnlo}

Having at hand the formula for the cross section in the \POWHEG
formalism allows us to discuss the alternative \MCatNLO approach on
the same footing.  The main idea in \MCatNLO is that the singular terms
$R_1\llsup{s}$ are taken to be {\em identical} to the subtraction
terms $S_1$.  They in turn are given by the convolution $S_1 =
B\otimes \llsplitP$, \ie by additionally identifying the universal
subtraction terms with the parton shower splitting kernels.  Therefore
\begin{eqnarray}
  \label{eq:matching:MC@NLO}
  \lld\llxsec\llsup{MC@NLO}&=&\lld\llPS_0
  \left[B(\llPS_0)+\alphaS V_1(\llPS_0)+\alphaS
    B(\llPS_0)\otimes\int\lld\llPS_{1|0}\llsplitP(\llPS_{1|0})\right]\nnb\\
  &&\quad\times\left[\llsud{0}(\llordmax,\llordcut)+
    \int_{\llordcut}\frac{\lld\llord_1}{\llord_1}
    \int\!\!\lld\llaux_1\,\frac{\alphaS}{2\pi}\llsplitP(\llaux_1)
    \llsud{0}(\llordmax,\llord_1)\right]\nnb\\
  &+&\lld\llPS_1\alphaS\left[R_1(\llPS_1)-
    B(\llPS_0)\otimes \llsplitP(\llPS_{1|0})\vphantom{\int}\right].
\end{eqnarray}
Again, undoing the integral in the second bracket, we obtain the first
parton shower splitting, and we can continue the shower as before. In
practice, two sets of events are given to the parton shower. The first
set contain Born-level states given by the first bracket in
\EqRef{eq:matching:MC@NLO}, where the integral in the second bracket
is simply undone by running the parton shower. The second set contains
events with one extra parton, where the parton shower is added using
suitable starting conditions. In this way the ${\cal O}(\alphaS)$
contribution of the parton shower is removed and is instead replaced
by the exact NLO result.

To ${\cal O}(\alphaS)$, this is equivalent to \POWHEG, however, we
note that the real emission matrix element is not exponentiated in the
Sudakov form factor (which is \llsud{} rather than \llsudb{}). Also, contrary
to \POWHEG, it is not guaranteed that the weights of the generated
states are positive definite, as one can easily imagine having
splitting functions that overestimate the real emission matrix
element, rendering the last bracket negative.

\mcsubsection{Tree-level multi-jet merging and CKKW}
\label{sec:matching-at-tree}

An alternative way of choosing the $R_1\llsup{s}$ term is to introduce
a cutoff, which we shall call the merging scale, $\llordms$ such that
$R_1\llsup{s}(\llPS_1)=R_1(\llPS_1)\times\Theta(\llordms-\llord(\llPS_1))$.
We then get a modified Born term compared to \EqRef{eq:matching:NLOBorn}
\begin{equation}
  \label{eq:matching:exclusiveborn}
  \tilde{B}(\llPS_{0}) =
  B(\llPS_{0})+\alphaS V_1(\llPS_{0})
  + \alphaS\int\lld\llPSmax{1|0}{\llordms} R_1(\llPS_{1}),
\end{equation}
which we immediately identify as the NLO expression for the exclusive
cross section with no partons above the scale \llordms, in the full
inclusive NLO cross section,
\begin{equation}
  \label{eq:matching:NLO4}
  \lld\llxsec\llsup{NLO}=
  \lld\llPS_0\tilde{B}(\llPS_{0})+
  \lld\llPS_1\alphaS R_1(\llPS_1)\Theta(\llord(\llPS_1)-\llordms).
\end{equation}
We also see that the second term is exactly what we would get from a
standard tree-level matrix element generator for the one-parton cross
section with \llordms as cutoff.

\mcsubsubsection{Merging for the first emission}

Ignoring the NLO-reweighting of the Born term for the moment, we can
now obtain a parton shower where the first emission is corrected to
the tree-level matrix element if it is above \llordms:
\begin{eqnarray}
  \label{eq:matching:CKKWfirst}
  \lld\llxsec\llsup{CKKW}&=&
  \lld\llPS_0B(\llPS_0)\left[\llsud{0}(\llordmax,\llordcut)
    \vphantom{\int\frac{R_1}{B}}\right.\nnb\\
    &&+\int_{\llordcut}\frac{\lld\llord_1}{\llord_1}
    \int\!\!\lld\llaux_1\,\frac{\alphaS}{2\pi}\llsplitP(\llaux_1)
    \Theta(\llordms-\llord_1)
    \llsud{0}(\llordmax,\llord_1)\nnb\\
    &&\left.+\int\lld\llPS_{1|0}\alphaS
    \frac{R_1(\llPS_1)}{B(\llPS_0)}\Theta(\llord(\llPS_{1|0})-\llordms)
    \llsud{0}(\llordmax,\llord_1)\right].
\end{eqnarray}
We see that the matrix element will fill the phase space above the
merging scale (jet production) and the parton shower will fill the
phase space below (jet evolution), effectively amounting to a phase
space slicing into two disjoint regimes. This is the basis of the
CKKW-based merging algorithms\cite{Catani:2001cc,Lonnblad:2001iq}.

A number of things are worth noting here:
\begin{enumerate}
\item The expression just exhibits the inclusive cross section for the
  first, hardest emission of a given Born configuration.  The effect
  of the phase space slicing is made manifest, and leads to the two
  emission terms.  The second of these emission terms, with
  $\llord>\llordms$, is in fact generated differently from how the
  equation above suggests: In practice this term gives rise to a
  separate term contributing to the inclusive sample, but with a
  Born-level matrix element for a one-parton emission process as seed
  rather than the initial Born matrix element.
\item The equation above exhibits a new feature, namely a violation of
  unitarity, \ie the square bracket does not integrate to one any more.
  As long as a reasonable range for the value of \llordms is chosen, this 
  does not have a big impact, though, and the total cross section of all
  contributions will be relatively stable with respect to changes in \llordms.  
\item The equation above explicitly shows that the logarithmic accuracy of
  the parton shower is preserved in the merging.
\end{enumerate}

\mcsubsubsection{Multi-jet merging}

If we again undo the integrals in \EqRef{eq:matching:CKKWfirst}, we
can as in \EqRef{eq:matching:PScorrfirst} continue the cascade below
$\llord_1$. However, as noted above, the second integral can then be
thought of as an additional Born-level contribution for a one-parton
emission process, for which we again can correct the first splitting. This
would give us
\begin{eqnarray}
  \label{eq:matching:CKKWsecond}
  \lld\llxsec_1\llsup{CKKW}&=&
  \lld\llPS_1\alphaS R_1(\llPS_1)\Theta(\llord(\llPS_{1|0})-\llordms)
  \llsud{0}(\llordmax,\llord_1)\times\nnb\\
  &&\left[\llsud{0}(\llord_1,\llordcut)
    \vphantom{\int\frac{R_1}{B}}\right.\nnb\\
    &&+\int_{\llordcut}\frac{\lld\llord_2}{\llord_2}
    \int\!\!\lld\llaux_2\,\frac{\alphaS}{2\pi}\llsplitP(\llaux_2)
    \Theta(\llordms-\llord_2)
    \llsud{0}(\llord_1,\llord_2)\nnb\\
    &&\left.+\int\lld\llPS_{2|1}\alphaS
    \frac{R_2(\llPS_2)}{R_1(\llPS_1)}\Theta(\llord(\llPS_{2|1})-\llordms)
    \llsud{0}(\llord_1,\llord_2)\right].
\end{eqnarray}
Clearly if we also have higher-order tree-level matrix elements,
$R_3,R_4,\ldots$, available we can now continue to correct also higher
parton multiplicities.

Further notes are in order.
\begin{itemize}
\item The Sudakov form factors can either be calculated analytically,
  as in the original CKKW scheme, or they can be generated by the
  shower itself as in CKKW-L, where the form factor is interpreted
  strictly as a no-emission probability.
\item If interpreted as no-emission probabilities, the Sudakov form
  factors are always below unity, and the reweighting can be
  implemented as a simple vetoing procedure (see
  \AppRef{mcmethods:veto}).
\item In the MLM and Pseudo-Shower schemes, the Sudakov form factors
  are approximated by allowing the shower to radiate all the way down
  to its cutoff, $\llordcut$, and this partonic state is then clustered
  with a jet algorithm with \llordms as resolution scale. The
  probability that these partonic jets are close to the original
  partons is then an approximation of the no-emission probability. The
  results are similar to those of CKKW(-L)\cite{Alwall:2007fs}, but it
  is not entirely clear how far the formal accuracy of the parton
  shower can be maintained.\footnote{See \eg the discussion in
    \cite{Lavesson:2007uu}.}
\item Equating the Sudakov form factors in \EqRef{Sudakov} with
  no-emission probabilities is only correct in final-state
  radiation. For initial-state emissions we must use the no-emission
  probability in \EqRef{ISRsudakov}, and it can be shown that this is
  related to the Sudakov form factor needed in the merging by a simple
  ratio of parton density function\cite{Krauss:2002up,Lavesson:2005xu}
  (see also \cite{Ellis:1991qj}), such that
  \begin{equation}
    \label{eq:matching-sudnoem}
    \llsud{i}(\llord_i,\llord_{i+1})=
    \frac{\pdf{}(x,\llord_{i})}{\pdf{}(x,\llord_{i+1})}\times
    \llsud{i}(\llord_i,\llord_{i+1};x).
  \end{equation}
  Hence, the procedure above needs to be amended with an extra
  reweighting with this PDF-ratio in the case of hadronic collisions.
\item We have not considered the running of \alphaS in the parton
  shower. In practice, the matrix-element generators will use a fixed
  \alphaS, and in the CKKW-based algorithms, an additional weight,
  $\prod\alphaS\llsup{PS}(\llord_i)/\alphaS$, is introduced.
\item If the parton shower is not ordered in hardness, or \kt, we
  cannot simply add a shower below the merging scale. Instead we must
  use the truncated, vetoed shower, described in
  \SecRef{sec:matching:truncated}, generalized to several hard
  emissions.
\end{itemize}

\mcsubsection{Multi-jet NLO merging}
\label{sec:nlo-merging}

If we look at \EqRef{eq:matching:CKKWfirst}, it is easy to see how we
can reintroduce the NLO corrections from \POWHEG in the CKKW-(L)
matching schemes. What is needed is to replace the splitting function
in the first integral with the same ratio $R_1/B$ as in the second,
and also reintroduce the corrected Sudakov form factors
$\llsud{0}\to\llsudb{0}$. If we then also reweight all states with the
dynamic $K$-factor $\bar{B}(\llPS_0)/B(\llPS_0)$, we arrive at the
so-called \MENLOPS procedure \cite{Hamilton:2010wh}, where the inclusive
cross section is correct to NLO, and the hardest emissions are
corrected with tree-level matrix elements.

Lately, there has also been some progress in combining parton showers
and matrix elements in such a way that also the cross sections for
multi-jet final states becomes correct to next-to-leading order.  The
scheme called NL$^3$, suggested in \cite{Lavesson:2008ah}, relies on
being able to generate Born states with $n$ extra partons above some
merging scale exactly according to the exclusive NLO cross section, in
the same way as in \EqRef{eq:matching:exclusiveborn}. From these
states a shower is then allowed to evolve below the merging scale. To
these event samples one can then add event samples generated according to
the standard CKKW-L procedure, but these are reweighted such that the
two first orders in \alphaS (corresponding to those in the NLO
samples) are subtracted, by carefully expanding out the Sudakov form
factors and the running of \alphaS.

The procedure is technically complicated and will not be described in
detail here, and so far it has only been implemented for \lleetoj. In
the end (returning to the language of \FigsRef{fig:matching-ordersPS} and
\ref{fig:matching-ordersNLO}) the procedure corresponds to correctly
taking into account the impact, both on shape \emph{and} cross
section, of all the blobs up to $\alphaS^n$ together with all the
$\alphaS^kL^{2k}$ and $\alphaS^kL^{2k-1}$ ones ($k>n$), for the
$n$ first real emissions above the cut, while all other emissions are
only correct to $\alphaS^nL^{2n}$ and $\alphaS^nL^{2n-1}$
(corresponding to filling all the half-filled blobs in
\FigRef[c]{fig:matching-ordersNLO}).

\mcsubsection{Summary}
\label{sec:matching-outlook}

\begin{itemize}
\item Fixed-order matrix elements and parton showers have different
  merits and shortcomings. They should be combined to get the best of
  both, for an optimal description of multi-parton states.
\item Tree-level matrix elements cannot be blindly combined with
  a parton shower. The former are inclusive in nature, while the latter
  produces exclusive final states.
\item Matrix elements must be supplemented with Sudakov form factors
  to give exclusive final states that can be combined with a parton
  shower.
\item Special care must be taken if the parton shower is not ordered
  in hardness. When adding such a parton shower to a
  matrix-element-generated state it must therefore be properly
  truncated and vetoed.
\item Combining next-to-leading-order matrix elements with parton
  showers for the first emission is now state of the art. For
  multi-leg matching and merging, the state of the art is still
  tree-level matrix elements.
\end{itemize}

Combining fixed-order matrix elements with parton showers is a very
active research topic, and is important for giving reliable precision
predictions for jet production from QCD. The last word is surely not
said yet, and we are looking forward to many new ideas and
improvements in the near future. In the long run it does not seem
inconceivable that we will have generators producing results that are
correct to next-to-next-to-leading order.

% Local Variables: 
% mode: LaTeX
% TeX-master: "../mcreview"
% End: 

\mcsection{PDFs in event generators}
\label{sec:pdfs-event-gener}

Parton Distribution Functions play a central role in event
generators, for the simulation of hard processes, parton showers 
and multiple parton interactions. The choice of PDF set therefore will 
influence both cross sections and event shapes.

To lowest order the function $f_i(x, \mu_F)$ describes the probability
to find a parton of species $i$ with a momentum fraction $x$ when a 
proton is probed at a scale $\mu_F$. This distribution cannot be predicted
from first principles, since it depends on the non-perturbative physics 
of the proton wave function. With an ansatz for the initial 
distributions at some low scale $\mu_{F0}$, the evolution towards larger 
scales is predicted by the DGLAP equations 
\cite{Gribov:1972ri,Dokshitzer:1977sg,Altarelli:1977zs}, however.
Different tunes have been made, by comparing an evolved ansatz with 
relevant data, \eg from Deeply Inelastic Scattering. Over the years
many such tunes have been presented, with increasing accuracy as newer
data have been added. Also the theoretical framework has seen some 
improvements. The CTEQ \cite{Lai:2010vv} and MRST/MSTW 
\cite{Martin:2009iq} collaborations have been especially diligent in 
regularly presenting updated tunes. These and others are available 
in the LHAPDF library \cite{Whalley:2005nh}.

Precision tests of QCD today normally involve comparisons with NLO 
matrix elements convoluted with NLO PDFs (\EqRef{Eq::Master_For_XSec}) 
both defined in the $\overline{\mathrm{MS}}$ renormalization scheme. In general 
this allows a greatly improved description of data, relative to LO 
results. But NLO expressions are not guaranteed to be positive definite, 
and do not have a simple probabilistic interpretation.
For PDFs, specifically, it is well known that the NLO gluon has a 
tendency to start out negative at small $x$ for small scales, and 
only turn positive by the QCD evolution towards larger $\mu_F$. In the 
MRST/MSTW sets the ansatz allows negative gluons, while the CTEQ ansatz
is constructed to be positive definite. Either way, the NLO gluon PDF
is very different from the LO one, also at larger scales. 
Specifically, it remains smaller in NLO than in LO at small $x$, which 
then typically is compensated by the NLO MEs being larger than the LO 
ones, by the presence of $\ln(1/x)$-enhanced terms.

In recent years the emphasis has been put on NLO PDFs, as offering the
best description of hard-process data. The problem is that generators
are largely of an LO character. That is, while several hard processes are 
implemented to NLO accuracy in some generators, notably in the MC@NLO 
and POWHEG frameworks, see \SecRef{sec:me-nlo-matching}, most
are still only 
provided at LO, as are all the standard PS and MPI models, \eg with 
LO splitting kernels in the showers. 
The latter two components additionally have most of their activity in 
the low-$p_{\perp}$ region, and the last one especially with low-$x$ 
gluons. A usage of NLO PDFs with LO MEs here would be strongly 
questionable, since this combination typically undershoots the 
interaction rate obtained both in LO and NLO calculations.

Therefore the norm is for LO generators to use the few LO PDF fits
that are still produced. In cases where the shape is more important
than the absolute normalization --- such as backwards evolution of ISR,
where it is ratios of PDFs that sets the evolution rate --- this may
be perfectly adequate. But for absolute cross sections NLO calculations 
usually lie above LO ones, and this enhancement is needed 
to obtain a good description of data. This introduces a tension in LO 
PDF fits: a data set that is mainly sensitive to a particular $x$ range
prefers to have more of the total momentum of the proton located in
that $x$ range, at the expense of other $x$ ranges. To address this 
issue, new Monte Carlo adapted PDFs have been presented, wherein the 
momentum sum rule $\sum_i \int_0^1 \xpdf{i}(x, \mu_F) \, \mathrm{d}x = 1$
is relaxed: MRST LO* and LO** \cite{Sherstnev:2007nd}, 
and CT09 MC1, MC2 and MCS \cite{Lai:2009ne}. By allowing the 
normalization to float, one typically finds a value 10 -- 15\% above 
unity, and the whole $x$ range 
can be enhanced without tension. This should certainly not be viewed 
as a breaking of momentum conservation, but as a way to include an 
approximate $K$ factor that can depend on the parton flavours and
momenta but not on the specific process.

Another trick introduced is to make use of pseudo-data, obtained by 
generating several different processes to NLO, but then fitting them
to PDFs within an LO context \cite{Lai:2009ne}. That way it is possible 
to obtain a more uniform coverage over a large $x$ range, and for 
different flavours. Additionally, of the above five sets, CT09 MCS 
does not relax the momentum sum rule, but instead optimizes
renormalization and factorization scales process by process.

Note that much of the case for the modified LO sets is based
on the LO PDF behaviour at small $x$ and $\mu_F$. In the opposite region,
large $x$ and $\mu_F$, differences are not  as clear-cut. Often the NLO
corrections to the MEs there correspond to rescaling by an approximately
constant factor. Then, if NLO PDFs provide a better shape than LO PDFs
in that region, the combination of LO MEs and NLO PDFs makes sense.
It is therefore possible to use one PDF set for the hard processes and 
another for the softer parton showers and multiple interactions. 

To conclude, the new PDFs offer the hope of improved descriptions 
of data within generators, and have already started to be used, both
by generator authors and by the experimental collaborations, as an
alternative to normal LO ones. The outcome is still not clear; studies
suggest that you can find distributions where the new sets do better 
than the traditional ones, and distributions where they do worse
\cite{Kasemets:2010sg}. There can be no doubt, however, that these
sets have put new tools in the hands of generator authors, and
opened the way for further developments of PDFs especially well suited
for generator applications.

\begin{itemize}
\item PDFs are used in generators for the description of hard processes,
showers and MPIs.
\item While NLO PDFs are appropriate for studies of hard processes,
especially when combined with NLO MEs, they are not well suited for
current LO shower and MPI models.
\item New tricks have been introduced to improve the usefulness of  
LO PDFs in generators, and new sets have been presented, but currently
there is no obvious winner.
\end{itemize}

% Local Variables: 
% mode: LaTeX
% TeX-master: "../mcreview"
% End: 

\mcsection{Soft QCD and underlying event physics}
\label{sec:minim-bias-underly}
Several distinct physics and modelling issues come under the 
heading of ``soft QCD'' and ``underlying event''. In \SecRef{sec:primkt}  
we discuss ``primordial \kT'', a
topic that lies on the intersection between parton showers and
soft QCD.
The rest of this section is  devoted to a discussion of
the different physics (sub-)processes that  contribute to the total
observed activity in hadron-hadron collisions. 
Thus, in \SecRef{sec:mbtypes}, we give a brief introduction to --- and
dictionary of --- the different 
QCD processes that form the dominant part of the total hadron-hadron
cross section, and to the origin of the so-called ``underlying event''
and the associated ``pedestal effect''.
In \SecRef{sec:mbmpi}, we then 
take a closer look at models based on multiple parton interactions
(MPI). \SecRef{sec:colrec} focuses on the particular issue of colour
reconnections. 
Finally, \SecRef{s_ue:pomerons} gives a very brief introduction to models
of diffraction, in particular models based on pomerons. 

For more specific details on the current implementations in each of
the main generators, see the individual descriptions in
\PartRef{sec:spec-revi-main}. 

\mcsubsection{Primordial \kT \label{sec:primkt}} 

In Monte Carlo models, the term ``primordial \kT'' is used to refer
collectively to any transverse momentum given to initial-state
partons \emph{beyond} that generated by the normal ISR shower
evolution described in \SecRef{sec:isr}.  

Physically, such additional momentum could come from several
sources, as follows:
\begin{enumerate}
\item Fermi motion of confined
partons inside their parent hadron, with a magnitude of order the
inverse hadron radius $\sim \LambdaQCD$.  
\item ``Unresolved'' ISR shower activity, coming from scales 
below the infrared cutoff.
\item Activity not accounted for, or incorrectly accounted for, 
by the particular shower model in question. In particular, this may be
relevant for parton evolution involving 
low parton momentum fractions $x$, as was briefly
discussed in \SecRef{sec:isr}.
\end{enumerate}
Of these, only Fermi motion is relatively straightforward.
It is also the only component that is genuinely
``primordial'', \ie intrinsic to the incoming hadron.
The others depend sensitively on issues that are
inherently ambiguous in the shower description: whether the
low-$\pT{}$ divergences in the parton shower are regulated by a sharp
cutoff or by a smooth 
suppression (and in what variable), how $\alphaS$ is treated close to
the cutoff, how the shower radiation functions and recoil effects 
behave, and whether any non-trivial low-$x$ effects are included. 

In the simplest possible treatment one lumps 
all these ``unresolved'' effects, regardless of origin, into 
a single number, called ``primordial \kT''. In each event, the
beam-collinear partons extracted from each of the original hadrons
can then be given a \pt\ of this magnitude, typically distributed
according to a Gaussian or similar distribution. (Since the beam
remnant must necessarily take up the recoil from this kick, 
an upper cutoff is usually also enforced, limiting the amount by
which the beam remnant is allowed to be kicked off axis by 
tails of this effect.)

However, even in this simplified case, the above discussion should serve
to illustrate that the value of the ``primordial \kT'' should not really be
perceived of as a universal constant. At the very least, the shower evolution
equations imply that it must have some  implicit dependence on the ISR cutoff --- 
if we increase the shower cutoff, for instance, 
the primordial \kT should increase slightly as well, to compensate for the now
missing shower activity in the region that has been cut
away. In current models, this scaling does not happen automatically,
but must be taken care of by retuning the primordial \kT if the ISR
cutoff (or any other parameter affecting the infrared regularization
of the initial-state shower) is changed. More troublingly, perhaps, 
since a lot of possible process-dependent physics effects have been
``swept under the rug'', there is also no strong reason why the same 
value of this primordial \kT should work equally well for all
processes or even in different phase space regions. 
Attempting to extract a value for it in several different, mutually
complementary, processes and regions, 
could therefore be a valuable input to guide future modelling. 

A next-to-simplest iteration can be obtained by letting 
the value of primordial \kT scale with the $Q^2$ of the hard
interaction \cite{Sjostrand:2004pf}. This generates a minimum of
process-dependence (\eg partons entering a soft QCD scattering can 
now be given a smaller primordial \kT than ones producing a $Z$
boson), but still does not really address the underlying physics. 

A first stab at a more physical
model was made in \cite{Gieseke:2007ad}, by including a
non-perturbative function that depends explicitly on the phase space
available for unresolved initial-state radiation, in
addition to a smaller and more
universal component to be modelled by a Gaussian. 
Although the data available at the time 
could not clearly differentiate this from the conventional models, 
the expectation is that this should generate a more realistic 
process- and collision-energy-dependence of the effective primordial
\kT.

A secondary modelling issue, relevant to the MPI models discussed in
the next subsection, is how much primordial \kT is assigned to
partons initiating multiple parton interactions, and 
how the associated recoil effects are distributed
among those initiators and the remnant. Typically, MPI initiators are
only assigned a primordial \kT of the order of Fermi motion, 
although this is a model-dependent statement that
may of course change, as models improve.

\begin{figure}[tp]
  \centering
    \label{fig:inline:drell-yan-intrinsic-kt-tvt}
    \includegraphics*[scale=0.55]{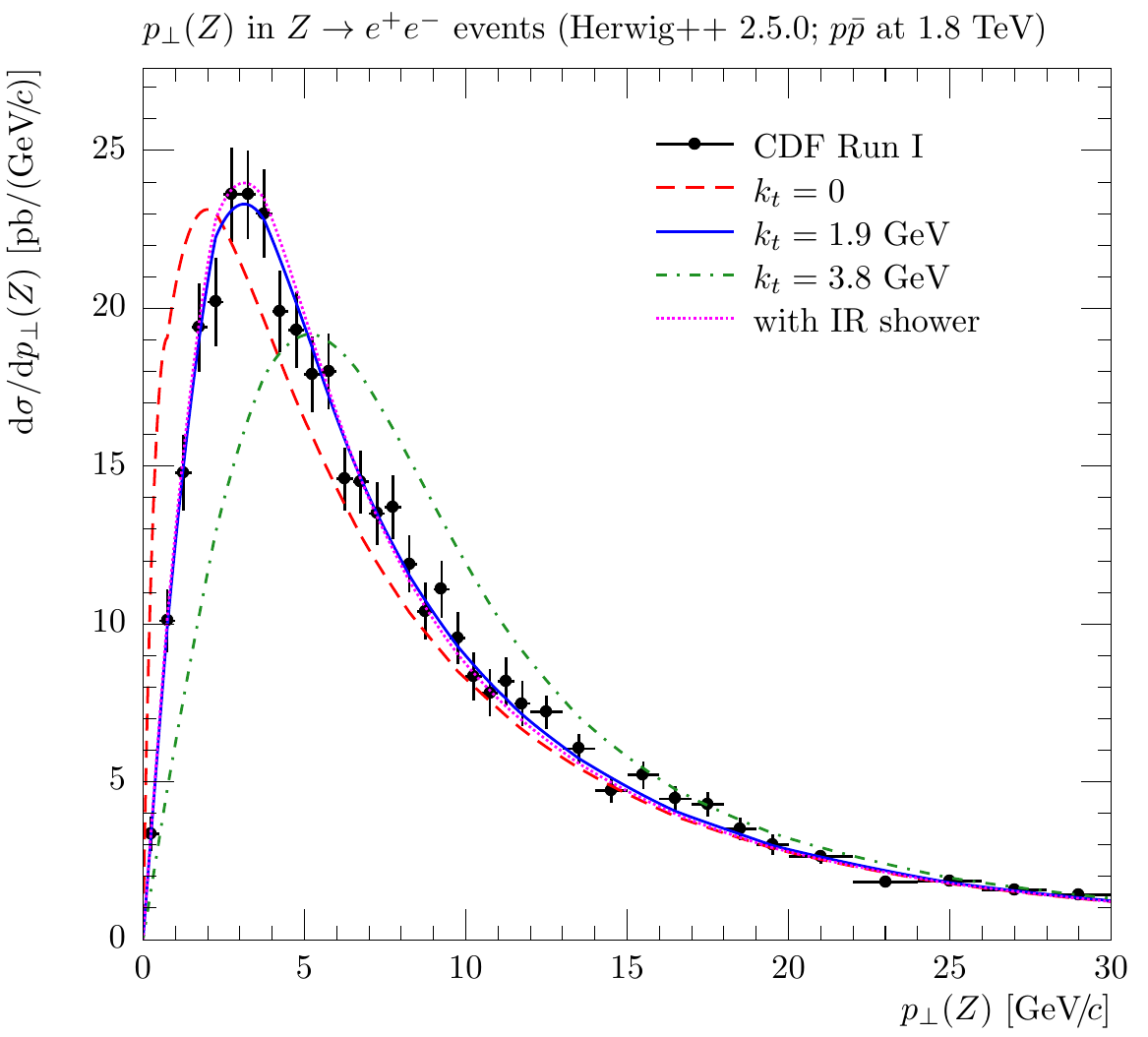}
    \includegraphics*[scale=0.55]{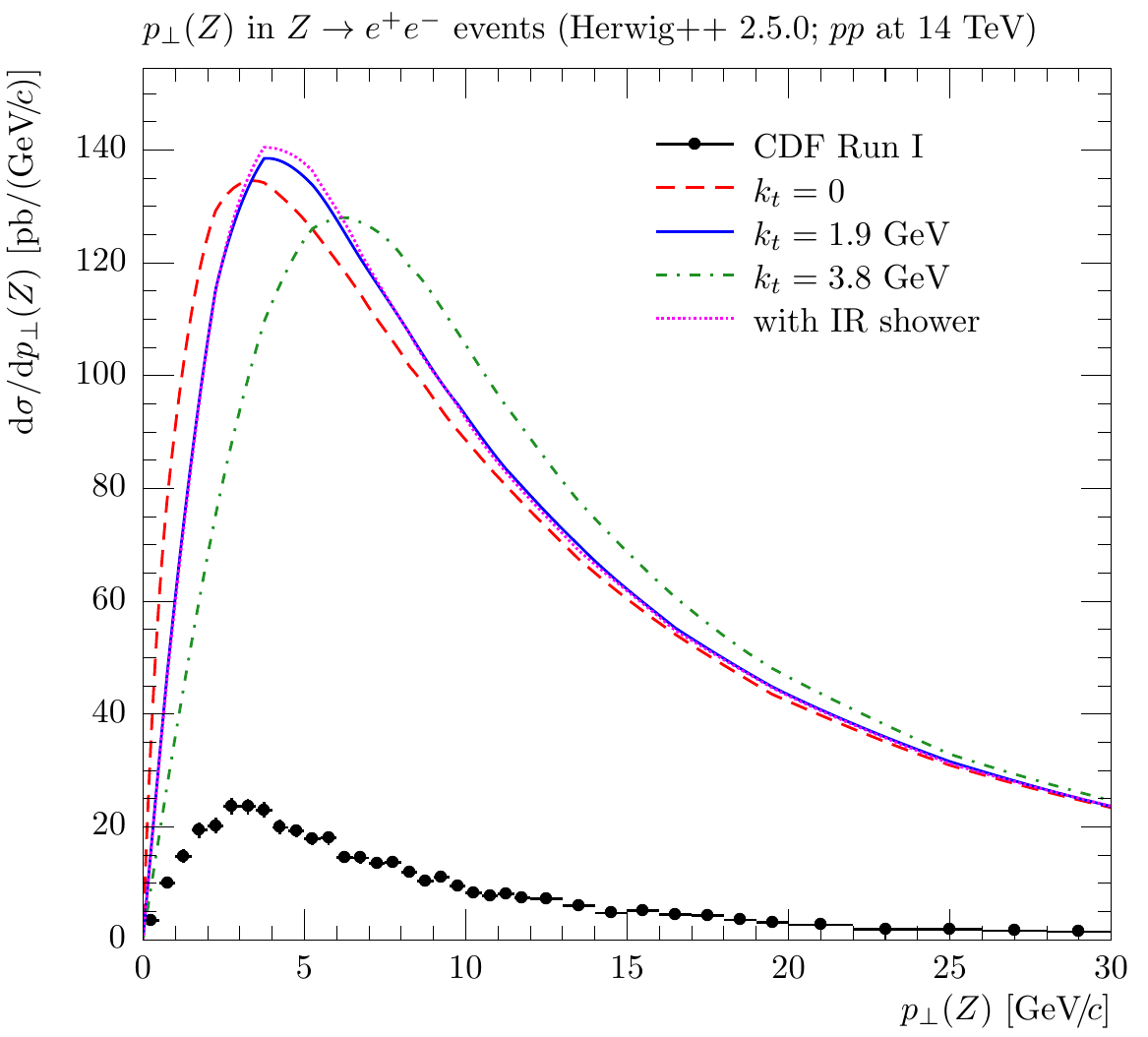}
  \caption{The low-\pt\ peak of the \pt\ distribution of lepton pairs
  in Drell-Yan events at the Tevatron, compared to CDF
  data \cite{Affolder:1999jh}. 
  A Monte Carlo model
  (\herwigpp) is shown with four different choices for the
  ``primordial \kT''. a) $p\bar p$ at 1.8 TeV b) $pp$ at
  14 TeV.
\label{fig:primkt}}
\end{figure}
\begin{figure}[tp]
  \centering
  \includegraphics[scale=0.7]{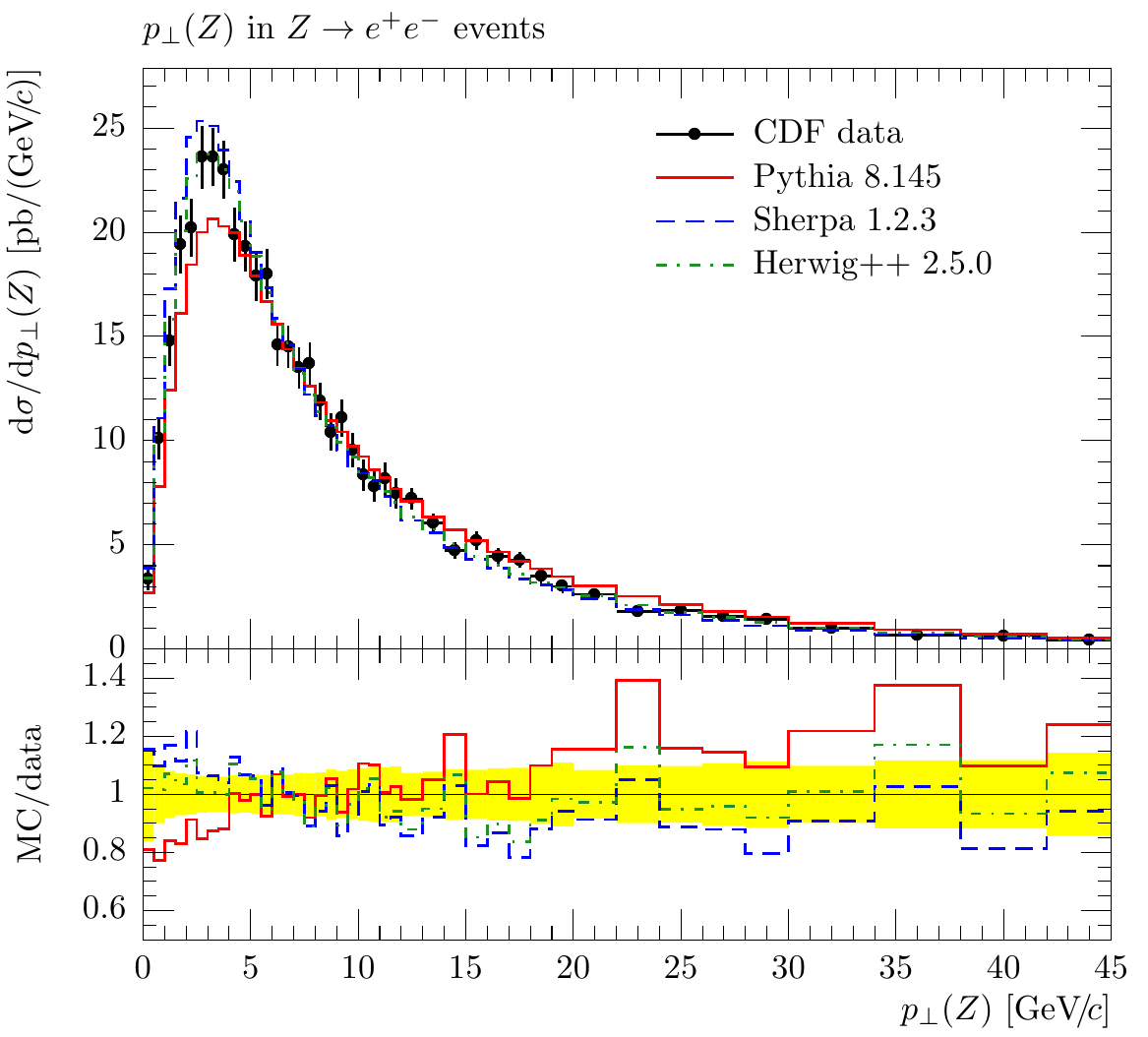}
  \caption{CDF 2000 $Z^0$ \pT peak \cite{Affolder:1999jh}.  The location
    of the peak is very sensitive to the degree of ``primordial \kT''
    smearing in the generator, and the higher-\pT region is affected by
    the parton shower. In all cases, the generators have been run with
    LO matrix elements, and the MC normalization is fixed to that of the
    data, to alleviate the requirement for an NLO cross-section.  An
    up-to-date version of this plot can be found at
    \url{http://mcplots.cern.ch/}.}
  \label{fig:cmp:intrinsic-kt}
\end{figure}

Empirically, the most important distribution for constraining the
magnitude of this effect is the \pt\ distribution of lepton pairs
in Drell-Yan events. The peak of this distribution is extremely
sensitive to infrared effects. In \FigRef[a]{fig:primkt}, we
compare the distribution measured by the CDF
experiment \cite{Affolder:1999jh} 
to  a Monte Carlo model (\herwigpp) with four different primordial-\kT
settings: $0$ GeV (off), 
          $1.9$ GeV (the default in \herwigpp), 
          $3.8$ GeV (twice the default), 
          and the IR-augemented shower
          model \cite{Gieseke:2007ad}. 
To illustrate how these predictions scale with  collider centre-of-mass energy,
keeping the $Q^2$ of the hard interaction fixed, we also include a
plot showing the \pt\ of Drell-Yan pairs in $pp$ collisions at 14 TeV
in \FigRef[b]{fig:primkt}; the distributions becomes broader, but the
peak position stays relatively constant. A comparison of different
generators on this distribution can be found
in \FigRef{fig:cmp:intrinsic-kt} in the comparisons section of the
review (\SecRef{sec:physics-areas-where}). 

It is also worth noting that, depending on the model, details of how
the transverse momenta generated by the initial-state parton shower
and the primordial component are combined, the latter can also have a
significant effect well above the peak region. In the more primitive
models, there is also the (probably artificial) possibility of a
``double-peak'' structure emerging at high energies, with a
higher-\pt\ peak generated by the perturbative shower and a low-\pt\
one by primordial \kT. 

As mentioned above, it is important to consider also
complementary distributions, involving different scales or $x$
values, to fully constrain this ambiguous component of Monte Carlo
models. Good examples here would be the Drell-Yan process at different $Q^2$
values or at different rapidities. A systematic comparison to
extractions in DIS could also be fruitful. 

\mcsubsection{Soft QCD processes\label{sec:mbtypes}}

\paragraph{Elastic and inelastic} Elastic scattering 
consists of  all reactions of the type 
\begin{equation}
A(p_A)B(p_B)\to A(p_A')B(p_B')~,
\end{equation}
where $A$ and $B$ are particles
carrying momenta $p_A$ and $p_B$, respectively. Specifically, 
the only exchanged quantity is momentum; all quantum numbers and
masses remain unaltered, and no new particles are produced. 
Inelastic scattering covers everything else, \ie  
\begin{equation}
 AB\to X \ne AB~,
\end{equation} 
where $X\ne AB$ signifies that one 
or more quantum numbers are changed, or more particles are
produced. The distinction between elastic and inelastic scattering is
physically observable and is therefore quantum mechanically
meaningful (see \SecRef{sec:phys-phil-behind} for further discussion
of this point). 
Thus, we divide the total hadron-hadron cross
section into two physically distinguishable components, 
\begin{equation}
\sigma_{\mathrm{tot}}(s) = 
\sigma_{\mathrm{el}}(s) +
\sigma_{\mathrm{inel}}(s)~, 
\end{equation}
where $s=(p_A+p_B)^2$ is the beam-beam centre-of-mass energy squared. 

\paragraph{Diffractive and non-diffractive}
If $A$ or $B$ are not elementary the inelastic final states may be
further divided into ``diffractive'' and ``non-diffractive''
topologies. This is a qualitative classification, usually based on
whether the final state looks like
the decay of an excitation of the beam particles 
(diffractive), or not (non-diffractive), or upon the presence of a
large rapidity gap somewhere in the final state which would separate such excitations.
There  are two schools of thought on how to specify this distinction more precisely: 
\begin{enumerate}
\item Use a theoretical model, whose 
different physics subprocesses can each be uniquely assigned as 
diffractive or non-diffractive. 
However, different models produce different final-state spectra,
and hence such a classification necessarily depends on
the model used to make it. Furthermore, if the 
model allows for events of both diffractive and non-diffractive origin
to populate the same phase space points, the
interference terms between them have no unique assignments and hence 
the classification cannot be made quantum mechanically meaningful, see
\SecRef{sec:phys-phil-behind} for a more general 
discussion of this issue. 
\item
Use one or more physical observables, which
guarantees that the definition will also be valid at the quantum level. 
In this case, the arbitrariness is instead reflected
in the fact that one has to choose what one means by a ``diffractive
topology'', at the level of a final-state observable, and this 
choice is without a unique ``correct'' answer. In general, one defines
diffractive topologies as 
events that contain large rapidity gaps in the activity,
consistent with (possibly multiple) decays of excited states, 
with ``large'' often taken to be 
somewhere in the range of 3--5 units of rapidity.
\end{enumerate}

\paragraph{Types of diffraction}
Given that an event has been labelled as diffractive either by a
theoretical model or by a final-state observable, we may distinguish
between three different classes of diffractive topologies, which it is
possible to distinguish between physically, at least in principle. 
In double-diffractive dissociation (DD) events, both of the beam particles are
diffractively excited and hence neither of them survive the collision
intact. In single-diffractive dissociation (SD) events, only one of the beam
particles gets excited and the other survives intact. The last 
diffractive topology is  central diffraction (CD),  in which 
both of the beam particles survive intact, leaving an excited system
in the central region between them\footnote{This latter topology also includes
so-called ``central exclusive production''~\cite{Khoze:2001xm}.}.
That is, 
\begin{equation}
\sigma_{\mathrm{inel}}(s) = 
\sigma_{\mathrm{SD} }(s)
+
\sigma_{\mathrm{DD}}(s) +
\sigma_{\mathrm{CD}}(s) + 
\sigma_{\mathrm{ND}}(s) ~, \label{eq:diff}
\end{equation}
where ``ND'' (non-diffractive, here understood not to include elastic
scattering) contains no gaps in the event
consistent with the chosen definition of diffraction. Further, 
each of the diffractively excited systems in the events labelled SD,
DD, and CD, respectively, may in principle consist of
several subsystems with gaps between them. \EqRef{eq:diff} may 
thus be defined to be exact, within a specific definition of
diffraction, even in the presence of multi-gap events. 
Note, however, that different
theoretical models almost always use different (model-dependent) definitions of
diffraction, and therefore the individual components in one model are
in general not directly comparable to those of another. It is
therefore important that data be presented at the level of physical
observables if unambiguous conclusions are to be drawn from them, see
\SecRef{sec:phys-phil-behind} for a more detailed discussion of this issue.
Monte Carlo models of diffraction will be discussed briefly in
\SecRef{s_ue:pomerons} below.

\paragraph{Minimum bias and soft inclusive physics}
The term ``minimum bias'' is an experimental term, used to define a
certain class of events that are selected with the minimum
possible selection bias, to ensure they are as inclusive as possible. This will
be discussed in more detail in \SecRef{sec:phys-phil-behind}.
In theoretical contexts 
the term ``minimum bias'' is often used with a slightly different
meaning: to denote specific (classes of) inclusive soft QCD
subprocesses in a given model. 
Since these two usages are not exactly identical, in this review we have chosen to 
reserve the term ``minimum bias'' to pertain strictly to
definitions of experimental measurements, and instead use  
the term ``soft inclusive physics'' as a generic descriptor for the
class of processes which generally dominate the various experimental
minimum bias measurements in theoretical models.

\begin{figure}[tp]
  \centering
  \subfigure[Transverse $N_\text{ch}$ at 900~GeV]{\label{fig:cmp:mpi-ue-atlas900-nch}\includegraphics[scale=0.7]{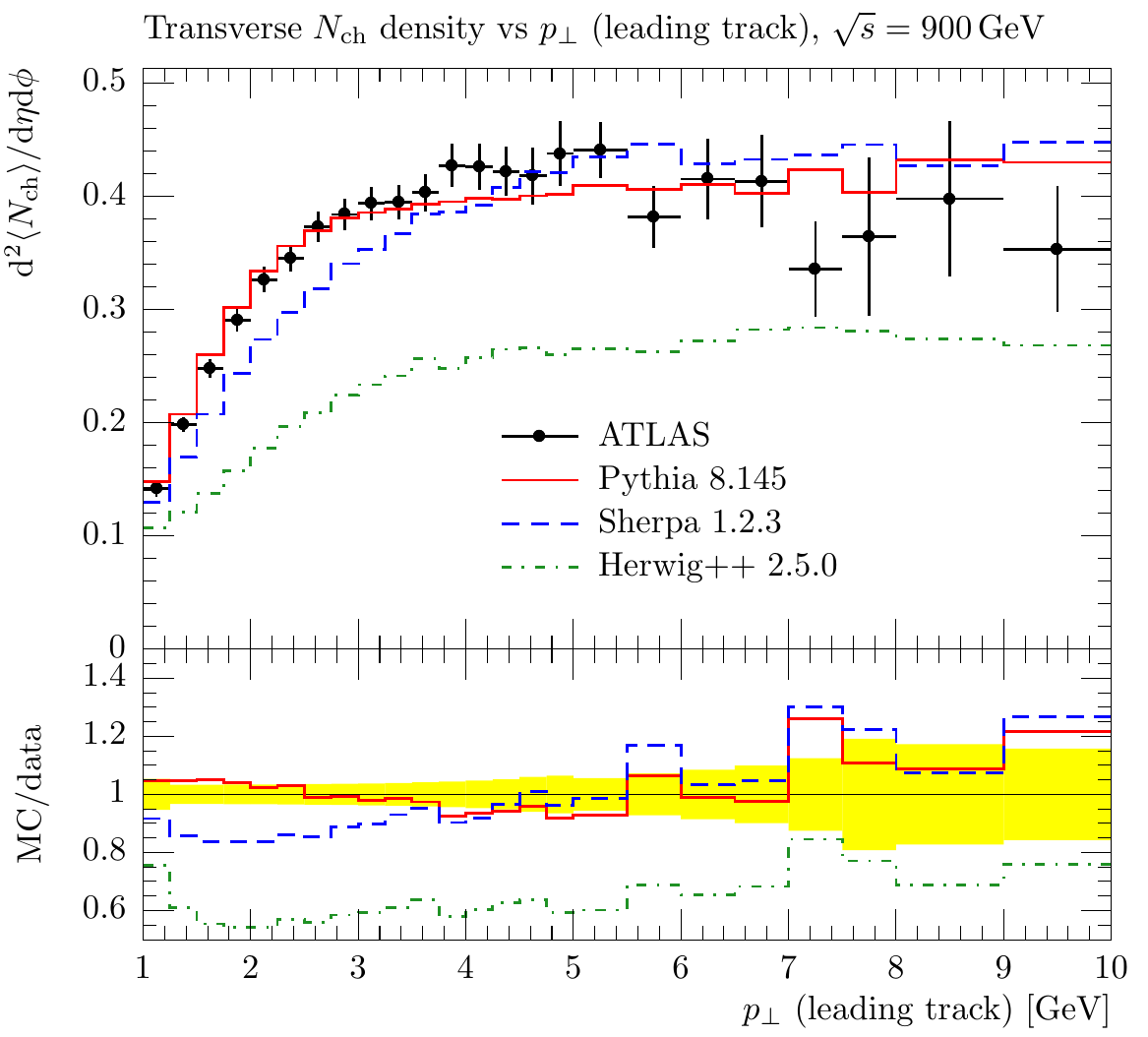}}
  \subfigure[Transverse $N_\text{ch}$ at 7~TeV]{\label{fig:cmp:mpi-ue-atlas7000-nch}\includegraphics[scale=0.7]{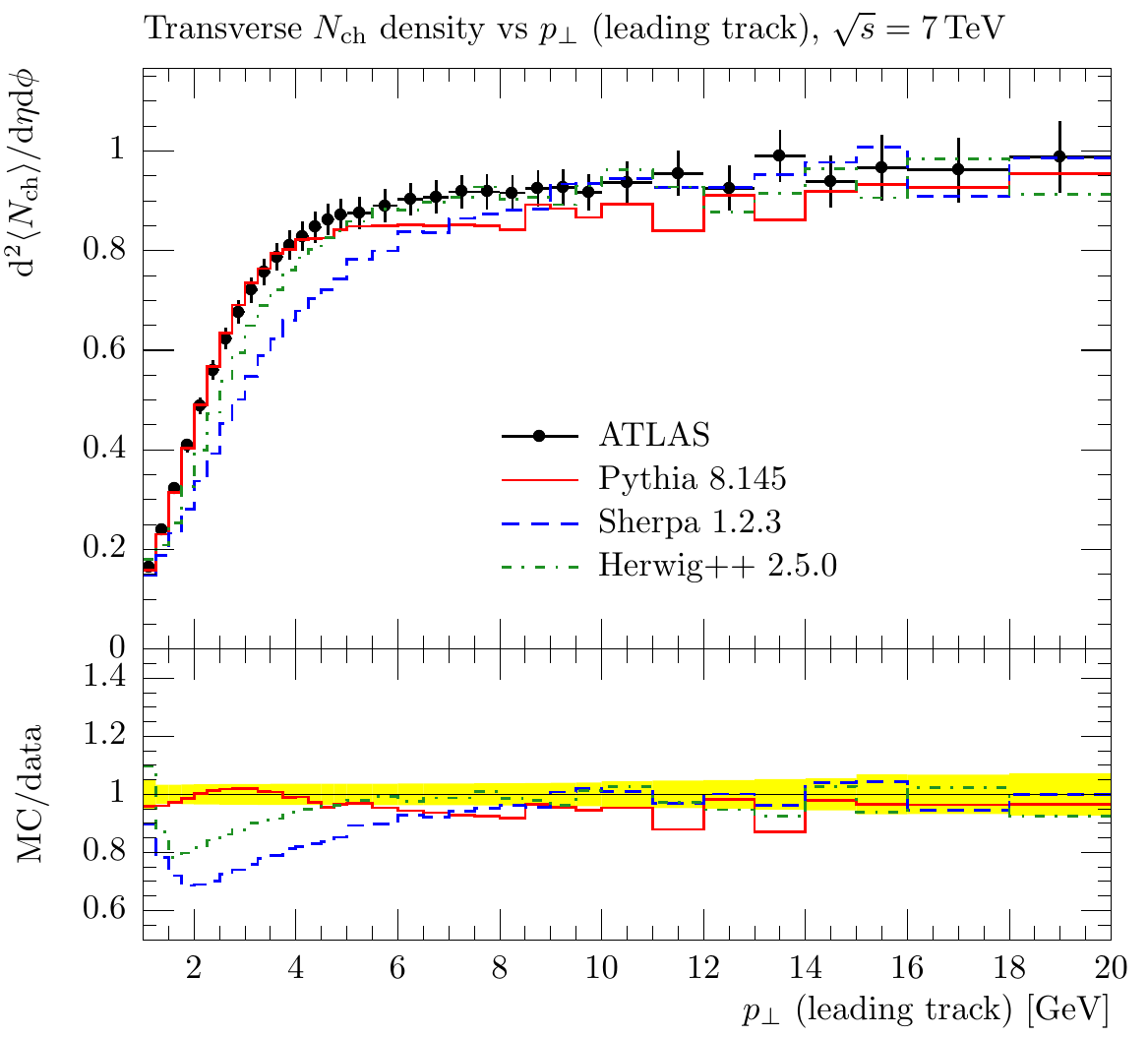}}
  \caption{ATLAS 900~GeV and 7~TeV underlying event observables, showing
    the dependence of MPI activity on the \pT of the leading charged
    particle in the event, with a charged particle \pT cut of
    $\pT > 500~\text{MeV}$, $|\eta| < 2.5$, $c\tau > 10\,\text{mm}$
    \cite{Aad:2010fh}. The MC description of
    these observables is dominated by the tuning of the MPI models, as
    described in \SecRef{sec:minim-bias-underly}. Up-to-date versions of
    these plots can be found at \url{http://mcplots.cern.ch/}.}
  \label{fig:cmp:mpi-ue-atlas-1}
\end{figure}

\begin{figure}[tp]
  \centering
  \subfigure[Transverse $\pT^\text{sum}$ at 900~GeV]{\label{fig:cmp:mpi-ue-atlas900-ptsum}\includegraphics[scale=0.7]{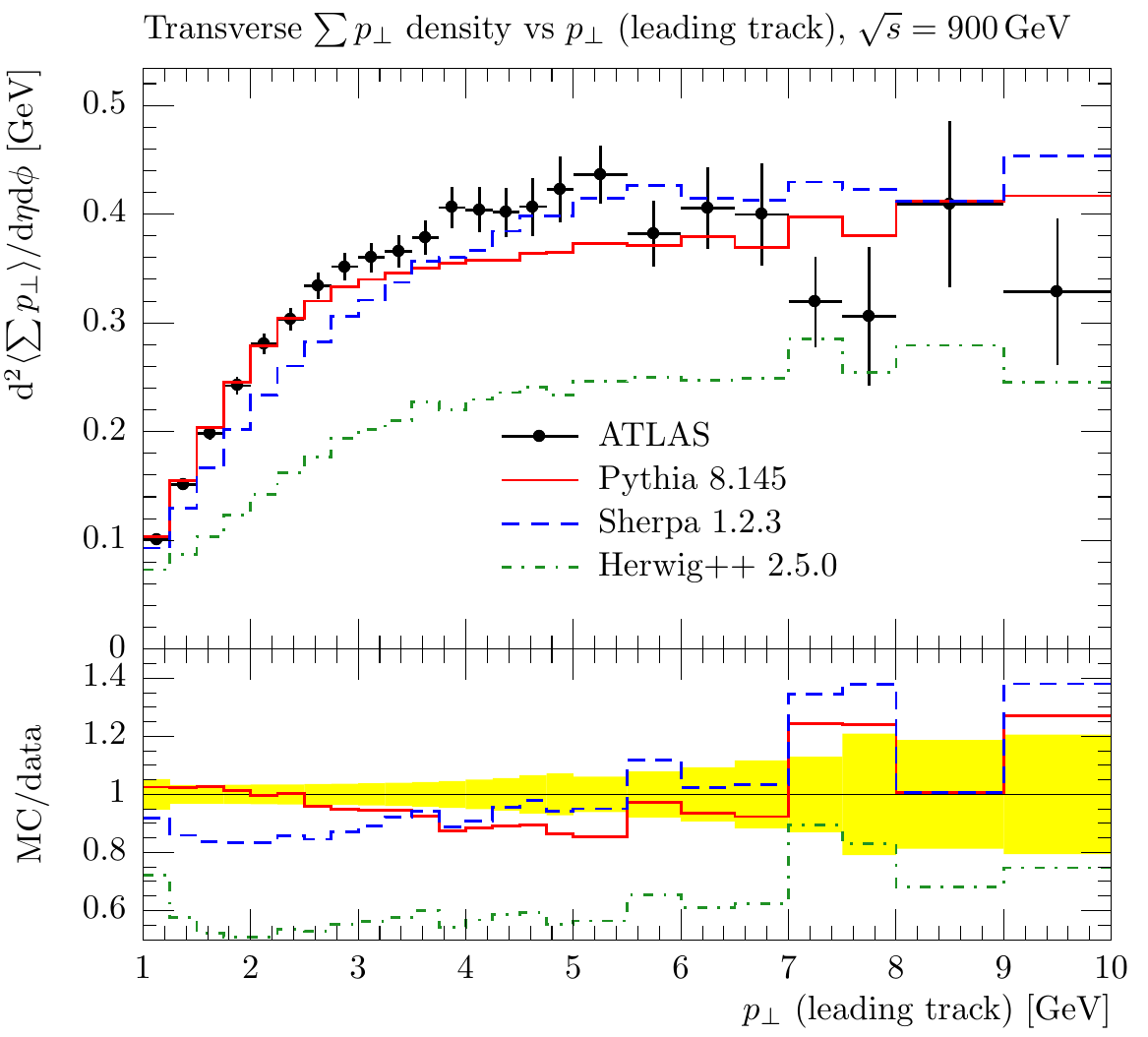}}
  \subfigure[Transverse $\pT^\text{sum}$ at 7~TeV]{\label{fig:cmp:mpi-ue-atlas7000-ptsum}\includegraphics[scale=0.7]{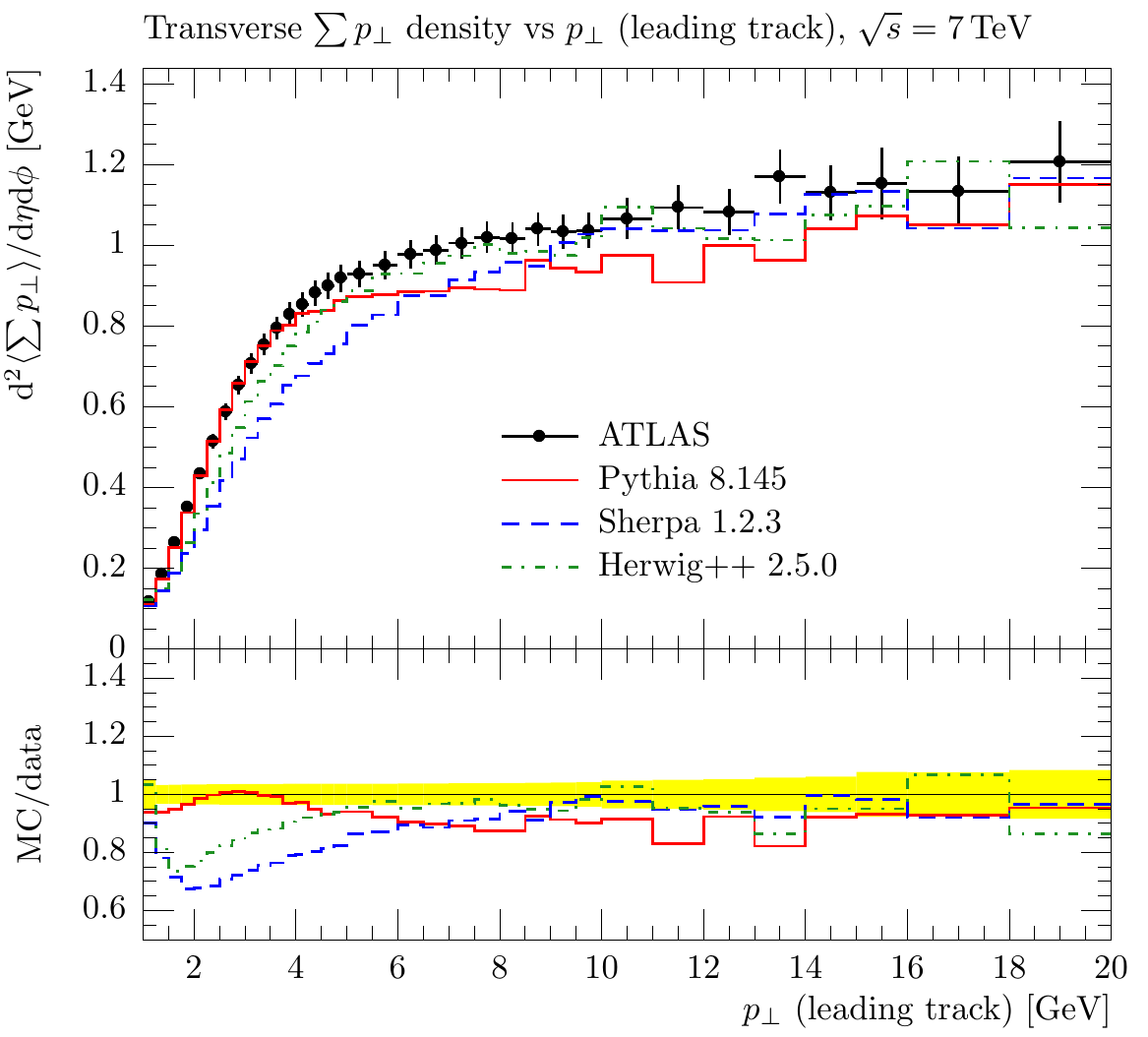}}
  \caption{ATLAS 900~GeV and 7~TeV underlying event observables, showing
    the dependence of MPI activity on the \pT of the leading charged
    particle in the event, with a charged particle \pT cut of
    $\pT > 500~\text{MeV}$, $|\eta| < 2.5$, $c\tau > 10\,\text{mm}$
    \cite{Aad:2010fh}. The MC description of
    these observables is dominated by the tuning of the MPI models, as
    described in \SecRef{sec:minim-bias-underly}. Up-to-date versions of
    these plots can be found at \url{http://mcplots.cern.ch/}.}
  \label{fig:cmp:mpi-ue-atlas-2}
\end{figure}

\begin{figure}[tp]
  \centering
  \subfigure[Transverse $\langle \pT \rangle$ vs. $N_\text{ch}$ at 900~GeV]{\label{fig:cmp:mpi-ue-atlas900-ptnch}\includegraphics[scale=0.7]{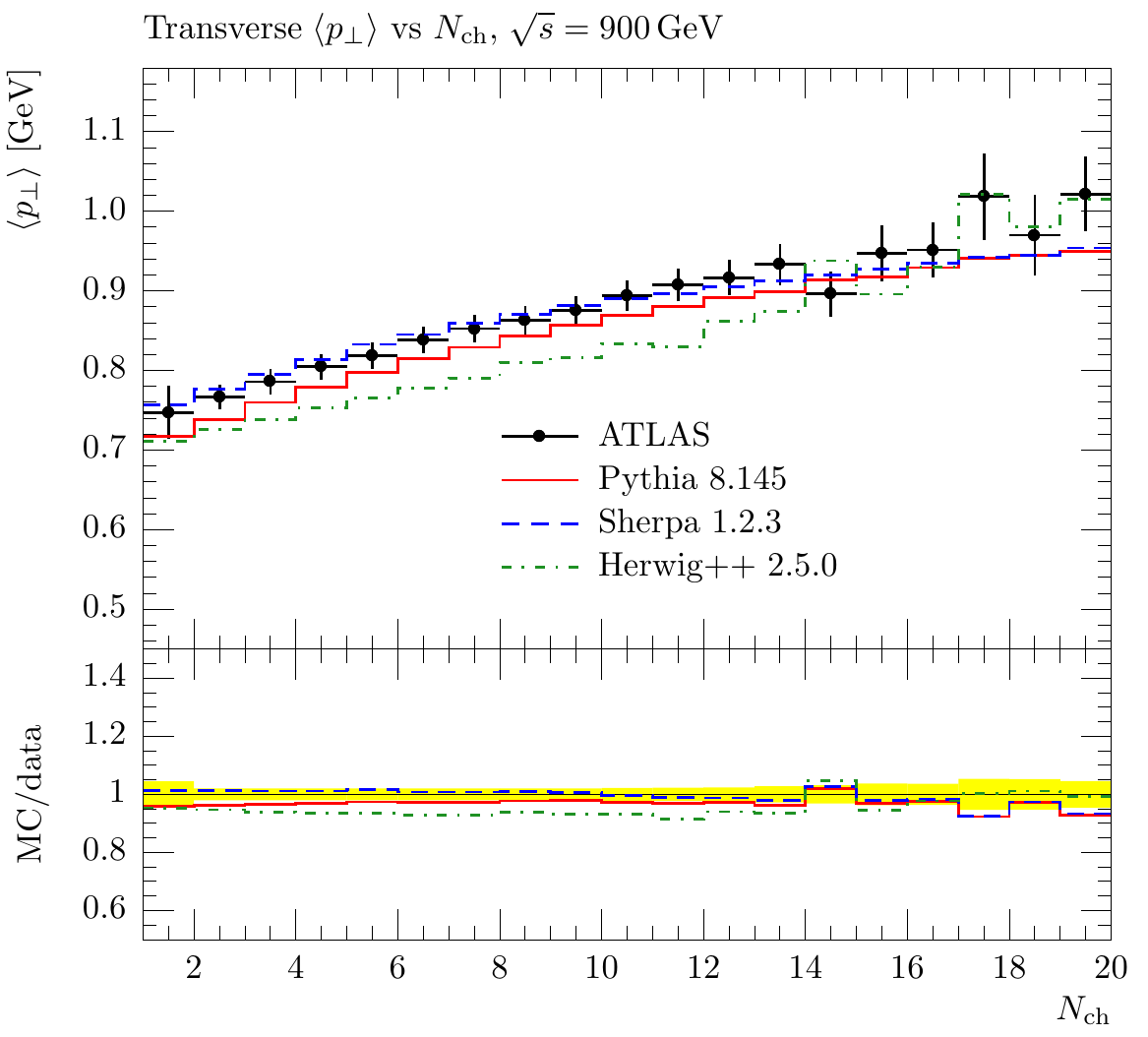}}
  \subfigure[Transverse $\langle \pT \rangle$ vs. $N_\text{ch}$ at 7~TeV]{\label{fig:cmp:mpi-ue-atlas7000-ptnch}\includegraphics[scale=0.7]{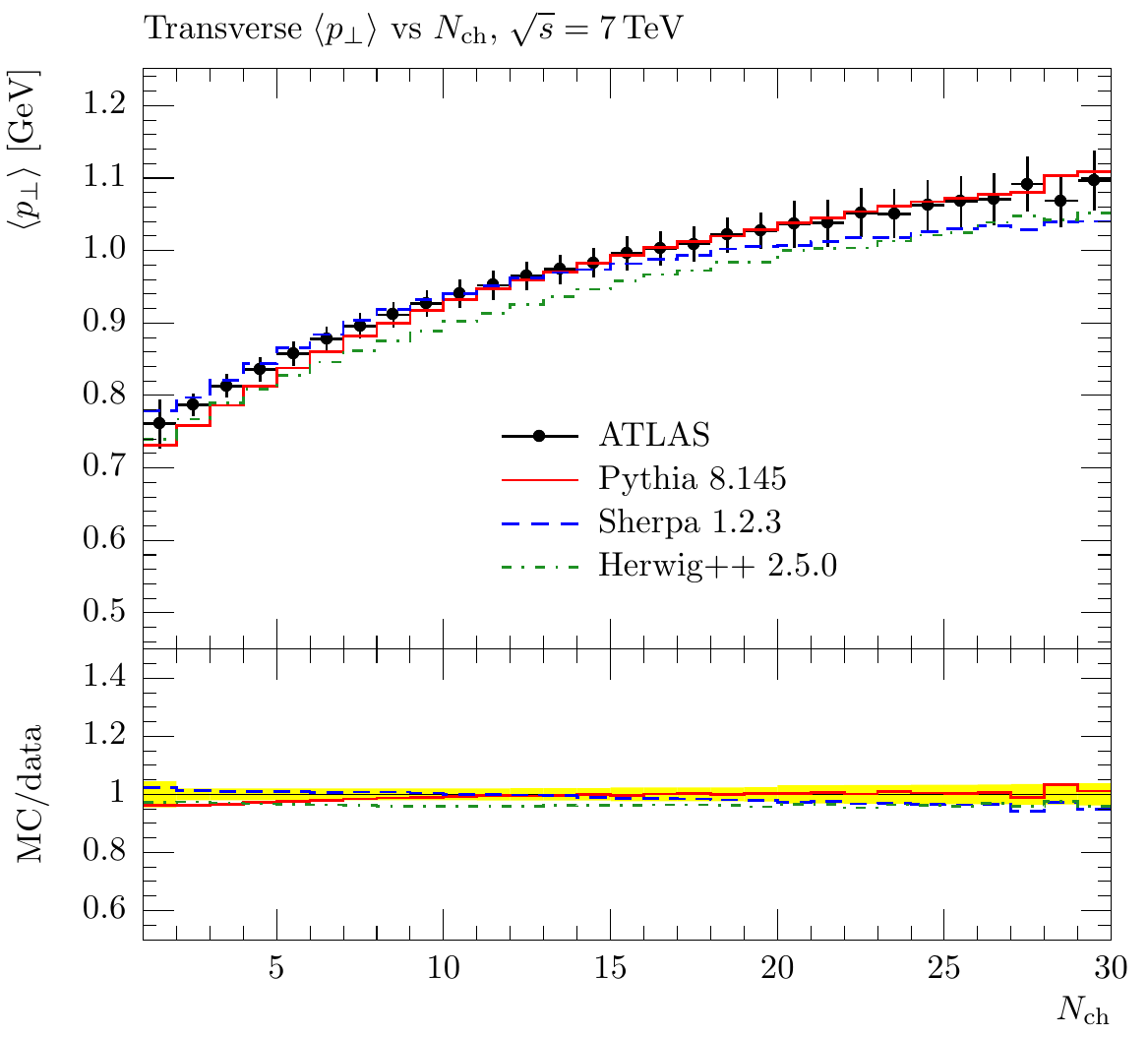}}
  \caption{ATLAS 900~GeV and 7~TeV underlying event $\langle \pT \rangle$
    vs. $N_\text{ch}$ correlation in the region transverse to the
    leading charged particle, with a charged particle \pT cut of
    $\pT > 500~\text{MeV}$, $|\eta| < 2.5$, $c\tau > 10\,\text{mm}$
    \cite{Aad:2010fh}. The MC description of
    these observables is dominated by the tuning of the MPI models, as
    described in \SecRef{sec:minim-bias-underly}. Up-to-date versions of
    these plots can be found at \url{http://mcplots.cern.ch/}.}
  \label{fig:cmp:mpi-ue-atlas-3}
\end{figure}

\paragraph{Underlying event and jet pedestals} 
In events containing a hard parton-parton interaction, 
the underlying event represents the additional
activity which is not directly associated with that interaction. 
There is some ambiguity in how one defines what is
``associated'' with the hard interaction, and what is not. Here, we
shall define the underlying event to represent the additional
activity \emph{after} all bremsstrahlung off the hard interaction
has already been taken into account. Specifically, initial-state
radiation off the hard interaction is \emph{not} included in our
definition of the underlying event. Note also that the underlying
event is usually much more active, with larger fluctuations, than
 soft-inclusive collisions at the same energy. 
This is called the ``jet pedestal'' effect (hard jets sit on top of a
higher-than-average ``pedestal'' of underlying activity), and is
interpreted as follows. When two hadrons collide at non-zero impact
parameter, high-$p_\perp$ interactions can only take place  
inside the overlapping region. 
Imposing a hard selection cut therefore statistically
biases the event sample toward more central collisions, which will also
have more underlying activity. The size of the pedestal, as a function
of leading track \pt, is illustrated in
\FigsRef{fig:cmp:mpi-ue-atlas-1}--\ref{fig:cmp:mpi-ue-atlas-3} in the
comparisons section of the review (\SecRef{sec:physics-areas-where}). 

\paragraph{Multiple interactions}
In a hadron-hadron collision more than one pair of partons may
interact, leading to the possibility of multiple interactions.
In Monte Carlo modelling contexts, the most striking and easily
identifiable consequence of multiple interactions is arguably
the possibility of observing several hard parton-parton 
interactions in one and the same hadron-hadron event\footnote{
Additional jet pairs produced in this way are sometimes
referred to as ``minijets'', and theoretically belong to a class of
perturbative  corrections called ``higher twist'', but in the interest
of maintaining a compact terminology, we shall here just call them MPI
jets.}. The main distinguishing feature of such jets is that they tend to form 
back-to-back pairs, with little total \pt. For comparison, jets from 
bremsstrahlung tend to be aligned with the direction of their
``parent'' partons. The fraction of multiple interactions that give
rise to additional reconstructible jets is, however, quite small (how
small depends on the exact jet definition used). 
Additional soft interactions, below the
jet cutoff, are much more plentiful, and can give significant
corrections to the colour flow and total scattered energy 
of the event. 
This affects the final-state activity in a more global way, increasing the
multiplicity and summed transverse energy, and contributing to the
break-up of the beam remnant in the forward direction. 

\begin{figure}[t]
\center
\includegraphics*[scale=0.75]{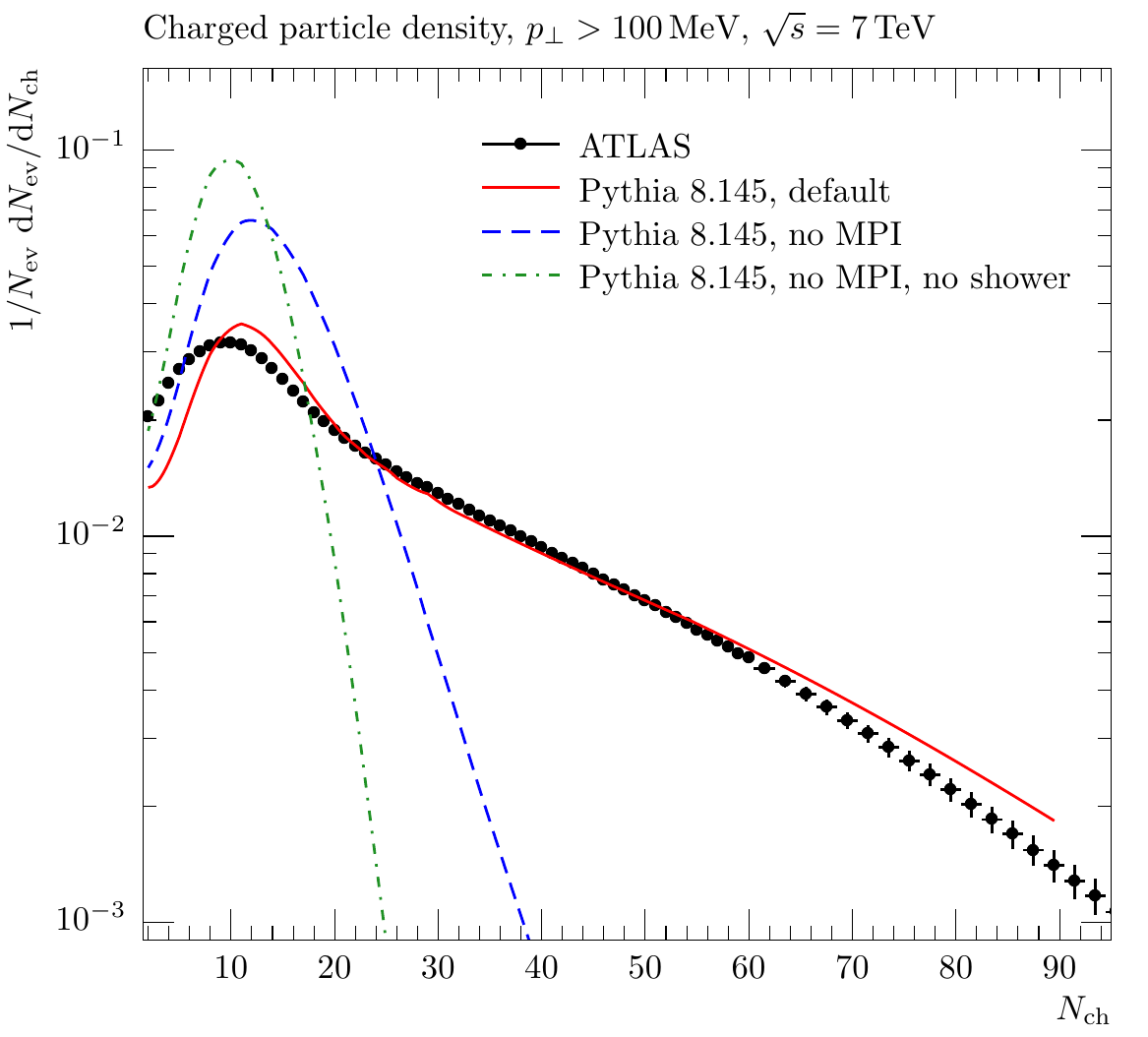}
\caption{Models with and without MPI and parton showers, compared to
the charged-particle multiplicity measured by the ATLAS
experiment \cite{Atlas:2010xx}, 
% The non-preliminary version: 
%experiment \cite{Collaboration:2010i}, 
for particles with $\pt>100$~MeV, $|\eta|<2.5$, and $c\tau > 10$~mm,
in events that contain at least two such particles. \label{fig:mbnch}}
\end{figure}
To illustrate this we include in \FigRef{fig:mbnch} a comparison
between an ATLAS minimum bias measurement of the charged-track
multiplicity at 7~TeV to a Monte Carlo model with and without MPI
switched on (curves labelled as ``default'' and ``no MPI'',
respectively). Clearly, the predicted multiplicity distribution
without MPI is far too narrow, regardless of whether parton showers
are included or not (curve labelled ``no MPI, no shower''). This by
itself is one of the strongest arguments that MPI must be included in
realistic models of soft-inclusive physics.

The possibility of multiple interactions has also been implicit
or explicit in many calculations of the total hadron-hadron cross
section.  Two recent and representative examples can be found in  
 \cite{Avsar:2008dn,Grau:2009qx}.  
The increase of the parton-parton cross section with CM energy is here
directly driving an increase also of $\sigma_{\mathrm{tot}}(s)$.   

The first detailed Monte Carlo model for 
perturbative MPI was proposed by Sj\"ostrand and van Zijl in 
\cite{Sjostrand:1987su}, and most modern implementations employ a
similar physical picture. Below, in section \SecRef{sec:mbmpi}, 
we therefore first summarize the main points of this basic framework,  
pointing out the differences between the currently existing models 
as we go along. 
Some useful additional references to the history and development of
 the subject of  multiple interactions also outside the Monte Carlo context 
can be found in the Perugia MPI workshop
 proceedings \cite{Bartalini:2010su} and in the mini-reviews contained
 in \cite{Sjostrand:2004pf,Gustafson:2007sb}.

\mcsubsection{Models based on multiple parton interactions (MPI) \label{sec:mbmpi}}
\mcsubsubsection{Basics of MPI \label{sec:mpibasics}}

Consider first the cross section for a \emph{single} parton-parton
scattering, \eg by $t$-channel gluon exchange (Rutherford scattering).
This process, and simple variations of it, make up the vast
majority of the total scattering
processes occurring between coloured particles, and it is thus on
this basic process that perturbative models of both soft inclusive and
underlying event physics are currently built.

An intuitive way of arriving at the idea of multiple interactions
is to view hadrons simply as ``bunches'' of incoming
partons. No physical law then  prevents several distinct pairs of partons
from undergoing scattering processes within one and the same hadron-hadron
collision.  The other key idea to bear in mind is that the exchanged QCD
particles are coloured, and hence such multiple interactions, even when soft, can cause non-trivial changes to the
colour topology of the colliding system as a whole, with potentially
major consequences for the particle multiplicity in the final state.

In the soft QCD region, the $t$-channel gluon propagator almost goes
on shell (reminiscent of the case of bremsstrahlung, described in detail in
\SecRef{sec:parton-showers}), causing the subprocess differential cross
section to become very large, behaving roughly as:
\begin{equation}
\mr{d}\hat{\sigma}_{2j} \propto \
\frac{\mr{d}t }{t^2} \ \sim \
 \frac{\drm\pPerp{}^2}{\pPerp{}^4}~, \label{eq:dpt4}
\end{equation}

\begin{figure}
\begin{center}
\vspace*{-30mm}\includegraphics*[scale=0.35]{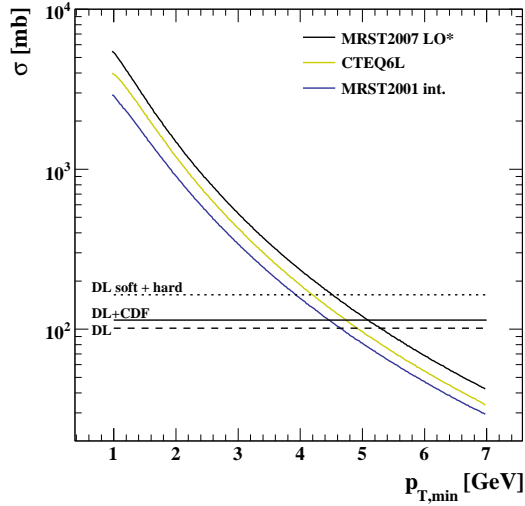}
\caption{The inclusive jet cross section calculated at LO for three
different proton PDFs, compared to various extrapolations
of the non-perturbative fits to the total pp
cross section at 14~TeV centre-of-mass
energy. From \cite{Bahr:2008wk}. \label{fig:sigma2to2}}
\end{center}
\end{figure}
An integration of this cross section from a lower cutoff
$\pPerp{\mr{min}}$ to $\sqrt{s}$, \
using the full (leading-order) QCD $2\to 2$ matrix elements
folded with some recent parton-density sets, is
shown in \FigRef{fig:sigma2to2}, for $pp$ collisions at 14~TeV
\cite{Bahr:2008wk}.
The solid curves, representing the calculated
cross sections as functions of $\pPerp{\mr{min}}$, are
compared to a few of the Donnachie-Landshoff (DL)
predictions \cite{Donnachie:1992ny,Donnachie:2004pi}
for the total $pp$ cross section $\sigma_{\mr{tot}}$,
shown as horizontal lines with  different dashing styles on the same
plot. Physically, the jet cross section can of course not
exceed the total $pp$ one, yet this is what appears to be happening
at scales of order 4--5~GeV in \FigRef{fig:sigma2to2}.
How to interpret this behaviour?

Recall that the interaction cross section is an inclusive
number. Thus, an event with two parton-parton interactions will count
twice in $\sigma_{2j}$ but only once in $\sigma_{\mr{tot}}$,
and so on for higher multiplicities. In the limit that all the
individual parton-parton interactions are independent and equivalent
(to be improved on below), we have
\begin{equation}
\sigma_{2j}(\pPerp{\mr{min}}) = \langle n \rangle(\pPerp{\mr{min}})
\ \sigma_{\mr{tot}} ~,
\end{equation}
with $ \langle n \rangle(\pPerp{\mr{min}})$ giving the
average of the distribution in the number of parton-parton interactions
above $\pPerp{\mr{min}}$ per hadron-hadron collision,
and that number may well be above unity.
This simple argument in fact expresses unitarity;
instead of the total interaction cross section diverging as $\pPerp{\mr{min}}\to 0$ (which would violate
unitarity), we have restated the problem so
that it is now the \emph{number of interactions per collision} that
diverges.

Two important ingredients remain to be introduced in order to fully
regulate the remaining divergence. The first is the correlation due to
energy-momentum conservation --- the interactions cannot use up more
momentum than is available in the parent hadron ---
which will suppress the large-$n$ tail of the na\"ive estimate above.
This is handled slightly differently in the various models on the market.
In the \pythia and \Sherpa models, the
multiple interactions are ordered in $\pPerp{}$, and the parton
distributions for each successive
interaction are explicitly constructed so
that the sum of $x$ fractions can never be greater than unity. In
the \herwigpp model, instead the uncorrelated estimate of
$\langle n \rangle$ above is used directly as an initial guess, but the actual
generation of interactions stop once the energy-momentum conservation
limit is exceeded (with the last ``offending'' interaction also
removed from consideration).

Even with this suppression taken into account, however, the number of
multiple interactions still grows uncomfortably fast as $\pPerp{\mr{min}}\to
0$. A second ingredient suppressing the number of interactions,
at low-\pPerp{} and $x$, is colour screening /
saturation. Screening and saturation both roughly correspond to
partons being unable to resolve each other as independent particles
at low scales, but the underlying physics pictures are slightly
different, as follows.

Screening is interpreted as an effect
of the wavelength $\sim$ $1/\pt$ of the exchanged particle becoming
larger than a typical colour-anticolour separation distance; it will
then only couple to an average colour charge that vanishes in the limit
$p_{\perp} \to 0$, hence leading to suppressed interactions.
This screening effectively provides an infrared cutoff for
MPI similar to that provided by the hadronization
scale for parton showers.
Saturation instead invokes explicit parton
recombination effects to reduce the growth of the parton
densities at low $x$. In either case, the product of cross section and parton densities is reduced. However, an
important modelling distinction is therefore that the reduction takes place
mainly as a function of \pt, for screening, whereas the $x$ variable
plays a central role in saturation arguments. As usual, the truth is
likely to be a combination of both. Since most of the models considered in
this review employ a screening-like cutoff, in \pt, we devote
a bit more space to this possibility here.

A first estimate of an effective lower cutoff due to colour screening would be
the proton size
\begin{equation}
\pPerp{\mr{min}} \simeq \frac{\hbar}{r_{p}} \approx
\frac{0.2~\mr{GeV}\cdot\mr{fm}}{0.7~\mr{fm}} \approx
0.3~\mr{GeV} \simeq \LambdaQCD\,, \label{eq:cutoff}
\end{equation}
but empirically this appears to be too low. In current models, one
replaces the proton radius $r_{p}$ in the above formula by a ``typical
colour screening distance'' $d$, \ie an average size of a region within
which the net compensation of a given colour
charge occurs. This number is not known from first principles, so
effectively this is simply a cutoff parameter, which can then just as
well be put in transverse momentum space.
The simplest choice is to introduce a step function
$\Theta(\pPerp{} - \pPerp{\mr{min}})$, such that the perturbative cross section
completely vanishes below the $\pPerp{\mr{min}}$ scale.
This is the procedure followed in the original \jimmy 
model \cite{Butterworth:1996zw} (an add-on to the Fortran
\herwig generator), whose predictions therefore have
a large dependence on the value of this parameter. The
\herwigpp model \cite{Bahr:2008dy} builds on this, and
improves it by adding a set of
``soft'' scatterings below $\pPerp{\mr{min}}$, with a $\pPerp{\mr{min}}$-dependent
cross section defined such that the two components add up to the
total (inelastic non-diffractive)
cross section \cite{Bahr:2008wk}, which should reduce the explicit
dependence on \pPerp{\mr{min}}.

Alternatively, one may note that the jet cross section is divergent
like $\alphaS^2(\pPerp{}^2)/\pPerp{}^4$, cf.~\EqRef{eq:dpt4},
and that therefore a factor
\begin{equation}
\frac{\alphaS^2(\pPerp{0}^2 + \pPerp{}^2)}{\alphaS^2(\pPerp{}^2)} \,
\frac{\pPerp{}^4}{(\pPerp{0}^2 + \pPerp{}^2)^2}
\label{eq:ptzerodampen}
\end{equation}
would smoothly regularize the divergences, now with $\pPerp{0}$ as the
free parameter to be tuned to data. This is the default in
the current \pythia and \Sherpa models. Note that,
since this merely represents a ``smoothed-out''
variant of the $\Theta$ function cutoff above, there is still a strong
dependence on the value of $\pPerp{0}$. This is thus one of the main ``tuning''
parameters in such models.

The parameters do not have to be energy-independent, however.
Higher energies imply that parton densities can be probed at smaller
$x$ values, where the number of partons rapidly increases. Partons
then become closer packed and the colour screening distance $d$
decreases. Just as the small-$x$ rise varies like some power of $x$
one could therefore expect the energy dependence of $\pPerp{\mr{min}}$ and
$\pPerp{0}$ to vary like some power of the centre-of-mass energy. Explicit toy simulations
\cite{Dischler:2000pk} lend some credence to such an ansatz, although with
large uncertainties. One could also let the cutoff increase
with decreasing $x$; this would lead to a similar phenomenology since
larger energies probe smaller $x$ values. Especially for models with
strong dependence on this cutoff, the uncertainty on this
energy and $x$ scaling of the cutoff is a major concern when
extrapolating between different collider energies.

As an alternative we therefore
note that the introduction of so-called unintegrated parton
densities, as used in the BFKL \cite{Kuraev:1977fs,Balitsky:1978ic}, CCFM
\cite{Ciafaloni:1987ur,Catani:1989sg}, and LDC
\cite{Andersson:1995ju,Andersson:1997bx,Kharraziha:1997dn}
approaches to initial-state radiation allows the possibility of
replacing the $\pPerp{\mr{min}}$ or $\pPerp{0}$ cutoff by parton densities that
explicitly vanish in the $\pPerp{} \to 0$ limit
\cite{Gustafson:2002jy}. This allows the possibility of an
alternative implementation of multiple interactions
\cite{Gustafson:2002kz} and may be a useful ingredient in future
phenomenological models, possibly in combination with more explicit
physical modelling of saturation effects.

\mcsubsubsection{Impact parameter dependence}
As mentioned in \SecRef{sec:mbtypes}, the so-called ``pedestal
effect'' (see also
\FigsRef{fig:cmp:mpi-ue-atlas-1}--\ref{fig:cmp:mpi-ue-atlas-3})
is partly driven by impact parameter dependence; in peripheral
collisions, only a small fraction of events contain any
high-\pt\ activity, whereas central collisions are more
likely to contain at least one hard scattering; a sample with a
high-\pt\ selection cut will therefore be biased towards small impact parameters.
The ability of a model to describe the shape of the pedestal (\eg to
describe both minimum bias data and underlying event distributions
simultaneously) is therefore related to its modelling of the
impact parameter dependence.
A related effect is that, also for a fixed selected \pt, events at
comparatively higher impact parameters should exhibit relatively less underlying
event and vice versa.

All the models discussed here contain an explicit treatment of
impact parameter, but we note that there are still
substantial simplifications made. Most importantly, the
impact parameter dependence is so far still assumed to be factorized from
the $x$ dependence, $\pdf{}(x,b)=\pdf{}(x)g(b)$, where $b$ denotes impact
parameter, a simplifying assumption that by no means should be treated
as inviolate, see \eg
\cite{Hagler:2007xi,Treleani:2007gi,Blok:2010ge}. Also, the
hadron-hadron impact parameter only enters in an averaged global
sense, not as a vector, and the individual MPI are not assigned
individual ``locations'' in transverse space.

In order to quantify the concept of hadronic matter overlap, one may
assume a spherically symmetric distribution of matter inside a
hadron at rest, $\rho(\mb{x}) \, \drm^3 x = \rho(r) \, \drm^3 x$.
The form of $\rho$ is a matter of some uncertainty, with various
more or less phenomenologically motivated choices available in
models. The options range from simple parameteric forms in
\pythia-based models, such as Gaussians, double Gaussians,
exponentials \cite{Sjostrand:1987su}, and forms interpolating between
them  \cite{Sjostrand:2004pf}, to a form based on the  electromagnetic
form factor in the \herwig-based ones
\cite{Forshaw:1991gd}. A possibility for future model refinements
thus lies in the input of more detailed information on the flavour-
or $x$-dependence of the transverse structure of the proton, \eg
obtained from sum rules, from analytic fits beyond the EM form factor,
or from lattice studies.

For a collision with impact parameter $b$, the time-integrated
overlap $\mathcal{O}(b)$ between the matter distributions of the
colliding hadrons is given by
\begin{equation}
\mathcal{O}(b) \propto \int \drm t \int \drm^3 x \, \,
\rho(x,y,z) \, \rho(x+b,y,z+t)     ~,
\end{equation}
where the necessity to use boosted $\rho(\mb{x})$ distributions
has been circumvented by a suitable scale transformation of the $z$
and $t$ coordinates, see \cite{Sjostrand:1987su}. The overlap function
$\mathcal{O}(b)$ is identical
to $A(b)$ in ``\jimmy notation''
\cite{Butterworth:1996zw,Bahr:2008dy}. It is
closely  related to the $\Omega(b)$ of eikonal models
(see, for example, \cite{Bourrely:2002wr,Borozan:2002fk,Treleani:2007gi}),
but is somewhat simpler in spirit.

The larger the overlap $\mathcal{O}(b)$ is, the more likely it is to
have interactions between partons in the two colliding hadrons.
In fact, to first approximation, there should be a linear relationship
\begin{equation}
\langle \tilde{n}(b) \rangle = k \mathcal{O}(b) ~,
\label{eq:bdepend}
\end{equation}
where $\tilde{n} = 0, 1, 2, \ldots$ counts the number of interactions
when two hadrons pass each other with an impact parameter $b$ and
$k$ is an undefined constant of proportionality, to be specified
below.

For each impact parameter, $b$, the number of interactions $\tilde{n}$
can be assumed to be distributed according to a Poissonian,
modulo momentum conservation, with the mean value of the Poisson
distribution depending on impact
parameter, $\langle \tilde{n}(b)\rangle$. If the matter
distribution has a tail to infinity (as, e.g., Gaussians do),
one may nominally obtain events with arbitrarily large $b$ values.
In order to obtain finite total cross sections, it is therefore
necessary to give a separate interpretation to the ``zero bin'' of the
Poisson distribution, which corresponds to ``no-interaction'' events.

In the \jimmy \cite{Butterworth:1996zw} and \herwigpp
\cite{Bahr:2008dy} models the part of the $pp$ cross section
containing hard scatters is calculated from the area
overlap function, the parton densities and the partonic cross section;
the ``no-interaction'' possibility is then accounted for as a
reduction of this cross section with respect to its value without
allowing for MPI. The \jimmy model stops here, considering only hard events,
and so it can only be applied to underlying event. As mentioned above,
the \herwigpp model also permits the possibility of soft scatters
(see also \SecRef{sec:herwigmpi}) and so can also be used to
simulate soft-inclusive physics.

In the framework of \cite{Sjostrand:1987su}, used by \pythia and
\Sherpa, the restriction to at least one perturbative scattering for soft
inclusive scatters implies that
the probability that two hadrons, passing each other
with an impact parameter $b$, will produce a real event is given by
\begin{equation}
\mathcal{P}_{\mr{int}}(b) =
\sum_{\tilde{n} = 1}^{\infty} \mathcal{P}_{\tilde{n}}(b) =
1 - \mathcal{P}_0(b) =
1 - \exp ( - \langle \tilde{n}(b) \rangle )
= 1 - \exp (- k \mathcal{O}(b) ) ~,
\end{equation}
according to Poisson statistics. The average number of
interactions per event at impact parameter $b$ is now
$\langle n(b) \rangle = \langle \tilde{n}(b) \rangle /
\mathcal{P}_{\mr{int}}(b)$, where the denominator comes from the
removal of hadron pairs that pass without interaction, \ie which do
not produce any events. While the removal of $\tilde{n}=0$ from the potential
event sample gives a narrower-than-Poisson interaction distribution
at each fixed $b$, the variation of $\langle n(b) \rangle$ with $b$
gives a $b$-integrated broader-than-Poisson interaction multiplicity
distribution.

Averaged over all $b$ the relationship $\langle n \rangle =
\sigma_{2j}/\sigma_{\mr{nd}}$ should still hold. Here, as before,
$\sigma_{2j}$ is the integrated interaction cross section for a given
regularization prescription at small \pt, while the inelastic
non-diffractive cross section $\sigma_{\mr{nd}}$ is taken from
parameterization \cite{Donnachie:1992ny,Donnachie:2004pi,Schuler:1993wr}.
This relation can be used to solve for the
proportionality factor $k$ in \EqRef{eq:bdepend}. Note that, since
now each event has to have at least one interaction, $\langle n
\rangle > 1$, one must ensure that $\sigma_{2j} >
\sigma_{\mr{nd}}$. The $\pPerp{0}$ parameter has to be chosen
accordingly small.

\mcsubsubsection{Perturbative corrections beyond MPI}
There are essentially two perturbative modelling aspects which go
beyond the introduction of MPI themselves. In particular, this concerns
\begin{enumerate}
\item Parton showers off the MPI.
\item Perturbative parton-rescattering effects.
\end{enumerate}

Without showers, MPI models would generate very sharp peaks for back-to-back
MPI jets, caused by unshowered partons passed directly to the hadronization
model. However, with the exception of the oldest \pythia~6 model
\cite{Sjostrand:1987su}, all of the models discussed in this
review do include such showers, and hence should exhibit more
realistic (\ie  broader and
more decorrelated) MPI jets --- although not much can be said concerning
their expected formal level of precision of course. A secondary effect is that a
showered interaction also generates a larger hadronic
multiplicity than an unshowered one. Therefore, a smaller total number
of MPI is needed when tuning models incorporating such showers.
More discussion of this tuning effect can be found in
\cite{Buttar:2008jx,Skands:2010ak}.
On the initial state side of the MPI shower issue the main questions
are whether and how correlated multi-parton densities
are taken into account (for a recent treatment of this issue see \eg
\cite{Gaunt:2009re} and references therein), and, as discussed
previously, how the showers are regulated at low \pt (or low $x$).
Although none of the Monte Carlo models currently
impose a rigorous correlated multi-parton evolution, all of them
include some elementary aspects. The most significant for
parton-level results is arguably momentum conservation, which is
enforced explicitly in all the models, although in slightly different
ways, as was discussed briefly above. The so-called ``interleaved'' models
\cite{Sjostrand:2004pf,Sjostrand:2004ef} attempt to go a step
further, generating an explicitly correlated multi-parton evolution
in which flavour sum rules can be imposed to conserve \eg
the total numbers of valence and sea quarks across interaction chains.

Perturbative rescattering in the final state can occur if
partons are allowed to undergo several distinct interactions, with
showering activity possibly taking place in between. This has so far
not been studied extensively, but a first
fairly complete model and exploratory study has been presented in the
context of \pythia~8 \cite{Corke:2009tk}.
The net effect there is a slight increase in the mean $p_\perp$ of the
partonic final states, but more dramatic signatures have not yet been
identified. In the initial state, parton rescattering effects may lead
to saturation (discussed briefly in \SecRef{sec:mpibasics}),
or to the incoming partons carrying enhanced
``primordial $k_\perp$'' values (discussed in
\SecRef{sec:primkt}). It could also produce  correlations
between different MPI initiators, in particular in colour
space. At the \emph{parton level}, however, such
colour correlations probably play a rather minor role.
This is suggested both by an exploratory study of ``intertwined''
multiple interactions \cite{Sjostrand:2004ef} (that is,
letting several partons from the same shower chain undergo
perturbative interactions, thus letting the perturbative shower
evolution generate their colour correlations)
which numerically found only very small effects, and by a
more heuristic argument; that the multiple interactions are likely to
be taking place slightly displaced from each other in
space-time. Their ``perturbative cross-talk'' should therefore be
suppressed by wave-function overlap
factors, which mean they should only be able to emit coherently at
rather small $p_\perp$ anyway.
Compared to the usual perturbative subleading-colour ambiguities
associated with parton shower Monte Carlos, this particular source of
colour space ambiguity should therefore not represent any significant
additional ambiguity \emph{at the parton level}. The interleaved
rescattering model of \cite{Corke:2009tk} is currently the only one to
address emissions inside colour dipoles spanned
\emph{between} different MPI subsystems.

\mcsubsubsection{Non-perturbative aspects\label{sec:mbmpi-np}}

Consider a hadron-hadron collision at a
resolution scale of about 1~GeV. The system of coloured partons emerging
from the short-distance phase (primary parton-parton interaction plus
parton-level underlying event plus beam-remnant partons)
must now undergo the transition to colourless hadrons.

In this context, it is useful to consider what happens to
infrared safe, and infrared sensitive, observables separately.
For infrared safe observables, such as
energy flow and jet observables, the parton flow in phase space
already gives quite a good
approximation, and hadronization only gives small corrections. (For
precision studies, these must of course still be taken into account, see \eg
\cite{Bhatti:2005ai,Dasgupta:2007wa}.)

Infrared sensitive observables, on the other hand, such as individual
hadron multiplicities and spectra are crucially dependent on the
parton-parton correlations in colour space, and on the properties and
parameters of the hadronization model used, pedagogical
descriptions of which can be found in \SecRef{sec:hadronization}.
Here, we concentrate on the specific issues connected with the
structure  of the underlying event in hadron collisions.

Keeping the short-distance parts unchanged, the colour structure
\emph{inside} each of the MPI systems is normally still described using
just the ordinary leading-colour matrix-element and parton-shower
machinery described in \SecsRef{sec:subprocesses} and
\ref{sec:parton-showers}. The crucial question, in the context of MPI,
is then how colour is neutralized \emph{between} different MPI
systems, including also the remnants. Since these systems can lie at
very different rapidities (the extreme case being the two opposite
beam remnants), the strings or clusters spanned between them can have
very large invariant masses (though normally low~\pt),
and give rise to large amounts of (soft)
particle production. Indeed, in the context of soft-inclusive physics,
it is precisely these ``inter-system'' clusters/strings which furnish the
dominant particle production mechanism (cf.\ again
\FigRef{fig:mbnch}), and hence their modelling
is an essential part of the infrared physics description.
For more on
the physics of the string and cluster hadronization models, see
\SecRef{sec:hadronization}.

\begin{figure}
\begin{center}
\includegraphics*[scale=0.7]{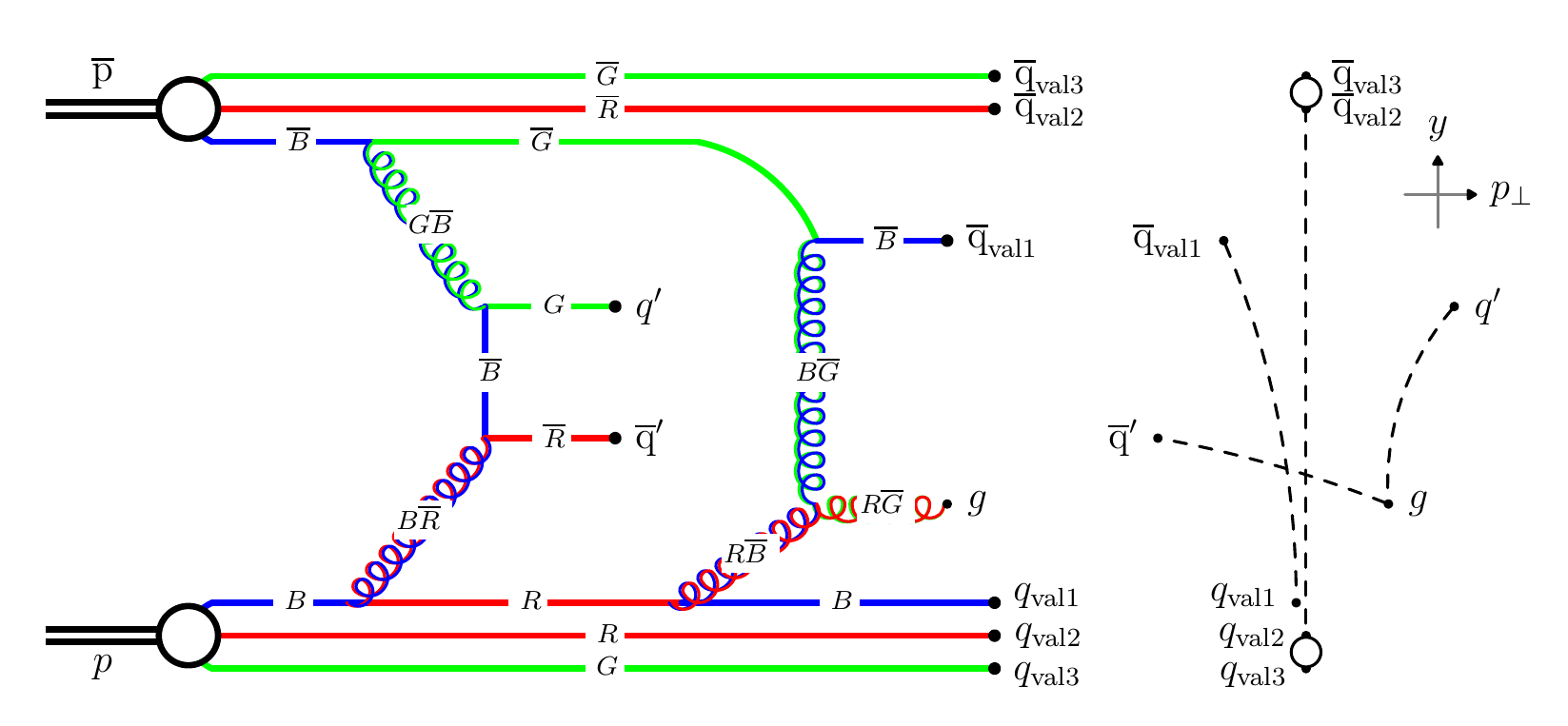}\\
\caption{Left: example of colour assignments in a $p\bar{p}$ collision with two
  interactions. Explicit colour labels are shown on each propagator
  line. Right: The string/cluster topology resulting from the colour
  topology on the left, with horizontal and vertical axes
  illustrating transverse momentum and rapidity, respectively. On the
  right, the two undisturbed valence quarks in each of the beam
  remnant are represented as ``diquarks'' carrying the (anti-)baryon
  number of the original beam particles. \label{fig:justring}}
\end{center}
\end{figure}

On the left-hand side of
\FigRef{fig:justring} we give a simple sketch of what a $p\bar{p}$
collision containing two distinct parton-parton interactions might
look like in (planar) colour space. The additional complications of parton
showers and further MPI have been suppressed, so
that we can focus entirely on the correlations \emph{between} the two
scatterings. Tracing the colour lines in this diagram and connecting
each colour-connected parton pair in the final state by a dashed line
results in the sketch shown to the right of the colour-flow
diagram. In current hadronization models, each of these dashed lines
would represent a string piece, or a cluster, as appropriate.
The vertical axis roughly represents rapidity, with the
beam-remnant partons at either extreme. The horizontal axis is
intended to illustrate $p_\perp$. Thus, the $q'\bar{q}'$ pair
from the primary parton-parton interaction
(furthest to the left in the colour-flow
diagram) are depicted at high $p_\perp$ and central rapidity, while the partons
from the secondary interaction, $\bar{q}_{\mr{val}1}g$
are depicted at larger rapidities, with
smaller transverse momenta.

\begin{figure}
\begin{center}\vspace*{2mm}%
\hspace*{-1mm}\includegraphics*[scale=0.7]{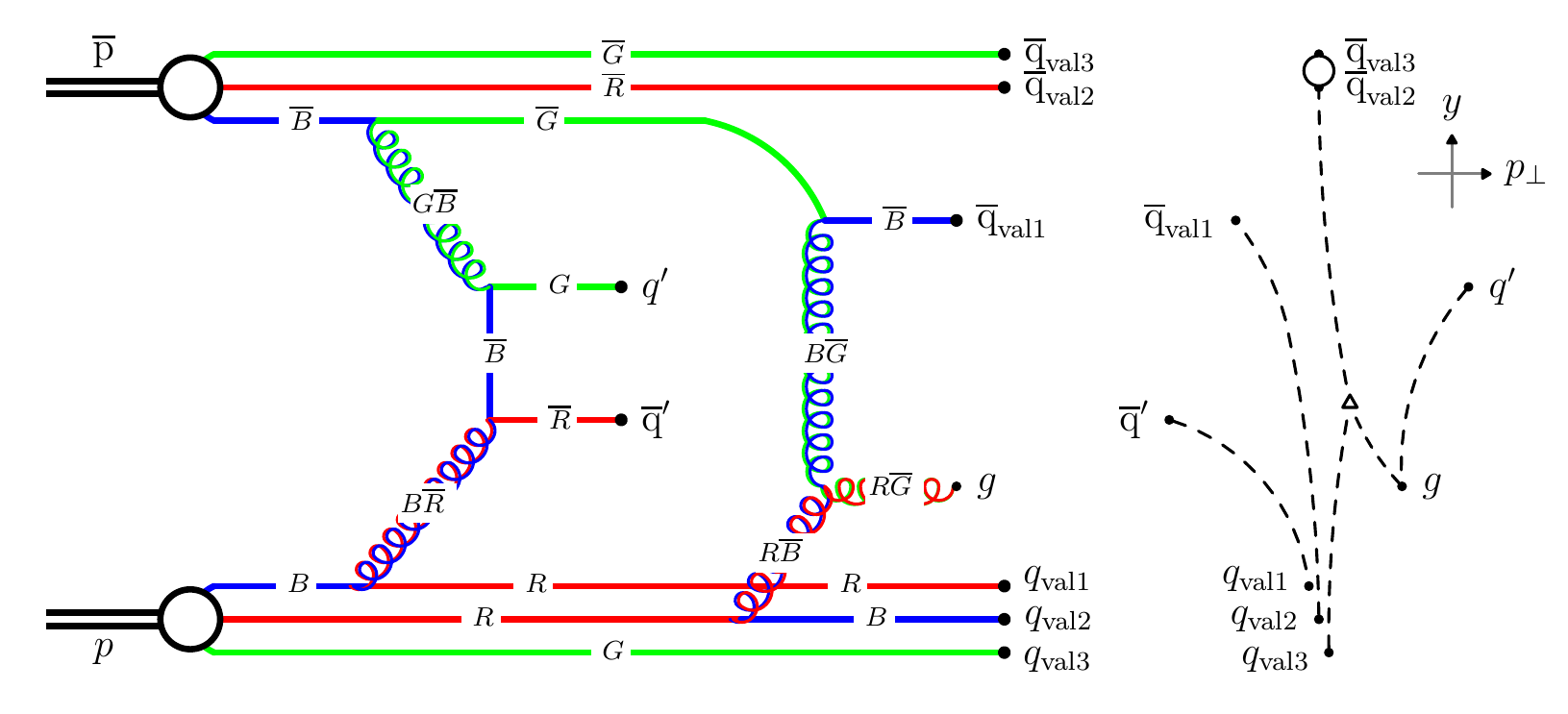}\\
\caption{The same momentum and perturbative colour-flow
configuration as in \FigRef{fig:justring}, but with the second gluon
extracted from the proton now attached to a different quark line than
that of the first gluon. The corresponding string topology reflects
the change and now has become more complicated.
 The baryon number of the proton beam has been ``liberated'', as represented by
  the $\Delta$ symbol towards the middle of the diagram.
\label{fig:justring2}}
\end{center}
\end{figure}

The ambiguity in the colour correlations can be illustrated by
comparing with \FigRef{fig:justring2} which shows exactly the same
perturbative momentum and colour-space configuration, but now with the
second gluon extracted from the proton attached to a different
beam-remnant quark line than that of the first gluon. The
corresponding string topology reflects the change and now has become
more complicated.
The baryon number of the proton beam has been ``liberated'', as represented by
the $\Delta$ symbol towards the middle of the graph. As an aside, this
also illustrates that the baryon number can migrate to the central
region despite all the valence quarks remaining in the beam
remnant; see \cite{Sjostrand:2004pf} for an explicit Monte Carlo model of this effect.
When adding more perturbative parton-parton interactions
this ambiguity grows, and there appears to be very little one can say
about it from perturbation theory.

In the \herwig-based models the colour correlations are set up
by first forcing the initiator of the primary parton-parton
interaction to be a valence quark at low $Q^2$. The initiators of the
subsequent MPI are then forced to be gluons,
which are attached at random to the primary-interaction valence quark
initiator. Since only one quark line is ``disturbed'' in this process,
the beam baryon number remains in the remnant, with the two
undisturbed valence quarks forming a ``diquark'' system
(see \SecRef{sec:hadronization}).

In the old \pythia model the colour flow of the primary
interaction is likewise taken as the basic skeleton onto which the
underlying event interactions are added. In that model, the MPI
final states can either be $q\bar{q}$ pairs, which form isolated
single string pieces, or $gg$ pairs, which either form a closed colour
loop or are inserted to give kinks on the existing primary-interaction
colour topology in a way that minimizes the total string length
(equivalent to minimizing a measure
of the classical potential energy). Again, since at most one quark in
the beam remnant is directly ``disturbed'', the beam baryon number
remains in the remnant. Empirically, the Tevatron data
appear to prefer almost exclusively $gg$ pairs that
minimize the string length. This is indicative of very strong colour
correlations between the MPI final states, although the model does not
address the physical origin of them. (We return to this point
below, under colour reconnections.)

In the new \pythia framework, described in \cite{Sjostrand:2004pf},
a more elaborate modelling of beam remnants was introduced, that
allowed the tracing of colour flow in more detail and also allowed
the beam baryon number to migrate away from the remnant. However,
also in this modelling context, empirical comparisons to the Tevatron
minimum bias data appear to require stronger colour correlations
between the MPI final states than those naively generated by the
model.

This brings us from colour connections, to colour
\emph{reconnections}, which we shall discuss in the next section.

\mcsubsection{Colour reconnections \label{sec:colrec}}
In a first study of colour rearrangements,
Gustafson, Pettersson, and Zerwas (GPZ) \cite{Gustafson:1988fs}
observed that, \eg in hadronic $WW$ events at LEP, colour interference
effects and gluon exchanges may cause `crosstalk' between the two
$W$ systems, leading \eg to uncertainties in the $W$ mass
determination.
In the GPZ picture, the corresponding changes occurred
already at the perturbative QCD level, leading to predictions
of quite large effects. Sj\"ostrand and Khoze (SK)
\cite{Sjostrand:1993rb,Sjostrand:1993hi}
subsequently argued against large perturbative effects
and instead considered a non-perturbative scenario in which
QCD strings can fuse or cut each other up (see \eg
\cite{Artru:1979ye}). These models resulted in
effects much smaller than for GPZ, leading to a predicted total uncertainty
on the $W$ mass from this source of $\sigma_{M_{W}}<40~\mathrm{MeV}$.

Subsequently, several alternative models have been proposed,
most notably by the Lund group, based on QCD dipoles
\cite{Gustafson:1994cd,Lonnblad:1995yk,Friberg:1996xc}, and by Webber based
on clusters \cite{Webber:1997iw}. Apart from $WW$ physics,
colour reconnections have also been proposed to model rapidity gaps
\cite{Buchmuller:1995qa,Edin:1995gi,Rathsman:1998tp,Enberg:2001vq}
and quarkonium production \cite{Edin:1997zb}.

Experimental investigations of colour reconnections at LEP
 \cite{Abbiendi:1998jb,Abbiendi:2005es,Schael:2006ns,
Abdallah:2006ve} were able to exclude at least the
most dramatic scenarios, such as GPZ and
extreme versions of SK with the recoupling strength parameter close to
unity, but more moderate scenarios have not been excluded.
Furthermore, in hadron collisions the initial state contains soft
colour fields with wavelengths of
order the confinement scale. The presence of such fields,
unconstrained by LEP measurements, could impact in a
non-trivial way the process of colour  neutralization
\cite{Buchmuller:1995qa,Edin:1995gi}. And finally,
the MPI produce an additional amount of displaced colour charges,
translating to a larger density of hadronizing systems. It is not
known to what extent the collective hadronization of such a system
differs from a simple sum of independent systems, as will also be briefly
mentioned in \SecRef{sec:hadronization} on hadronization.

A new generation of colour-reconnection toy models have therefore been
developed specifically with soft-inclusive and underlying event
physics in mind \cite{Sandhoff:2005jh,Skands:2007zg,Skands:2010ak},
and also the cluster-based \cite{Webber:1997iw} and Generalized-Areal-Law
\cite{Rathsman:1998tp} models have been revisited in that context.
Although still quite crude, these models do appear to be able to
describe significant features of the Tevatron and LHC data, such as
the distribution of the mean \pt\  of charged particles vs. the number of charged particles, $\langle \pt \rangle(N_{\mr{ch}})$.
\begin{figure}[t]
\center
\includegraphics*[scale=0.75]{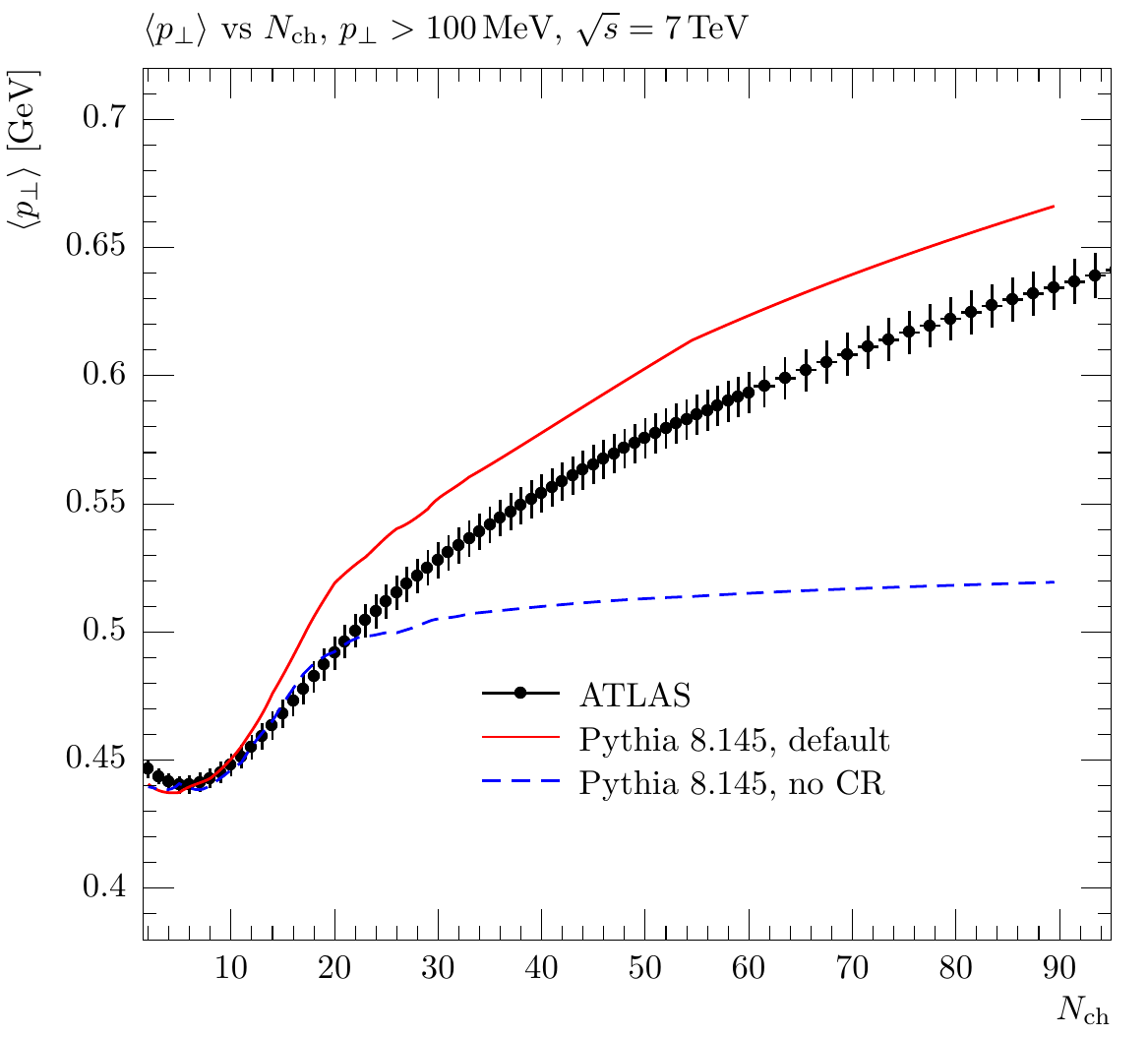}
\caption{Models with and without colour reconnections compared to
the $\langle \pt \rangle (N_{\mr{ch}})$ distribution measured by the ATLAS
experiment \cite{Atlas:2010xx}, 
% The non-preliminary version: 
%experiment \cite{Collaboration:2010i}, 
for particles with $\pt>100$~MeV, $|\eta|<2.5$, and $c\tau > 10$~mm,
in events that contain at least two such particles. \label{fig:mbptofnch}}
\end{figure}
To illustrate this, we include in \FigRef{fig:mbptofnch} a comparison
between an ATLAS minimum bias measurement of the $\langle \pt \rangle
(N_{\mr{ch}})$ distribution at 7~TeV and two Monte Carlo models,
with and without colour reconnections switched on (curves labelled as
``default'' and ``no CR'', respectively). Without colour
reconnections, the predicted  $\langle \pt \rangle
(N_{\mr{ch}})$ distributions appear to rise too slowly with $N_{\mr{ch}}$.

It is nonetheless clear
that the details of the full fragmentation process in hadron-hadron
collisions are still far from completely understood.

\mcsubsection{Diffraction and models based on pomerons \label{s_ue:pomerons}}

Essentially, the MPI-based models discussed above start from
 perturbatively calculable cross sections and attempt to extend these
 to low $p_\perp$ by a combination of resummations of soft
 perturbative effects and  explicit modelling of
 non-perturbative effects.

It is also possible to start from a non-perturbative standpoint,
using unitarity to relate elastic and inelastic scattering processes
 through the optical theorem. In this language, the total cross
 section is driven by the exchange of ``reggeons'' and ``pomerons''
 --- colour-singlet fluctuations with leading
$f\bar{f}$ and $gg$ contents, respectively ---
 with the latter dominating at high energies.

In this picture diffractive events originate from collisions between
an effective flux of such (virtual) colour-singlet objects within the beam
particles, leading to a characteristic spectrum that varies roughly
like $dM^2/M^2$, with $M$ the invariant mass of the diffractively
excited system. (For
multiple diffraction, the behaviour is a product of such factors.)
An important modelling aspect
is whether the pomerons are considered to have a substructure
themselves. If so, partonic collisions between pomeron
fluctuations can generate high-mass diffractive processes
such as diffractive dijets and other hard central exclusive
processes. If not, only the  $dM^2/M^2$ spectrum is present.

Inelastic events are understood in terms of cut pomerons, which
furnishes a relation between diffractive and non-diffractive
scattering that is absent in the MPI-based models discussed
above. The relation between MPI, diffraction, and pomerons is usefully
discussed in \cite{Treleani:2007gi}.

Translated into the terminology used in this review,
each cut pomeron corresponds
to the exchange of a soft gluon, which results in two `strings'
being drawn between the two beam remnants. Uncut pomerons give
virtual corrections that help preserve unitarity. A variable number
of cut pomerons are allowed, which furnishes the equivalent of
MPI in this language. However,
note that cut pomerons were originally viewed as purely soft objects,
and so generated only transverse momenta of order
$\LambdaQCD$, unlike the multiple interactions considered
above. In \textsc{Dtujet} \cite{Aurenche:1994ev},
\textsc{Phojet} \cite{Engel:1994vs,Engel:1995yda} and \textsc{Dpmjet}
\cite{Ranft:1994fd,Roesler:2000he}, however,
also hard interactions have been included, so
that the picture now is one of both hard and soft pomerons, ideally
with a smooth transition between the two.
The three programs all make use of the Lund string hadronization
description, however, and hence share the fundamental properties and
ambiguities of this part of the
modelling with the string-based MPI models discussed
above.

\mcsubsection{Summary}
\begin{itemize}
\item Several parton-parton interactions can occur within
  a single hadron-hadron collision. This is called multiple
  parton interactions (MPI).
\item The hard perturbative tail of MPI is approximately proportional
  to $\drm\pt^2/\pt^4$. This tail produces additional observable
  (pairs of) jets  in the  underlying event. The oft repeated mantra
  that the underlying event is non-perturbative is thus a misconception.
\item Most MPI \emph{are} relatively soft, however, and do not lead to easily
  identifiable additional jets. Instead, they contribute to building
  up the total amount of scattered energy and cause colour exchanges
  between the remnants, thereby increasing the number of particles
  produced in the hadronization stage.
\item Hadron-hadron collisions at small impact parameter
  have a higher number of MPI than peripheral ones. What the enhancement
  factor is depends on the shape of the hadron transverse mass
  distribution.
\item A hadron-hadron collision with a large number of MPI has a
  higher probability of containing a hard jet than one with few
  MPI. This produces the ``pedestal effect''.
\item The number of MPI is regulated by colour screening and
  saturation effects. The detailed behaviour of this regularization,
  including its dependence on collider centre-of-mass energy, is
  poorly known and represents one of the main uncertain / tunable
  aspects of the models.
\item In addition to the MPI $2\to 2$ scatterings,
  realistic models must also incorporate showers off the MPI, to
  describe the broadening and decorrelation of MPI jets.
\item It is also possible to include perturbative rescattering
  effects, but this is so far not available in all models.
\item At the non-perturbative level, the assumed structure of the beam
  remnant can be important, \eg affecting the event structure at large
  rapidities and migration of the beam baryon number in rapidity.
\item There are significant ambiguities concerning colour-space
  correlations, in particular \emph{between} the various MPI
  systems. In current models some amount of \emph{colour
    reconnections} appear to be necessary to properly describe
  minimum bias and underlying event data.
  This is probably the most poorly understood part
  of the modelling, however, and is associated with significant
  uncertainties.
\item The distinction between the  diffractive and non-diffractive
  components of the total inelastic hadron-hadron cross section is
  fundamentally ambiguous and must be interpreted with care, as discussed
  in \SecRef{sec:mbtypes}. This issue is also discussed in
  \PartRef{sec:comp-gener-with}, there from the point of view of
  measurements.
\item Diffractive processes are typically modelled as a separate class
  of processes driven by the exchange of so-called pomerons. These can
  be viewed either as purely soft objects or as having an internal
  partonic substructure. The latter provides a mechanism for
  generating high-mass diffractive processes such as diffractive dijets.
\end{itemize}
% Local Variables:
% mode: LaTeX
% TeX-master: "../mcreview"
% End:

\mcsection{Hadronization}
\label{sec:hadronization}
\mcsubsection{Definition and early developments}

In the general context of QCD studies, the term ``hadronization'' has
been used with somewhat different meanings. 
In the present context it refers to the specific
model used in an event generator for the transition from the partonic
``final'' state to a complete representation of the actual hadronic
final state.  We should emphasize that this is a transition for which we
still have only models, albeit inspired by QCD, because the only
available rigorous approach to non-perturbative hadronic phenomena, lattice
QCD, is formulated in Euclidean space-time and therefore cannot deal
with inherently Minkowskian processes like the time-evolution of
partons into hadrons.

Other ``hadronization'' meanings exist. When quantities that
are calculable within perturbative QCD, for example hadronic event
shapes in $e^+e^-$ annihilation, are compared with experimental data,
there are discrepancies that are commonly ascribed to ``hadronization
corrections''.  They are often estimated and corrected for by comparing
the hadron-level prediction of an event generator with a parton-level
result computed at the end of parton showering.\footnote{We strongly
  deprecate the correction of experimental data to ``parton level''
  by this method. See \SecRef{sec:phys-phil-behind} for discussion of
  this point.}  However, such a
parton-level quantity is not really comparable to the result of a
perturbative calculation, certainly not at fixed order, nor even when
resummed to all orders, as the shower result depends on the scale and
details of the cutoff that terminates it.  The origin of the
discrepancies is instead generic non-perturbative contributions
that do not depend on the detailed mechanism of hadron
formation.  They can be parameterized as power-suppressed terms
related to the so-called infrared renormalon ambiguity of the perturbation
series~\cite{Webber:1994cp,Korchemsky:1994is,Dokshitzer:1995zt,Dokshitzer:1995qm}.

Another use of the term hadronization is to describe the
parameterization of single-particle distributions from hard processes
in terms of fragmentation functions -- for recent reviews
see~\cite{Amsler:2008zzb,Albino:2008gy}.
This is analogous to the parameterization of PDFs for incoming hadrons,
and indeed similar factorization properties allow the scale dependence
of the distributions to be predicted once they have been fitted at
some scale, for example in $e^+e^-$ annihilation.  However once again
such analyses do not illuminate the mechanism of hadron formation.

The earliest models for hadronic final states were based 
on isotropic or longitudinal phase space. On their own, these are 
mainly of interest at very low energies, and are not discussed 
further. The first model that points the way to current approaches 
is the Artru--Mennessier one \cite{Artru:1974hr}, which manages to 
pioneer both string and cluster concepts. The code had a 
number of limitations, however, and never had a practical impact.

In contrast, the Field--Feynman model \cite{Field:1977fa} 
a few years later kickstarted the whole field of hadronization
studies by Monte Carlo simulation. It introduced an iterative 
recipe for the construction of realistic jets. By considering each
outgoing parton separately, it became possible to write 
``independent fragmentation'' generators for $e^+e^-$ physics 
\cite{Hoyer:1979ta,Ali:1979em}.\footnote{We note here that historically the terms
  ``fragmentation'' and ``hadronization'' have been used
  interchangeably, and we follow that usage in this Section, although in other
  contexts fragmentation can refer to the whole process of parton
  showering plus hadron formation.} These models soon became outdated
since the independent fragmentation framework is not Lorentz invariant, 
is not safe under collinear emissions of partons, and also suffers 
from other deficiencies \cite{Sjostrand:1984iy}.

The two main hadronization classes in current use are the string 
and cluster ones. These are described in the following. The main 
difference is that the former transforms partonic systems
directly into hadrons, while the latter employs an intermediate 
stage of cluster objects, with a typical mass scale of a few GeV.

\mcsubsection{String model}
\label{sec:string-model}

An early string fragmentation model is that of Artru and Mennessier, 
introduced above. The most sophisticated and well-known string 
model is the Lund one, however. Its development began in 1977,  
followed by the first primitive Monte Carlo implementation in 1978. 
The core framework was complete by 1983 \cite{Andersson:1983ia,%
Andersson:1998tv}. Thereafter many different additions and 
alternatives have been studied, but only 
a few of them are available in the standard implementation in the 
\pythia event generator \cite{Sjostrand:2006za,Sjostrand:2007gs}.
It is this core Lund string framework that is presented here.  

\begin{figure}
\includegraphics[width=\textwidth]{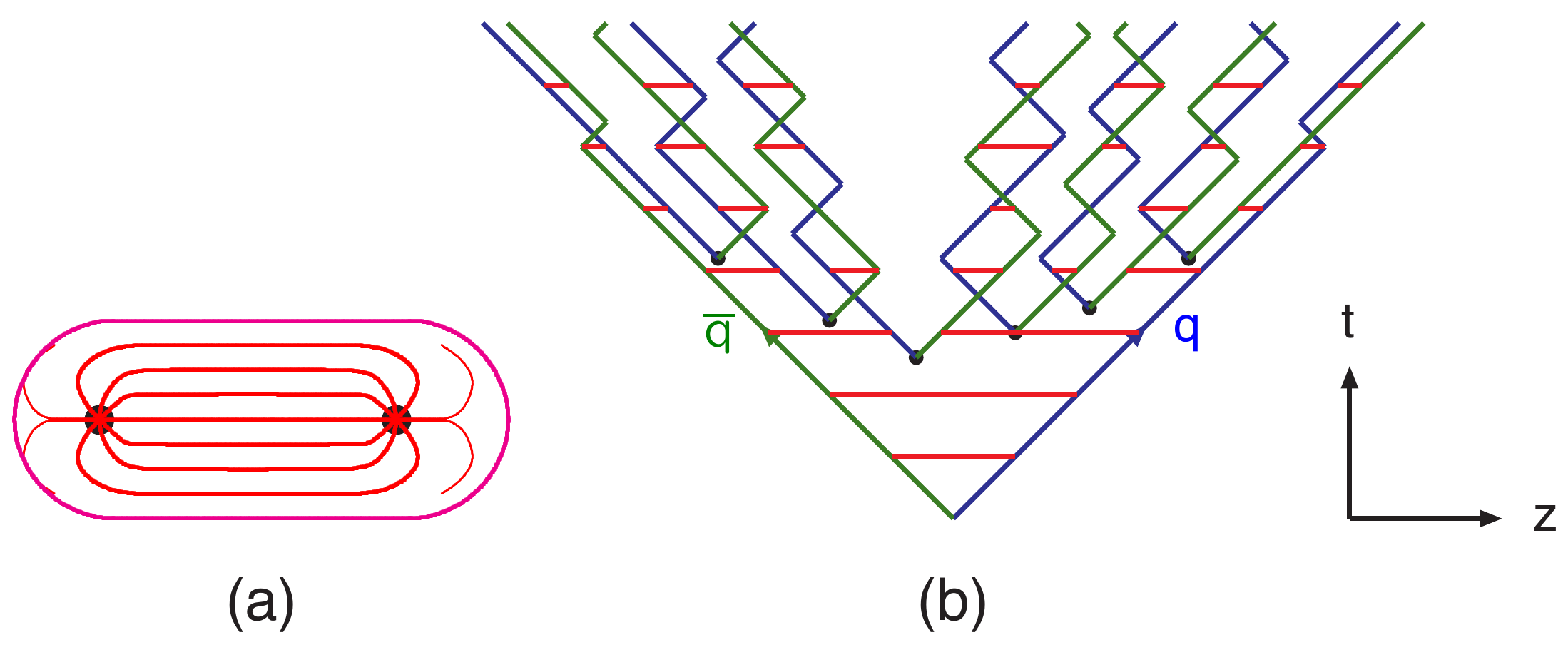} 
\caption{(a) A flux tube spanned between a quark and an antiquark.
(b) The motion and breakup of a string system, with the two 
transverse degrees of freedom suppressed (diagonal lines are 
(anti)quarks, horizontal ones snapshots of the string field).
\label{fig:stringone}}
\end{figure}

In QCD, a linear confinement is expected at large distances. This
provides the starting point for the string model, most easily
illustrated for the production of a back-to-back $q\overline{q}$ 
pair, \eg in $e^+e^-$ annihilation events. As the partons move 
apart, the physical picture is that of a colour flux tube being 
stretched between the $q$ and the $\overline{q}$, 
\FigRef[a]{fig:stringone}. The transverse dimensions of the tube 
are of typical hadronic sizes, roughly 1~fm. If the tube is assumed 
to be uniform along its length, this automatically leads to a 
confinement picture with a linearly rising potential, 
$V(r) = \kappa r$. From hadron mass spectroscopy the string 
constant $\kappa$, \ie the amount of energy per unit length, 
is known to be 
$\kappa \approx 1~\mathrm{GeV/fm} \approx 0.2~\mathrm{GeV}^2$.

This picture is also supported by lattice QCD calculations in the 
quenched approximation, \ie with a gluonic field but no dynamical 
quarks. At small distances an additional Coulomb term is required,
but the assumption of the Lund model is that this term can be 
neglected in the overall production pattern of hadrons. Its 
influence would be felt in the properties of the individual  
hadrons, such as wave functions and masses, however.

In order to obtain a Lorentz covariant and causal description of the 
energy flow due to this linear confinement, the most straightforward 
approach is to use the dynamics of the massless relativistic string with 
no transverse degrees of freedom. The mathematical, one-dimensional 
string can be thought of as parameterizing the position of the axis 
of a cylindrically symmetric flux tube.  The expression ``massless'' 
relativistic string is somewhat of a misnomer: $\kappa$ effectively 
corresponds to a ``mass density'' along the string.

Now consider a simple $q\overline{q}$ two-parton event further.
As the $q$ and $\overline{q}$ move apart from the creation vertex, 
say along the $\pm z$ axis, the potential energy stored in the string 
increases, and the string may break by the production of a new 
$q'\overline{q}'$ pair, so that the system splits into two 
colour-singlet systems $q\overline{q}'$ and $q'\overline{q}$. 
These two systems move apart, and a widening no-field region opens up
in between, \FigRef[b]{fig:stringone}. For simplicity the quarks are 
shown as massless, so they move with the speed of light. If the 
invariant mass of either of these systems is large enough, further 
breaks may occur, and so on until only ordinary hadrons remain. 
Typically, a break occurs when the $q$ and the $\overline{q}$ ends 
of a colour singlet system are 1--5~fm apart in the $q\overline{q}$ 
rest frame, but note that the higher-momentum particles at the 
outskirts of the system are appreciably Lorentz contracted.

At the end of the process, the string has broken by the creation of a 
set of new $q_i\overline{q}_i$ pairs, with $i$ running from 1 to
$n-1$ for a system that fragments into $n$ primary hadrons (\ie
hadrons before secondary decays). Each hadron is formed by the quark 
from one break (or an endpoint) and the antiquark from an adjacent break:
$q\overline{q}_1$, $q_1\overline{q}_2$, $q_2\overline{q}_3$, \ldots, 
$q_{n-1}\overline{q}$.

The space--time picture of string motion, \eg in 
\FigRef[b]{fig:stringone}, can be mapped onto a corresponding 
energy--momentum picture by noting that the constant string tension 
implies that the quarks obey 
\begin{equation}
\left| \frac{\mathrm{d}E}{\mathrm{d}z} \right| =
\left| \frac{\mathrm{d}p_z}{\mathrm{d}z} \right| = 
\left| \frac{\mathrm{d}E}{\mathrm{d}t} \right| =
\left| \frac{\mathrm{d}p_z}{\mathrm{d}t} \right| = \kappa ~.
\label{eq:stringtension}
\end{equation}
It follows that a hadron formed between vertices $1$ and $2$ has 
$E = \kappa\Delta z = \kappa(z_1 - z_2)$ and 
$p_z = \kappa\Delta t = \kappa(t_1 - t_2)$ 
\cite{Andersson:1983ia}. The different breaks are spacelike separated, 
$(\Delta t)^2 - (\Delta z)^2 < 0$, \ie they occur ``independently'' 
of each other in a causal sense. Nevertheless two adjacent
breaks are constrained by the fact that the string piece created
by them has to be on the mass shell for the hadron being produced:
$m_{\perp}^2 = m^2 + p_x^2 + p_y^2 = E^2 - p_z^2 = %
\kappa^2((\Delta z)^2 - (\Delta t)^2)$,
\FigRef[a]{fig:stringtwo}. Here transverse mass is introduced, since
it is this quantity that becomes relevant for the $(E, p_z)$ and 
$(t, z)$ pictures, rather than normal mass, once transverse momentum 
fluctuations are introduced, see below. The total probability for an 
event to be formed can therefore be written as the product of $n-1$ 
breakup vertex probabilities times $n$ delta functions for the 
(transverse) hadron masses.

\begin{figure}
\includegraphics[width=\textwidth]{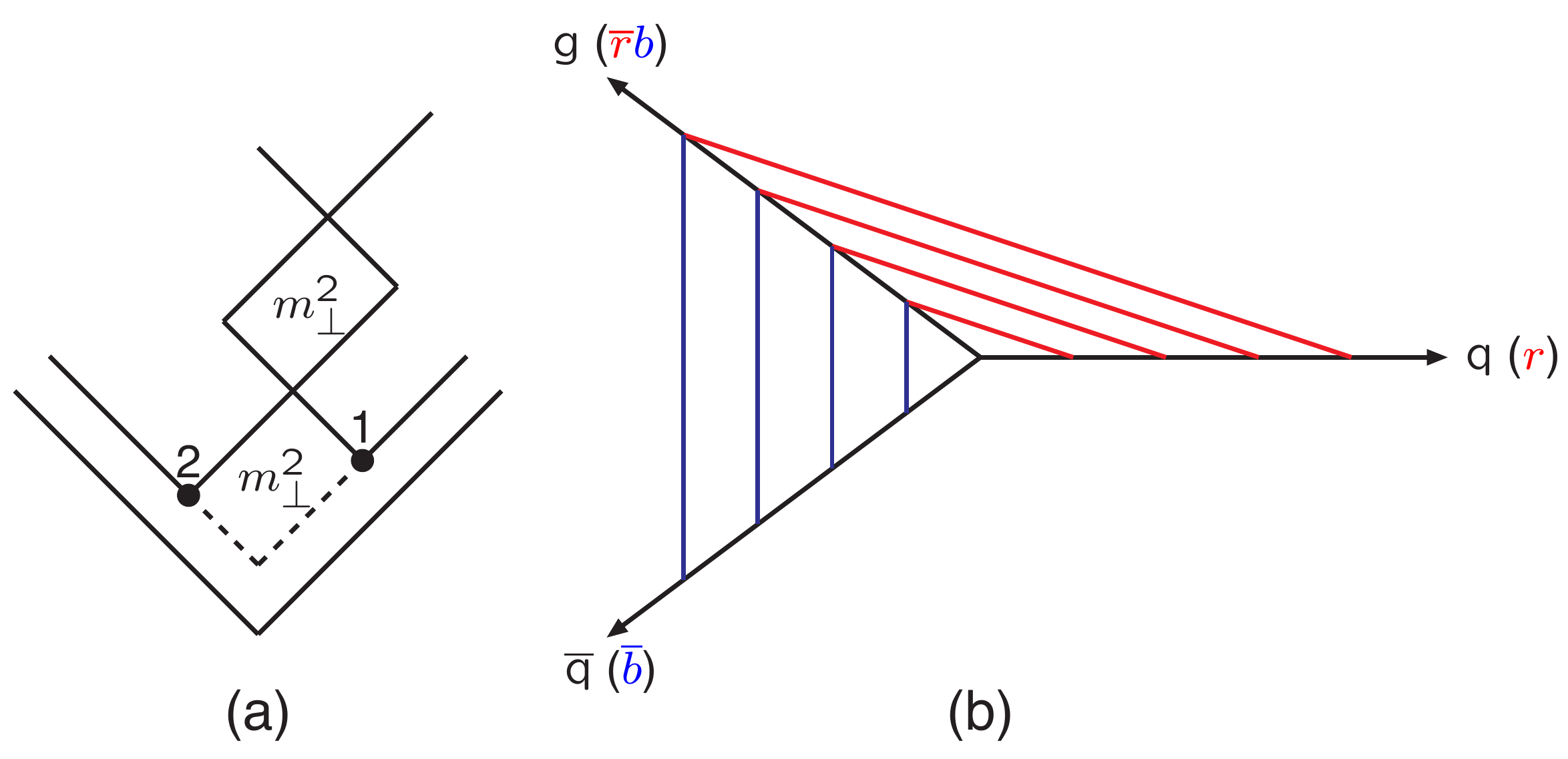} 
\caption{(a) conditions on nearby string breaks;
(b) string motion in three-jet $q\overline{q}g$ events
\label{fig:stringtwo}}
\end{figure}

Technically such an approach would be cumbersome. Fortunately 
an iterative procedure can be used to give the same result.
Since there is no natural ordering, one is free to consider the 
breaks in any order. For instance, one can start
at the $q$ end of the system and iterate ``left'' towards the 
$\overline{q}$ end. Alternatively, one can start at the 
$\overline{q}$ end and iterate the other way, towards ``right''. 
Either approach should give the same overall answer,
``left--right symmetry''. Focusing on the production of a single
hadron between vertices 1 and 2, \FigRef[a]{fig:stringtwo},
the requirement reads $\mathcal{P}(1) \, \mathcal{P}(1 \to 2) =
\mathcal{P}(2) \, \mathcal{P}(2 \to 1)$, where $\mathcal{P}(i)$
is the probability to reach vertex $i$ by iteration from left/right
and $\mathcal{P}(i \to j)$ the probability to take a step from 
vertex $i$ to vertex $j$.  

The solution to this equation can be written in terms of a 
fragmentation function $f(z)$, where $z$ is the fraction of 
the remaining lightcone momentum that the new hadron takes,
with $E \pm p_z$ fraction for iteration to the left/right:
\begin{equation}
f(z) \propto \frac{1}{z} \, (1 - z)^a \, 
\exp\left( - \frac{b m_{\perp}^2}{z} \right) ~,            
\label{eq:stringfz}
\end{equation}
where $a$ and $b$ are two free parameters (for the derivation
see~\cite{Andersson:1983jt}).\footnote{A more complicated
  expression, with different $a$ parameters for different flavours, is
  possible in principle, but only studied for baryon production.} 

As a by-product, the derivation of $f(z)$ also gives the probability 
distribution in invariant time $\tau$ of $q'\overline{q}'$ breakup 
vertices. In terms of $\Gamma = (\kappa\tau)^2$, this distribution is
$\mathrm{d}\mathcal{P}/\mathrm{d}\Gamma \propto  \Gamma^a \exp(-b \Gamma)$,
with the same $a$ and $b$ as above. In a given event, the connection 
between adjacent $\Gamma$ values is given by the formula
\begin{equation}
\Gamma_2 = (1 - z) \, \left( \Gamma_1 + \frac{m_{\perp}^2}{z} \right) ~, 
\label{eq:stringgamma}
\end{equation}
where $\Gamma_1$ is the ``old'' and $\Gamma_2$ is the ``new'' value 
obtained after taking a step $z$ for the production of a hadron 
with transverse mass $m_{\perp}$. The initial values at the $q$ and 
$\overline{q}$ ends of the system are $\Gamma_0 = 0$. Note that
$a = 0$ corresponds to a pure exponential decay as a function of 
the area swept out by the string, exactly as in the Artru--Mennessier 
model, but that correlations between adjacent $\Gamma_i$ values is  
affected by the difference in mass spectra between the Lund and 
Artru--Mennessier models. 

Heavy quarks, \ie charm and bottom, are not produced at new string 
breaks (see below), but may be at the endpoints of a string. 
Unlike massless quarks, heavy quarks
do not move along straight lines, which implies a
changed area swept out by the string field. The argument of an 
exponential decay with area then leads to a modified shape 
\cite{Bowler:1981sb}
\begin{equation}
f(z) \propto \frac{1}{z^{1 + b m_Q^2}} \, (1 - z)^a \, 
\exp\left( - \frac{b m_{\perp}^2}{z} \right) ~,
\label{eq:stringfzsheavy}
\end{equation}
where $m_Q$ is the heavy-quark mass. 

The $f(z)$ formulae above, for the breakup of a system into a hadron and
a remainder-system, strictly speaking only apply when the mass of the
remainder-system is large. In a Monte Carlo program, it is therefore
necessary to introduce a special procedure to cover the production of
the last two particles. This contains no new physics, but has just to be
constructed so that the place where one selects to ``patch up'' the
fragmentation from the $q$ end with that from the $\overline{q}$ one 
looks as closely like any other as is possible. In addition, steps 
are taken from the left and right ends of the system at random, so that 
the matching procedure is not applied at the same place in all events.

A related but different issue is what to do with a low-mass string,
which occasionally may occur as part of an event with several separate 
strings. An attempt is then made to form an exclusive two-body state, 
with orientation preferentially along the string axis. If the string mass
is too small for this to 
work, there is a possibility to let a small string collapse into one 
single hadron. To put this hadron on the mass shell, some shuffling of 
energy and momentum with other partons in the event is then necessary. 
This machinery thus has some similarities with the cluster fragmentation 
approach, but is in practice only used for a small fraction of the
total particle production.  

In a colour field a $q'\overline{q}'$ pair, where the
$q'$ and $\overline{q}'$ have no mass or transverse momentum, 
can classically be created in one point and then be pulled apart 
by the field. If the quarks have mass or transverse momentum, 
however, they must classically
be produced at a certain distance so that the field energy between
them can be transformed into the transverse mass $m_{\perp}$. Quantum
mechanically, the quarks have to be created in one point and then 
must tunnel out to the classically allowed region. The production 
probability for this tunnelling process is proportional to
\begin{equation}
\exp( -\pi m_{\perp}^2 / \kappa) =  \exp( -\pi m^2 / \kappa) \,
\exp( -\pi p_{\perp}^2 / \kappa) ~.
\label{eq:stringtunnel}
\end{equation}
  
The factorization of the transverse-momentum and the mass terms leads
to a flavour-independent Gaussian spectrum for the 
$q'\overline{q}'$ pairs. Since the string is assumed to have 
no transverse excitations, this $p_{\perp}$ is locally compensated 
between the quark and the antiquark of the pair, and
$\langle p_{\perp q}^2 \rangle = \sigma^2 = \kappa/\pi 
\approx (250~\mathrm{MeV})^2$. Experimentally a number closer to 
$\sigma^2 \approx (350~\mathrm{MeV})^2$
is required, which could be explained as the additional effect of 
soft-gluon radiation below the shower cutoff scale. That radiation 
would have a non-Gaussian shape but, when combined with the ordinary 
fragmentation $p_{\perp}$, the overall shape is close to Gaussian, 
and is parameterized correspondingly in the program. Hadrons receive 
$p_{\perp}$ contributions from two $q'\overline{q}'$ pairs and have 
$\langle p_{\perp h}^2 \rangle = 2 \sigma^2$.
  
The formula also implies a suppression of heavy quark production,
$u : d : s : c \approx 1 : 1 : 0.3 : 10^{-11}$. Charm and heavier quarks
hence are not expected to be produced in the soft fragmentation.
The suppression of $s\overline{s}$ production is left as a free 
parameter in the program, but the experimental value agrees 
qualitatively with theoretical prejudice.
  
A quark and an antiquark may combine to produce either a pseudoscalar
or a vector meson. From counting the number of spin states
one would expect the relative probability for pseudoscalar : vector
to be 1 : 3. This should be modified by wave function effects, as
manifested \eg in the mass splitting between pseudoscalar and vector
mesons, bringing the ratio closer to $1 : 1$. The production of higher 
resonances is assumed to be low in a string framework. The four 
$L = 1$ multiplets are implemented, but are disabled by default, largely
because several states are poorly known and thus may result in a 
worse overall description when included. 
  
The simplest scheme for baryon production is that, in addition to
quark--antiquark pairs, also antidiquark--diquark pairs are occasionally 
produced in the field, in a triplet--antitriplet representation.
Such an assumption does not imply that a diquark should be considered as
a single excitation of an elementary field, only that the soft
chromoelectric field effectively acts on a diquark as if it were a single unit.
Due to the large uncertainty in the definition of diquark masses, the
tunnelling formula cannot be used directly to predict the expected rate of
diquark production. Rather, from data a relative probability for diquark
to quark production is determined to be $qq / q \approx 0.1$. Further
parameters are needed to pin down the rates of individual diquarks, 
again for lack of well-defined diquark masses.

A more general framework for baryon production is the so-called
popcorn model, in which diquarks as such are never produced, but
rather baryons appear from the successive production of several 
$q'\overline{q}'$ pairs. Part of the time, the end result will be 
exactly the same $B\overline{B}$ situation as above, \ie with an 
adjacent baryon $B$ and antibaryon $\overline{B}$ sharing a 
diquark--antidiquark pair. However, further possibilities of the type 
$BM\overline{B}$, $BMM\overline{B}$, etc., can occur,
where a varying number of mesons $M$ are produced in between 
the baryon and antibaryon. The $B$ and $\overline{B}$ then have just 
one $q'\overline{q}'$ pair in common, rather than two. In its present 
form, the program generates $B\overline{B}$ and $BM\overline{B}$ 
configurations with roughly equal probability, while $BMM\overline{B}$ 
and even longer meson chains are neglected.

A given quark--diquark pair may combine to produce either a spin $1/2$
(``octet'') or a spin $3/2$ (``decuplet'') baryon. Again higher resonances 
are neglected. A very important constraint is the fact that a baryon is 
a symmetric state of three quarks, neglecting the colour degree of 
freedom. When a diquark and a quark are joined to form a baryon, it is 
therefore necessary to weight the different flavour and spin states 
by the probability that they form a symmetric three-quark system.

So far only the simplest possible system, $q\overline{q}$, has been
considered. If several partons are moving apart from a common origin, 
the details of the string drawing become more complicated. For a 
$q\overline{q}g$ event, a string is stretched from the $q$ end 
via the $g$ to the $\overline{q}$ end, \FigRef[b]{fig:stringtwo}, 
\ie the gluon is an energy- and momentum-carrying kink on the string. 
It is assigned an incoherent sum of one colour charge and one anticolour 
one. 

As a consequence of the gluon having two string pieces attached, the
ratio of gluon/quark string forces becomes 2, a number that can be
compared with the ratio of colour charge Casimir operators, $\Nc/C_F =
2/(1-1/\Nc^2) = 9/4$. In this, as in several other respects, the
string model can therefore be viewed as a variant of QCD where the
number of colours $\Nc$ is not 3 but infinite (\cf\SecRef{sec:large-nc-limit}).

Note that the factor 2 above does not depend on the kinematical 
configuration: a smaller opening angle between two partons corresponds 
to a smaller string length drawn out per unit time, but also to an 
increased transverse velocity of the string piece, which gives an 
exactly compensating boost factor in the energy density per unit 
string length. 

In an event with several gluons, these will still appear as kinks on 
the string between the $q$ and $\overline{q}$ ends. It is also possible 
to have a closed gluon string, \eg in $\Upsilon \to ggg$ decays.

One of the key predictions of the string model is that, in
$q\overline{q}g$ events, the $qg$ and $\overline{q}g$ angular 
regions should receive enhanced particle production, while the 
$q\overline{q}$ one should be depleted. This arises as a consequence
of having fragmenting string pieces boosted into the former two
regions, but not into the latter one. It was confirmed by JADE 
\cite{Bartel:1983ii}, which inspired the Leningrad study of 
perturbative coherence in such events \cite{Azimov:1986sf}. 
It led to the picture of linked dipoles driving a 
perturbative evolution \cite{Gustafson:1986db,Gustafson:1987rq}, 
\SecRef{sec:dipoles}.  

One of the key virtues of the string fragmentation approach is that it
is collinear and infrared safe. That is, the emission of a collinear or 
soft gluon disturbs the overall string motion and fragmentation 
vanishingly little in the small-angle/energy limit 
\cite{Sjostrand:1984ic}. Therefore the choice of lower cutoff scale 
for parton showers is not crucial: letting the shower evolve to 
smaller and smaller scales just adds smaller and smaller wrinkles 
on the string, which still maintains the same overall shape.% 
\footnote{In a generator implementation there are technical 
complications, however, and also an increasing time consumption, 
implying that it does not pay to take things to the extreme.}

\begin{figure}
\hspace*{0.1\textwidth}
\includegraphics[width=0.8\textwidth]{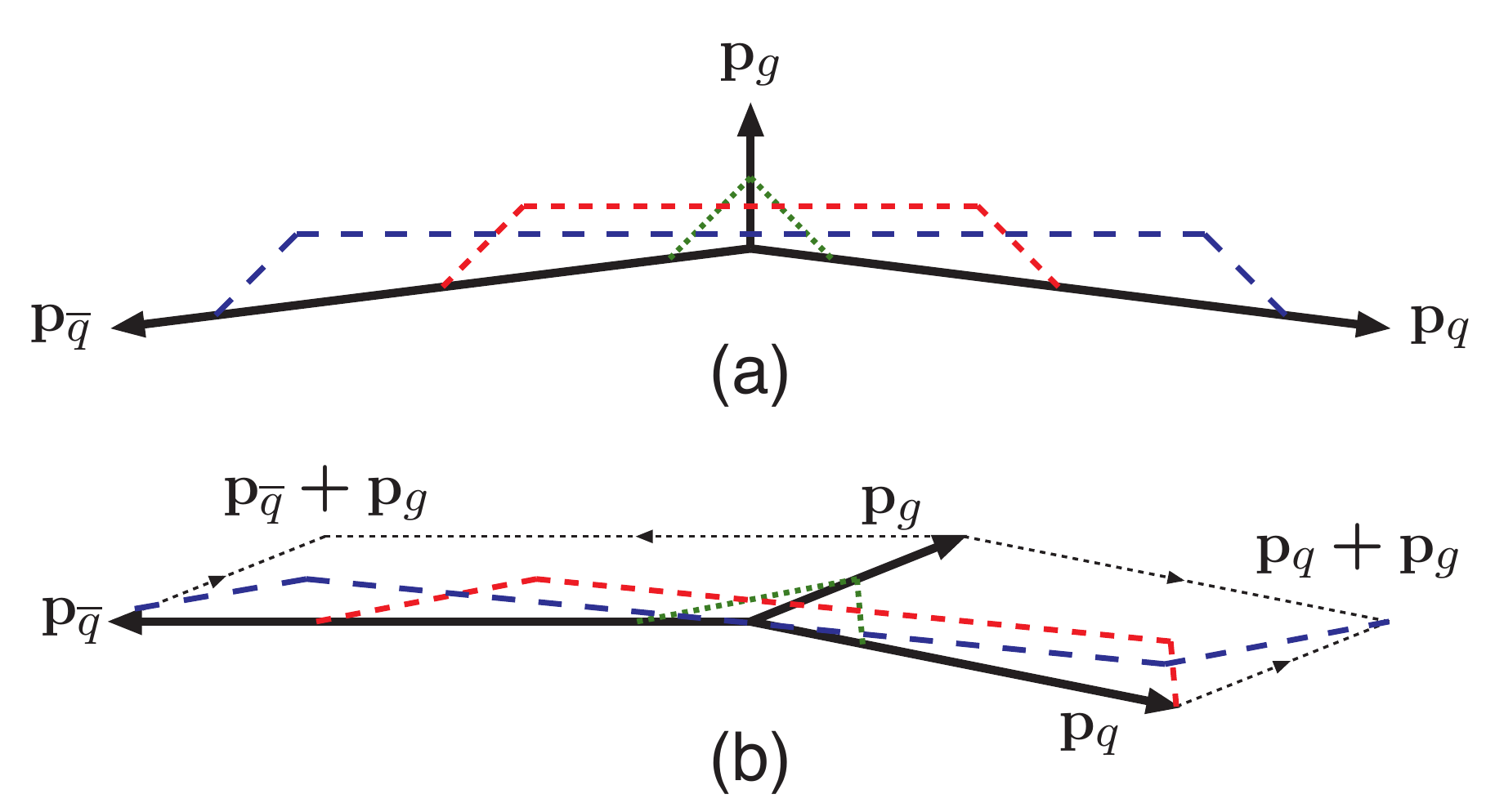} 
\caption{Snapshots in time of the string motion for (a) soft and
(b) collinear gluon emission.
\label{fig:stringthree}}
\end{figure}

To understand this point a bit better, consider the string motion
without any fragmentation. A three-jet $q\overline{q}g$ event initially 
corresponds to having a string stretched from the $q$ via the $g$ to 
the $\overline{q}$, \ie two string pieces, \FigRef[b]{fig:stringtwo}.
In the string piece between the $g$ and the $q$ ($\overline{q}$), 
$g$ four-momentum is flowing towards the $q$ ($\overline{q}$) end,
and $q$ ($\overline{q}$) four-momentum towards the $g$ end. When 
the $g$ has lost all its energy, the $g$ four-momentum continues 
moving away from the middle, \ie where the $g$ used to be, and 
instead a third string region is formed there, consisting of inflowing 
$q$ and $\overline{q}$ four-momentum.
  
For an energetic gluon it takes a long time for the gluon to lose its
energy, and by then hadronization is already well under way. For a 
low-energy gluon, on the other hand, the third string region 
appears early, and the overall drawing of the string becomes fairly 
two-jetlike, \FigRef[a]{fig:stringthree}. In the limit of vanishing 
gluon energy, the two initial string regions disappear, and 
the ordinary two-jet event is recovered. Also for a collinear gluon, 
\ie $\theta_{qg}$ (or $\theta_{\overline{q}g}$) small, the stretching 
becomes two-jetlike, \FigRef[b]{fig:stringthree}. In particular, the 
$q$ string endpoint first moves out a distance $\mathbf{p}_q / \kappa$, 
and then a further distance $\mathbf{p}_g / \kappa$, a first half 
accreting gluon four-momentum and a second half re-emitting it.
The end result is, approximately, that a string is drawn out as if
there had only been a single parton with energy 
$|\mathbf{p}_q + \mathbf{p}_g|$, such that the simple two-jet event 
again is recovered in the limit $\theta_{qg} \to 0$.
These discussions for the three-jet case can be extended to the 
motion of a string with an arbitrary number of intermediate gluons. 
  
The generalization of the left--right symmetry requirement to the
fragmentation of multiparton configurations is not completely unique. 
A sensible physics ansatz is that the distribution in 
invariant time of breakup vertices should not depend on the exact 
shape of the string. The $z$ variable no longer has any simple physical 
interpretation, but \EqRef{eq:stringfz} and 
\EqRef{eq:stringgamma} taken together still provide a valid recipe 
for the relationship between adjacent $\Gamma$ values. These $\Gamma$ 
values themselves are always well defined and, if taken together with
the constraint of hadrons being on mass shell, uniquely define the 
position of each breakup vertex.

Until now only the string model as such has been introduced. It is 
useful to briefly put it into the context of the generation of a 
complete event in hadronic collisions, a topic discussed in 
\SecRef{sec:mbmpi-np}. String fragmentation is almost at the end of the
generation chain, only followed by particle decays. It is applied when
a number of partons have already been produced,
by a combination of hard processes and parton showers, and with
coloured leftover beam remnants. Since a remnant cannot be described 
by perturbative means, special rules are needed to assign colours to 
the individual partons in it, and to describe how the available 
remnant momentum is split between them. Some of these rules are 
based on common sense, such as that the average valence quark ought to 
carry more momentum than the sea one, but our ignorance also leads to 
some arbitrary decisions. With the remnant subdivided, and with colours 
traced in the large-$\Nc$ limit, the string topology can be constructed. 
The event that way can be split into a number of string systems. 
Mainly these will be open strings, with quarks or diquarks at the 
endpoints and with gluons in between, but closed gluon loops are 
also possible. 

One special issue is that of baryon number flow if, say, (the colour 
of) two of the valence quarks of a proton are kicked out in different 
directions. For situations like these a three-quark system can be 
viewed as a Y-shaped topology: three strings, each with a quark at 
one endpoint, and at the other coming together in a junction. 
Each of the strings can break by the production of new 
$q'\overline{q}'$ pairs, but at the end of the process there will be one
unmatched $q'$ nearest to the junction in each string, and these 
three together give a baryon. Thus the baryon number is ``carried''
by the junction, and the balance between the different string pieces
pulling on the junction largely determines the net motion of this
baryon \cite{Sjostrand:2002ip}. Depending on the overall colour 
topology of an event it thereby becomes possible to transport 
the baryon number over large rapidity distances.

One potential weakness of the string model is that it is formulated
in terms of the fragmentation of one single string in isolation,
as you may expect it to be in $e^+e^-$ annihilation events. 
If several strings are produced, they fragment independently of each 
other. For hadronic collisions the MPI framework is likely to lead to 
a picture where several strings overlap in space and time during the 
fragmentation process, however, especially at high collision energies. 
It is not unthinkable that this leads to collective phenomena bordering 
on those of the quark--gluon plasma expected in heavy-ion collisions, 
which then are not modelled by the standard string framework. 
For instance, a dense hadronic gas could lead to non-negligible
rescattering corrections.

Finally, in addition to the basic ideas presented here, the string 
concept has been used as a starting point for various extensions, 
say for colour reconnections in $W^+W^-$ events (see
\SecRef{sec:mbmpi-np}) or Bose--Einstein effects among identical
mesons. Such effects are not included in simulations by default.

\mcsubsection{Cluster model}
\label{sec:cluster-model}
The cluster model of hadronization is based on the so-called
preconfinement property of parton showers, discovered by Amati and
Veneziano~\cite{Amati:1979fg}.  They showed that the colour structure of the
shower at any evolution scale $Q_0$ is such that colour singlet
combinations of partons (clusters) can be formed with an
asymptotically universal invariant mass distribution.  Here
`universal' means dependent only on $Q_0$ and the QCD scale $\Lambda$,
and not on the scale $Q$ or nature of the hard process initiating the
shower, while `asymptotically' means $Q\gg Q_0$.  If in addition
$Q_0\gg\Lambda$, then the mass distribution of these colour singlet
clusters, together with their ($Q$-dependent) momentum and
multiplicity distributions, can be computed
perturbatively~\cite{Amati:1979fg,Bassetto:1979vy}.  It turns out that the mass distribution
is power-suppressed at large masses, the mean cluster
multiplicity $\abr{n}$ rises faster than any positive power of $\ln Q$,
and the asymptotic multiplicity distribution is a universal function
of $n/\abr{n}$ (KNO scaling~\cite{Koba:1972ng}).

\begin{figure}
\begin{center}\includegraphics[width=8cm]{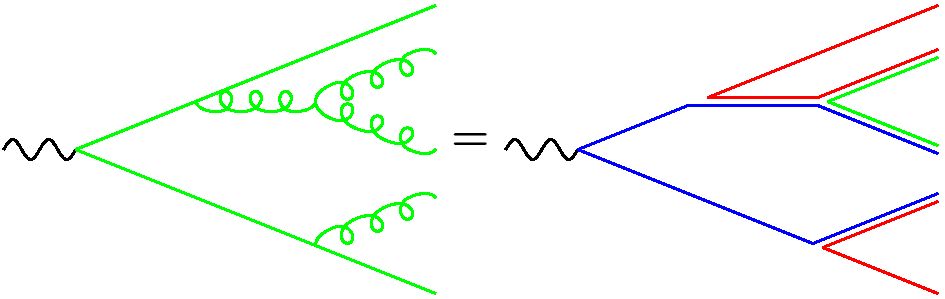}\end{center}
\caption{Colour structure of a parton shower to leading order in
  $\Nc$.\label{fig:planar}}
\end{figure}

The preconfinement mechanism can be seen most simply in the limit
of a large number of colours $\Nc$ (\cf\SecRef{sec:large-nc-limit})\footnote{The proof of preconfinement is
  valid for any value of $\Nc$.   However, as in the string model,
  subleading terms of order $1/\Nc^2$ are neglected in the cluster
hadronization model.}.   To leading order in $\Nc$, the gluons in
the shower can be represented by pairs of colour-anticolour lines
that are connected at vertices (see \FigRef{fig:planar}).  Then each
colour line at the low-scale end of the shower is connected to an
anticolour partner line at the same scale.  In this limit the
colour structure of the shower can be drawn on a plane, such that
these colour-anticolour partners are adjacent.  Adjacent partners can
form colour singlets, whereas non-adjacent lines have a vanishing
probability of doing so as $\Nc\to\infty$.   Furthermore, adjacency
tends to imply closeness in phase space, leading to the suppression of large
masses and an asymptotically universal mass distribution of adjacent
objects.  

A model of hadronization based on preconfinement was first proposed by
Wolfram~\cite{Wolfram:1980gg} and incorporated into an event generator for $e^+e^-$
annihilation by Field, Fox and Wolfram~\cite{Fox:1979ag,Field:1982dg}.  The key idea
was to enforce non-perturbative splitting of gluons into
quark-antiquark pairs at the shower cutoff scale $Q_0$.  Then
adjacent colour lines become quark-antiquark pairs that can form
physical clusters with mesonic quantum numbers.  For low values of the
cutoff, the typical cluster invariant masses will be low and the hadrons
from the decay of each cluster will be spread over a limited region of
phase space.  This leads naturally to a distribution of final-state
hadrons closely connected to that of partons at the cutoff scale,
\ie to local parton-hadron duality~\cite{Azimov:1984np,Azimov:1985by}.

The enforced gluon splitting corresponds to an effective enhancement
of the $g\to q\bar q$ vertex, which would be expected to reduce or
even reverse the running of the QCD coupling at low scales.  Thus this
mechanism also agrees, at least qualitatively, with the notion of a finite
effective low-scale value of $\alpha_S$, which is suggested by studies
of hadronization corrections to event shapes\cite{Dokshitzer:1995zt,Dokshitzer:1995qm,Dasgupta:2003iq}
 and jet profiles~\cite{Dasgupta:2007wa}.
 
Another intriguing hint of enhanced gluon splitting is the high
yield of soft photons in hadronic $Z^0$ decays~\cite{Abdallah:2005wn,Abdallah:2010tk}, which cannot be
explained in terms of radiation from perturbatively produced quarks
plus bremsstrahlung from initial-state leptons and final-state hadrons.
This suggests a nonperturbative phase in which many charged particles
are formed and propagate for significant times before hadronization,
as might happen between gluon splitting and cluster formation.

Once the mechanism of gluon splitting has been adopted, a number of
issues need to be addressed in building a quantitative model of
hadronization.  First of all, what should be the momentum distribution
and flavours of the quarks produced in gluon splitting?  The
momentum distribution is not a major issue as long as the
light flavours are treated as having constituent quark masses, $m_{u,d}\sim
300$~MeV, $m_s\sim 450$~MeV.  Then the effective gluon mass at the end
of showering, $m_g\sim Q_0\sim 1$~GeV, is close to the threshold for splitting,
and the kinematic range of quark momentum is too small for it to have 
much effect on the momenta and masses of the clusters.

On the other hand the flavour distribution in gluon splitting will clearly
be important.  At such low scales, kinematics will ensure that heavy flavours are
forbidden and strangeness suppressed.   However, these flavours can be
pair-produced in cluster decays, as can baryon-antibaryon pairs.
Baryons might also come from gluon splitting into light
diquark-antidiquark pairs. Parameters can be introduced to tune the
yields of different flavours, but generally the effects of kinematics
work quite well as a first approximation. 

\begin{figure}
\begin{center}\includegraphics[width=8cm]{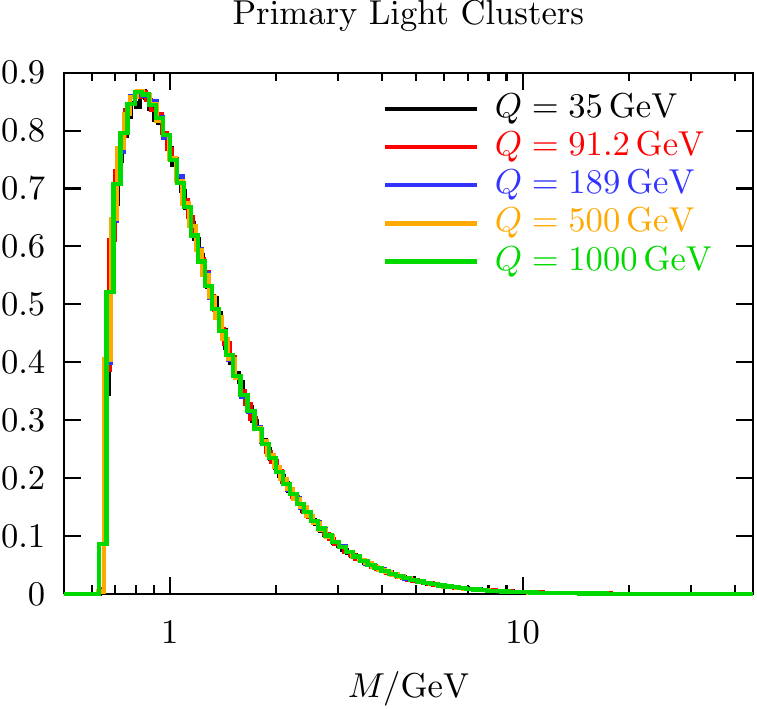}\end{center}
\caption{Invariant mass distribution of colour-singlet clusters in \herwig.\label{fig:cluster}}
\end{figure}

The cluster mass distribution in $e^+e^-$ annihilation into
light-quark pairs, obtained from the \Herwig\ event generator, is shown in
\FigRef{fig:cluster} for a wide range of  centre-of-mass energies.
One sees clearly the universality of the distribution and the typically
low scale of cluster masses, determined by the shower cutoff,  $Q_0\sim 1$~GeV.

Once these clusters have been formed, how should they decay into the
observed hadrons?  The typical cluster masses are low enough for them
to be treated as a smoothed-out spectrum of excited mesons, in which
case quasi-two-body decay into less excited states seems to be
preferred by Nature.   Assuming that matrix element effects tend to average out,
the simplest model is then to select at random among all two-body decay
channels allowed by flavour and kinematics, with probabilities
proportional to the available phase space for each, including spin
degeneracy.  The preference for decay directly to the lowest-mass
states is somewhat offset by the larger number of spin states amongst
excited hadrons.  One must be careful to include full flavour
multiplets of hadrons of each spin and parity, in order not to bias
the flavour selection. This can entail some guesswork about the masses
and decays of excited heavy-flavour hadrons that are not yet well established.

This basic model of cluster decay comes surprisingly
close to fitting the hadron distributions observed in jet
fragmentation, with virtually no free parameters other than the shower
cutoff.  The multiplicities of different flavours of mesons and
baryons are determined by their masses and spins, and their
transverse momenta relative to the jet axis are naturally limited by
the phase space available in cluster decay. The characteristic
differences between quark and gluon jets, with the latter having
softer hadron spectra, higher multiplicities and wider profiles, all come
from the higher rate of parton showering from gluons; apart from
leading flavour effects, the relative proportions of different hadron species
in all types of jets should be universal.

For a more refined description of hadronic final states, at the level
demanded nowadays from event generators, the basic cluster model
described above requires some adjustments.  The sharp
transition from perturbative to non-perturbative physics at the
cutoff scale tends to over-suppress heavy states such as
multiply-strange (\eg $\Omega^-$) and charmed baryons.  A smoother
transition over a range of scales would clearly be more physical.
Light-flavour baryon production
comes mainly from the decay of mesonic clusters into baryon-antibaryon
pairs.   While this gives about the right multiplicities, it tends to
produce pairs that are too closely correlated in rapidity.  The
requirement that each cluster should produce at least two hadrons leads
to single-hadron distributions that are too low at high fractions of the
jet momentum.  This can be improved somewhat by allowing low-mass
clusters to form a single hadron, transferring some four-momentum
to a `nearby' cluster.  In reality the parton showering mechanism
is not dominant at very high momentum fractions and more exclusive
processes must take over.

 A related problem is the treatment of
high-mass clusters.  Although the cluster mass distribution is
mostly confined to low values, there is always a high-mass tail, as
seen in \FigRef{fig:cluster}, coming from events with very little
parton showering.   Again, exclusive hadron production
would in fact predominate over these highly Sudakov-suppressed
parton configurations.  In the absence of such contributions, special
treatment of high-mass clusters is required, as  the model of
isotropic phase-space decay is clearly unreasonable.  The prescription
commonly adopted is sequential binary fission, preserving the
orientation along the axis defined by the constituent partons of the
original cluster, until the sub-cluster masses fall below some value,
typically  $3-4$~GeV, after which the standard phase-space decay is
resumed.  The treatment of high-mass clusters thus becomes close to
that of the string model, and indeed a smooth merging of the two
models at intermediate cluster masses would seem well worth investigation. 

The two cluster hadronization models in wide
use~\cite{Webber:1983if,Winter:2003tt},
incorporated in the \Herwig\ and \Sherpa\ generators respectively, differ in
their detailed treatment of these issues but follow the basic approach
outlined above.  The reader is referred to the papers cited, the relevant sections
below, and the generator manuals for further details.

An interesting recent development is the application of a statistical
hadronization model to cluster decay~\cite{Bignamini:2010jy}.  In this
way one may generate the spectra, multiplicities and flavour
composition of the produced hadrons with few free parameters, while
retaining the limited transverse momenta and jet structure which are
difficult to explain if a statistical approach is applied directly to the whole
final state.

As in the case of the string model, the basic cluster model does
not include interaction between clusters, except for some
momentum transfer to permit light ones to form single hadrons.
There are optional schemes for colour reconnection between clusters,
for example in $W^+W^-$ hadronic decays, as discussed in
\SecRef{sec:mbmpi-np}.  There could be additional collective effects
when the cluster density becomes high.

\mcsubsection{Summary}
\begin{itemize}
\item Hadronization cannot be calculated from first principles, but has
  to be modelled. The two most commonly used model classes are the string
  and cluster ones. 
\item The string model is based on the assumption of linear confinement.
\item A quark corresponds to an endpoint of a string, and a gluon to a
  kink on it, with partons ordered in colour along the string.
\item The string offers a very predictive framework for how its 
  space--time motion and breakup translates into an energy--momentum
  distribution of the primary hadrons. 
  This framework also applies for complicated multiparton 
  configurations, and has been successfully tested in $e^+e^-$ collisions.
\item The main (known) weakness of the string model is that there are 
  many parameters related to flavour properties, which ultimately have 
  to be pinned down from data itself.
\item The cluster hadronization model is based on the preconfinement
  property of parton showers, which leads to colour-singlet parton
  clusters with a universal mass distribution at low scales.
\item Cluster hadronization starts with non-perturbative splitting of
  gluons into quark-antiquark (and possibly diquark-antidiquark) pairs.
  Clusters are then formed from colour-connected pairs. 
\item Most clusters undergo quasi-two-body sequential phase-space
  decay.  The limited cluster mass spectrum naturally leads to limited
  transverse momenta and suppression of heavy flavour, strangeness and
  baryon production.
\item The decay of heavier clusters requires a more string-like
  initial stage of anisotropic decay into lighter clusters.
\item When combined with angular-ordered parton showers, the cluster
  model gives a fairly good overall description of high-energy
  collider data, usually slightly less good than the string model but
  with fewer parameters.
\item The busy environment of high-energy hadronic collisions could lead to
  nontrivial collective effects, currently not simulated either in string
  or in cluster models. 
\end{itemize}

% Local Variables: 
% mode: LaTeX
% TeX-master: "../mcreview"
% End: 

\mcsection{Hadron and tau decays}
\label{sec:hadron-decays}

 Following the hadronization phase of event generation a number of unstable
 hadrons are produced, which must be decayed into particles that are
 stable on collider timescales.
 This is an important part of the event
 simulation, because the observed final-state hadrons result
 from a convolution of hadronization and decay, so that a particular set
 of tuned hadronization parameters is applicable only in combination
 with a particular decay package.

 Simulating hadron decays involves non-trivial modelling.
 At first glance it might seem that all the information
 needed to simulate these decays is readily available in the Particle Data Group~(PDG)'s
 Review of Particle Physics~\cite{Amsler:2008zzb}, but the information
 on particle properties in the PDG is often insufficient and numerous choices have to be made.
 This is particularly true for members of excited meson multiplets, excited mesons containing
 heavy~(bottom and charm) quarks, and baryons containing heavy quarks.
 The number of choices which have to be made
 increases as more excited meson and baryon multiplets
 are added to the simulation.

 The first choice that must be made is which hadrons to include in the simulation.
 This choice is generator specific and is closely connected with
 the tuning of hadronization parameters.  In the cluster model in particular it
 is important that all the light members\footnote{The hadrons containing only up, down and strange quarks.}
 of a multiplet are included, as the absence of members can lead to isospin or $SU(3)$ flavour
 violation at an unphysical rate. All the general-purpose event
 generators include the lightest pseudoscalar, vector, scalar, even and odd charge conjugation
 pseudovector, and tensor multiplets of light mesons. In addition, some excited
 vector  multiplets of light mesons are often included. Usually the mesons containing
 a single heavy quark, or heavy quarks with different flavours,
 from the same multiplets are included, although particularly
 for bottom mesons the properties of the excited mesons are taken from theoretical models
 rather than the PDG. A large number of states containing $c\bar c$ or $b\bar b$ have been
 observed and usually most of these states are included in the simulation, with the
 exception of some recently discovered particles for which the quark model interpretation
 is unclear. While a large number of mesons are normally included, usually only the
 lightest octet, decuplet and singlet baryons are present, although both \herwigpp and
 \Sherpa now include some heavier baryon multiplets. Although a number of
 baryons containing two heavy quarks have been observed these are not generally
 included in the standard generators as their production is rare.

 Having selected the hadrons to use in the simulation, the choice of which decay modes
 to include and how to simulate them is closely related. Consider the example
 of the $a_1$ meson, which decays to three pions, $a_1\to\pi\pi\pi$, where
 the dominant contribution takes place via an intermediate rho meson. If
 we choose
 to use a simple simulation without matrix element or off-shell effects, then
 this decay is best simulated as $a_1\to\rho\pi$ followed by the decay of the
 rho meson to two pions. However, if a matrix
 element for the decay is included it is better to generate the three-body
 decay including
 the effect of the intermediate rho, and other suppressed contributions,
 in the matrix element for the decay.

 Historically, the standard generators included few matrix elements for
 hadron decays and at best used a na\"ive Breit-Wigner smearing of the masses
 of the particles. More sophisticated simulation of hadronic decays
 was then performed using specialized external packages such as
 \evtgen~\cite{Lange:2001uf} for hadron decays and
 \tauola~\cite{Golonka:2003xt,Jadach:1993hs,Jadach:1990mz} for tau decays.
 This still holds true for \pythiaeight, while \herwigpp and \sherpa
 now include much better
 simulation of hadronic, and particularly tau lepton, decays. This was primarily
 motivated by the need to provide a better description of spin effects in tau decays.
 The perturbative production mechanism of the tau can have observable
 effects on its decay properties, which can be used to probe the properties of
 the Higgs boson and particles in BSM models. This is facilitated by using
 the same approach for both the perturbative and non-perturbative
 decays.

  The decays of different types of hadron, and the tau lepton, are simulated
  in a variety of different ways.
  The light mesons and baryons which decay via the weak
  interaction typically have long lifetimes and therefore these decays do not
  need to be simulated in high energy collisions.
  The remaining strong and electromagnetic decays of the
  light mesons are normally simulated using simple matrix elements based
  on parity and charge conjugation invariance.  
  It is important that modes with relatively low branching
  ratios, for example pion Dalitz decay $\pi^0\to e^+e^-\gamma$, are included
  as although they rarely occur for a single particle they can
  contribute significantly
  given the large rate for the production of light mesons.

  The simulation of light baryon decays is often the most primitive part of the simulation,
  particularly in the external decay packages, as these were originally developed
  to simulate events at the B-factories, where baryons are rarely produced. 
  The new hadron decay models  have
  significant improvements for the simulation of baryon decays, typically using
  simple matrix elements based on the relevant conservation laws in the same way as for
  the light mesons.

  While not a hadron, due to its mass the tau lepton primarily decays
  semi-leptonically to a tau neutrino and a small number of light
  mesons.\footnote{The semi-leptonic branching ratio of the tau lepton
  is approximately 65\% with the remaining 35\% being fully leptonic decays.}
  This can be simulated as the decay of the tau lepton to its associated
  neutrino and a virtual $W$ boson. The matrix element can be written as
\begin{equation}
\mathcal{M} = \frac{G_F}{\sqrt{2}}\,L_\mu\,J^\mu,\qquad
L_\mu       = \bar{u}(p_{\nu_\tau})\,\gamma_\mu(1-\gamma_5)\,
        u(p_{\tau}),
\label{eqn:taudecay}
\end{equation}
  where $p_\tau$ is the momentum of the $\tau$ and $p_{\nu_\tau}$ is the momentum of the
  neutrino produced in the decay. The information on the decay products of the
  virtual $W$ boson is contained in the hadronic current, $J^\mu$.
  These currents are calculated either for low-energy effective theories or
  fits to experimental data. The currents for a large number of
  decays, from both modern theoretical models and experimental fits, are included
  in the most recent simulations of
  tau decay~\cite{Golonka:2003xt,Jadach:1993hs,Jadach:1990mz,Grellscheid:2007tt,Gleisberg:2008ta}.
  In some hadron decay models these currents are also used to simulate
  the weak decay of heavy mesons and baryons in the na\"ive factorization
  approximation~\cite{Wirbel:1985ji,Bauer:1986bm}.

  In recent years there has been a lot of interest in the decays of charm and, especially,
  bottom mesons motivated by the study of the CKM matrix and CP violation at the B-factories
  and the Tevatron. This has led to the development of detailed simulations, in particular the
  \evtgen package~\cite{Lange:2001uf}, for these decays. However, this 
  package mainly concentrates on the simulation of $B^0$--$\overline{B}^0$ mixing and rare
  $B$-meson decays that are of interest for the study of CP violating phenomena. 

  While a large number of inclusive decay modes of the weakly decaying mesons containing a 
  single charm or bottom quark have been observed, the branching ratios for these modes
  are insufficient to account for all the decays. The simulation of these decays
  therefore uses a combination of:
\begin{itemize}
\item a number of inclusive, generally low multiplicity, decays
      simulated using either a phase-space distribution, or matrix elements based on 
      na\"ive factorization or experimental fits;
\item partonic decays of the heavy quark, for example $b\to c\ell^-\bar\nu_\ell$, followed
      by the hadronization of the partonic final state including the spectator quark
      using the hadronization models described in
      \SecRef{sec:hadronization} to simulate the remaining observed decay modes.
\end{itemize}
  This approach for the simulation of heavy meson decays 
  is sufficient in most collider physics applications.
  However, the simulation of the oscillations of $B^0_d$ and $B_s^0$ mesons
  and CP violation in $B$-meson mixing and decays are needed for both
  $B$-physics studies and some other applications that are sensitive to 
  mixing phenomena.
  
 \herwigpp and \pythiaeight include the oscillation of neutral
  $B$-mesons using the probability for the meson to oscillate into its antiparticle
  before it decays.
  \sherpa and \evtgen use a more sophisticated simulation including
  CP-violating effects and, for common decay modes of the neutral meson
  and its antiparticle, the interference between the direct decay and
  oscillation followed by decay.

  While a number of decay modes of the weakly decaying charm baryons are known,
  very few weak decays of bottom baryons have been observed and, with the exception
  of the $\Lambda_c^+$, only ratios of the branching ratios are known. The simulation
  of the decays of the weakly decaying heavy baryons therefore uses a 
  very small number of inclusive modes together with partonic decays for
  the majority of the decays.

  A number of excited charm, and in the recent years, bottom mesons
  have been observed, although the properties of a number of the excited
  bottom mesons are uncertain.
  The strong and electromagnetic decays of excited bottom
  and charm mesons are normally
  treated in the same way as the decays of the light mesons, \ie
  using simple matrix elements based on the relevant conservation laws. 

  While a number of charm baryons which decay via either electromagnetic or
  strong interactions have been observed, only the $\Sigma_b$ and $\Sigma_b^{*}$ bottom
  baryons, decaying via the electromagnetic and strong
  interactions respectively, have been observed.
  In general the baryons containing a single heavy quark required to 
  complete the octet and decuplet baryon multiplets are included, although
  for many of the strongly decaying particles the masses and decay modes are based
  on theoretical models or the properties of the corresponding charmed baryons,
  rather than experimental results.

  The decay rates of bottom- and charm-onium resonances to $\ell^+\ell^-$ and various
  partonic final states can be computed in terms of the quarkonium wavefunction, which 
  is calculated in various models. As knowledge of the wavefunction is only needed to
  compute the width, which is taken from experimental results in event generators, the
  matrix elements can be used to simulate the decays of quarkonium states. In practice
  the simulation of the exclusive decays of these resonances is usually supplemented
  with the inclusion of a number of observed low multiplicity decay modes in
  a similar way as for weakly decaying charm and bottom hadrons.

  \evtgen, \herwigpp and \sherpa include spin correlations between different decays
  in all hadron decays
  where matrix elements are used to calculate the distributions of the decay
  products, whereas \pythiaeight only includes correlations in certain decay chains.
  All the simulations include at least the generation of the masses of
  unstable particles according to the Breit-Wigner distribution, with improvements
  in some simulations for particles where new decay modes become kinematically
  accessible close to the particle's mass.

In summary:
\begin{itemize}
\item the simulation of hadron decays is based on a combination of
      experimental results and theoretically motivated assumptions which
      are required in order to generate exclusive events;
\item the modern simulations of hadron decay are sophisticated, including matrix elements
      for many modes and spin correlations;
\item given the close relationship between the hadron decay and hadronization models
      care should be taken when changing the hadron decay model, unless the
      hadronization parameters are retuned.
\end{itemize}

% Local Variables:
% mode: LaTeX
% TeX-master: "../mcreview"
% End:

\mcsection{QED radiation}
\label{sec:qed-radiation}
The simulation of electromagnetic radiation in general-purpose event generators
uses one of two approaches. The most common is to use the same parton
shower algorithm that was used for the simulation of QCD radiation. Indeed this 
is the preferred option for processes where the emission of
both QCD and QED radiation is possible. The simulation
of QED radiation proceeds in a similar way as for QCD radiation, with the evolution
partner selected according to the charge, rather than the colour flow. This
can cause problems in some processes where there are destructive contributions
that would be suppressed by $1/\Nc^2$ in QCD, but which are leading in QED.
Despite these problems this is the most common approach in Monte Carlo simulations
as both QED and QCD radiation can be generated at the same time.
This interleaving of both types of radiation in one shower gives
interesting phase space competition effects and could be used
to shed light on the parton shower ordering
variable\cite{Seymour:1994bx}.

An alternative to the parton shower is the Yennie-Frautschi-Suura (YFS) formalism~\cite{Yennie:1961ad}
which proceeds by exponentiating the full eikonal distribution for
soft photon emission, below a cut-off,
together with the corresponding virtual corrections, given by the YFS form factor.
In this approach, starting with the production of
$n$ particles, the cross section with the radiation of
an additional $n_\gamma$ photons can be written as 
\begin{eqnarray}
\sigma  
&=& 
\displaystyle{\frac{(2\pi)^4}{2\hat{s}}
\prod_{i=1}^{n} \frac{{\rm d}^3p_i}{(2\pi)^32p_i^0}
|\overline{\mathcal{M}}|^2
\delta^4\left(l_1+l_2-\sum_{i=1}^np_i-\sum_{i=1}^{n_\gamma}k_i\right)}
\label{eqn:YFSmaster} \\
&&\displaystyle{\times
\sum^\infty_{n_\gamma=0}\frac1{n_\gamma!}\prod_{j=1}^{n_\gamma}\int 
\frac{{\rm d}^3k_j}{k_j^0}\tilde{S}_{\rm total}(k_j)}
 e^{Y_{\rm total}(\Omega)},
\nonumber
\end{eqnarray}
where $p_i$ are the momenta of the outgoing particles, $k_i$ are those
of the outgoing photons, $l_i$ those of the incoming partons and
 $|\overline{\mathcal{M}}|^2$ is the 
spin summed/averaged matrix element for the leading-order process.
  The total dipole radiation function is
\begin{equation}
\tilde{S}_{\rm total}(k) = \sum_{i=0}^n\sum_{j=1,j>i}^n \frac{\alpha Z_i\theta_iZ_j\theta_j}{4\pi^2}
\left(\frac{p_i}{p_i\cdot k}-\frac{p_j}{p_j\cdot k}\right)^2,
\end{equation}
  where $Z_{i,j}$ is the charge of the $i,j^{\rm th}$ particle in units 
  of the positron charge and \mbox{$\theta_{i,j}=+1(-1)$} if the $i,j^{\rm th}$ particle is
  outgoing~(incoming).

 The total YFS form factor~\cite{Yennie:1961ad}, $Y_{\rm total}(\Omega)$,
  is a sum of contributions from pairs of charged particles:
\begin{equation}
Y_{\rm total}(\Omega) = \sum_{i=0}^n\sum^n_{j>i} Y_{ij}(p_i,p_j,\Omega),
\end{equation}
  where $\Omega$ is used to symbolically indicate the dependence
  on the infrared cutoff on the photon energy.
 The YFS form factor for a pair of charged pairs is given by
\begin{equation}
Y_{ij}(p_i,p_j,\Omega) = 2\alpha
\left(\mathcal{R}e B_{ij}(p_i,p_j)+\tilde{B}_{ij}(p_i,p_j,\Omega)\right).
\end{equation}
  The real emission piece, $\tilde{B}_{ij}$, is
\begin{equation}
\tilde{B}_{ij}(p_i,p_j,\Omega) =
\frac{Z_i\theta_iZ_j\theta_j}{8\pi^2}\int_0^{\left|\mathbf{k}\right|<\omega}
\frac{{\rm {d}}^3k}{\left|\mathbf{k}\right|}
\left(\frac{p_i}{k\cdot p_i}-\frac{p_j}{k\cdot p_j}\right)^2,
\end{equation}
  where $\omega$ is the upper limit on the photon energy.
  The virtual piece does not depend on the cutoff and is given by
\begin{equation}
B_{ij}(p_i,p_j) = -\frac{iZ_i\theta_iZ_j\theta_j}{8\pi^3}
\int {\rm d}^4k \frac1{k^2}
\left(
\frac{2p_i\theta_i-k}{k^2-2k\cdot p_i\theta_i}+
\frac{2p_j\theta_j+k}{k^2+2k\cdot p_j\theta_j}\right)^2.
\end{equation}

The standard technique to generating photons according
to \EqRef{eqn:YFSmaster} works in two stages.
First, the distribution is generated according to the leading-order result 
in which each photon is produced independently.
A correction weight is then applied in order to give exactly the
distribution in \EqRef{eqn:YFSmaster}.
The major advantage of this technique is that because the distribution
used to generate the additional photons is known analytically, higher
order corrections can be included exactly. It is this feature which allowed the construction of high precision 
Monte Carlo simulations for LEP 
physics~\cite{Jadach:1988gb,Jadach:2000ir,Jadach:1999vf,Placzek:2003zg,Jadach:2001mp,Jadach:2001uu}
and is included in \sherpa for initial-state photon radiation in lepton collisions~\cite{Schalicke:2002ck}.

In the general-purpose event generators the parton shower
approach is used in the majority of perturbative processes, where
both QCD and QED radiation must be generated. However, both
\herwigpp~\cite{Hamilton:2006xz} and \sherpa~\cite{Schonherr:2008av}
include the simulation of QED radiation using the YFS formalism
in cases where no QCD radiation is possible, \ie
for the leptonic decays of $W^\pm$ and $Z^0$ bosons, hadron and tau decays. 
In particular, the latter two applications simplify the decay tables
considerably since many decay modes are produced by adding photons to
simpler modes.
In the previous generation of Monte Carlo simulations the production
of QED radiation in particle decays was normally simulated using an interface to the PHOTOS 
program~\cite{Barberio:1993qi,Barberio:1990ms,Golonka:2005pn}. This program
is based on the collinear approximation for the radiation of 
photons together with corrections to reproduce the correct result in the soft
limit~\cite{Barberio:1993qi,Barberio:1990ms}. Recently it has been improved 
to include the full next-to-leading order QED corrections for certain 
processes~\cite{Golonka:2005pn}. However, given the inherent problems
with interfacing to external programs, the superior accuracy of the YFS
formalism and the ability to systematically improve it, 
in \herwigpp and \sherpa the YFS approach is preferred.

In summary:
\begin{itemize}
\item QED radiation can be simulated using either a parton shower or
      YFS based approach;
\item historically the parton shower approach has been more common in
      general-purpose event generators and is still used when both QED and
      QCD radiation is possible;
\item both \herwigpp and
      \sherpa now use the YFS formalism for the simulation of
      QED radiation in particle decays.
\end{itemize}
% Local Variables: 
% mode: LaTeX

\mcsection{BSM in general-purpose generators}
\label{sec:bsm-general-purpose}
  We do not know what kind of physics beyond the Standard Model may be encountered
  at the LHC;  if any is found, a variety of new physics models will need to be considered in 
  order to determine its exact nature. Despite the large number of
  models, they can be split into two broad 
  classes:\footnote{There are some scenarios such as Little 
                    Higgs~\cite{ArkaniHamed:2002qx,ArkaniHamed:2002qy}
                    and Leptoquark models which are intermediate 
                    between the two cases with only a small number of
                    additional particles.}
\begin{enumerate}
\item models that contain either new effective operators which 
      modify the cross sections and distributions for Standard Model
      processes, or only a few new particles which are generally
      produced as resonances, \eg the 
      ADD~\cite{ArkaniHamed:1998rs,Antoniadis:1998ig} or
      Randall-Sundrum~\cite{Randall:1999ee} extra-dimensional models;
\item models that contain a large number of new particles, often new partners
      for each Standard Model particle, which can be produced in a variety of
      ways at the LHC and then decay, for example the Minimal Supersymmetric
      Standard Model~(MSSM), Universal Extra Dimension
      models~(UED)~\cite{Appelquist:2000nn,Cheng:2002ab} or
      Little Higgs models with T-parity~\cite{Low:2004xc,Hubisz:2004ft}.
\end{enumerate}
  In general the first class of models are relatively simple to simulate with only
  minor changes to the Standard Model production processes that are
  present in all general purpose event generators. The simulation of the second
  class of models is more complicated.
  There are two approaches that 
  have been adopted to simulate these models:
\begin{enumerate}
\item the production of the new heavy particles is simulated first, usually using a
      leading-order $2\to2$ scattering process, followed by the subsequent decay of
      the heavy particles, which often leads to long decay chains as the 
      heavier BSM particles cascade decay into lighter ones;
\item a high multiplicity matrix element including all the final-state partons
      is used to simulate the process including the decays of any unstable
      heavy particles.
\end{enumerate}
  The first approach has the advantage of both computational simplicity
  and being able to easily simulate the fully inclusive BSM signal. However,
  while the second approach is more computationally expensive it has the advantage
  of correctly treating unstable intermediate particles and any correlation effects.

  Methods have therefore been developed to allow all the correlation 
  effects to be retained, in the approximation that only resonant diagrams
  are included and all interferences are neglected, while still simulating
  the production and decay of heavy particles 
  separately~\cite{Collins:1987cp,Knowles:1987cu,Knowles:1988hu,Knowles:1988vs,Richardson:2001df}. Generally
  when using such methods the masses of the heavy particles
  are smeared using the Breit-Wigner distribution, although more sophisticated
  techniques have been developed~\cite{Gigg:2008yc}. Currently \herwigpp and \pythiaeight use
  the first of the above approaches, with \herwigpp using the methods of
  Refs.~\cite{Richardson:2001df,Gigg:2007cr,Gigg:2008yc} 
  to include spin correlations and off-shell effects.
  This also has the advantage that QCD radiation from new coloured
  particles can be simulated more easily.
  As \sherpa includes
  a sophisticated matrix element generator, it currently uses the second approach.

  Historically models of new physics were implemented directly in the Monte Carlo
  event generators by hard coding the production and decay matrix elements.
  In recent years this has changed, with both \sherpa and \herwigpp using a method
  where the production processes and decays are automatically calculated from
  the Feynman rules, for arbitrary processes in \sherpa and for $2\to2$ scattering
  processes and $1\to2$ or 3 decays in \herwigpp. It has also become increasingly
  common to use an external matrix element generator interfaced via the Les
  Houches Accord~\cite{Boos:2001cv,Alwall:2006yp}
  to simulate the hard scattering process. The most recent 
  development is the \FeynRules~\cite{Christensen:2008py} package which
  can automatically calculate the Feynman rules in a given model from the Lagrangian
  in a form that can be used by a matrix element generator. \sherpa already uses
  this approach to allow a large range of models to be simulated and work is
  in progress to use it with \herwigpp.

  While in most cases models of new physics only require the simulation of
  the hard process, any subsequent decays, and the QCD radiation from the heavy particles,
  recently a number of more exotic models have been proposed where
  the nature of the new physics leads to changes in other parts of the Monte
  Carlo simulation. In general this occurs when the new model involves colour 
  structures which do not occur in the Standard Model. 
  Three situations have arisen.

  Firstly, in R-parity violating SUSY models baryon number can be violated
  by a new operator which 
  couples three particles in the fundamental, or anti-fundamental, representation
  of $SU(3)_C$ via the total antisymmetric tensor $\epsilon^{ijk}$. This
  can be considered as a junction where three
  colour lines meet. This presents a problem both in the selection of 
  the colour partners for the parton shower evolution and later in the hadronization
  stage. The simulation of these models, with the angular-ordered parton
  shower and cluster hadronization model~\cite{Gibbs:1994cw,Dreiner:1999qz} and
  in the string model~\cite{Sjostrand:2002ip}~(see \SecRef{sec:string-model}),
  has been studied in detail with
  the selection of the colour partners for the perturbative radiation being done 
  at random from among the potential partners, and after the shower a 
  special treatment of three partons colour connected to the 
  junction.
  
  Secondly, in hidden valley
  models~\cite{Strassler:2006im,Strassler:2006qa} there are particles
  that are charged under a new strongly interacting gauge group.
  In these models, radiation of the gauge bosons of the new strong force
  by the new particle must be simulated both in the parton shower phase,
  as in~\cite{Carloni:2010tw}, and in the subsequent hadronization.
  
  Finally, some models have recently been proposed in which there
  are particles in representations of the $SU(3)_C$ group of the strong
  force other than those we know how to simulate (the fundamental and adjoint),
  for example particles in the sextet 
  representation~\cite{Chen:2008hh,Han:2009ya,Berger:2010fy}. The simulation
  of these particles is not currently possible in any of the general-purpose
  event generators.

  In summary:
\begin{itemize}
\item most BSM models can be simulated either by incorporating changes to 
      Standard Model production processes or by adding the production
      and decay of the new particles in the specific model;
\item in general the production and decay of new particles are simulated
      separately in order to generate exclusive production processes;
\item if new colour structures are present the parton shower and
      hadronization phases must also be modified.
\end{itemize}
 
% Local Variables: 
% mode: LaTeX
% TeX-master: "../mcreview"
% End: 

\mcpart{Specific reviews of main generators}
\label{sec:spec-revi-main}
In this part we briefly review the MCnet event generators,
referring back to \PartRef{sec:revi-phys-behind} for the physics
involved and the modelling options implemented in them.  The first to
be discussed, \ariadne, has proved highly successful for $e^+e^-$ and
$ep$ physics but is still under development for hadron-hadron
collisions, as explained in \SecRef{sec:ariadne}.  The next two,
\herwigpp and \pythiaeight, are new C++ generators based on earlier
Fortran programs, while \sherpa is a wholly new C++ generator; all of
these three are already extensively used for LHC physics, although
still being actively improved as discussed in their respective Sections.

\mcsection{Ariadne}
\label{sec:ariadne}
\mcsubsection{Introduction}%

The \Ariadne program \cite{Lonnblad:1992tz} was the first parton
shower generator to implement a dipole cascade. It uses the colour
dipole model by Gustafson \lletal
\cite{Gustafson:1986db,Gustafson:1987rq,Andersson:1988gp}, where gluon
emissions are modelled as coherent radiation from two colour-connected
partons.

For final-state radiation the \Ariadne cascade is rather similar to
any other dipole-based cascade, such as the ones described in
\SecsRef{sec:pythia8} and~\ref{sec:sherpa}. In \llee annihilation
into quarks, the first gluon emission is given by a dipole splitting
function identical to the exact differential cross section for
$e^+e^-\to q\bar{q}g$. Hence the matrix element matching
  described in \SecRef{sec:matching-first-ps} is automatically
  included.  The next emission will then either come from the dipole
between the quark and the gluon or from the dipole between the gluon
and the antiquark, with a trivial generalization for subsequent
emissions. The only difference in the subsequent emissions is the
colour factors and the non-singular behaviour of the splitting
functions, which in the soft and collinear limits coincide with the
standard Altarelli--Parisi splitting functions in \EqRef{DGLAP}.

The recoils in the emissions are taken by both emitting partons in the
radiating dipole. In this way all partons can be put on-shell in each
step of the cascade. The way the recoils in the transverse direction
are distributed between the radiating partons differs somewhat between
different types of dipoles. For example in quark--gluon dipoles, the quark
takes the full transverse recoil, so that the neighbouring dipole on
the gluon side is minimally disturbed, while for a gluon--gluon dipole
the transverse recoil is distributed so as to minimize the sum of the
squared transverse momenta of the emitters. Apart from these recoils,
all dipoles are treated independently.

The emissions are ordered (using appropriate Sudakov form factors) in
a Lorentz-invariant transverse momentum defined as
\begin{equation}
  \label{eq:ariadne-pt}
  \llipt^2=\llsdip(1-x_1)(1-x_2),
\end{equation}
where \llsdip is the squared invariant mass of the dipole, and
$x_i=2E_i/\sqrt{\llsdip}$ are the scaled energies of the partons after
the emission in the dipole rest frame. The transverse momentum is also
used as the scale in \alphaS. With an invariant definition of rapidity
\begin{equation}
  \label{eq:ariadne-y}
  \llirap=\frac{1}{2}\ln\frac{1-x_1}{1-x_2}
\end{equation}
it can easily be shown that the dipole splitting function is well
approximated by $\llD(\llipt,\llirap)\propto d\llirap\, d\ln\llipt$,
so apart from the running of \alphaS, the emission probability is
essentially flat in the $(\ln\llipt,\llirap)$ plane, where the
available phase space is given as an approximately triangular region,
$\ln(\llipti{\max}/\llipt)\lesssim|\llirap|$.

For final-state emissions, \Ariadne also includes the
$\llg\to\llqrk\llqbar$ splitting, by simply dividing the normal
Altarelli--Parisi splitting function between the two dipoles to which
the gluon is connected \cite{Andersson:1989ki}.

\mcsubsection{Hadronic collisions}%
What makes \Ariadne truly unique is the handling of radiation in
collisions where there are incoming hadrons. In a normal parton shower
one would then apply a backwards evolution of initial-state
splittings, and in more recent dipole shower implementations such as
those in \pythiaeight and \sherpa (see \SecsRef{sec:pythia8} and
\ref{sec:sherpa}) dipoles are defined between \eg incoming and
outgoing partons in the hard interaction. The \Ariadne program, in
contrast, uses the so-called Soft Radiation Model
\cite{Andersson:1988gp}, where there are dipoles between the hadron
remnants and the partons from the hard interactions.

Consider the process of deeply inelastic \llep scattering. We can view
it as a quark being kicked out of the proton by the virtual
photon. The quark carries colour, while the corresponding anti-colour
is continuing with the proton remnant down the beam pipe. From a
semi-classical viewpoint we then would have a large dipole spanned
between the struck quark and the proton remnant, and we could argue
that this dipole would radiate gluons in the same way as a dipole
between a \llqrk and \llqbar in an \llee-annihilation.

The difference between \llee and \llep is that in the former case the
emitting \llqqbar-pair is essentially point-like, while the proton
remnant in the \llep case is an extended object with about the same
size as the proton itself. And, just as the emission of
small-wavelength photons from an extended electric dipole antenna is
suppressed, one can argue that high-\llipt emission of gluons in the
proton direction should be suppressed \cite{Andersson:1988gp}. In
\Ariadne this is implemented by assuming that in any emission from a
dipole connected to a hadron remnant, only a fraction
\begin{equation}
  \label{eq:ariadne-softsup}
  a=\left(\frac{\mu}{\llipt}\right)^\alpha
\end{equation}
of the remnant energy is available. Here $\mu$ is the inverse
(transverse) size of the remnant (typically around 1~GeV), and
$\alpha$ is a parameter related to the dimensionality of the remnant
(1 would correspond to a string-like remnant, and 2 to a disc --- the
default value is 1). This gives a sharp cutoff in the phase space
allowed for gluon radiation, but optionally also some emission outside
this region is allowed with a power suppressed tail (in \llipt).

One can relate this suppression to the ratios of parton densities
which enter the initial-state splittings in a conventional backward
evolution shower (\cf \EqRef{eq:matching-sudnoem});
 however, especially in the remnant direction at
small-$x$, the suppression in the \Ariadne case is much less
severe. This shows up very distinctly in the case of forward jet rates
at HERA (see \eg \cite{Aktas:2005up}), where \Ariadne gives a much
higher jet rate than conventional cascades, in better agreement with
data. Comparing the Sudakov form factors one can see that the dipole
shower in \Ariadne resums some large logarithms of $1/x$, similarly to
what is done in BFKL evolution \cite{Rathsman:1996jc}.

There is one additional peculiarity in \Ariadne related to the proton
remnant. As only a part of the remnant takes part in the emission,
only that part of it will receive a recoil. This means that there will
be an extra gluon produced which is given some transverse
momentum. This gluon will, however, also be suppressed by an
additional Sudakov form factor corresponding to the probability that
no standard emission would produce a higher \llipt.

In \eg \llW boson production in hadronic collisions, where there are no
final-state coloured partons in the hard subprocess, the initial
dipole will be spanned between the two remnants. In this case the
recoil from emissions must also be shared with the \llW boson, which
is done in a way described in \cite{Lonnblad:1995ex}.

As in the final-state cascade, the initial-state $\llg\to\llqqbar$
splitting is not naturally described in terms of dipole
radiation. Instead this is included as a standard backward-evolution
step in a conventional initial-state shower. Also, the initial-state
emission of a quark in a $\llqrk\to\llg\llqrk$ splitting may be
included in the same way.

\mcsubsection{The \Ariadne program and the LHC}

The \Ariadne program was initially written in Fortran, to be run
together with the Fortran version of \Pythia, simply replacing the
\Pythia parton shower with the dipole cascade. In principle it can be
used to generate \lhc events, but some care must be taken when
generating processes with incoming gluons, such as Higgs production,
since the initial-state emission of quarks in the $\llqrk\to\llg\llqrk$
splitting mentioned above was not implemented in the Fortran version.

The \Ariadne program is currently being rewritten in \Cpp using the
\thepeg framework \cite{Lonnblad:2006pt} (see section
\SecRef{sec:thepeg}) and should soon be publicly available for
generating dipole cascades for any Standard Model process. The aim is
that it then will also be easily merged with matrix element generators
using the CKKW-L algorithm (see
\SecRef{sec:matching-at-tree}). Possibly it will also include
NLO-merging (see \SecRef{sec:nlo-merging}), but this is conditional on
whether the concept of recoil gluons, described briefly above, can be
reformulated in a way that is compatible with a proper \alphaS
expansion of the dipole emissions.

Finally it should be noted that the \Cpp version of \Ariadne is also
used in conjunction with a new initial-state evolution model
\cite{Avsar:2005iz,Avsar:2006jy,Avsar:2007xg} based on Mueller's
dipole evolution \cite{Mueller:1993rr,Mueller:1994jq,Mueller:1994gb}
formulated in impact-parameter space. The new program, called \llDIPSY
\cite{Flensburg:2010xx} is mainly intended to generate soft QCD
events, and can do so for both hadron collisions and heavy ion
collisions.

% Local Variables: 
% mode: LaTeX
% TeX-master: "../mcreview"
% End: 

\mcsection{\herwigpp and \thepeg}
\label{sec:herwig++--thepeg}
\mcsubsection{\gensectionintro}

Historically, \herwigpp is based on the event generator \Herwig 
(Hadron Emission Reactions With Interfering
Gluons), which was first published in 1986 \cite{Marchesini:1987cf} and was
developed throughout the era of LEP, with the latest major release
version 6.5.10 \cite{Corcella:2000bw,Corcella:2002jc} in
2005\footnote{Version 6.5.20, released in 2010, contains bug fixes.}. From the
beginning it has featured angular ordered parton showers to take colour
coherence effects into account.  The cluster hadronization model it uses
(\SecRef{sec:cluster-model}) was developed at the same time.

\herwig was written  in \fortran, but
with the advent of the LHC it was decided to freeze its development and
develop a new generator, with the same strengths as the old program, in
\cpp.  The idea was to not just rewrite the generator but to introduce
physics improvements whenever they seemed necessary and feasible.  The
new generator, \Herwigpp, was first released only for $e^+e^-$
annihilation in 2003 \cite{Gieseke:2003hm}.  Since then it was
further developed into a complete event generator for collider
physics \cite{Bahr:2008tf}, with the current version 2.4.2 released in 2009.
The code and its physics features are fully documented in the main
reference \cite{Bahr:2008pv}, which will be updated as the code
develops continually. Some distinctive physics features of \Herwigpp are:
\begin{itemize}
\item Automatic generation of hard processes and decays with full spin
  correlations for many BSM models.
\item Matching of many hard processes at NLO with the POWHEG method
  built in. 
\item Angular ordered parton showers. 
\item Cluster hadronization.
\item Sophisticated hadronic decay models, particularly for bottom
  hadrons and $\tau$ leptons. 
\item Hard and soft multiple partonic interactions to  model the
  underlying event and soft inclusive interactions. 
\end{itemize}
We will describe the most important details of the physics models in the
remainder of this section.

\mcsubsection{ThePEG}
\label{sec:thepeg}

\Herwigpp is distributed as a comprehensive collection of plugin
modules to \thepeg, the Toolkit for High
Energy Physics Event Generation \cite{Lonnblad:2006pt}. 
\thepeg provides all the infrastructure that
is necessary to construct an event generator, handling \eg random
number generation, the event record, tuneable parameter settings, and
most importantly, a mechanism to plug in physics implementations for
all steps of event generation. \Herwigpp provides such a set of
plugins and comes with several complete generator setups for $e^+e^-$,
$ep$ and hadron-hadron collisions. 

\thepeg's core component is the Repository, which holds the relations
between all the different modules involved in a generator run and
their tuneable parameter settings. It can be controlled through a
simple command language in plain text, which is used to set up the
modules involved in a generator run.  Using such files at run time,
the user can override any of the default parameters that \Herwigpp
comes with; no recompilations are necessary to change parameters, or to
switch between physics models, different matrix elements or analyses.

\thepeg provides a reader for the Les Houches Accord event format
\cite{Alwall:2006yp} to read in parton-level events for further processing, an
output module for HepMC events \cite{Dobbs:2001ck}, as well as a native
interface to Rivet \cite{Buckley:2010ar}, which avoids the overhead of having
to pipe events through text files. Additionally, \thepeg can be linked
to LHAPDF \cite{Whalley:2005nh} to get direct access to any PDF sets that are
available there.

The repository plugin structure allows for easy inclusion of user-defined
modules. Any \cpp object that inherits from the respective base
classes in \thepeg can be used transparently in addition to, or instead
of, one of the default plugins. Any user code can be loaded at
runtime as dynamically linked libraries. This allows modification
of the program's behaviour without having to recompile the 
main program or needing to edit the core libraries.
They can therefore always be installed centrally, possibly as part of a larger framework.

\mcsubsection{\gensectionhard} 
Three main mechanisms for
simulating hard processes are available 
in \herwigpp. First, there is a large set of
hand-coded matrix elements for the most common subprocesses for hadron,
lepton and DIS collisions. They are written using a reimplementation
of the HELAS helicity amplitude formalism \cite{Murayama:1992gi}, which allows
the spin correlations to be carried forward to the remaining event
simulation consistently. Second, \Herwigpp also contains a generic
matrix element calculator for $2 \to 2$ processes, mainly used for BSM
physics, which automatically determines the permitted diagrams for a
set of given external legs from a list of active vertices. The third
source of hard subprocesses is \thepeg's Les Houches reader, which
allows parton-level events with any number of legs to be read from
external sources.

%%% \paragraph{diffraction}

For several processes, \Herwigpp incorporates the full NLO corrections in
the parton shower \cite{Hamilton:2008pd,Hamilton:2009za,Hamilton:2010mb}
using the POWHEG formalism (\SecRef{sec:powheg}). An implementation of
the CKKW merging scheme for tree-level multi-jet events
(\SecRef{sec:matching-at-tree}) will be included in an upcoming
release \cite{Hamilton:2009ne}.

\mcsubsection{BSM physics}\label{sec:hwbsm}
The simulation of BSM physics in \herwigpp
\cite{Gigg:2007cr,Gigg:2008yc} makes extensive use of \thepeg's plugin
architecture. Each model is implemented in a model class, which
holds the relevant new parameters, and a list of Feynman rules for its
vertices. Based on this information, all possible production and decay
matrix elements with up to four external legs are constructed and can
be selected in the text-based input files. 
\Herwigpp currently provides models to simulate processes in the MSSM
\cite{Haber:1984rc,Gunion:1984yn} and NMSSM \cite{Ellwanger:2009dp}
scenarios with an SLHA \cite{Skands:2003cj,Allanach:2008qq}
 file reader to provide the relevant parameters,
a model for universal extra dimensions
\cite{Cheng:2002iz,Appelquist:2000nn},
an implementation of
Randall-Sundrum \cite{Randall:1999ee}
and ADD-type gravitons \cite{ArkaniHamed:1998rs,Antoniadis:1998ig},
 as well as a model of transplanckian
scattering \cite{Giudice:2001ce}.

The production and decay matrix elements are all calculated using
helicity amplitude techniques so that spin correlations between the
production and decay of unstable particles can be generated using the
approach of \cite{Richardson:2001df}, as described in
\SecRef{sec:bsm-general-purpose}.  This ensures that \Herwigpp can
generate the spin correlations for individual decay chains in a
computationally efficient way, while still allowing the simulation of
inclusive BSM signals. The efficiency comes at the expense of
neglecting interference effects with other decay chains leading to the
same final state.  Off-shell effects -- including the suppression of
decay modes close to threshold -- are simulated using the approach of
\cite{Gigg:2008yc}, which includes the running width of the unstable
particle in the denominator of the Breit-Wigner propagator and in the
calculation of the production matrix element for the particle.

\mcsubsection{\gensectionshower}

The parton shower in \Herwigpp is based on a new evolution variable
$\tilde q$ \cite{Gieseke:2003rz}, motivated from the branching of gluons
off heavy quarks \cite{Catani:2002hc}.  This is one of the possible
choices in \EqRef{eq:evolchoices}.  As in \Herwig, the evolution in this
variable ensures the angular ordering of parton shower emissions,
to take colour coherence effects into account 
(see \SecRef{parton-shower:initial-conditions}). In addition to the treatment
of mass effects in the splitting functions and the showering of
coloured particles in BSM models,
 the shower differs from \Herwig's implementation in the way it
 fills the
available phase space for emissions.  Considering only the first gluon
emission from a $q\bar q$ pair, the new
variable fills the soft gluon emission region of phase space 
without any overlap between the parton showers. 

A so-called dead region is still present in the phase space, as in
\Herwig, but is filled by either a hard matrix element correction or by
higher order emissions.  Potential discontinuities in the emission phase
space at the transition from the parton shower to the hard emission
region are avoided by applying a so-called soft matrix element
correction: the emission rates in the parton shower overestimate the
rates one would obtain from a full matrix element calculation. For each
parton shower emission, the overestimated rate is then corrected down to
the matrix element by a veto (see \AppRef{mcmethods:veto}) 
which reflects the relative emission
probability between parton shower and matrix element, respectively.  In
\Herwigpp there are matrix element corrections for $e^+e^-\to q\bar q$,
Drell-Yan production of vector bosons in hadronic collisions, $gg\to
h^0$ and for top decays.  In order to fill the phase space smoothly, it
should be noted that the starting scales of the parton shower are
adjusted to the values that are given by the requirement of
colour coherence, see
\EqRef{eq:psinitialconditions}.  In \herwigpp, these
initial conditions cannot be
altered, \eg by raising the initial evolution scale. 
Before the parton shower generates emissions, all heavy unstable
particles in the partonic
final state, which typically have a very narrow width, are decayed.  
All intermediate coloured lines are then also showered.
These decays are done for the Higgs particle,
electroweak gauge bosons, top quarks, and also \eg
supersymmetric particles (see \SecRef{sec:hwbsm}).

\mcsubsection{\gensectionMPI}
\label{sec:herwigmpi} 

The default model for the simulation of
underlying
event physics is a model for multiple partonic interactions
\cite{Bahr:2008dy}.  Both hard and soft multiple partonic interactions
are taken into account.  The hard interactions are modelled as hard QCD
$2\to 2$ scatters with a transverse momentum above the cutoff value
$\pt^{\rm min}$.  The hard scattering centres are thought to be
spatially distributed within the proton similarly to the charge,
as measured in elastic electron--proton scattering, leading to the
dipole form factor.  However, a different width for the
distribution of colour charges, quarks and gluons, parameterized by the inverse
radius $\mu^2$, is allowed. 
In the region $0<\pt<\pt^{\rm min}$, soft scatters are generated
with a transverse momentum distribution that has a Gaussian form, has an
integral given by the soft parton-parton cross section, and is
continuously matched with the perturbative distribution at $\pt=\pt^{\rm min}$
\cite{Borozan:2002fk}. These constraints are sufficient to uniquely specify
this distribution. The soft partons' longitudinal momentum
distribution is taken to be like that of the low-energy sea,
\ie $\xpdf{}(x) \approx \text{flat}$.
The spatial distribution of soft colour charges, given by a
parameter $\mu_{\rm soft}^2$, is allowed to be different from the hard ones as
otherwise it was not possible to obtain a model that is consistent with Tevatron
data on multiple scattering \cite{Bahr:2008wk}.  The soft cross
section and $\mu_{\rm soft}^2$ are fixed from measurements of the total
cross section and the elastic slope parameter, if available, or
parameterizations of them otherwise.

After initial studies, good agreement with Tevatron underlying event
measurements from Run I and II were found \cite{Bahr:2008dy}.  With
the availability of first LHC data, e.g\ \cite{Aad:2010rd,Atlas:2010xx}
it became clear that the model suffered from the lack of a colour
reconnection mechanism, which will be included in an upcoming release.
It gives a very satisfactory description of these hard underlying event
data.

\mcsubsection{\gensectionhadronize}

\Herwigpp uses the cluster hadronization model, described in
\SecRef{sec:cluster-model}.  Its first step is a non-perturbative gluon
splitting, where each gluon splits
isotropically in its rest frame into a $q\bar q$ pair of
one of the three lightest flavours. We stress that at this stage all
partons are treated as non-perturbative objects and acquire a
constituent mass.  The value of the gluon mass in particular is one of
the important model parameters.  After cluster formation we are left
with a small number of heavier clusters of mass $M$, that will fission
in binary sequential decays, whenever the condition
\begin{equation}
  \label{eq:clfission}
  M^p \geq M_{\rm max}^p +(m_1+m_2)^p
\end{equation}
is fulfilled, where $m_{1, 2}$ are the masses of the constituent partons
of the cluster and $M_{\rm max}$ and $p$ are the main parameters of the
cluster hadronization model, chosen independently for light,
charmed and bottom clusters.  Once a cluster is split, a new
particle-antiparticle pair of quarks or diquarks is taken out of the
vacuum, chosen with adjustable weights.  The kinematics of the new clusters
preserve the original directions of the constituent particles and depend
on whether they contain a perturbative parton or a beam remnant.  Once
clusters fall below the limit of \EqRef{eq:clfission}, they
decay isotropically in their rest frames 
into pairs of hadrons.  The hadron species
are determined according to available phase space and phenomenological
weights for flavour multiplets.  As heavier baryons tend to be
suppressed in this approach \cite{Kupco:1998fx}, the choice between a
baryonic or non-baryonic decay is made before the hadron
species are selected.  In some cases clusters will turn out to be too
light to decay into a pair of hadrons; they will decay into a
single light particle instead and share some momentum with a 
cluster close by in spacetime.
For any cluster that contains a parton from the original hard
process, \eg a bottom quark, the resulting heavy meson retains the
original parton direction in the cluster rest frame, up to some Gaussian
smearing.

In addition to the hadronization of partonic final states, the
implementation of the model in \Herwigpp can also handle stable coloured
particles or baryon number violating vertices, which both occur in BSM models.

As with all tuneable parameters, a detailed list and description can be found
in the manual \cite{Bahr:2008pv}.

\mcsubsection{\gensectiondecay}

The decays of both fundamental particles and unstable hadrons in
\Herwigpp are modelled in the same framework, using either a general
matrix element based on the spin structure of the decay, or with a
specific matrix element for important decay modes, with a particular
emphasis on baryon decays. This allows for a
sophisticated treatment of off-shell effects, the treatment of excited
baryonic multiplets, and for example the easy
integration of the semileptonic $\tau$ lepton
decays~\cite{Grellscheid:2007tt}. Spin correlation effects are
included fully for the decays of all unstable
particles~\cite{Richardson:2001df} and are
consistent with the preceding stages of event generation all the way
back to the production matrix element. QED radiation in decays is
simulated using the YFS formalism
\cite{Hamilton:2006xz} (see \SecRef{sec:qed-radiation}). 
All the decay matrix elements have been extensively tested against
external packages where available, and are in full agreement. 

The particle properties such as masses, widths, lifetimes, decay modes
and branching ratios that are used in \Herwigpp can be found
in the online interface to its database of particle properties at
\begin{center}
\href{http://www.ippp.dur.ac.uk/~richardn/particles/}{\tt
  http://www.ippp.dur.ac.uk/$\sim$richardn/particles/}
\end{center}

\mcsubsection{Outlook}

\Herwigpp in its current version (2.4.2) is 
superior to its \fortran predecessor in almost all aspects of
physics simulation and is the recommended version for new
studies. Once this is true without any exceptions the version 
number will move to 3.0.  The program package, together
with \thepeg can be found at 
\begin{center} 
\texttt{http://projects.hepforge.org/herwig/}
\end{center}

% Local Variables: 
% mode: LaTeX
% TeX-master: "../mcreview"
% End: 

\mcsection{\pythiaeight}
\label{sec:pythia8}

\mcsubsection{\gensectionintro}

\pythia is a general-purpose event generator. It has been used 
extensively for $e^+e^-$, $ep$ and $pp/p\overline{p}$ physics, 
\eg at LEP, HERA and the Tevatron, and during the last 20 years 
has probably been the most used generator for LHC physics studies.
As a building block it has also been used in heavy-ion
physics and cosmic-ray physics.
 
The history of the Lund family of event generators began with 
\jetset \cite{Sjostrand:1982fn,Sjostrand:1982am,Sjostrand:1985ys,
Sjostrand:1986hx} in 1978, which later was merged into 
\pythia \cite{Bengtsson:1982jr,Bengtsson:1984yx,Bengtsson:1987kr,
Sjostrand:1993yb,Sjostrand:2000wi,Sjostrand:2006za}. Over the years 
many new physics models, especially for perturbative and 
nonperturbative QCD, have been developed and tested in parallel 
with the respective code. Thus the \pythiasix generator is the 
product of over thirty years of progress, but some of the code 
has not been touched in a very long time. New options have been 
added, but old ones seldom removed. The basic structure has been 
expanded in different directions, well beyond what it was once 
intended for, making it rather cumbersome by now.

{}From the outset, all code was written in Fortran~77. For the
LHC era, the experimental community made the decision to discontinue
Fortran and move heavy computing to \cpp. Therefore it was logical
also to migrate \pythia to \cpp, and in the process clean up and 
modernize various aspects. A first attempt in this direction was 
the \pythiaseven project \cite{Lonnblad:1998cq,Bertini:2000uh}, 
however, early on this was redirected to become a generic 
administrative structure, and renamed \thepeg (see \SecRef{sec:thepeg}).

\pythiaeight is a clean new start, to provide a successor to \pythiasix. 
It is a completely standalone generator, but several optional hooks 
for links to other programs are provided. Work on it began in 2004, 
and the first fully operational version (8.100) was released in 2007 
\cite{Sjostrand:2007gs}. It is not yet as well tested and tuned as 
\pythiasix, and therefore not as much used, although a slow shift is 
underway. Since priority has been to be ready in time for LHC startup,
some topics have not yet been addressed. Other parts of the \pythiasix 
were deemed obsolete and are permanently dropped.

Here follows a very brief summary of the current \pythia~8.1 program.
Much of the physics is the same as documented in the \pythia~6.4 manual 
\cite{Sjostrand:2006za} and the literature quoted there, with some relevant 
later updates \cite{Corke:2009tk,Corke:2010zj,Navin:2010kk,
Kasemets:2010sg,Carloni:2010tw,Corke:2010yf}. 
A complete manual also comes with the code distribution.
 
The physics summary below is split into core processes, the further 
perturbative evolution, and hadronization. An introduction to the 
code structure completes this outline.  

\mcsubsection{\gensectionhard}

Currently the program is only set up to handle collisions either between 
hadrons, such as $p$, $\overline{p}$, $\pi^{\pm}$ and $\pi^0$, or between 
same-generation leptons. That is, $pp$, $p\overline{p}$ and $e^+e^-$ 
beam combinations can be used,  but currently not $ep$, 
$\gamma\mathrm{p}$ or $\gamma\gamma$.

\pythia contains an extensive list of hardcoded subprocesses, over 200, 
that can be switched on individually. These are mainly $2 \to 1$ and 
$2 \to 2$, some $2 \to 3$, but no multiplicities higher than that. 
Consecutive resonance decays may of course lead to more final-state 
particles, as will parton showers. A brief summary of the main sets 
of subprocesses is as follows:
\begin{itemize}
\item hard  QCD processes, giving two high-$p_{\perp}$ partons;
\item $t$-channel exchange of a $\gamma^*/Z^0$ or $W^{\pm}$, also giving
  two high-\pt partons;
\item prompt-photon production with one or two photons in the final state;
\item a single $\gamma^*/Z^0$ or $W^{\pm}$ gauge boson, a pair of gauge 
bosons, or a gauge boson together with a parton; 
\item charmonium and bottomonium in the colour singlet and octet models;
\item top and a hypothetical heavy fourth generation;
\item Higgses within and beyond the Standard Model;
\item Supersymmetry (in progress); and
\item other exotic physics with new gauge bosons, left--right symmetry,
leptoquarks, excited fermions, hidden valleys, or extra dimensions. 
\end{itemize}

The subprocess cross sections have to be convoluted with PDFs to obtain
the event rates. Several proton PDFs are hardcoded in \pythia for ease 
of use and speed. These include
\begin{itemize}
\item traditional LO sets such as GRV 94L, CTEQ 5L, 6L and 6L1, 
and MSTW 2008 LO; 
\item the newer-style Monte-Carlo-adapted modified LO sets MRST LO* and 
LO**, and CT09 MC1, MC2 and MCS; and 
\item two central members of NLO sets, namely MSTW 2008 NLO and
CTEQ 66.00; 
\end{itemize}
see \SecRef{sec:pdfs-event-gener}.  
Further sets are available through an interface to the LHAPDF library.
It is possible to use separate PDF sets for the hard interaction, on one
hand, and for the subsequent showers and MPIs, on the other. Specifically, 
NLO sets are only intended to be used for hard subprocesses.

Obviously this list is far from complete, in terms of what will be
required at the LHC. Furthermore \pythia does not have automatic code 
generation for new processes, unlike some other generators. 
The intention is that \pythia should be open to external input to the 
largest extent possible, however. That way specialists from many areas 
can contribute hard subprocesses, which thereafter are handled further 
by the normal \pythia machinery.
\begin{itemize}
\item If an external program can generate a Les Houches Event File
\cite{Alwall:2006yp}, this can easily be read in by \pythia. 
A large number of programs can do just that. This includes 
general-purpose matrix-element programs, such as MadGraph or 
CompHep/CalcHep, and ready-made collections of processes, such as 
ALPGEN or AcerMC, see \SecRef{sec:subprocesses}. It also includes 
several processes implemented in the POWHEG approach to NLO calculations,
see \SecRef{sec:me-nlo-matching}. 
\item It is also possible to have a runtime link to \cpp or Fortran
programs, using the Les Houches Accord \cite{Boos:2001cv} structure 
to transfer information between the programs.
\item You can implement your own hard process inside a class derived from 
a \pythia base class, send in a pointer to it, and then let \pythia
handle the generation exactly as if it were an internal process. 
Notably MadGraph~5 will provide a facility whereby the complete code 
for such a class can be written automatically, ready to be linked.  
\end{itemize}  
\pythia is also open to input from other sources, such as the 
SUSY Les Houches Accord \cite{Skands:2003cj,Allanach:2008qq}.

\mcsubsection{Soft processes}

The so-called soft processes are elastic, single and double diffractive, 
and nondiffractive, see \SecRef{sec:mbtypes}. Together they are intended to 
offer an inclusive description of the total $pp$ cross section, with the 
exception of some of the rare (and even hypothetical) processes that
are better simulated separately. Thus the inelastic event sample includes 
high-$p_{\perp}$ physics as a tail of the low-$p_{\perp}$ one, in a 
consistent way, as provided by the MPI framework. To be precise, 
``soft'' events contain an inclusive production of standard QCD 
$2 \to 2$ processes, prompt photons, charmonia and bottomonia, low-mass 
Drell-Yan pairs, and $t$-channel $\gamma^*/\mathrm{Z}^0/\mathrm{W}^{\pm}$ 
exchange, in their expected proportions, with the MPI approach ensuring
that several of them can occur in the same event. One can alternatively use 
the same list as exclusive ``hard'' processes, if one is only interested in 
the high-$p_{\perp}$ tail, where a generation of the complete cross section 
would be inefficient.

Nondiffractive events provide the bulk of the inelastic cross section,
\ie what is observed in central detectors. Inside \pythia it is also 
referred to as the {\it minbias} component, but it does not have a one-to-one
overlap with the experimental definition of minimum bias,
see the warning in \SecRef{sec:mbtypes}. The nondiffractive
component is expected to provide an even bigger fraction of the events
that contain a hard process. Therefore, if the user requests an exclusive
hard process, currently \pythia would always simulate the underlying 
event as being of the nondiffractive type.

Diffractive events are handled in the Ingelman--Schlein picture
\cite{Ingelman:1984ns}, wherein single diffraction is viewed
as the emission of a pomeron pseudoparticle from one incoming proton,
leaving that proton intact but with reduced momentum, followed by the 
subsequent collision between this pomeron and the other proton. The
pomeron is, to first approximation, viewed as a glueball state with
the quantum numbers of the vacuum, but by QCD interactions it will also
have a quark content. The pomeron--proton collision can then be handled 
as a normal hadron--hadron nondiffractive event, displaying the same
structure with MPI, ISR, FSR and the rest. Double diffractive
events contain two pomeron--proton collisions. 

Elastic scattering by default only includes strong interactions
\cite{Schuler:1993wr}, but it is possible to switch on the QED 
Coulomb term and interference as well \cite{Bernard:1987vq}.  

\mcsubsection{The perturbative evolution}

The \pythiaeight showers are ordered in transverse momentum 
\cite{Sjostrand:2004ef}, both for ISR and for FSR. Also MPIs are ordered 
in $p_{\perp}$ \cite{Sjostrand:1987su}.
This allows a picture where MPI, ISR and FSR are interleaved in one 
common sequence of decreasing $p_{\perp}$ values. 
This is most important for MPI and ISR, since they are in direct 
competition for momentum from the beams, while FSR mainly 
redistributes momenta between already kicked-out partons.
The interleaving implies that there is one combined evolution equation
\begin{eqnarray}
\frac{\mathrm{d} \mathcal{P}}{\mathrm{d} p_{\perp}}&=& 
\left( \frac{\vphantom{\left(\right)} \mathrm{d}\mathcal{P}_{\mathrm{MPI}}}%
{\mathrm{d} p_{\perp}}  + 
\sum   \frac{\vphantom{\left(\right)} \mathrm{d}\mathcal{P}_{\mathrm{ISR}}}%
{\mathrm{d} p_{\perp}}  +
\sum   \frac{\vphantom{\left(\right)} \mathrm{d}\mathcal{P}_{\mathrm{FSR}}}%
{\mathrm{d} p_{\perp}} \right)
\nonumber \\ 
 & \times & \exp \left( - \int_{p_{\perp}}^{p_{\perp\mathrm{max}}} 
\left( \frac{\vphantom{\left(\right)} \mathrm{d}\mathcal{P}_{\mathrm{MPI}}}%
{\mathrm{d} p_{\perp}'}  + 
\sum   \frac{\vphantom{\left(\right)} \mathrm{d}\mathcal{P}_{\mathrm{ISR}}}%
{\mathrm{d} p_{\perp}'}  +
\sum   \frac{\vphantom{\left(\right)} \mathrm{d}\mathcal{P}_{\mathrm{FSR}}}%
{\mathrm{d} p_{\perp}'} 
\right) \mathrm{d} p_{\perp}' \right) 
\label{eq:pythiacombinedevol}
\end{eqnarray}
that probabilistically determines what the next step will be.
Here the ISR sum runs over all incoming partons, two per
already produced MPI (including the hard process), the FSR sum runs over 
all outgoing partons, and $p_{\perp\mathrm{max}}$ is the $p_{\perp}$ of the 
previous step.

Starting from a large $p_{\perp}$ scale of the hard process, the 
decreasing $p_{\perp}$ of \EqRef{eq:pythiacombinedevol} can be viewed 
as an evolution towards
increasing resolution; given that the event has a particular structure
when activity above some $p_{\perp}$ scale is resolved, how might that 
picture change when the resolution cutoff is reduced by some infinitesimal
$\mathrm{d} p_{\perp}$? That is, let the ``harder'' features of the event 
set the pattern to which ``softer'' features have to adapt.
It does not have a simple interpretation in absolute time;
all the MPIs occur essentially simultaneously (in a simpleminded
picture where the protons have been Lorentz contracted to pancakes),
while ISR stretches backwards in time and FSR forwards in time. 

\mcsubsection{\gensectionshower}

The initial- and final-state algorithms are partly based on a 
dipole-type approach to recoils, see \SecRef{sec:dipoles}, but with 
some modifications. 

In the simplest case, consider a colour dipole stretched between 
two final-state partons. The emission off such a dipole can be 
associated with either of the two ends, approximately in proportion 
to the respective $1/Q^2$ propagator, which gives a smooth transition 
across phase space for having it associated with either end, 
say $(1 \pm \cos\theta)/2$ in the soft-gluon limit.
By this classification the radiator end is the one that branches 
into two, which implies changed kinematics, to be compensated by 
the recoiler end. 

If the colour dipole is stretched between a final and an initial 
parton, radiation off  the final end has to be compensated by a 
changed momentum for this incoming parton. The subdivision of 
radiation from the two dipole ends is also somewhat more delicate 
in this case, and necessitates the introduction of a special damping 
factor on emission from the final end. 
 
ISR is described by backwards evolution \cite{Sjostrand:1985xi}, 
wherein branchings are constructed from the hard subprocess, back to 
the shower initiators. In each step the whole previously generated
partonic system takes the recoil of the newly emitted parton.
This applies whether the radiating initial dipole end has a 
colour partner in the initial or in the final state. In the latter 
case the pure dipole picture dictates that only this one partner
should take the recoil, but such a picture would not give the correct 
resummation behaviour \eg for $Z^0$ production, see \SecRef{sec:dipoles},  
and is therefore rejected. Emissions are allowed partially to line up 
in azimuthal angle by colour flow, however, which retains some memory 
of the dipole structure.

The evolution variable is closely related to the transverse momemtum
of a branching, but is not identical with it. Instead the 
lightcone-motivated relationships $p_{\perp\mathrm{evol}}^2 = (1-z)Q^2$ 
for ISR and $p_{\perp\mathrm{evol}}^2 = z(1-z)Q^2$ for FSR are used to 
define the space- or time-like virtuality $Q^2$ of the off-shell 
intermediate parton, given the chosen $p_{\perp\mathrm{evol}}$ 
and $z$. When kinematics are actually reconstructed, Lorentz
invariant expressions for $z$ are being used, based on ratios of 
invariant masses, which leads to a kinematical 
$p_{\perp\mathrm{kin}} \leq p_{\perp\mathrm{evol}}$.
Specifically, as a function of emission angle, $p_{\perp\mathrm{kin}}$
peaks at $90^{\circ}$, whereas $Q^2$ and hence $p_{\perp\mathrm{evol}}$ 
keeps on rising with angle.

Both QCD and QED emissions are allowed, and fully interleaved. 
Currently allowed branchings in the shower are $q \to qg$,
$g \to gg$, $g \to q\overline{q}$, $q \to q \gamma$,
$\ell \to \ell \gamma$ and, for FSR only, $\gamma \to \ell^+ \ell^-$
and $\gamma \to q\overline{q}$. 

Many resonance decays involve full matching to NLO QCD matrix 
elements. For production, however, all internally implemented 
processes are LO only. Production of $\gamma^*/Z^0/W^{\pm}$ is
matched to the real-emission corrections, so as to obtain the 
NLO $p_{\perp}$ spectrum, but without the NLO $K$-factor.
Showers have been constructed so that they, by default, have a 
sensible behaviour over the full phase space, all the way up to 
the kinematical limit, for a wide range of processes, but they 
will not be perfect. 

Final-state showers have a sharp lower cutoff, that should define 
the transition to hadronization. For ISR it is also possible to
use a sharp cutoff, but a valid alternative is a smooth turn-off
related to what is done for MPI below.

\mcsubsection{\gensectionMPI}

MPI modelling has traditionally been a hallmark of \pythia.
The framework is extensively described in \SecRef{sec:mbmpi},
and here only the basic principles are recapitulated, to put
them in context.

The perturbative $2 \to 2$ QCD cross section, which is dominated by 
$t$-channel gluon exchange, diverges roughly like 
$\mathrm{d} p_{\perp}^2 / p_{\perp}^4$. But this is based on the 
assumption of free incoming states, which is not the case when 
partons are confined in colour-singlet hadrons. Screening by nearby 
opposite colour charges will dampen the interaction of gluons with a 
large transverse wavelength. This is introduced by reweighting the 
interaction cross section by the factor in \EqRef{eq:ptzerodampen},
where $p_{\perp 0}$ is a free parameter in the model. To be more precise, 
it is the physical cross section $\mathrm{d} \sigma / \mathrm{d} p_{\perp}^2$ 
that needs to be regularized, \ie  the convolution of
$\mathrm{d} \hat{\sigma} / \mathrm{d} p_{\perp}^2$ with the two 
parton densities. One is thus at liberty to associate the screening 
factor with the incoming hadrons, half for each of them, instead of 
with the interaction. Such an association also gives a recipe to 
regularize the ISR divergence, as already noted.

The $p_{\perp 0}$ parameter can be energy-dependent,
since higher energies probe partons at smaller $x$, where the 
parton density increases and thereby the colour screening distance
decreases. An ansatz $p_{\perp 0} \propto E_{\mathrm{CM}}^{\epsilon}$
is therefore assumed, with some small power $\epsilon$. 

The spatial shape of the proton determines the balance between 
peripheral and central collisions, as reflected for example in the width 
of the multiplicity distribution. Several different shapes are available,
starting with a simple Gaussian ansatz.

Rescattering has been implemented, \ie the possibility of one parton 
scattering several times. So far no good experimental signals have been 
found for it, and it is off by default.

For dedicated studies of two low-rate processes in coincidence, two
hard interactions can be set in the same event, by a somewhat 
simplified duplication of the normal hard-process selection machinery. 
There are no Sudakov factors included for these two interactions, 
similarly to normal events with one hard interaction.

Rescaled parton densities are defined after each interaction, 
that take into account the nature of the previous partons extracted
from the hadron. This guarantees energy--momentum--flavour conservation.

Currently there is only one scenario for colour reconnection in the 
final state, see \SecRef{sec:mbmpi-np}, in which there is a certain 
probability for the partons of two subscatterings to have their 
colours interarranged in a way that reduces the total string length. 
(This is intermediate in character between the original strategy 
\cite{Sjostrand:1987su} and the more recent ones \cite{Skands:2010ak}.) 

At the end of the perturbative stage, a number of leftover partons
are found in the proton beam remnants, with colour connections to
the scattered partons, see \SecRef{sec:mbmpi-np}. Primordial $k_{\perp}$'s 
are introduced both for the scattering subsystems and the remnants, 
colours are assigned to connect the subsystems and the remnants with 
each other, and leftover longitudinal momentum is split between the 
remnant partons. When necessary, the junction approach is used to 
keep track of the baryon number, see \SecRef{sec:string-model}.

\mcsubsection{\gensectionhadronize}

Hadronization is based solely on the Lund string fragmentation
framework \cite{Andersson:1983ia,Sjostrand:1984ic}, 
\SecRef{sec:string-model}, which is at the origin of the \jetset 
program and thus of \pythia.

Particle data have been updated in agreement with the 2006 PDG
tables \cite{Yao:2006px}.  Some updated charm and bottom decay 
tables have been obtained from the DELPHI and LHCb collaborations.

The BE$_{32}$ model for Bose--Einstein effects \cite{Lonnblad:1997kk} 
has been implemented, but is not on by default. It does a reasonable
job with $e^+e^-$ data but not so well for hadronic collisions.   

\mcsubsection{Program structure and usage}

The \pythiaeight homepage is at
\begin{center} 
\texttt{http://www.thep.lu.se/}$\sim$\texttt{torbjorn/Pythia.html}
\end{center}
and from there you can download the most recent version as a gzipped
tar file, which also includes documentation as well as several example
main programs illustrating different ways in which \pythiaeight can be
used and linked. The documentation can also be accessed directly from the
\pythiaeight homepage. 

It is possible to perform analyses of the event record inside the 
main program. Alternatively events can be output to the HepMC
format, from which they can be studied further, or sent on to
detector simulation programs like Geant.

\mcsubsection{\gensectionoutlook}

\pythiaeight by now offers a complete replacement of \pythiasix
for essentially all aspects related to LHC physics studies, and in 
many respects contains improved physics models and new features. 
While development of \pythiasix has stopped, and new subversions
will only be prompted by bug fixes, \pythiaeight is being further
improved and extended in several directions. Experimental usage is 
still lagging behind, but interest is picking up, so one should 
expect a gradual phaseover during the next few years. 

% Local Variables: 
% mode: LaTeX
% TeX-master: "../mcreview"
% End: 

\mcsection{\Sherpa}
\label{sec:sherpa}

\mcsubsection{\gensectionintro}
\label{Sec:intro_sherpa}
\Sherpa is a general-purpose event generator, capable of simulating
the physics of lepton-lepton, lepton-hadron, and hadron-hadron collisions 
as well as photon induced processes. Unlike the programs
 \Ariadne, \Herwig and \Pythia, it was constructed from the beginning in 
\cpp, and in contrast to the \cpp versions of those programs some of the
physics modules (such as the old parton shower, encoded in 
\Apacic, or the matrix element generator \Amegic) were established
before the actual event generation framework.  The construction paradigm
of the \Sherpa framework can be summarized as follows:
\begin{itemize}
\item {\em emphasis on strict modularity of physics modules}\\
  In fact the organization is such that physics modules are only
  connected through relatively unspecific event phase handlers, which in turn
  call interfaces to the underlying physics modules.  These interfaces
  are constructed such that they can connect to various independent
  modules performing the same tasks.  A prime example in \Sherpa is
  the treatment of hard matrix elements, where various ME generators
  (see below) are available to \Sherpa, but all of them accessible 
  through one and the same {\tt Matrix\_Element\_Handler}.  This allows 
  a comparably simple replacement of outdated modules, for instance 
  the old \Apacic parton shower.  
\item {\em bottom-up approach}\\
  The event organization within \Sherpa is kept as simple as possible.
  In particular, there are no abstract overheads for possible 
  event phases when there exists no corresponding physics module yet.
\end{itemize}
Traditionally \Sherpa's main focus is on the perturbative event phase; 
\Sherpa is a frontrunner in the automated generation 
of tree-level matrix elements and hosts two fully-fledged ME generators with
highly advanced phase-space integration methods.  In recent years, the scope
there has widened to also include infrastructure to support the calculation
of cross sections at NLO accuracy, by providing automated subtraction methods.
In addition, the cornerstone of \Sherpa's event simulation, from the beginning,
was the multijet merging described in \SecRef{sec:matching-at-tree}.
Only quite recently the description of parton showering in \Sherpa has been 
improved by the inclusion of a parton shower based on Catani-Seymour
subtraction \cite{Schumann:2007mg} and the development of a true dipole 
shower \cite{Winter:2007ye}, the latter still awaiting full incorporation 
into the framework.  Similarly, in the beginning hadronization in \Sherpa
was performed through an interface to the Fortran version of \Pythia and
only in recent years a new independent implementation of the cluster
hadronization idea \cite{Winter:2003tt}, see \SecRef{sec:cluster-model}, 
has been added. Other additions include a complete model of hadron and 
$\tau$ decays and QED final-state radiation \cite{Schonherr:2008av} and 
a simulation of the underlying event based on the multiple-parton 
scattering ideas of \cite{Sjostrand:1987su}.

\mcsubsection{\gensectionhard}
\label{Sec:hard_process_sherpa}
\paragraph{Tree-level matrix-element generators}

Processes simulated by \Sherpa are selected by defining initial and final 
states of the hard subprocess, generated by the matrix element generator chosen, 
see below.  These initial and final state also define the particles that 
will actually appear in the event record, following the philosophy outlined 
in \SecRef{sec:phys-obj-mc-truth}.\footnote{
  It should be noted that in \Sherpa projections on intermediate states
  in various schemes (narrow width or propagator, both with full spin
  correlations) are also available; these intermediate states, however, will
  typically {\em not} appear in the event record.
}
To generate the cross sections for the hard subprocesses, \Sherpa provides 
two built-in matrix-element generators, \Amegic~\cite{Krauss:2001iv} and 
\Comix~\cite{Gleisberg:2008fv}, as well as facilities for hard-coded matrix 
elements, see \AppRef{sec:app_mcs}.

\Amegic is a Feynman diagram based generator that constructs tree-level 
amplitudes and suitable phase-space mappings from given sets 
of interaction vertices. The Feynman diagrams then get translated into 
helicity amplitudes using an algorithm similar to the one described in 
\cite{Kleiss:1985yh,Ballestrero:1992dv} and extended to include also 
spin-two particles in \cite{Gleisberg:2003ue}. The list of supported 
physics models covers:
\begin{itemize}
  \item the complete Standard Model,
  \item extension of the SM by a general set of anomalous triple and
    quartic gauge couplings \cite{Appelquist:1980vg,Appelquist:1993ka},
  \item extension of the SM by a single complex scalar
	\cite{Dedes:2008bf},
  \item extension of the SM by a fourth generation,
  \item extension of the SM by an axigluon
	\cite{Pati:1975ze,Hall:1985wz,Frampton:1987dn,Frampton:1987ut,Bagger:1987fz},
  \item Two-Higgs-Doublet Model,
  \item Minimal Supersymmetric Standard Model \cite{Hagiwara:2005wg}, 
  \item ADD model of large extra dimensions
    \cite{ArkaniHamed:1998rs,Antoniadis:1998ig}.
\end{itemize}
Other new physics models can easily be invoked by providing model 
representations generated with the \FeynRules 
program~\cite{Christensen:2008py,Christensen:2009jx}. Based on the 
information of all Feynman diagrams contributing to a given process,
\Amegic automatically constructs suitable phase-space mappings. For the
actual integration all contributing channels are combined in a self-adaptive 
multi-channel integrator, see \AppRef{Sec:PS_ME}, which automatically
adjusts to the relative  importance of the single phase-space maps to
minimize the variance. The efficiency of the integrator is further 
improved by applying the self-adaptive \Vegas \cite{Lepage:1980dq} 
algorithm on single phase-space maps.

\Comix is especially suited for the simulation of highest-multiplicity 
processes. This generator is based on an extension of the colour-dressed 
Berends-Giele recursive relations 
to the full Standard Model, see \AppRef{Sec:TL_ME}. Within \Comix any 
four-particle vertex of the Standard Model is decomposed into 
three-particle vertices. This leads to a significantly improved 
performance for large final-state multiplicities, compared to \Amegic. 
The summation (averaging) over colours in QCD and QCD-associated processes 
is performed in a Monte Carlo fashion and colour-ordered amplitudes can 
therefore be computed. Following the reasoning of~\cite{Duhr:2006iq}, 
the colour-flow basis is employed throughout the code. As discussed 
in~\cite{Maltoni:2002mq}, this yields a certain correspondence between 
the large-$\Nc$ limit employed in parton-shower simulations and full QCD 
results, which is especially useful in the context of a merging with the 
parton shower, see \SecRef{sec:me-nlo-matching}.

\paragraph{Next-to-leading order event generation}
\label{Sec:NLO_ME_Sherpa}
The \Amegic matrix-element generator has the further functionality to 
construct dipole-subtraction terms and their integrals over the one-parton 
emission phase space in the Catani--Seymour formalism \cite{Catani:1996vz}
for arbitrary Standard Model processes \cite{Gleisberg:2007md}. 
When supplemented with corresponding one-loop amplitudes, using the 
Binoth-Les-Houches-Accord \cite{Binoth:2010xt} interface structure, \Sherpa 
is capable of generating parton-level events at next-to-leading order
precision, see \SecRef{sec:subprocesses:NLOcross_sections}. This framework
has for example been used to evaluate the QCD NLO corrections to $W/Z+3$jets
\cite{Berger:2009zg,Berger:2009ep,Berger:2010vm} and $W+4$jets 
\cite{Berger:2010zx} production, with the loop amplitudes obtained from 
{\sc BlackHat} \cite{Berger:2008sj}. In \cite{Binoth:2009wk} 
the NLO corrections to $ZZ+$jet production have been calculated for the 
first time, relying on {\sc Golem} \cite{Binoth:2005ff} for the 
generation of the loop-amplitude expressions. 

Besides the implementation of the Catani--Seymour dipole subtraction 
method, facilitating parton-level event generation at NLO, \Sherpa also 
provides the possibility to generate hadron-level events at NLO accuracy 
using the \POWHEG algorithm to combine NLO matrix elements with the \Sherpa 
parton shower. This is achieved in a completely process-independent way, 
using a reformulation of the original \POWHEG method, which was presented 
in~\cite{Hoeche:2010pf}. The respective \POWHEG generator is based 
on the same principles as the internal parton-shower module described in 
the following section.

\mcsubsection{\gensectionshower}
\label{Sec:shower_sherpa}
\Sherpa's default parton-shower algorithm, first presented in 
\cite{Schumann:2007mg}, is based on the Catani--Seymour 
dipole factorization formalism \cite{Catani:1996vz,Catani:2002hc}.
The underlying key idea is to derive the corresponding shower splitting 
operators from the four-dimensional unintegrated dipole-subtraction terms 
by performing the large-$\Nc$ limit and summing and averaging over all 
spin degrees of freedom. Accordingly, one arrives at a completely factorized 
approximation for the real-emission process in terms of the underlying Born 
channel times a sum of suitable splitting operators that correctly account
for (quasi-)collinear and soft emissions. 

The emerging shower picture corresponds to sequential splittings of dipoles 
where, in the Catani--Seymour formulation, a dipole is made up of the actual 
parton that is supposed to split and a well-defined spectator parton that 
is colour-connected to the emitter. Four dipole configurations have to be 
considered, classified by the emitter/spectator being either in the final 
(F) or initial (I) state; FF, FI, IF and II. All dipole configuration are 
treated on an equal footing and as a consequence there is no formal distinction 
between initial- and final-state parton showers. Successive emissions are 
ordered in terms of the invariant transverse momentum between final-state 
splitting products or with respect to the emitting beam particle. At present 
the Catani--Seymour dipole shower in \Sherpa implements all QCD splittings 
in the Standard Model and the MSSM as well as QED photon emissions 
\cite{Hoeche:2009xc}.

In its original formulation presented in \cite{Schumann:2007mg}, the recoil 
strategy for the various types of dipole splittings closely followed the 
choice of the Catani--Seymour formalism \cite{Catani:1996vz,Catani:2002hc}.
However, when considering initial-state splittings this can lead to the
situation that only the first splitting of an initial--initial dipole 
transfers transverse momentum to the rest of the event. As an intuitive 
example, consider the shower evolution of a Drell-Yan event, which starts from just initial-initial dipoles. In the extreme case the gauge boson 
would get a finite recoil from the first splitting only, clearly at odds 
with the resummation of associated large logarithms. In 
\cite{Platzer:2009jq,Hoeche:2009xc} alternative, crossing symmetric, 
recoil strategies were presented that avoid this peculiar feature. 
For the \Sherpa implementation, \cite{Carli:2010cg} studied the impact 
of different recoil strategies in the context of deep-inelastic lepton 
scattering events.

The shower formulation based on Catani--Seymour dipole factorization 
offers two substantial advantages with respect to traditional parton showers,
which help to facilitate the merging with fixed-order matrix-element 
calculations:
\begin{itemize}
\item Due to the notion of specific spectator partons, four-momentum 
conservation is maintained locally, while only a single external
particle, the spectator, takes the recoil when the splitting parton
goes off-shell. This is important for the construction of a backward 
clustering algorithm based on the parton shower in the spirit of 
\cite{Hoeche:2009rj}.
\item The parton-shower model inherently respects QCD soft-colour coherence. 
By construction in Catani--Seymour factorization, the eikonal factor associated 
with soft gluon emission off a colour dipole, used to derive the angular ordering 
constrained in conventional parton showers, is exactly mapped onto two CS dipoles,
which only differ by the role of emitter and spectator.
\end{itemize}

\mcsubsection{Matrix-element parton-shower merging}
\label{Sec:meps_sherpa}
One of the key features of \Sherpa is a generic implementation of the 
technique for combining tree-level matrix elements with parton showers
that was presented in~\cite{Hoeche:2009rj}, see \SecRef{sec:me-nlo-matching}. 
The method was extensively tested and validated for multijet production in 
$e^+e^-$ and hadron-hadron 
collisions, as well as deep-inelastic scattering processes, a scenario where 
event generators based on collinear factorization assumptions 
are unreliable due to a lack of matrix elements with sufficiently 
high final-state multiplicity~\cite{Carli:2010cg}.
An extension of the merging algorithm, which simulates hard QED radiation 
in a democratic approach, \ie on the same footing as QCD radiation, 
was implemented in \Sherpa and reported in~\cite{Hoeche:2009xc}. 
It yields excellent agreement with existing experimental data on prompt
photon production at both $e^+e^-$ and hadron colliders.
Although the novel merging technique implemented in recent versions of \Sherpa
has yielded significant improvements over the original CKKW algorithm,
in the sense that results are more accurate and stable, the CKKW approach
itself was already employed in former versions of \Sherpa with 
great success~\cite{Krauss:2004bs,Krauss:2005nu,Gleisberg:2005qq,Alwall:2007fs}.

In order to realize the ME+PS merging, \Sherpa makes use of its two 
internal tree-level matrix-element generators \Amegic and \Comix. Soft and 
collinear parton radiation is simulated by means of the internal parton 
shower. It should be noted that \Sherpa implements its matrix-element 
parton-shower merging in a modular way, distributing only necessary tasks to 
the matrix-element and parton-shower generators and handling all cross-module
interaction in the overall framework. This means in particular that the
matrix-element generator is only used to identify possible parton-shower
histories in the matrix elements by testing for respective subamplitudes
in the Feynman diagrams. The parton shower supplies information about 
the weight associated with a backward clustering that would reduce the 
actual partonic final state to the respective subamplitude. If external 
parton showers or matrix-element generators are provided by the user,
they must be capable of performing these operations. If so, they can in turn
be employed for automatic matrix-element parton-shower merging without 
any further adjustments of the \Sherpa framework, 
see also \SecRef{Sec:interfaces_extensions_sherpa}.

The \MENLOPS algorithm for merging lowest-multiplicity NLO matrix-elements
with higher-order tree-level contributions as presented independently 
in~\cite{Hamilton:2010wh} and~\cite{Hoeche:2010kg} is fully implemented
in the \Sherpa generator~\cite{Hoeche:2010kg}. It relies on the internal
generic \POWHEG generator described in \SecRef{Sec:NLO_ME_Sherpa}, 
which drives the lowest-multiplicity simulation and interfaces to the 
Catani--Seymour dipole shower to generate additional parton radiation.

\mcsubsection{\gensectionMPI}
\label{Sec:mpi_sherpa}
The multiple-interactions model used in \Sherpa closely follows the original 
ideas of~\cite{Sjostrand:1987su}. There are however important details where
the approach deviates from the formalism in \Pythia. Secondary interactions 
undergo parton-shower corrections in \Sherpa, but the evolution does not 
interleave parton showers and additional hard scatterings.
Care must then be taken when combining ME+PS merging with the modelling of 
multiple interactions. It is vital that the parton showers related to secondary 
collisions do not alter the initial jet spectra of the hard process. This can 
be achieved by a special jet veto procedure, which is described in some detail
in~\cite{Gleisberg:2008ta}.

The modelling of beam remnants in \Sherpa is realized in such a way that only 
a minimal set of particles (quarks and diquarks, the latter as carriers of 
baryon number) is produced in order to reconstruct the constituent 
flavour configuration of an incoming hadron. The distribution of colour in the 
remnants is guided by the idea of minimizing the relative transverse momentum 
of colour dipoles spanning the outgoing partons.  When including multiple 
parton interactions in the simulation, it is not always possible to accomplish 
free colour selection in the hard process and minimization of relative transverse 
momenta simultaneously.  In such cases the colour configurations of the 
matrix elements are kept but the configuration of the beam remnants is shuffled 
at random until a suitable solution is found.

In addition to the issues related to colour neutralization with the
beam remnants, all shower initiators and beam partons obtain a
primordial~\kt, see \SecRef{sec:primkt}. When tuning this
distribution, for example by using the data shown in \FigRef{fig:cmp:intrinsic-kt}, 
the mean and width parameter values obtained are typically rather
small (about $0.5-1.0$~GeV).

\mcsubsection{\gensectionhadronize}
\label{Sec:hadronization_sherpa}
The idea underlying \Ahadic, \Sherpa's module dealing with hadronization, 
is to take the interpretation of clusters as excited hadrons very literally, 
to compose clusters out of all possible flavours including diquarks 
and to have a flavour-dependent transition scale between clusters and 
hadrons.   This results in converting only the very lightest clusters 
directly into hadrons, whereas slightly heavier clusters experience a 
competition between either being converted into heavy hadrons or decaying 
into lighter clusters. For all decays, QCD-inspired, dipole-like 
kinematics are chosen. In somewhat more detail, in \Ahadic, the hadronization 
of quarks and gluons proceeds as follows:
\begin{itemize}
\item Firstly, all gluons are forced to decay into quark or diquark-pairs,
  $q\bar q$ or $d\bar d$, and all remaining partons are brought on 
  constituent mass shells.  Recoils are compensated mainly through 
  colour-connected particles.
\item Subsequent decays of heavy clusters are modelled by first emitting 
  a gluon from the $q\bar q$ pair and then splitting this gluon again.
\item In all non-perturbative decays ($g\to q\bar q$ and cluster decays) the 
  transverse momentum is limited to be smaller than a parameter 
  $\pt^{\rm max}$, typically of the order of the parton-shower cutoff 
  scale, and $\pt^2$ is chosen according to 
  $\alphaS(\pt^2+p_0^2)/(\pt^2+p_0^2)$, invoking a second parameter
  $p_0$ \footnote{There is also the option to use a non-perturbative 
    $\alphaS$ coupling, agreeing with a measurement from the GDH sum rule
    \cite{Deur:2008rf}.}.
  In this picture lighter-flavour pairs are preferentially produced due to 
  available phase space; this is supplemented by weight parameters.
\item The decays of clusters into hadrons are determined by various weights 
  including flavour wave functions, phase-space factors, flavour and 
  hadron-multiplet weights, and other dynamical measures.  
\end{itemize}

\mcsubsection{\gensectiondecay}
\label{Sec:decays_sherpa}

\Sherpa's hadron decay module is quite exhaustive, with approximately 
200 decay tables (one for each particle) consisting of more than 2500 
decay channels. Each of them is modelled by isotropic decay
and the branching ratio, but on top of that, spin-dependent
matrix elements and even form factors can be included.  This leaves \Sherpa
in a situation where for some decays various form factor models
are available\footnote{There is a plethora of sources and models, 
  for instance HQET \cite{Neubert:1993mb,Caprini:1997mu,Richman:1995wm}, 
  quark-model predictions~\cite{Isgur:1988gb,Scora:1995ty,Goity:1994xn} or
  QCD sum rules\cite{Ball:2004ye,Ball:2004rg,Ball:2007hb,Aliev:2007uu},
  all for heavy meson decays.  In addition, form factor models for $\tau$
  decays based on the Kuhn-Santamaria parameterization \cite{Kuhn:1990ad},
  or on Resonance Chiral Theory 
  \cite{Weinberg:1978kz,Gasser:1983yg,Gasser:1984gg,Ecker:1988te}
  have also been included.}, 
while for others even the branching ratios are not well known and have to be 
estimated from symmetry principles and phase-space arguments.  

In addition to the simulation of individual decays, non-trivial quantum 
effects are also modelled in \Sherpa, including spin correlations in 
sequential decays and CP violation introduced by mixing phenomena or their 
interplay with direct CP violation in decays.  For the latter, \Sherpa allows 
the user to include separate decay tables for particles and antiparticles.  

For QED FSR, the \Photons module \cite{Schonherr:2008av} is invoked, 
which employs the YFS formalism (see \SecRef{sec:qed-radiation} )
allowing for a systematic improvement of the 
eikonal approximation order-by-order in the QED coupling constant. 
${\cal O}(\alpha)$ corrections are
included for a number of processes, among them decays of vector particles
into leptons, leptonic $\tau$ decays and some $B$ decays.  
At present the module is only capable of handling single-particle initial 
states, \ie particle decays possibly including QED FSR off the hard 
process.  In contrast to some other implementations, however, it can deal with
decays involving more than two charged particles.  

\mcsubsection{Interfaces and extensions}
\label{Sec:interfaces_extensions_sherpa}
\paragraph{Interfaces provided}
\Sherpa supports most of the commonly used standard interfaces for information input or output:
\begin{itemize}
\item parameters and interactions of new physics models can be 
incorporated through \FeynRules generated input files \cite{Christensen:2009jx},
\item the spectra of supersymmetric models can be provided in the form of 
SLHA files \cite{Skands:2003cj,Allanach:2008qq},
\item to link external parton densities the LHAPDF package is supported,
\item hard-process configurations generated with either \Amegic or \Comix 
can be output in the Les-Houches-Event-File format,
\item one-loop amplitudes can be invoked using the Binoth-Les-Houches-Accord 
outlined in \cite{Binoth:2010xt},
\item fully showered and hadronized events can be output in the HepMC 
\cite{Dobbs:2001ck} or HepEvt format. 
\end{itemize}

\paragraph{Extending \Sherpa}
Extensions of \Sherpa can be provided in various ways. 
The easiest would certainly be to enhance the functionality of
an existing module of the program, thus providing the code with
the capability to, for example, simulate reactions in a new physics
scenario. The most challenging, but nevertheless available option
would be to supply a complete new module to the event-generation
framework, encoding for example an alternative underlying event 
model. Considerable modifications of the core framework of \Sherpa
will only be necessary if an extension of the program requires 
cross-module interaction that has not been foreseen and therefore
has not been implemented yet. In such cases, users are strongly 
encouraged to coordinate their efforts for implementing extensions
with the authors of \Sherpa. In most cases, however, existing 
structures will suffice to satisfy the needs for possible enhancements.

\Sherpa provides the option of loading most possible extensions of the
program package at runtime, using dynamically linked libraries.
This mechanism is especially convenient to use in bigger software
frameworks, where the \Sherpa core library itself is just part of
a larger event generation and analysis framework. It also allows
the user to install \Sherpa in a predefined location and to provide an 
extension of the program without altering its core modules.

For most of its extensions \Sherpa employs a so-called ``getter''
mechanism to identify possibly available external sources.
This means that an extension module is registered with the \Sherpa
instance at load time of its shared library, using a predefined
protocol associated with the physics task of the extension module.
An example would be an externally supplied parton shower,
which registers using its name (``Apacic'', for example) using the 
parton-shower identification protocol. At runtime, users can then 
specify this name in the input card to enable the respective shower
model in event generation.

The following extensions can currently be supplied to \Sherpa using 
``getter'' methods:
\begin{itemize}
\item {\em Analysis programs}\\
Both the Rivet library and the HZTool library are interfaced 
using such extensions of \Sherpa. These interfaces are distributed 
with the \Sherpa package itself.
\item {\em Parton distribution functions}\\
Despite most PDFs being available nowadays within the LHAPDF library, 
it might, in some cases, become necessary to interface the code for
a dedicated PDF. \Sherpa provides this option and supplies, for example,
in-house interfaces to photon PDFs.
\item {\em Matrix elements}\\
While there is little need to extend \Sherpa with tree-level matrix 
elements, this option is nevertheless provided to allow implementation 
of special matrix elements, for example for upsilon production.
Additionally, \Sherpa provides the option of including external NLO virtual
matrix elements, which can then be combined with automatically generated
Born-level, real-emission and subtraction terms.
\item {\em New physics models}\\
Even though the usage of the \FeynRules program package and its interface 
to \Sherpa is strongly encouraged, \Sherpa provides the option of implementing
a new-physics scenario directly. The corresponding vertices will then be
available for both internal matrix-element generators, \Amegic and \Comix.
\item {\em Helicity-amplitude building blocks}\\
New helicity-amplitude building blocks might become necessary for exotic BSM scenarios.
They can be provided for both internal matrix-element generators, 
\Amegic and \Comix. Of course they will have a different underlying 
structure in each case.
\item {\em Matrix-element generators}\\
If necessary, a complete external matrix-element generator can be supplied.
Note, however, that it must also satisfy the requirements imposed by the
possibility of merging matrix-element level events with parton showers.
This means in particular that it must provide a clustering algorithm
that identifies allowed parton-shower histories in tree-level matrix 
elements.
\item {\em Parton-shower generators}\\
If necessary, a complete external parton-shower model can be supplied.
Such a parton shower must, however, comply with \Sherpa's rules for 
matrix-element parton-shower merging, \ie it must provide a related
algorithm for computing the branching probability leading to tree-level 
matrix-element final states.
\item {\em Hadron decayers}\\
As already hinted at above, \Sherpa already provides quite an extensive
library for hadron decays.  They can be further extended by providing
form factors or ``skeleton'' matrix elements in the \Hadrons package.
\end{itemize}

\mcsubsection{\gensectionoutlook}
As with any other event generator, it is hard to conceive that \Sherpa will 
ever reach a state of ``perfection'', where nothing is left to be done.  
However, for the near future, a number of enhancements are foreseen:
\begin{itemize}
\item Work on the automated implementation of the 
  \POWHEG algorithm, including non-trivial colour configurations and 
  new-physics processes, will be finalized.  Equipped with this tool, the path towards a 
  multijet merging at NLO seems to be viable.  
\item A second, independent parton-shower formulation is ready to be fully
  included into the framework.  This will allow systematic comparison of
  parton-shower effects with two independent modules in the same framework,
  a huge step forward.
\item A new model for the simulation of soft inclusive physics and the
  underlying event, based on the multichannel eikonal approach of
  \cite{Ryskin:2009tj} is under way.  It will supplement or replace the
  old model, based on MPI.  
\end{itemize}

The current and future releases of the \Sherpa package as well as the most recent documentation can be found at
\begin{center} 
\href{http://projects.hepforge.org/sherpa/}{\tt http://projects.hepforge.org/sherpa/}
\end{center}

% Local Variables: 
% mode: LaTeX
% TeX-master: "../mcreview"
% End: 

\mcpart{The use of generators}
\label{sec:comp-gener-with}

\mcsection{Physics philosophy behind phenomenology and generator validation}
\label{sec:phys-phil-behind}
As discussed in \SecRef{sec:general-introduction}, Monte Carlo simulations are used in
various ways when performing measurements in particle physics experiments that require
the comparison of theoretical predictions with data.  Similarly, measurements of SM
processes in data provide important input to Monte Carlo tools. They provide validation
of theoretical predictions and allow free parameters to be tuned.

There are some basic philosophies that experimentalists using Monte Carlo tools
and making experimental measurements potentially useful for their
validation should be aware of.  These are discussed in this section.

\mcintrosubsection{Physical observables and Monte Carlo truth}
\label{sec:phys-obj-mc-truth}

When simulating a process with a Monte Carlo event generator it is
important to make the distinction between ``Monte Carlo truth'' and
``physical observables'' (see also the discussion of Monte Carlo truth
contained in the contribution by Buckley et al.\ in \cite{Butterworth:2010ym}).

It is often desired to specify a process in terms of intermediate
objects as well as initial and final states, for example lepton
pair hadroproduction via production and decay of a $Z^0$ boson. However,
the intermediate objects are not physical observables,
and in practice it is not always possible to classify the process in
this way.
In particular one must keep in mind that such a classification is only exact in the
limit that all quantum interference effects can be neglected.
Thus, although it may be convenient to model a
double-slit experiment by shooting particles either through slit S
(signal) or slit B (background), that distinction, as it stands, is not  quantum
mechanically meaningful when both slits are open.\footnote{``Background''
here refers only to fundamentally {\it irreducible} background, which
can produce the same final states as the signal.} Likewise, soft
bremsstrahlung in particular depends strongly on interference effects
(coherence, see \SecRef{sec:parton-showers} on parton showers),
and hence the assignment of radiation as coming off this or that
parton is inherently ambiguous.
The one fail-safe way to make sure a distinction
is quantum mechanically meaningful to all orders is well known: to
classify an event according to the values of specific physical
observables (such as where the photon struck the
actual screen, in the case of the double-slit experiment).

The \Sherpa event generator (see \SecRef{sec:sherpa}) goes so far as to insist that a process is
defined in terms of initial and final states, such that it is
not possible for a user to access any intermediate objects.
All possible contributing subprocesses, as well as any interference terms between
them, are then included in the calculation.
While other event generators do allow the user to specify the process
of interest in more detail, users should be aware of the possible limitations.
In addition, when an experimental measurement is  performed it should  be
presented in an unambiguous way, in terms of physical observables.

\mcsubsection{Making generator-friendly experimental measurements}
\label{sec:mcfriendlyobs}

For a measurement to be useful in the context of the development and
improvement of Monte Carlo models, it must be well-defined in terms of
the observed initial- and final-state particles, rather than in terms of
intermediate unstable particles or a particular type of process.
This is also a desirable attribute for any physical measurement to have
meaning beyond a particular theoretical framework. Indeed, quantities
not defined in terms of physical observables run the risk of not being
quantum-mechanically meaningful.

The philosophy advocated here is that the data are ``golden'', and any
dependence on current theoretical tools should be minimized to ensure
the longevity and usefulness of the experimental result. Therefore
corrections and extrapolations to different regions of phase-space using
a Monte Carlo or other theoretical prediction should be minimized. This
should not be confused with correcting for detector effects. In fact it
is required that the effects of the detector (resolutions, efficiencies)
are removed to within some stated systematic uncertainty, or at least
quantified in terms of systematic uncertainties.

The initial state is generally the colliding beams, which are well
known, although in some cases, for example almost-on-shell photons, they
may be treated as a quasi-real initial state.  The final state can be
more problematic. In general the best approach is to define all
particles with proper lifetimes beyond some cut~\cite{Buttar:2008jx}
(typically 30~ps) as being the ``stable'' final state of the event, and
derive all event properties, cross section fiducial regions and so on
from these. The result should be stated in terms of final-state
particles within the acceptance of the detectors, without extrapolations
into regions that are not measured. Statistics allowing, it is even
better to split up the observed phase space into a few complementary
regions, and quote the result for each separately, which can provide a
non-trivial cross check on the ability of the models to interpolate
among those regions.

If any theory-based corrections are applied (for example QED radiative
corrections) they need to be clearly stated and quantified and the
result without the correction should also be stated, since in principle
these corrections would be included in the ``ideal'' Monte Carlo.

These guidelines are best illustrated with different examples of common
measurements at collider experiments.

\paragraph{Measurements of charged-particle distributions (``minimum bias")}

Typically distributions of charged particles, such as charged particle
multiplicity, transverse momentum \pt\ and pseudo-rapidity $\eta$ distributions are made. These measurements
are often referred to as ``minimum bias'', because the idea is usually
to be as inclusive as possible and include the distributions of events
in which it is known that an inelastic collision occurred.  This is
inferred from the detection of final-state particles (other than the
incoming ones). Following the philosophy previously described, such a
measurement should be made within a well-defined region of phase space.
For example if a detector can reconstruct tracks from charged particles
in the region $|\eta| < 2.5$ and \pt\ $>$ 100~MeV, then the result
should be expressed in terms of charged particles with the same
(or tighter) kinematic cuts.  Furthermore, if the distributions are
normalized to the total number of events in the sample (as is often the
case) it should be well defined what is meant by an event. For example
an event could be defined as ``any event with at least one charged
particle with $|\eta| < 2.5$ and \pt\ $>$ 100~MeV''. This is a
definition that can easily be reproduced at the generator level.
Normalizing to all events from a $pp$ collision is not well defined
experimentally, as some minimal experimental criterion is required to
detect a collision. The only way to correct distributions for events
with no particles within the acceptance of the detector is to use a
theoretical model or ad hoc extrapolation, which does not give any extra
information from the data, and in fact contaminates the result with the
model that is used to perform the correction.

Another important point for this type of analysis is the definition of
the final state. As previously stressed, the result should be given in
terms of final-state particles only. No claims should be made about the
type of process that produced this final state, as it is not possible to
state unambiguously what the process was.  Historically many such
measurements have been made for non-single-diffractive events.  The reasoning is usually the use of
a double-armed trigger, which selects events based on the presence of
forward particles on both sides of the detector. These triggers are
typically inefficient for single-diffractive events, where one of the
colliding hadrons remains intact, resulting in a void of activity on one
side of the detector.  The distributions are often corrected for the
remaining single-diffractive contribution, using a given Monte Carlo model. These
models are very poorly constrained and unreliable, resulting in model
dependent corrections with systematic uncertainties that are very
difficult to quantify. A preferable approach is to leave the
distributions uncorrected for a certain type of process.

To further suppress diffraction, one would instead add more requirements
(more ``bias'') on the final state, such as the presence of more than
one track in the fiducial region or the absence of large rapidity
gaps in the event.

This particular example raises a more subtle issue, which is correcting
for detector effects such as the trigger described previously. It is
preferable to use a trigger that is as inclusive as possible, and highly
efficient with respect to the event sample definition. It is also highly
desirable to correct for the trigger efficiency using a data-driven
approach. Relying on a model to correct for a trigger that does not
overlap in phase-space with the particles being measured can also lead
to unreliable results, as the prediction of the particle distributions
in the region outside the measurement acceptance is model dependent.
Again, the systematic uncertainties on these corrections can be large
and very difficult to quantify. Alternatively the trigger signal should
be corrected for detector effects and converted into a hadron-level
definition to be included in the event definition discussed above (\eg
at least one charged particle in the region  $3.0 < |\eta| < 5.0$ with
\pt\ $>$ 50 MeV).

\paragraph{Measurement of the $\ell^+\ell^-$ transverse momentum
distribution in $Z^0/\gamma* \rightarrow \ell^+\ell^-$ events}

An interesting measurement for constraining QCD initial-state radiation
predictions is the \pt\ of the $Z^0$ boson.  There are features of such a
measurement that illustrate many of the issues introduced above. The measurement that is actually made is of the di-lepton
($\ell^+\ell^-$) \pt, not the $Z^0$ \pt, as it is the final-state leptons
that are detected. Correcting back to the $Z^0$ \pt\ traditionally
involves two steps, both of which should be avoided. Firstly, QED
radiation of photons from the final-state leptons would have to be
corrected for, as the photons will carry off some fraction of the lepton
and hence the  $Z^0$ \pt. This involves using a model of QED radiation
that is likely to have uncertainties and may neglect interference
between photons emitted in the initial and final states. It may be
argued that the experimental measurement should be comparable to a
theoretical prediction that does not itself include the effects of QED
radiation. Of course the ideal prediction should include all effects,
but in reality this is not always practical. In this case the result
should be given both with and without the QED corrections. Only
presenting results corrected for QED effects implies that the best Monte Carlo
generators must switch off part of the true process in order to compare
to the data that has been corrected with another (potentially less
accurate)  model -- which is an unnecessarily complicated procedure and
in fact reduces the accuracy of the experimental result.

The correction of QED radiation is a subtle issue and
should be treated with some caution.  In the case of final-state
electrons, the energy and hence \pt\ measurement is typically made in a
calorimeter. Any final-state photons that are emitted in a direction
collinear to the electrons may, if they do not convert into $e^+e^-$
pairs, end up in the same calorimeter cell(s) and hence can be
indistinguishable from the electrons. Thus their energy will be
automatically ``clustered'' back into the energy of the electron. Wide-angle
radiation will of course not be included in the energy
measurement.  Depending upon
the detector, it may therefore be preferable to define the electron in
terms of a cone of electromagnetic particles, analogous to a hadronic jet. The size
of this cone might be experiment specific, but it would be well defined
and easily reproducible at the final-state particle level in a
generator.
See the contribution of Buckley et al.\ in \cite{Butterworth:2010ym}
for a detailed discussion of this issue.
Defining the final state in this way is also theoretically more reliable as
the required corrections are smaller.

In the case of final-state muons the momentum is typically measured by a
tracker that measures the curvature of charged particle tracks in a magnetic
field. In this case the photons do not contribute to the particle
momenta and only the di-muon momentum is measured.
However, it should be noted that the effect of enhanced collinear radiation from muons is much less than that from electrons due to their larger mass.

The second correction required to ``measure'' the $Z^0$ \pt\ is for the
contribution from the virtual photon propagator and for $Z^0/\gamma^*$
interference terms. It is not possible to experimentally distinguish
these contributions, although the $Z^0$ contribution can be greatly
enhanced by making a cut on the invariant mass of the lepton pair in a
window around the $Z^0$ boson mass, \eg 66 $< M_{\ell\ell} <$
116~GeV. Again, this definition is unambiguous and reproducible by any
Monte Carlo generator and should therefore be preferred over claims of a process
involving a particular propagator.

Another important issue that can be demonstrated in this example is the
correction for the acceptance of the final-state leptons.  The
measurement can only be performed on leptons that fall within the
acceptance; the result should therefore be presented in terms of leptons
that pass certain kinematic cuts, \eg \pt\ $>$ 25~GeV and $|\eta| <
2.0$. An attempt to correct for the regions of phase space not measured
can result in large extrapolations, using a certain prediction with
its associated limitations and uncertainties. Again, this adds no
information to the measurement and in fact reduces its
accuracy, reliability and usefulness for validation and tuning.

\paragraph{Jet cross sections}

Jets are designed to reflect and be sensitive to short-distance physics,
but they are composed of hadrons.  An example of poor practice which is
fortunately by now almost extinct in current experimental measurements
is the correction of jets to some ``parton-level final state''. While
hadronization and other soft corrections do need to be evaluated in
order to compare to perturbative QCD calculations, they are now
typically applied to the theory rather than data, and in any case the
data are almost invariably presented first in terms of final-state
particles, even if later corrected in such comparisons. The soft QCD
physics used in such corrections is typically the least theoretically
constrained aspect of a given Monte Carlo program. Cases where, for example, the
underlying event is corrected for at the same time as pile-up, or where
even, bizarrely, the data are corrected directly from a detector level
distribution to some ``leading order'' partonic state, are essentially
useless for any theory comparison except possibly to the particular
version of the particular Monte Carlo generator used to make the correction. Note that there are
methods of correcting for, or reducing the effects of, underlying event which
are well defined and model-independent, see for example~\cite{Cacciari:2009dp}.

\mcsubsection{Evaluation of MC-dependent systematic errors}

Even if a measurement is defined in a model-independent way, as
described above, there will still in general be model dependence in the
corrections applied to remove or quantify detector effects, since
detector response is often evaluated using Monte Carlo tools.  In addition it is
often necessary to use Monte Carlo predictions of signal and background rates
or kinematic distributions in order to extract the significance of a
signal or place limits in the absence of a signal.

As a general rule, experimentalists should use all Monte Carlo generators that
simulate the process of interest and potential backgrounds. While it is
not always sufficient to take the difference between the result of two
different generators or tunes as a systematic uncertainty, it is useful
for getting an idea of the limitations of and differences between the
predictions.

Often it is necessary to unfold a distribution, both for detector
inefficiencies and resolutions.  In some cases, such as a low detector
efficiency that is localized in $\phi$, the azimuthal angle around the
beam-pipe, using a Monte Carlo to model the inefficiency will be very reliable.
There is no physical significance to this region, so if the detector is
well modelled and if the generator models the physics well in other
$\phi$ regions, the modelling of the inefficiency will be robust.  In
other cases the underlying physics of the model used to do the
unfolding can have a significant effect on the result.  There are many
different possible techniques for unfolding, which depend to varying
degrees on the underlying model. The dependence on the model can be
reduced by \eg reweighting the Monte Carlo events to match the data for relevant
kinematic distributions.
Note that it is not necessarily only the distribution that is being
unfolded that is relevant for the unfolding and sometimes unfolding in
two dimensions, with a second distribution that is chosen because it
is strongly correlated with the size of the correction, can be useful.

 Residual uncertainties can be determined by
comparing different Monte Carlo models or tunes, as long as the differences
between the tunes are sufficiently large to cover the difference
between Monte Carlo and data in relevant distributions.

A common Monte Carlo systematic is the modelling of a detector response to QCD
jets. In general this depends upon the details of the detector and upon
the fragmentation of the jets (charge-to-neutral ratio, energy
partition, etc). The model dependence of a calibration may be tightly
constrained by requiring that the simulation describes quantities such
as the number of charged particles near jets, or the energy flow around
jets, satisfactorily.

Finally, there are often MC-dependent systematic uncertainties
associated with the modelling of the background and/or signal rates and
kinematic distributions.  Such uncertainties can be evaluated and
minimized in the manner described above, with the constraining
requirements coming from comparisons between data and Monte Carlo in control
regions. In addition, if any relevant measurements from the same or other
experiments provide constraints on the Monte Carlo predictions used to extract
the signal, these should be used in the assessment of the systematic
uncertainties.

In general how to assess MC-dependent systematic uncertainties depends
on the specific analysis. However, it is good practice to consider all
possible constraints from data, whether it be control regions in the
same measurement or different results. In addition the limitations of
the Monte Carlo generator used should be understood, and different generators and tools
should be considered.

% Local Variables:
% mode: LaTeX
% TeX-master: "../mcreview"
% End:

\mcsection{Validation and tuning}
\label{sec:rivet-professor}

Validation in the context of Monte Carlo generators means confronting a model with all
relevant data that it claims to be able to describe. It is essential that the
validation is global, because the model should describe the underlying physics
and not just parameterize the data, otherwise it would not have any predictive
power. In this sense validation is important for developing models as well as
for debugging both code and physics models. Tuning means adjusting the free
parameters of the model within their allowed ranges to improve the description
of the relevant data.

\mcsubsection{Generator validation and tuning strategies}
\label{sec:tuningstrategy}

As mentioned, generator validation must simultaneously consider a range of
observables to be meaningful and predictive beyond the observables considered.
The choice of observables must also be limited according to the model being
considered: poor description of an observable whose responsible process is not
modelled conveys little information.

For tuning, similarly, a range of observables is required for predictivity and
to obtain a generally usable single set of parameters. Again, depending on the
suitability of observables to the model being tuned, it may or may not be
\emph{possible} to describe all data simultaneously: this in itself may be a
useful result for model development. The optimization of MC parameters to the
chosen observables may be performed manually -- guided by the expected physical
behaviour of the models -- or by a more automated method driven by the quality
of the fit to data. In both approaches, some sampling of the parameter space is
typical to ascertain the generator behaviour in response to parameter changes.
The allowed ranges of parameter values in this sampling typically span all
values for which the underlying physical picture is valid, although scans of
more restricted ranges are usually necessary to produce a final tune.

The choice of reference data is important since all simulations lack some known
physics effects. Generator tuning should primarily optimize phenomenological
simulation aspects, and not make up for shortcomings in modelling of event
aspects that should be robustly described by calculable QCD. For example, tuning
a Monte Carlo generator that contains only a $2 \to 2$ scattering matrix element to
high jet-multiplicity data will tend to distort the parton shower and underlying
event in an attempt to make up for the lack of higher multiplicity matrix
elements. However, there are modelling aspects that do not fall neatly into
either a perturbative or non-perturbative definition, \eg the primordial~\kt as
discussed in \SecRef{sec:primkt}. The parameters of these models are perfectly
valid for use in tuning -- with appropriate care. For example, while a
primordial~\kt width may have no numerical upper limit in the generator
implementation, a tuned value located far into the perturbative regime would be
an abuse of the model and suggest deficiencies elsewhere.

As general-purpose event generators contain models for many processes, most have
of order 15 or more tuning parameters. This defines a parameter space whose
dimensionality is far too high for comprehensive exploration, even with an
automated sampling method. The practical consequence is that factorization of
the parameters into minimal sets suitable for each group of observables has been
found to be important. Hence, tuning of generators usually occurs in several
distinct stages, in the following order:
\begin{itemize}
\item {\em Hadronization and final-state fragmentation:} The flavour and kinema\-tic
  structure of the final-state shower and hadronization mechanisms are assumed
  to be universal between \ee and hadron colliders. As \ee
  observables (\eg event shapes and identified particle rates and \pt spectra)
  may be described by a generator without first requiring a reasonable tune of
  initial-state hadron collider effects, these are typically used to tune final-state
  shower and hadronization parameters. Flavour structure and kinematics
  may themselves be factorized to some extent, perhaps with iteration.

  Some typical parameters for tuning of fragmentation kinematics are the
  \alphaS/\LambdaQCD values and IR-cutoff for the final-state shower, the string
  tension and fragmentation function parameters for string hadronization models,
  and the gluon constituent mass and cluster momentum smearing in cluster
  hadronization models. Light and heavy quark fragmentation kinematics are often
  treated separately, which permits further factorization to charm- and
  bottom-specific observables without compromising the statistically dominant
  light fragmentation. Tuning of flavour parameters in hadronization -- for
  string hadronization in particular -- introduces an extra collection of
  parameters for, \eg, enhancement and suppression of strangeness/charm/beauty,
  $\eta/\eta'$ and baryon fractions. A final semi-distinct group of parameters
  may be available for adjusting the admixtures of different orbitally excited
  hadron states: whether these are considered in tuning depends on the purpose
  at which the tuning is aimed.

  The recent availability of identified particle data from hadron colliders such
  as RHIC and the LHC is of interest from the point of view of hadronization
  tuning, but violates the desirable feature of not requiring a viable
  initial-state effect tune before beginning. At present, such data have not
  been included in tunings: they are, however, of great interest for validating
  the assumption that hadronization parameters tuned to \ee observables will
  remain valid in a hadron collider environment.

\item {\em Initial-state parton shower:} Once a reasonable tune of final-state
  parameters has been obtained, the typical next step is to tune the
  initial-state (space-like) parton shower parameters. The reason for tuning
  this before the soft QCD effects is that we desire the shower to be tuned
  to observables with little MPI/beam-remnant contamination, and then use the
  full flexibility of the heavily-parameterized MPI machinery to make the final
  best fit to data. This way, we avoid the danger of absorbing effects which
  should be perturbatively describable into the relatively unconstrained MPI
  modelling.

  Some typical observables for initial-state shower tuning are dijet azimuthal
  decorrelations (from the Tevatron and the LHC, with concern that $2 \to 2$ matrix
  elements are not abused to include third hard jet contributions) and hadron
  collider jet shapes. Typical parameters are the shower IR-cutoff, the shower
  \alphaS/\LambdaQCD, and perhaps a scaling factor for the \alphaS evaluation
  scale and the starting scale for the parton cascade. The philosophy of what
  shower parameters are available for tuning varies according to generator: some
  permit use of multiple \alphaS definitions, while others insist that the same
  values be used throughout the generator, perhaps based on the value specified
  in the PDF.

\item {\em MPI and beam remnant effects:} As discussed above, since MPI modelling is
  the element of Monte Carlo modelling least constrained by \textit{ab initio} QCD
  calculation, it is left untuned until the final stage. There may be many
  parameters in MPI models -- essentially all modelling aspects described in
  \SecRef{sec:mbmpi} can introduce one or more parameters.  The key parameters
  common to most eikonal MPI models are the \pPerp{\mr{min}} cutoff/regulator
  for perturbative $2 \to 2$ scattering, the parameterization of the scaling of
  this cutoff with collision energy, the hadronic matter distribution/overlap,
  and any parameters relating to colour-reconnection of either strings or
  clusters. The primordial~\kt width is often considered as part of this tuning
  step, as it may affect soft QCD observables as well as the peak region of the
  $Z^0$ \pt spectrum. As MPI models generate multiple scattering from low-$x$
  gluons extracted from the beam-remnants, they are profoundly affected by the
  choice of PDF. Hence, distinct MPI tunings are required for each PDF. The most
  obvious parameter affected by a change of PDF is \pPerp{\mr{min}}: when using
  a PDF with a large low-$x$ gluon fraction, the MPI model will require more
  screening of the divergent partonic cross section than for a PDF with a
  smaller amount of soft gluon. Accordingly, tunes with PDFs such as
  LO*~\cite{Sherstnev:2007nd} which have a lot of low-$x$ gluon tend to have
  higher \pPerp{\mr{min}} values than tunes of the same generator using \eg the
  CTEQ6L1~\cite{Pumplin:2002vw} PDF.

  The observables for MPI tuning are minimum bias and underlying event data from
  as many hadron colliders as possible. As a key feature of soft QCD modelling
  is the scaling of MPI activity with the collider center-of-mass energy, a wide
  range of collision energies is desirable. Experimental tunings may place
  emphasis on the collider of most interest -- currently the LHC -- for the
  purpose of best describing the soft QCD backgrounds to hard-process simulations
  at that collider. To date the most comprehensive MPI tunes have included data
  from the CERN S\ppbar{S}, RHIC, the Tevatron, and the LHC. HERA data have not yet
  been included in hadron collider MPI tuning, but whether a single tune can
  describe both \ep and \pp/\ppbar data is a strong check on the domain of validity of
  the generator's MPI model~\cite{Carli:2010jb}. LHC results on identified
  particle distributions in minimum bias and underlying event data will provide
  another test of currently unprobed model details.
\end{itemize}

The \rivet~\cite{Buckley:2010ar} package for MC generator validation and the
\professor~\cite{Buckley:2009bj} system for generator tuning have become
established tools in both the collider theory and experiment communities. Their
strength is in systematically verifying event simulations and optimizing their
parameters, where required and physically sensible. Both tools are
described in the following sections.

\mcsubsection{\rivet}
\label{sec:rivet}

\rivet is a Monte Carlo \emph{validation} tool: it encodes MC equivalents of an ever
more comprehensive set of high-energy collider analyses which are particularly useful
for testing the physics of MC generators. \rivet does not itself produce
generator tunings, but provides a standard set of analyses by which to verify
the accuracy of a given generator with a given tuning. These analyses are based
upon a set of calculational tools that make writing of new analyses by either
phenomenologists or experimentalists relatively straightforward.

Several fundamental design principles have been derived from the experience on
\rivet's predecessor system, \hztool~\cite{Bromley:1995np,Waugh:2006ip}, and
from iteration of the \rivet design:
\begin{itemize}
\item No generator steering: \rivet relies entirely on being provided, by
  unspecified means, with events represented by the \hepmc~\cite{Dobbs:2001ck}
  event record.
\item No generator-specific analyses: official \rivet analyses are specifically
  not allowed to use the generator-specific portions of the supplied event
  records. Apart from a few very limited exceptions, all analyses are based
  solely on physical observables, \ie those constructed from stable particles
  (those with \hepmc status 1) and physical decayed particles (those with status
  2). This approach is fully compatible with the approach to robust generator
  phenomenology discussed in \SecRef{sec:mcfriendlyobs}.
\item \rivet can be used as a \cpp library to be interfaced with generator
  author or experiment analysis frameworks, as a Python module for construction
  of higher-level tools (for example, much of the \rivet documentation is
  generated this way), or as a command line tool (which itself makes use of the
  Python interface). This exemplifies a general philosophy to keep the tools
  simple and flexible, rather than constrain \rivet's applicability to a
  pre-defined collection of specific tasks.
\end{itemize}

Internally, \rivet analyses are based on a comprehensive set of calculational
tools called \emph{projections}, which perform standard computations such as jet
algorithms (using \fastjet~\cite{Cacciari:2006sm}), event shape observables, and
a variety of other common tasks. Use of projections allows, \eg
\begin{itemize}
\item simplification of analysis code;
\item encapsulation of complexities arising from the ban on use of event record
  internal entities (the summation of photon momenta around charged leptons
  during vector boson reconstruction is a good example);
\item and efficiency gains over pure library functions, via a complex (but
  hidden) system of automatic projection result caching.
\end{itemize}

Users can write their own analyses using the \rivet components and use them via
the \rivet programming interface (API) or command-line tool without re-compiling \rivet, due to use of
an analysis ``plugin'' system. Separation between generator and \rivet on the
command-line is most simply achieved by using the \hepmc plain text
\texttt{IO\_GenEvent} format via a UNIX pipe (a.k.a. FIFO): this avoids disk
access and writing of large files, and the CPU penalty in converting event
objects to and from a text stream is in many cases outweighed by the
general-purpose convenience. For generator-specific use of \rivet, the
programmatic interface allows \hepmc objects to be passed directly in code,
without this computational detour. While this method requires some up-front
integration into generator frameworks, eliminating the temporary conversions to and
from a plain text format provides a significant performance gain, and the API
control gives more flexibility, \eg in use of resulting histograms, than is
possible with the command-line tool. A sister tool,
\agile~\cite{Buckley:2007hi}, is provided for convenient control of several
legacy \fortran-based generators, while the \mcnet generators either feed a
\hepmc text stream into \rivet, or in several cases use the direct programmatic
interface.

Reference data for the standard analyses are included in the \rivet package as a
set of XML files in the AIDA~\cite{AIDA} format. After several years of
re-development as part of the CEDAR~\cite{Buckley:2007hi} project, the
\hepdata~\cite{Buckley:2010jn} database of HEP experimental results is used to
directly export data files usable by \rivet from its Web interface at
\url{http://hepdata.cedar.ac.uk/}: this can be used by anyone developing new
analyses based on papers in \hepdata. Analysis histograms are directly booked
using the reference data as a binning template, ensuring that data and MC
histograms are always maximally consistent.

% \rivet is in use within the MC generator development community, particularly in
% general-purpose shower MC programs, and the LHC experimental community, for MC
% validation and MC analysis studies which do not require detector simulation.

The most recent version of \rivet at the time of writing is 1.4.0. This release
focuses on quality control of official analyses, and was largely driven by
requirements of LHC experiment MC tuning studies, by validation requirements of
\mcnet ME/PS merging algorithm developments, and by increasingly wide use of
weighted events to cover disparate phase space regions in single MC
runs. Development of \rivet has also driven much of the feature development and
bug-fixing in \hepmc in recent years, in particular improving the treatment of
physical units, propagation of cross section information in the event record
(supported by all \mcnet generators), and a more complete event weighting
system.

With the increasing user demand for \rivet functionality, major effort has been
devoted to making the command-line tools and post-processing scripts intuitive,
comprehensive and bug-free. The emphasis on usability also led to making \rivet
analyses ``self-documenting'': each analysis has a structured set of metadata
specifying name, authors, run conditions, a description, etc., which is used
(via the Python interface) to provide interactive help, HTML documentation, and
a reference section in the \rivet manual.

The next major stage of development is the upgrade of \rivet's histogramming and
data analysis code, which is currently rather basic. The developing data
analysis library will be of general-purpose usefulness, but will provide some
features particularly useful for MC validation and tuning analyses, such as
parallel handling of event weight vectors for integrated event re-weighting, and
non-contiguous histogram binning as required by several experimental analyses.
The upgrade will also enable statistically accurate combination of runs,
allowing for greater parallelization of \rivet analyses that require large
event statistics: this major development will mark \rivet version 2.0.0.

\mcsubsection{\professor}
\label{sec:professor}

The \professor system builds on the output of MC validation analyses such as
those in \rivet, by optimizing generator parameters to achieve the best possible
fit to reference data. The main description of \professor's details is found in
reference~\cite{Buckley:2009bj}, so we will only summarize it here. The most
recent version of \professor at the time of writing is 1.1.0.

Fundamentally, generator tuning is an example of the more general problem of
optimizing a very expensive function with many parameters: the volume of the
space grows exponentially with the number of parameters and the CPU requirements
of even a single evaluation of the function mean that any attempt to scan the
parameter space will fail for more than a few parameters. Here, the expensive
function is running a generator with a particular parameter set to recreate a
wide range of analysis observables, using a package such as \rivet. The approach
adopted by \professor is to parameterize the expensive function based on a
non-exhaustive scan of the space: it is therefore an approximate method, but its
accuracy is systematically verifiable and it is currently the best approach available.

The MC parameterization is generated by independently fitting a function to each
of the observable bin values, approximating how they vary in response to changes
in the parameter vector. One approach to fitting the functions would be to make
each function a linear combination of algebraic terms with $n$ coefficients
$\alpha_i$, then to sample $n$ points in the parameter space. A matrix inversion
would then fix the values of $\alpha_i$. However, use of a pseudoinverse for
rectangular matrices allows a more robust coefficient definition with many more
samples than are required, with an automatic least-squares fit to each of the
sampled ``anchor points'': this is the method used by \professor. By aggregating
the parameterizations of all the observable bins under a weighted goodness of
fit (GoF) measure a numerical optimization can be used to create an ``optimal''
tune. The GoF currently used in \professor is a heuristic function based on a
\chisq, but augmented with inclusion of all available errors -- as opposed to
the traditional Pearson definition which uses the number of MC events in each
bin as the sole uncertainty measure in the denominator.  In practice, many
different semi-independent (sampled with replacement) combinations of MC runs
are used to provide a systematic handle on the degree of variation expected in
tunes as a result of the inputs, to avoid the problem that a single
``maximum-information'' tune may not be typical of the parameter space.

The \professor tools have been used in tuning of several generators, including
the \mcnet ones already featured, particularly for the hadronization and soft
QCD multiple-scattering aspects of event generation, where theory is least
predictive and generators have most free parameters. Initial studies focused on
the \pythiasix MC generator~\cite{Sjostrand:2006za}, as this had already been
the focus of a CDF tuning campaign and was well understood. It was found that
the parameterization method worked well in all cases, and a range of systematic
methods and tools were developed to check the accuracy of the approximations,
such as line-scans through the parameter space. Parameter spaces and observables
with discontinuous behaviours -- \eg some aspects of cluster hadronization --
remain problematic for any method that assumes smooth parameterizations.
Several approaches exist to handle this, including parameter transformation, use
of separate parameterizations in distinct regions, and manually avoiding tuning
across such discontinuities. As discussed in \SecRef{sec:tuningstrategy},
factorization of the total parameter space into block diagonal tuning stages is
required: with \professor, \oforder{10} parameters at a time has been found to
be a practical maximum.

Much of the effort in constructing a generator tune is now focused on the
development of a set of fit weights for the observables in a tune: different
applications may wish to place different emphases on different observables,
\eg LHC vs.\ Tevatron data, or underlying event vs.\ minimum bias data. Once a
set of weights has been chosen, it is a matter of logistics to create equivalent
tunes for different PDFs: this permits a more accurate measure of the systematic
effect of PDF choice than was previously possible. In the particular case of MPI
model tuning to soft QCD observables, this approach has shown that much of the
effect of PDF changes can be absorbed into typical MPI model parameter
choices.

The intermediate parameterizations have proven useful in their own right: the
\texttt{prof-I} GUI tool provides interactive visualization of observable
responses to parameter changes and is useful for MC developers as a
model-exploration and debugging tool. The usefulness of fast parameterization is
not limited to MC generators, and \professor has been used for other studies
from extra-solar planets to exploration of supersymmetric model phenomenology.
As the LHC era matures, the demand for ``new'' tunes will naturally reduce -- to
be replaced with a need for more accurate assessments of systematic
uncertainty. \professor will remain a useful tool for this purpose, as the MC
parameterizations can be used for construction of tune-error estimates. In one
approach, parameter points sampled from the multi-dimensional Gaussian
distribution defined by the numerical minimizer covariance matrix are mapped
into observables using the parameterizations, defining error bands for given
statistical confidences. Alternatively, the same covariance matrix can be used
slightly differently to construct Hessian ``error tunes'', or
``eigentunes''. The latter approach is already in use by LHC experimental
collaborations to improve the accuracy of MC-derived modelling systematics for
detailed LHC physics studies.

% Local Variables:
% mode: LaTeX
% TeX-master: "../mcreview"
% End:

\mcsection{Illustrative results}
\label{sec:physics-areas-where}
In this section we show results from the \herwigpp, \pythiaeight and \sherpa MC
generators described in \PartRef{sec:spec-revi-main}, compared to data from a
variety of collider experiments from \lep to the \lhc. In all these plots, the
versions and tunes shown are: for \herwigpp, a pre-release copy of version 2.5.0
with the default tune to the MRST~LO$**$ PDF; for \pythiaeight, version 8.145
with tune 4C and the \cteq{6L1} PDF; and for \sherpa, version 1.2.3 with the
default tune and the \cteq{6L1} PDF. All the analyses shown are in \rivet.

It should be emphasised here that the generators have not yet been optimally
tuned to LHC data overall, or indeed to any of the particular plots shown.  The
intention here is rather to give an ``existence proof'' of output from the
programs.  Therefore the success or otherwise of a generator in fitting the data
should not at this stage be taken as a true measure of its performance.  As
stated in the figure captions, as tuning progresses the plots will be updated
and archived at \url{http://mcplots.cern.ch/}.

\FigRef{fig:cmp:intrinsic-kt} shows the $Z^0$ transverse momentum
distribution in $p\bar p$ collisions at $\sqrt s=1.8$ TeV, compared
with CDF data~\cite{Affolder:1999jh}.  As discussed in \SecRef{sec:primkt}, the
position and shape of the peak in this distribution is sensitive to
the modelling of non-perturbative effects generically termed
``primordial \kT''.

\begin{figure}[tp]
  \centering
  \subfigure[Charged multiplicity]{\label{fig:cmp:mpi-minbias-atlasnch}\includegraphics[scale=0.7]{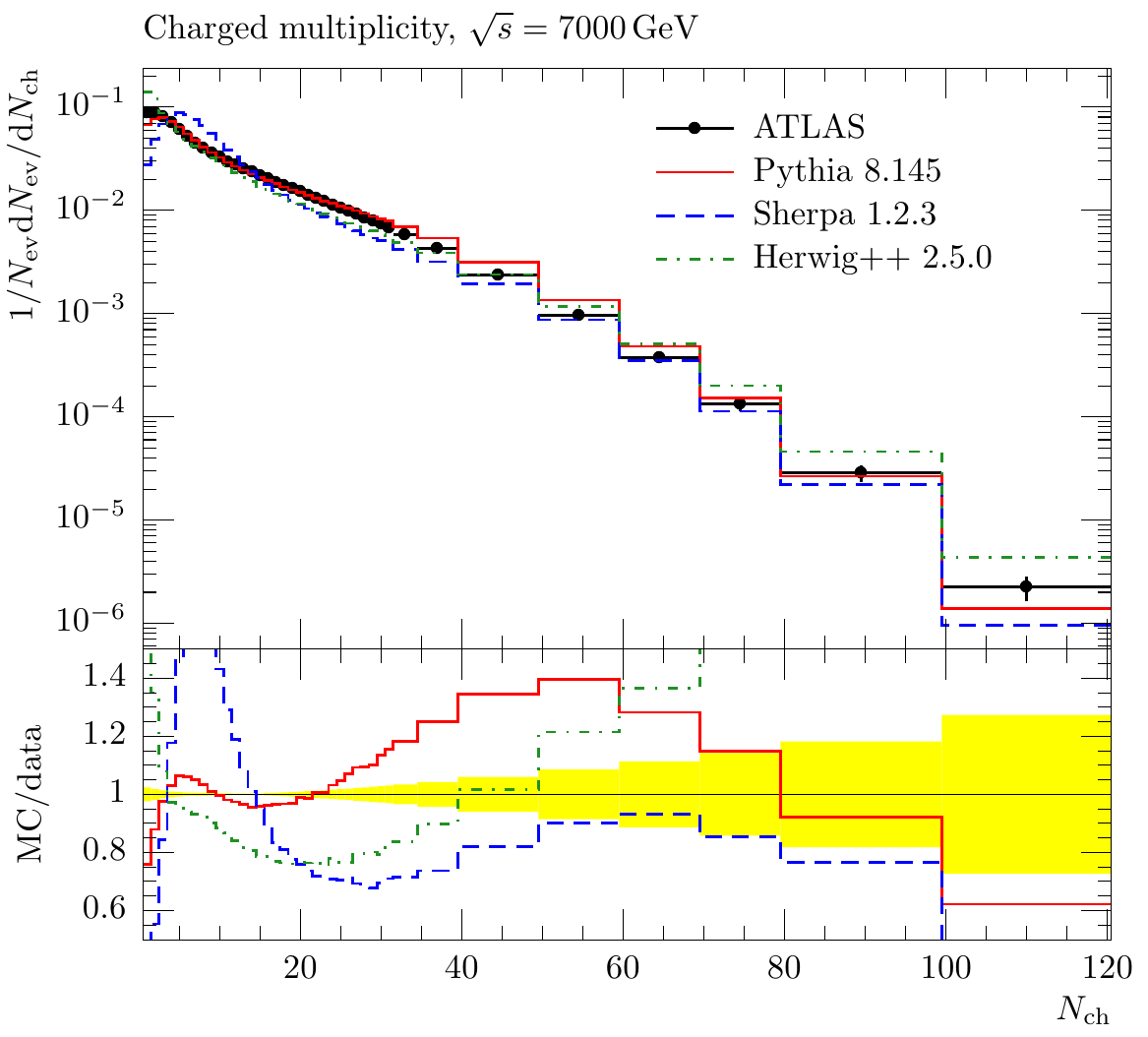}}
  \subfigure[$\langle \pT \rangle$ vs. $N_\text{ch}$]{\label{fig:cmp:mpi-minbias-atlasptnch}\includegraphics[scale=0.7]{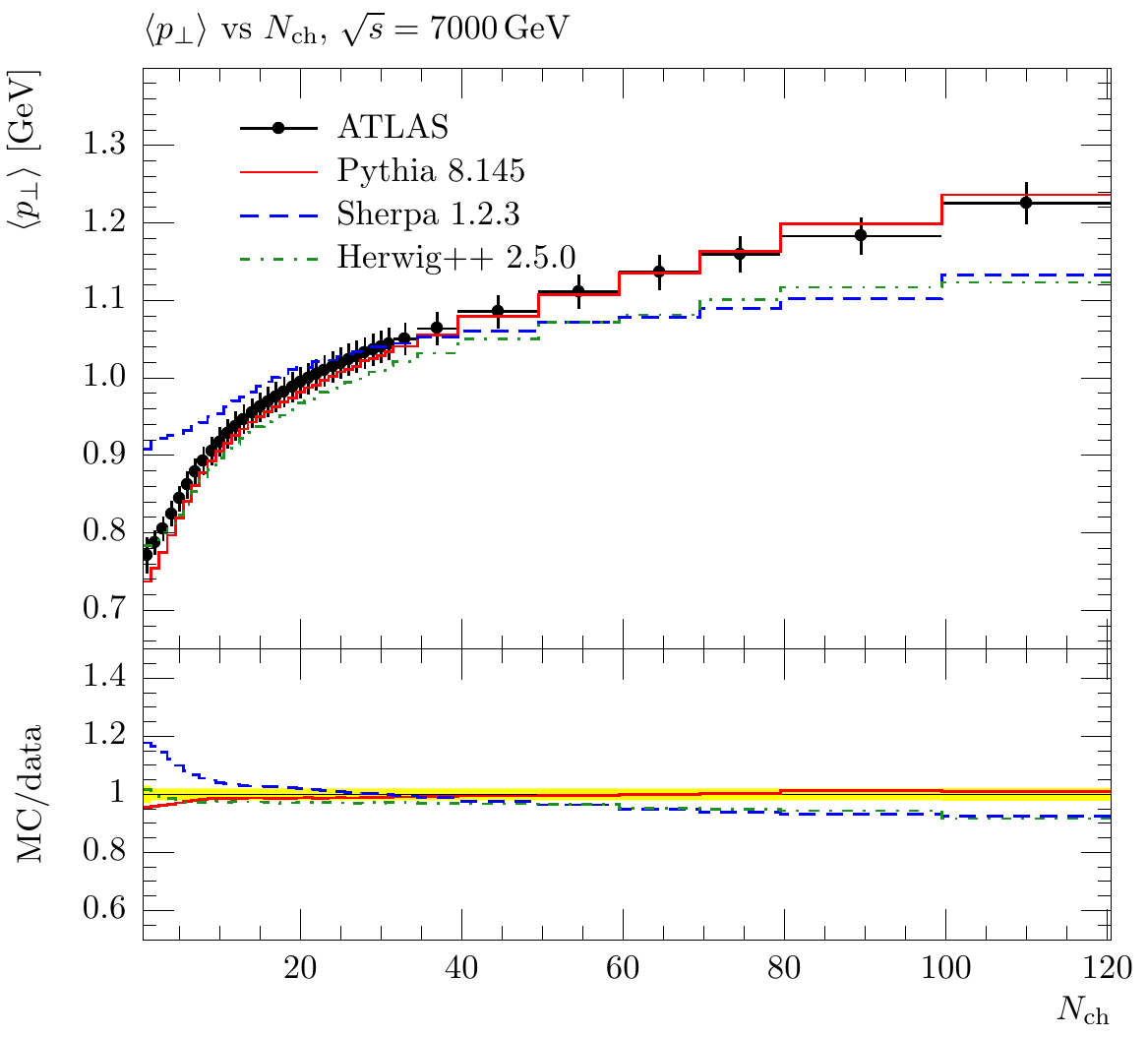}}
  \caption{ATLAS minimum bias charged particle distributions at 7~TeV,
    with a charged particle \pT cut of $\pT > 500~\text{MeV}$,
    $|\eta| < 2.5$, $c\tau > 10\,\text{mm}$
    \cite{Atlas:2010xx}. The MC description of these observables is
    dominated by the tuning of the MPI models: the inclusive charged
    multiplicity is dependent on the level of MPI activity, and the
    correlation between $\langle \pT \rangle$ and $N_\text{ch}$ is
    affected by colour reconnection, as described in
    \SecRef{sec:minim-bias-underly}. Up-to-date versions of these plots
    can be found at \url{http://mcplots.cern.ch/}.}
  \label{fig:cmp:mpi-minbias}
\end{figure}

In \FigRef{fig:cmp:mpi-minbias} we show results on soft QCD processes
compared with ATLAS minimum bias data~\cite{Atlas:2010xx}.

\begin{figure}[tp]
  \centering
  \subfigure[CDF~Run~2 transverse $\pT^\text{sum}$ in leading jet events at 1960~GeV]{\label{fig:cmp:mpi-ue-cdf2-ptsum}\includegraphics[scale=0.7]{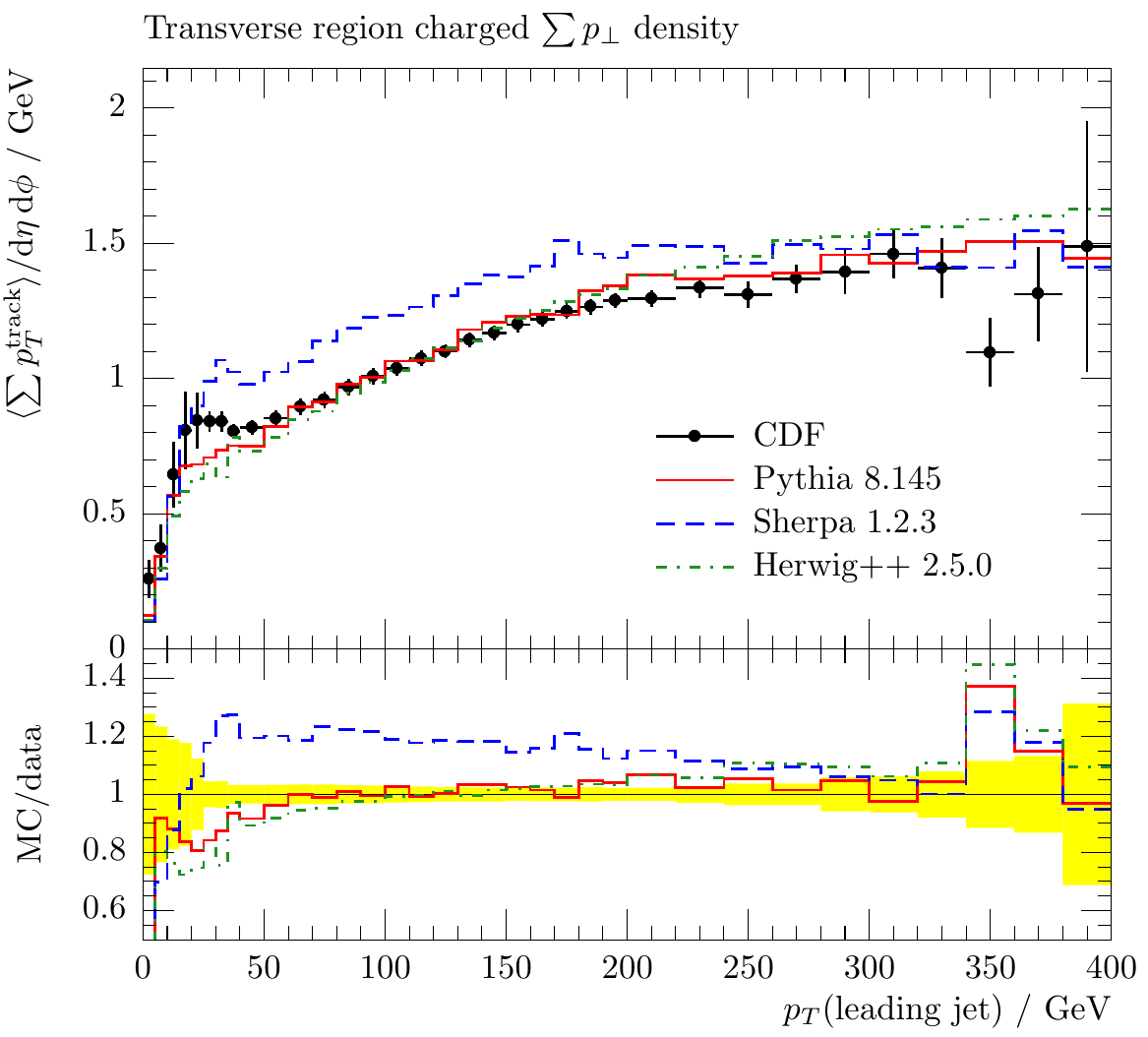}}
  \subfigure[CDF~Run~2 towards $\pT^\text{sum}$ in Drell-Yan events at 1960~GeV]{\label{fig:cmp:mpi-ue-cdf2-ptsum-drellyan}\includegraphics[scale=0.7]{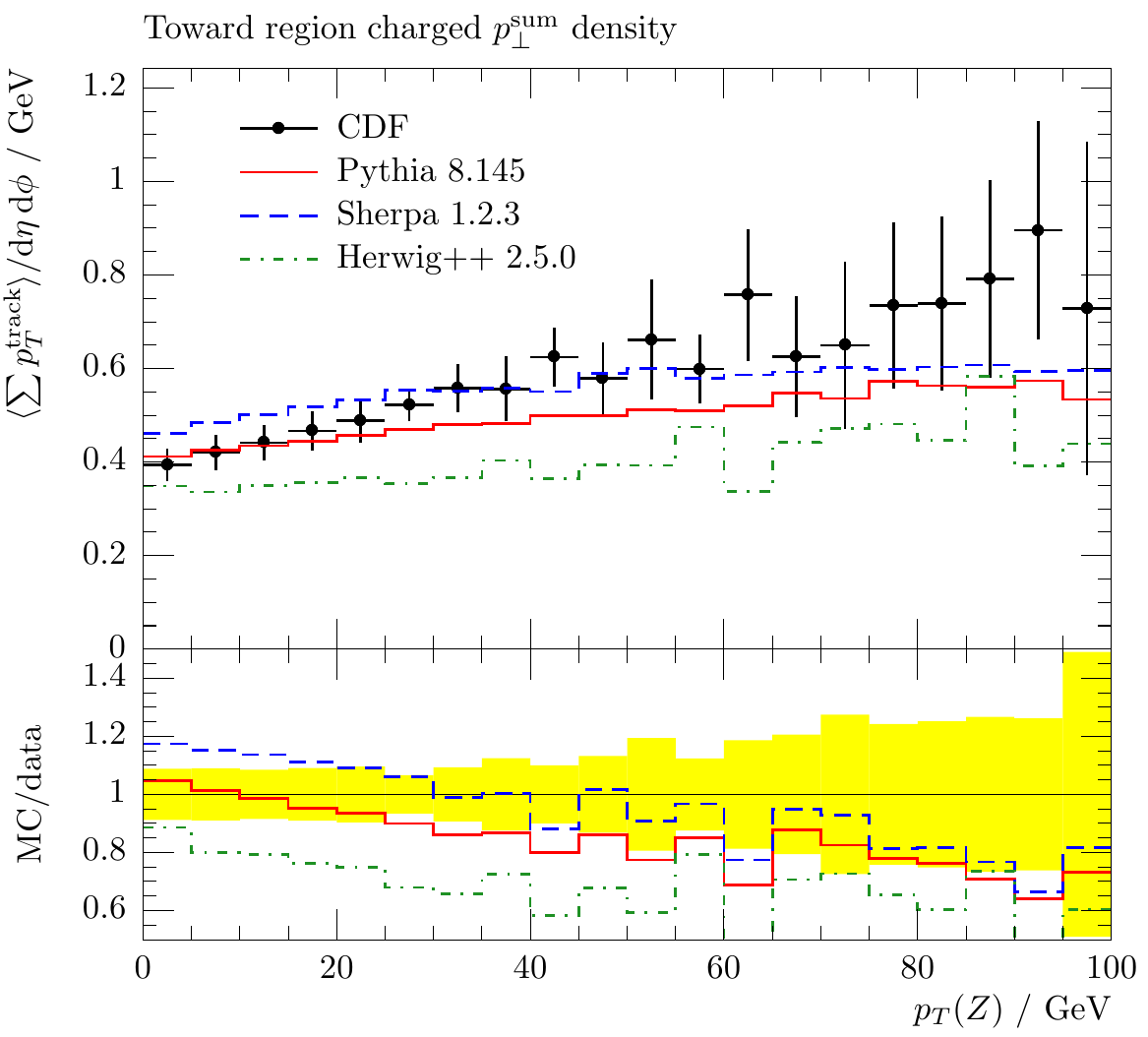}}
  \caption{CDF Run~2 underlying event profile observables: the $\pT^\text{sum}$
    is shown in the transverse region for leading jet events, and the towards
    region in Drell-Yan events \cite{Aaltonen:2010rm}. Up-to-date
    versions of these plots can be found at \url{http://mcplots.cern.ch/}.}
  \label{fig:cmp:mpi-ue-cdf}
\end{figure}

\FigsRef{fig:cmp:mpi-ue-atlas-1}--\ref{fig:cmp:mpi-ue-atlas-3} show
observables relevant to the underlying event, discussed in
\SecRef{sec:minim-bias-underly}, compared to ATLAS data at 900 GeV
  and 7 TeV~\cite{Aad:2010fh}.  Various indicators of event activity
  are measured in the transverse region, \ie at $60^\circ-120^\circ$ in
  azimuth, relative to the leading-\pt\ charged particle.
\FigRef{fig:cmp:mpi-ue-cdf} shows similar results for the Tevatron, in
the transverse region relative to the leading jet, and in the towards
region, \ie closer than $60^\circ$ in azimuth, relative to the $Z^0$
direction in Drell-Yan events, compared to CDF data~\cite{Aaltonen:2010rm}.

\begin{figure}[tp]
  \centering
  \subfigure[Summed out-of-event-plane $\pT$ \cite{Barate:1996fi}]{\label{fig:cmp:eeshapes-ptout}\includegraphics[scale=0.7]{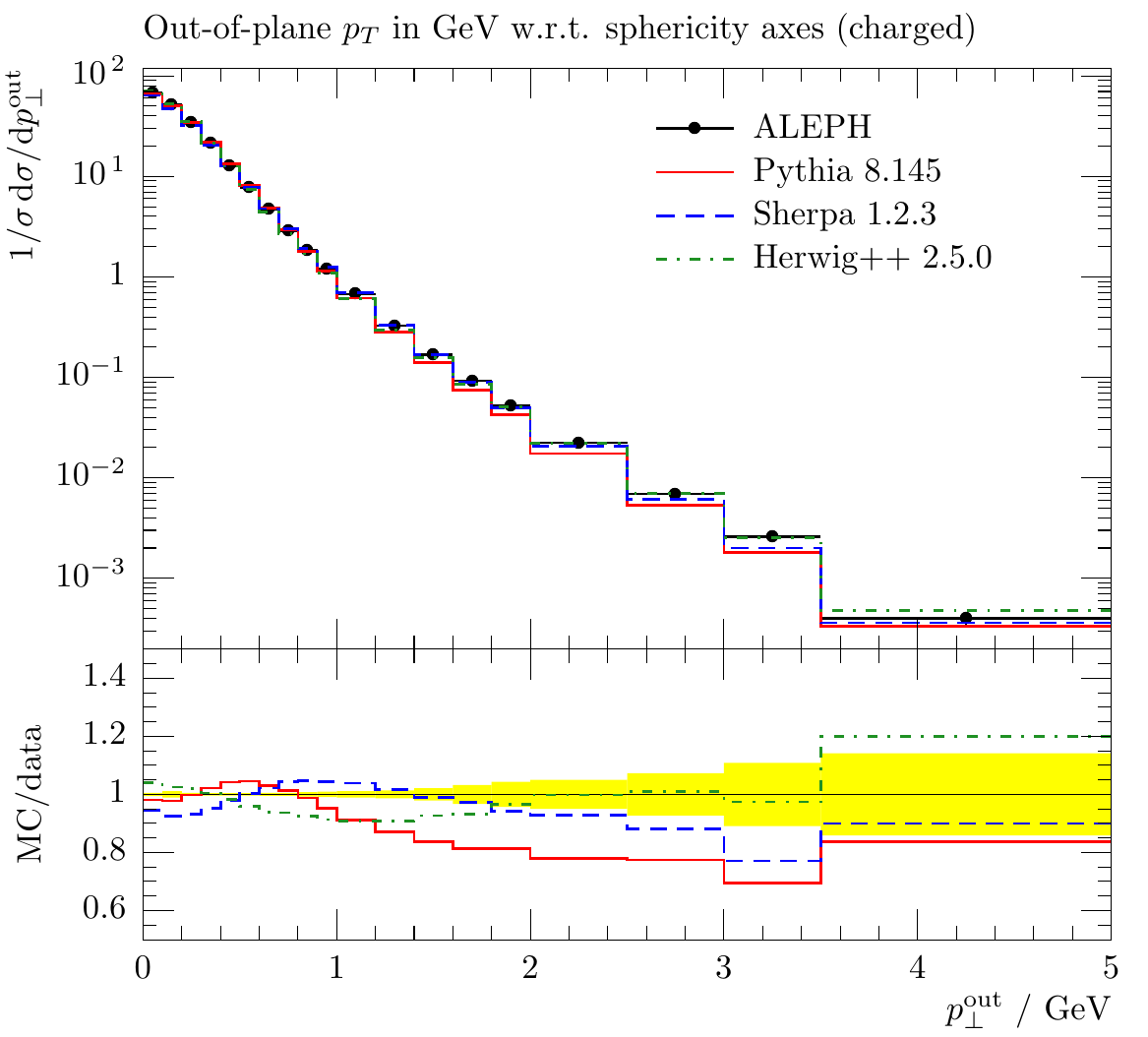}}
  \subfigure[$1-\text{Thrust}$ \cite{Heister:2003aj}]{\label{fig:cmp:eeshapes-thrust}\includegraphics[scale=0.7]{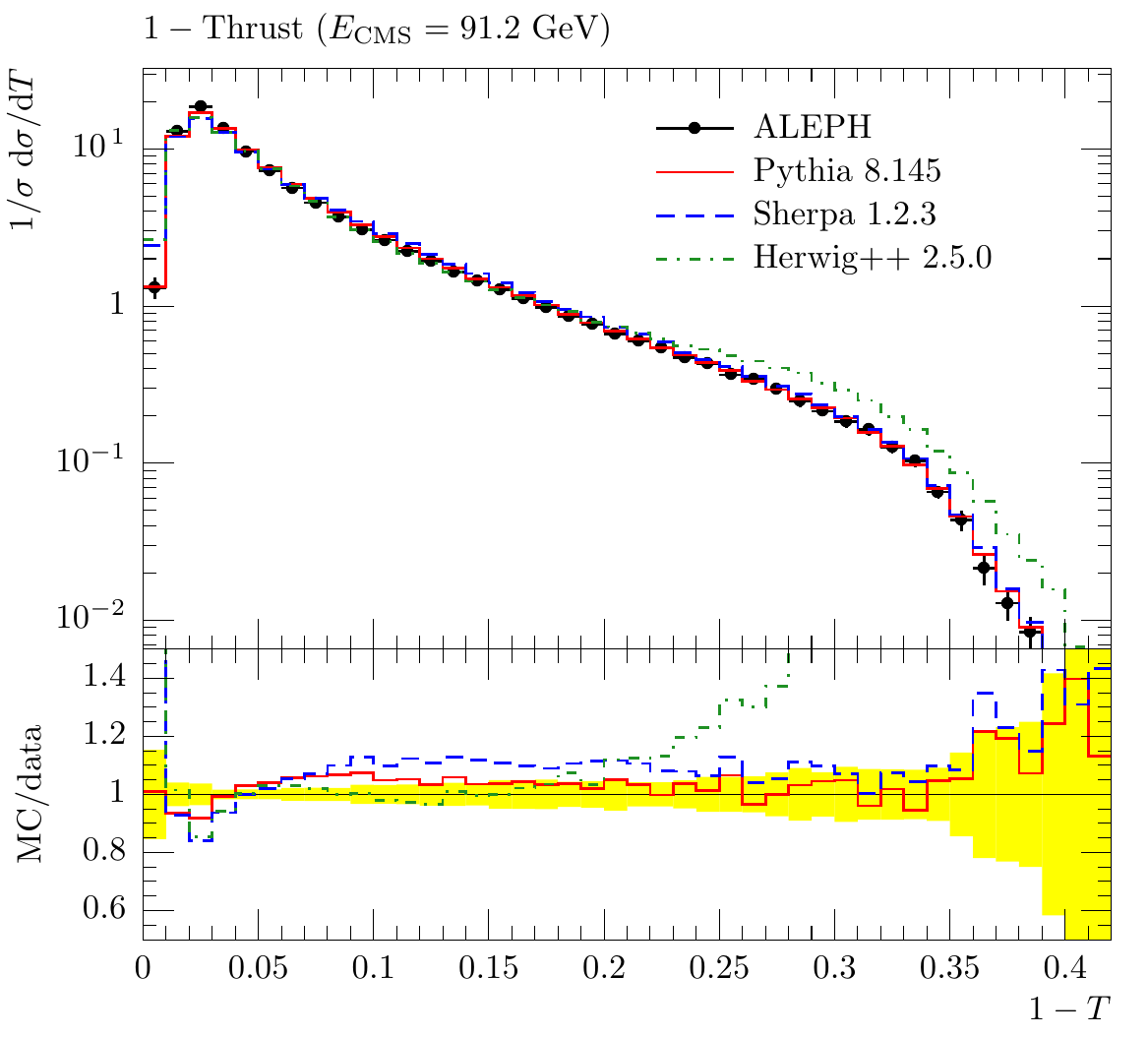}}\\
  \caption{\ee event shapes measured by ALEPH at 91\,GeV \cite{Barate:1996fi,Heister:2003aj}. Up-to-date versions of these plots can be found at \url{http://mcplots.cern.ch/}.}
  \label{fig:cmp:eeshapes}
\end{figure}

\begin{figure}[tp]
  \centering
  \subfigure[Differential 3-jet rate]{\label{fig:cmp:eeshapes-jety23}\includegraphics[scale=0.7]{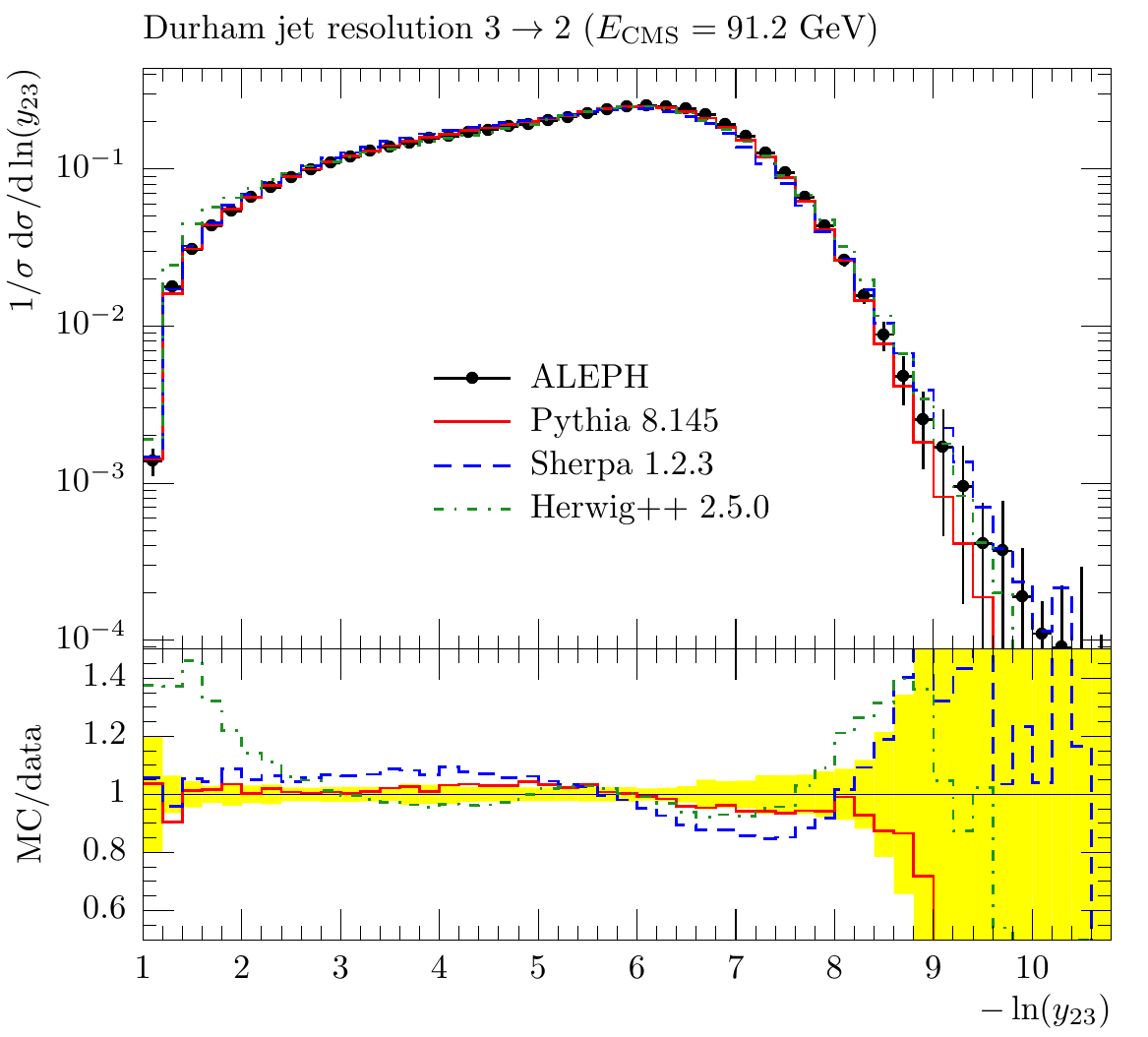}}
  \subfigure[Differential 5-jet rate]{\label{fig:cmp:eeshapes-jety56}\includegraphics[scale=0.7]{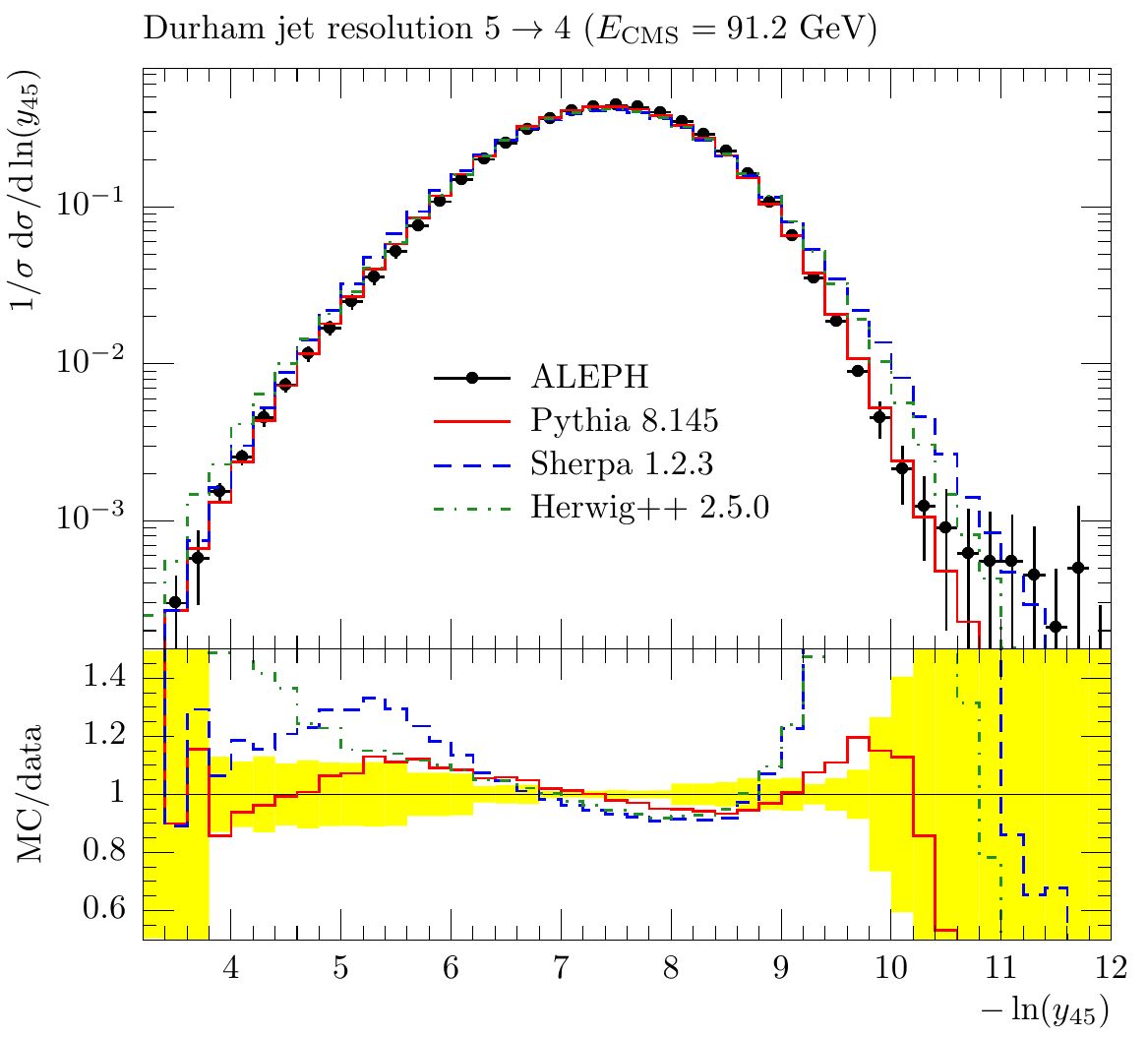}}
  \caption{\ee differential jet rates measured by ALEPH at 91\,GeV \cite{Heister:2003aj}.  The quantity $y_{n-1,n}$ is
    the value of the \kt-jet resolution at which $n$ jets are just
    resolved.  Up-to-date versions of these plots can be found at
  \url{http://mcplots.cern.ch/}.}
  \label{fig:cmp:eejetrates}
\end{figure}

\begin{figure}[tp]
  \centering
  \subfigure[D\O{} dijet azimuthal decorrelation \cite{Abazov:2004hm}]{\label{fig:cmp:dijet-azimdecorr}\includegraphics[scale=0.7]{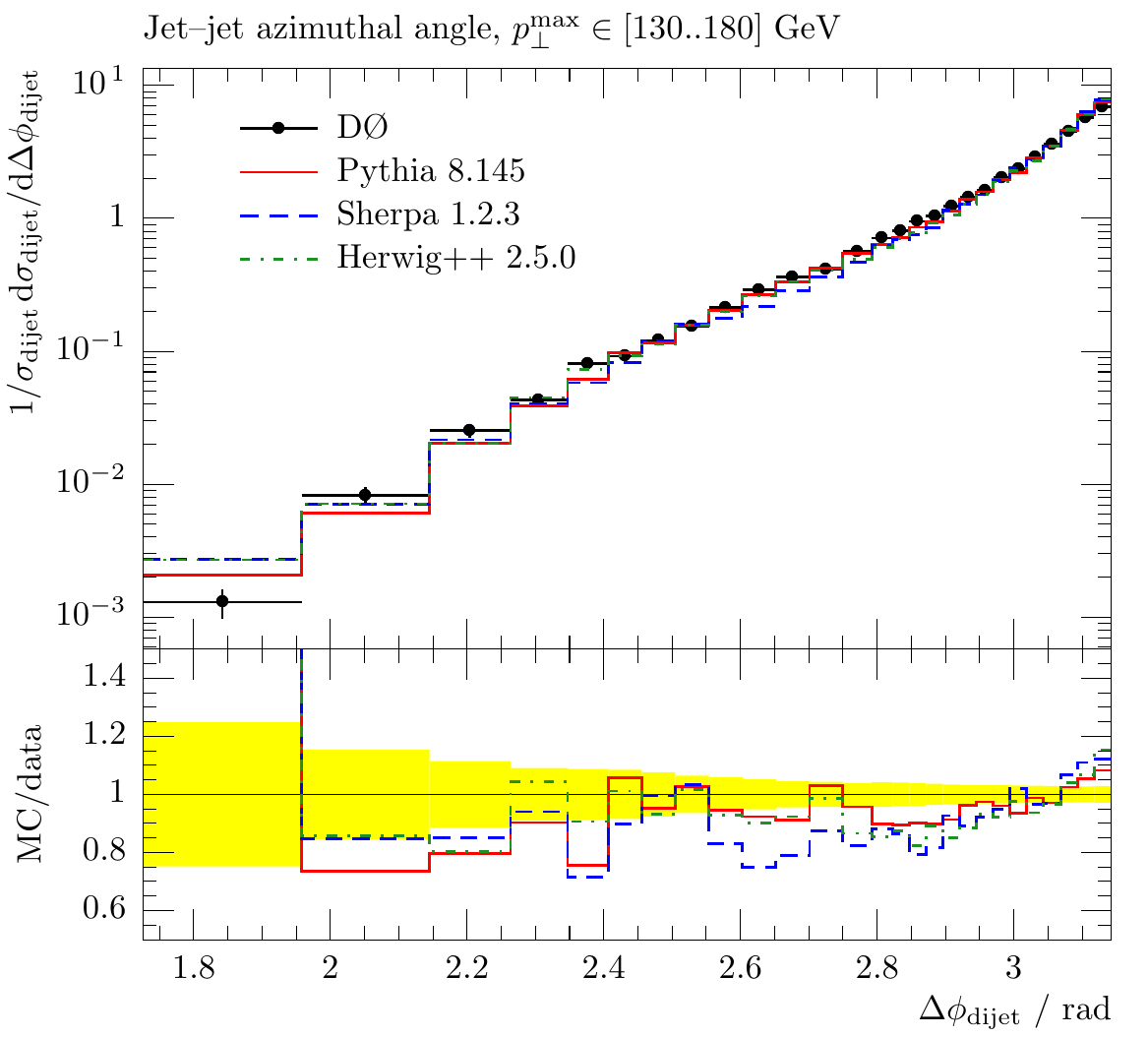}}
  \subfigure[Hadron collider jet shapes: CDF \cite{Acosta:2005ix}]{\label{fig:cmp:hadronic-jetshapes}\includegraphics[scale=0.7]{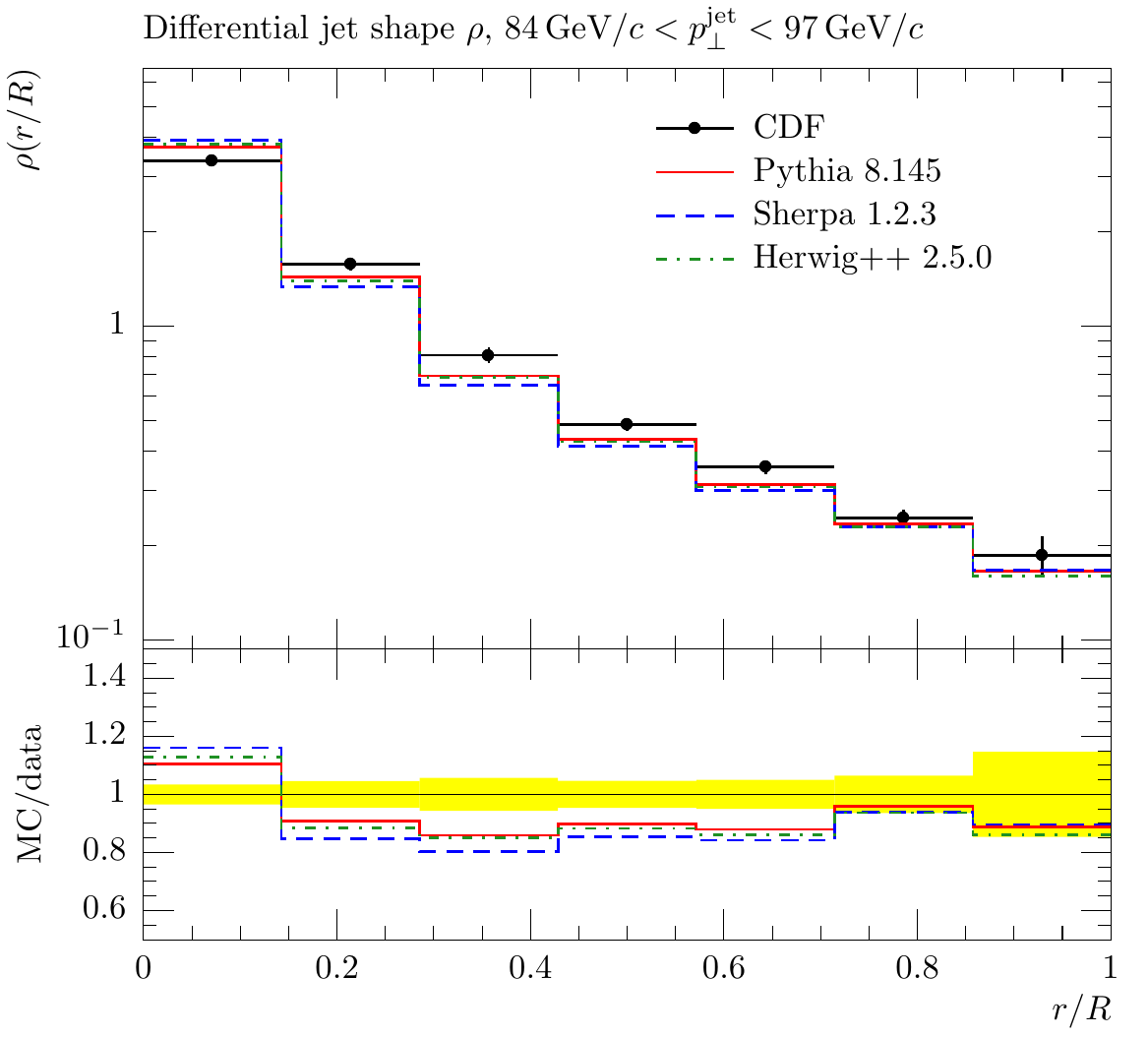}}
  \caption{Hadron collider shower-sensitive observables: dijet azimuthal
    decorrelation and jet shapes measured by D\O{} and CDF in Run~2 \cite{Abazov:2004hm,Acosta:2005ix}.
    The azimuthal decorrelation is a measure of
    the influence of three-jet configurations and shower emissions in disrupting
    a purely back-to-back two-parton configuration. Jet shapes measure the
    distribution of (transverse) momentum as a function of radius within jets.
    Up-to-date versions of these plots can be found at
    \url{http://mcplots.cern.ch/}.}
  \label{fig:cmp:ppshapes}
\end{figure}

\begin{figure}[tp]
  \centering
  \subfigure[LEP/SLD identified particle multiplicities \cite{Amsler:2008zzb}]{\label{fig:cmp:multis}\includegraphics[scale=0.7]{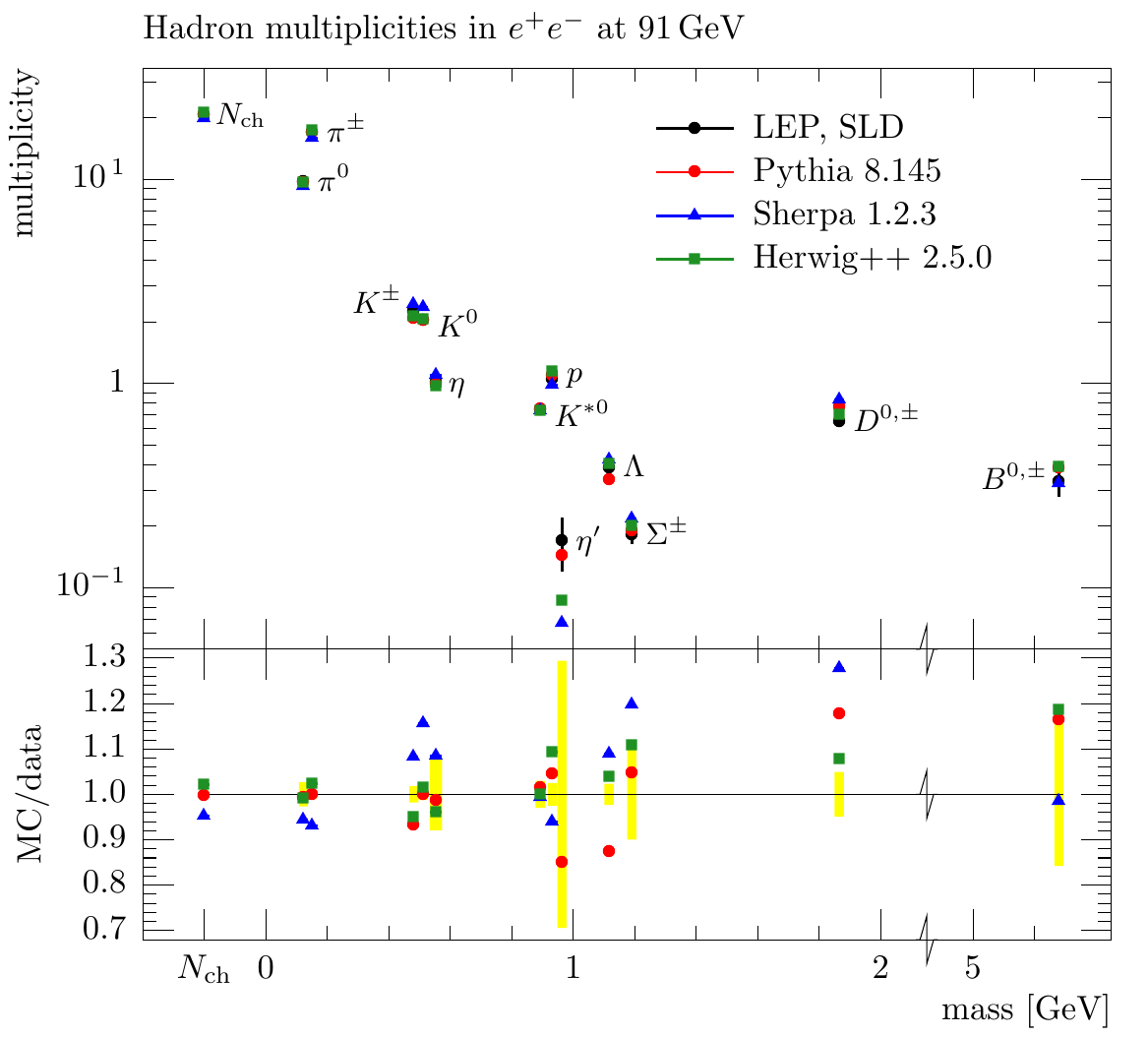}}
  \subfigure[STAR identified particle mean $p_\perp$ \cite{Abelev:2006cs}]{\label{fig:cmp:particle-mean-pt}\includegraphics[scale=0.7]{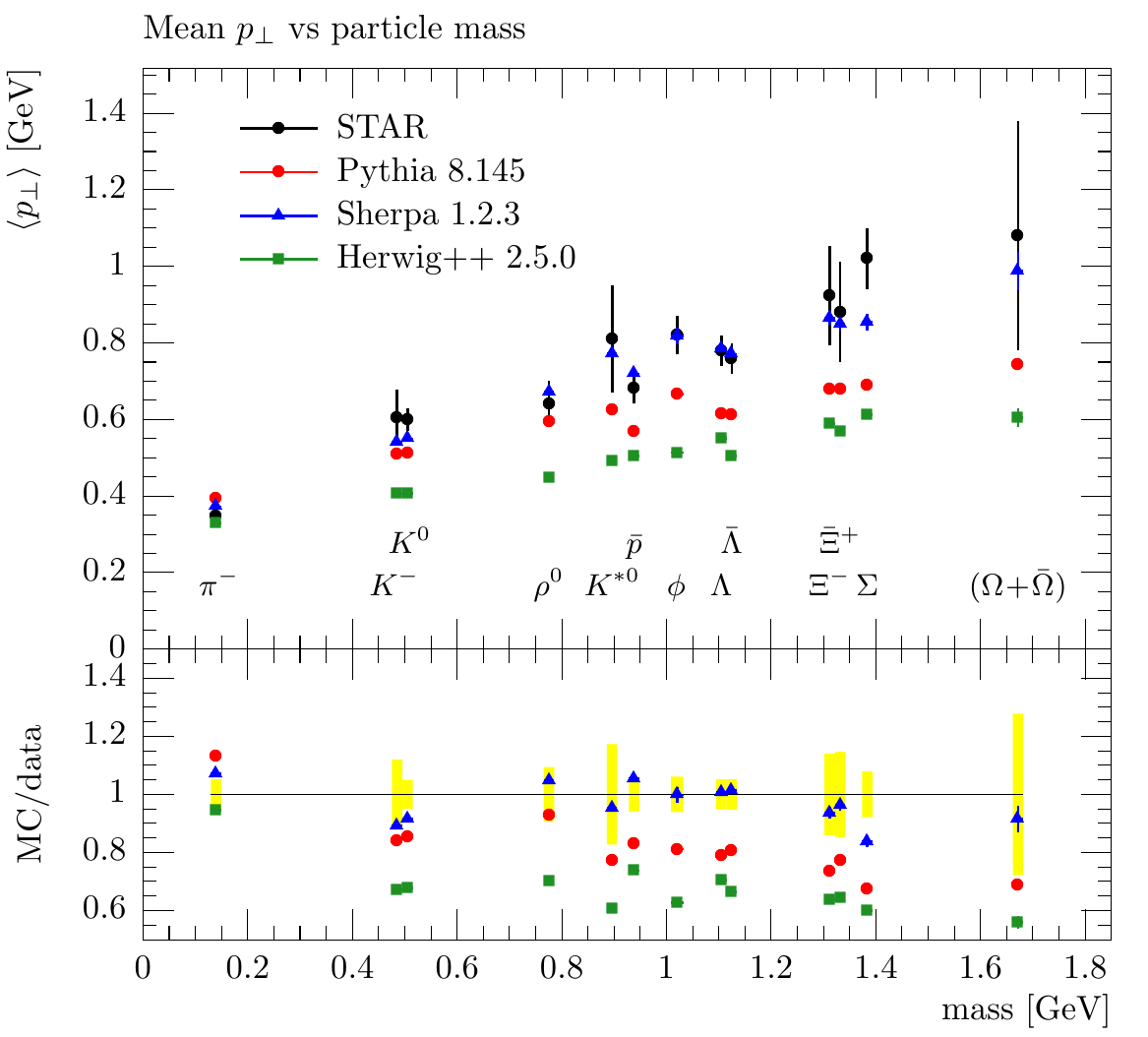}}
  \caption{Identified hadron multiplicities in $e^+e^-$ collisions at
    the $Z$ peak and $\langle \pt \rangle$ vs. particle mass in $pp$ collisions
    at 200\,GeV. These observables are determined primarily by the tuning
    of the hadronization models, both the flavour and kinematic aspects, but the
    overall multiplicities are also strongly dependent on the tuning of the
    parton showers (and MPI models, for the hadron collider observables).
    Up-to-date versions of these plots can be found at
    \url{http://mcplots.cern.ch/}.}
  \label{fig:cmp:idparticle-rates}
\end{figure}

\begin{figure}[tp]
  \centering
  \includegraphics[scale=0.7]{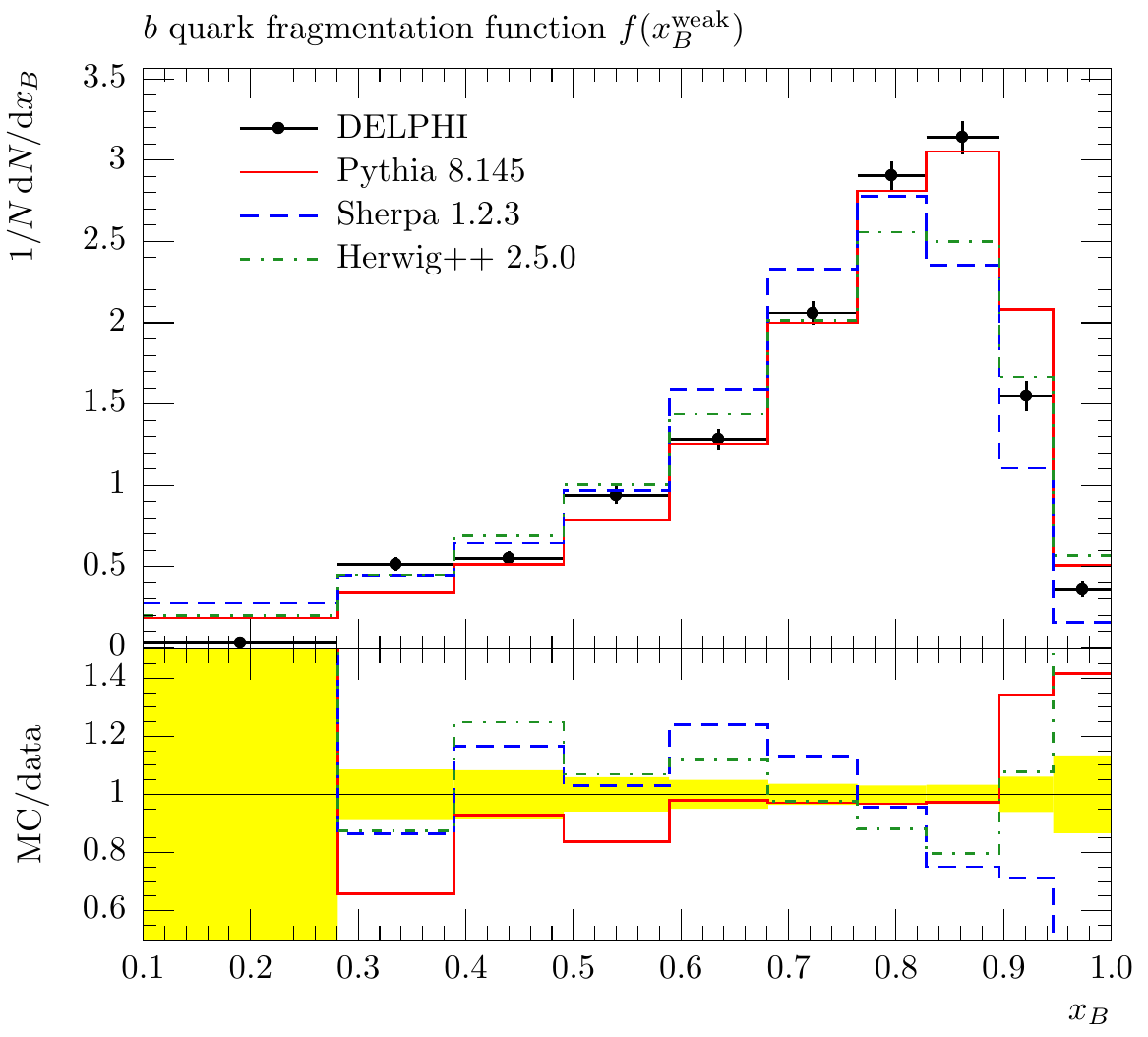}
  \caption{DELPHI $B$ fragmentation function $x_B = 2 E_B / \sqrt{s}$
    for weakly decaying $b$ hadrons \cite{Barker:2002}. Most Monte Carlo
    models apply a special fragmentation function treatment to heavy
    quarks, but this observable is not entirely decoupled from light
    quark fragmentation parameters. An up-to-date version of this plot
    can be found at \url{http://mcplots.cern.ch/}.}
  \label{fig:cmp:bfragfunction}
\end{figure}

Some results on final states in $e^+e^-$ annihilation at the $Z^0$
peak are shown in
\FigsRef{fig:cmp:eeshapes}--\ref{fig:cmp:eejetrates}, together with
ALEPH data~\cite{Barate:1996fi,Heister:2003aj}.  Other plots relevant to jet
fragmentation are shown in
\FigsRef{fig:cmp:ppshapes}--\ref{fig:cmp:bfragfunction}.

\begin{figure}[tp]
  \centering
  \includegraphics[scale=0.7]{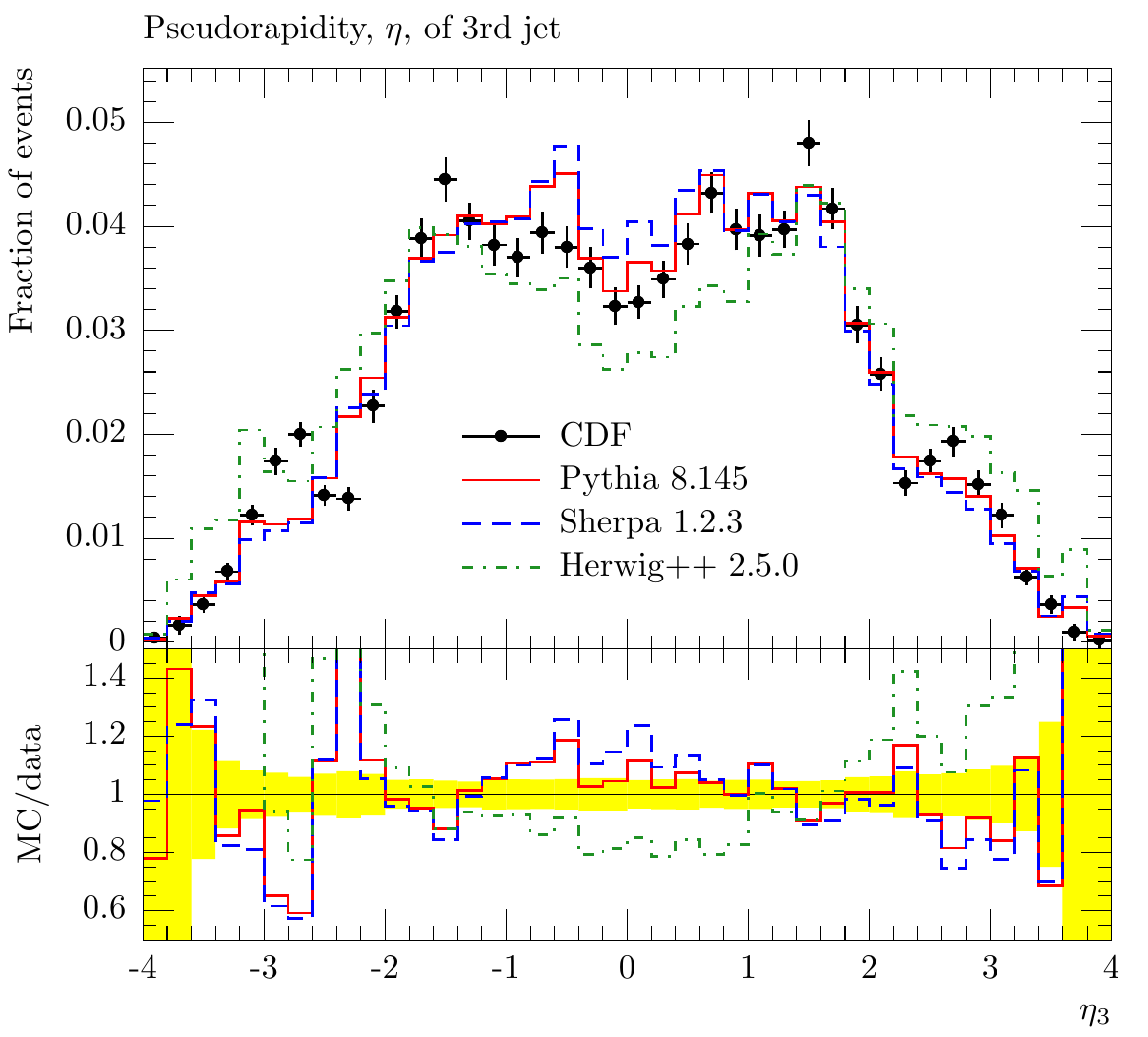}
  \caption{CDF's evidence for colour coherence in $p\bar{p}$ collisions at
    $\sqrt{s} = 1.8$~TeV \cite{Abe:1994nj}. The pseudorapidity of the
    third jet is plotted, uncorrected for detector effects, with the
    \herwigpp, \pythiaeight and \sherpa Monte Carlo generators. All
    these generators include colour coherence effects via either angular-ordered
    parton showers (\herwigpp) or transverse-momentum-ordered dipole showers
    (\pythiaeight, \sherpa),  and hence correctly exhibit a dip at central
    rapidity. The correlated fluctuations between MC samples are due to
    statistical errors on the detector-smearing correction factors taken
    from the CDF paper. An up-to-date version of this plot can be found
    at \url{http://mcplots.cern.ch/}.}
  \label{fig:cmp:cdf-colour-coherence}
\end{figure}

Finally \FigRef{fig:cmp:cdf-colour-coherence} shows results from
the MCnet generators on the colour-coherence test discussed in
\SecRef{parton-shower:initial-conditions} (already displayed in
\FigRef{fig:CDFcoherence} for the earlier generator \pythiasix),
again compared with CDF data~\cite{Abe:1994nj}.

% Local Variables:
% mode: LaTeX
% TeX-master: "../mcreview"
% End:

\clearpage
\part*{Acknowledgements}\label{sec:acknowledgements}
\addcontentsline{toc}{section}{Acknowledgements}
This work was supported by the European Union as part of the EU Marie
Curie Research Training Network MCnet (MRTN-CT-2006-035606).  We are
most grateful to the members of the MCnet network for their dedication
and hard work, which underlie many of the new developments reviewed here.

We thank Andr\'e Hoang for permission to use his lectures at
the 2009 MCnet summer school as the basis for
\AppRef{sec:top-quark-masses} and for his and Iain Stewart's comments on
the manuscript.

JMB's work is supported in part by
a Royal Society Wolfson Research Merit Award. 
SG acknowledges support from the Helmholtz Alliance ``Physics at the
Terascale''. 
SH's work was supported by the US Department of Energy 
under contract DE--AC02--76SF00515.
LL and TS acknowledge support from the Swedish Research Council
(contract numbers 621-2007-4157, 621-2008-4252 and 621-2009-4076).
ELN's work is supported in part by
a Royal Society University Research Fellowship. 
MHS's work is supported in part by an IPPP Associateship and in part by
a Royal Society Wolfson Research Merit Award.
SS acknowledges support by the German Federal Ministry of Education and 
Research (BMBF). 
BW acknowledges the support of a Leverhulme Trust Emeritus Fellowship.


\appendix
\mcpart{Appendices}
\mcsection{Monte Carlo methods}
\label{sec:mc-methods}
\mcsubsection{Generating distributions}
\label{mcmethods:dist}  
We give here a very brief review of the numerical methods used in
Monte Carlo event generators. The basic requirement of such a
generator is to produce a set of representative points in the phase
space of the process under study, in such a way that the density of points
follows the probability distribution predicted for that process. The
simplest case is that of a single variable $x$ to be distributed in
the region $[x_{\rm min},x_{\rm max}]$ with probability distribution
proportional to $f(x)\geq 0$.  Then if $R\in[0,1]$ is a uniform
pseudo-random number, we want to generate $x$ such that
\beq
\int_{x_{\rm min}}^{x} f(x')\,\done x'= R \int_{x_{\rm min}}^{x_{\rm max}}f(x')\,\done x'\;.
\eeq
If the indefinite integral of $f(x)$ is a known function $F(x)$ then
this is equivalent to solving
\beq\label{eq:mcFx}
F(x) = R\,F(x_{\rm max}) +(1-R)\,F(x_{\rm min})\;.
\eeq
If the inverse function $F^{-1}$ is known, the problem is solved.
Otherwise, the solution can often be obtained quite fast numerically,
because the positivity of $f(x)$ ensures that $F(x)$ is monotonic.

If the indefinite integral of $f(x)$ is not known, or if the method of
numerical solution is too slow, then the {\it hit-or-miss} method can
be used.  Here we suppose that a function $g(x)\geq f(x)$ on the
interval $[x_{\rm min},x_{\rm max}]$ has a known indefinite integral
$G(x)$ that can be inverted or solved for $x$. Then we generate the
distribution according to $g(x)$ and accept the resulting point with
probability $f(x)/g(x)$, \ie if $f(x)>R'g(x)$ where $R'\in[0,1]$ is
another uniform pseudo-random number.  In particular we can choose a
constant $g(x)=g_u\geq\max\{f(x),x\in [x_{\rm min},x_{\rm max}]\}$,
generate points uniformly as
\beq
x = R\,x_{\rm max} +(1-R)\,x_{\rm min}\;,
\eeq
and accept those points that satisfy $f(x)>R'g_u$.  However this may
be very inefficient (many points may be rejected) if $f(x)$ is very
non-uniform or if $g_u$ is chosen too large.

\mcsubsection{Monte Carlo integration and variance reduction}
\label{mcmethods:integ}   
In reality the phase space is multi-dimensional.
Then it is important to appreciate that the Monte Carlo method is based on the
concept of an integral as an average.  Suppose we have a matrix element-squared
$f({\bf x})$ which is a function of the $n$-component vector ${\bf x}$, and
we want to integrate it over a region $V$ of ${\bf x}$-space, for
example to compute a cross section:
\beq
I[f] = \int_V \done^n x\,f({\bf x})\;.
\eeq
Standard methods of integration (Simpson's, Gaussian, \ldots) are too
laborious and/or inaccurate for $n$ large (say $n>3$).  However, if
$N$ points $\{{\bf x}_i,i=1,\ldots,N\}$ are distributed (pseudo-)randomly in
$V$,  then the central limit theorem of statistics tells us that
the {\em mean value} of $f$ on those points is an unbiased estimator of the integral,
\beq
I[f]\simeq\abr{f} = \frac 1N \sum_{i=1}^N f({\bf x}_i)\;,
\eeq
and that the estimated error $E[f]$ on this evaluation is given by
the {\em variance} of $f$,
\beq
\mbox{Var}(f)=\abr{(f-\abr{f})^2} = \abr{f^2}-\abr{f}^2\;,
\eeq
as
\beq
E[f]= \sqrt{\frac{\mbox{Var}(f)}{N-1}}\;.
\eeq
Thus the error decreases as the inverse square root of the number of
points, independent of the dimensionality of the integral.
Furthermore, for a given number of points, the error will be less
if the variance of the integrand is small.

The variance can be reduced by a change of variables that ``flattens''
the integrand.  Consider the mapping  ${\bf x}\to {\bf y}({\bf x})$ with
Jacobian
\beq
\left|\frac{\partial({\bf y})}{\partial({\bf x})}\right| = g({\bf x})\;.
\eeq
Then
\beq
I[f] =  \int_{V'} \done^n y\,\frac{f({\bf x})}{g({\bf x})}\;,
\eeq
where $V'$ is the region in ${\bf y}$-space corresponding to $V$ in ${\bf x}$-space.
If $h=f/g$ is a function with less variance than $f$ itself then the error will be
reduced by distributing points uniformly in ${\bf y}$-space. This is
known as {\it importance sampling}. To obtain
a set of points distributed according to $f({\bf x})$,
as desired for an event generator, we can now apply the hit-or-miss
method, accepting points with probability $h/h_{\rm lim}$, where
$h_{\rm lim}$ is an upper bound on the value of $h$ in $V'$. The Monte
Carlo efficiency, as measured by the fraction of points accepted,
$\abr{h}/h_{\rm lim}$, will usually also be increased as a result of
the variance reduction.

 In some applications, for example NLO cross section calculations, the
 integrand $f({\bf x})$ can contain integrable singularities. Although
 these would give a finite result if they were calculated
 analytically, their variance is divergent and hit-or-miss Monte Carlo
 will fail to converge. Variance reduction, with a carefully chosen
 generated distribution $g({\bf x})$, becomes mandatory in such cases.

More sophisticated methods for variance reduction, such as {\it
stratified} or {\it multichannel sampling}, are also applied in Monte
Carlo generators, particularly when dealing with matrix elements that
have sharp peaks due to resonance production or matrix element singularities
close to the physical region, as discussed in \AppRef{sec:app_mcs}.

If it is difficult to arrive at an acceptable efficiency by reducing
the variance of the integrand and/or finding a good upper bound on it,
one may wish to resort to generating {\it weighted events}.  In that
case the phase-space points $\{{\bf x}_i\}$ (or $\{{\bf
y}_i\}$ if some variance reduction has been achieved) are used to
represent events, but each event has a different weight $f_i$ (or
$h_i$) when contributions to observables are computed.  In that case
one has to take account of the variance of the weights when computing
error on observables.  That means, for example, that one must keep track of
the sum of the squared weights as well as the weights contributing to
each bin of a histogram.  In contrast the error for the {\it
unweighted events} obtained from hit-or-miss is just given by the
square root of the number of events in the bin.

An example of a situation in which weighted events can be useful is in the
study of jet hadroproduction, where the distribution of jet transverse
energy $E_T$ falls very rapidly, roughly as $E_T^{-5}$.  If events are
generated according to the relevant hard subprocess matrix elements multiplied
by $\pt^5$, where $\pt$ is the transverse momentum of the hardest
final-state parton in the subprocess, and then weighted by
$\pt^{-5}$, event properties can be explored over the full range of
jet $E_T$ without generating huge event samples.

For further details of general Monte Carlo methods
see, for example,\ \cite{James:1980yn} and the relevant section of the
Review of Particle Physics~\cite{Nakamura:2010zzi}.

\mcsubsection{Veto method}
\label{mcmethods:veto}
It often happens in event generators that one wishes to generate an
ordered sequence of values $\{q_i\}$ of some variable $q$, for example
the evolution variable of a parton shower, according to a distribution
function with a rather complicated form, in this case the relevant
Sudakov form factor.  The {\it veto method}, a variant of
hit-or-miss, is a useful way of achieving this.

Suppose that, given $Q$, we wish to generate $0<q_1<Q$ such that the
probability that $q_1<q$ is $F(q)/F(Q)$, where $F(q)$ is a monotonically increasing
function with $F(0)=0$, \eg the Sudakov form factor $\Delta(Q^2,q^2)$ in
\EqRef{Sudakov}\footnote{Strictly speaking, \EqRef{Sudakov} requires
  $q^2>2Q_0^2$ where $Q_0>0$ and $\Delta(Q^2,2Q_0^2)>0$.  We consider
  here the case that $Q_0\to 0$, and discuss below the effect of
  $Q_0>0$.}  In simple cases we can do this as in
\EqRef{eq:mcFx}, by solving the equation $F(q_1) = R
F(Q)$ where $R\in[0,1]$ is a uniform pseudo-random number.  However,
if $F(q)$ is too complicated for this, but its derivative $f(q)=\done
F/\done q$ is known, and we can find a simpler monotonic function
$G(q)\geq 0$ with derivative $g(q)$ such that $f(q)/F(q)<g(q)/G(q)$ for
$q<Q$, we can proceed as follows:
\begin{enumerate}
\item Solve $G(q') = R\, G(Q)$ for $q'$, where $R$ is a random number
as above.
\item If  $f(q')/F(q')>R'\, g(q')/G(q')$,  where $R'$ is another random number, set $q_1=q'$.
\item Otherwise {\it veto} this choice of $q_1$, \ie  set $Q=q'$ and
go back to step 1 to find $q''<q'$.
\end{enumerate}
To see that this generates the correct probability
$P(q_1<q)=F(q)/F(Q)$, we note first that the probability distribution
of $q'$ from step 1 is $dP/dq'=g(q')/G(Q)$, and the probability of
vetoing $q'$ is
\beq
P_{\rm veto}(q') = 1-\frac{f(q')G(q')}{F(q')g(q')}\;.
\eeq
Now the probability of finding $q_1<q$ with no veto is
\beq
P(q_1<q)_{\rm 0-veto} = \frac{G(q)}{G(Q)}\;,
\eeq
while the probability of finding $q_1<q$ after one veto is
\bea
P(q_1<q)_{\rm 1-veto} &=& \int_q^Q\done q'\frac{g(q')}{G(Q)}
P_{\rm veto}(q')\frac{G(q)}{G(q')}\nonumber\\
&=& \frac{G(q)}{G(Q)}\int_q^Q\done
q'\left[\frac{g(q')}{G(q')}-\frac{f(q')}{F(q')}\right]\nonumber\\
&=& \frac{G(q)}{G(Q)}\left[\ln\frac{G(Q)}{G(q)}-\ln\frac{F(Q)}{F(q)}\right]
\;.
\eea
Similarly the probability of finding $q_1<q$ after two vetoes is
\beq
P(q_1<q)_{\rm 2-veto}
=\frac 1{2!}\frac{G(q)}{G(Q)}\left[\ln\frac{G(Q)}{G(q)}-\ln\frac{F(Q)}{F(q)}\right]^2
\;,
\eeq
where the $1/2!$ comes from the fact that the vetoes are ordered,
$q''<q'<Q$.  Summing over all numbers of vetoes gives an
exponential series,
\bea
P(q_1<q) &=& \sum_{n=0}^\infty
\frac
1{n!}\frac{G(q)}{G(Q)}\left[\ln\frac{G(Q)}{G(q)}-\ln\frac{F(Q)}{F(q)}\right]^n\nonumber\\
&=&\frac{G(q)}{G(Q)}\exp\left[\ln\frac{G(Q)}{G(q)}-\ln\frac{F(Q)}{F(q)}\right]
=\frac{F(q)}{F(Q)}\;,
\eea
as required.

As a simple example, suppose we have an upper bound $a >f(q)/F(q)$
for all $q<Q$.  Then we can take $G(q)=\exp(aq)$, so that step 1 gives
$q'=Q+(\ln R_1)/a$, and veto $q'$ if $f(q')/F(q')<a R_2$. As in simple
hit-or-miss, the method remains valid but becomes less efficient (more
vetoes) if $a$ is larger than necessary.

Once a value of $q_1$ has been accepted, $q_2$ can be generated by
repeating steps 1--3 with $Q$ replaced by $q_1$, and so on to
create a decreasing ordered sequence $\{q_i\}$.  In the case of a
parton shower, the sequence terminates at $q_n$ when the step 1 with
$Q=q_n$ produces a value of $q'$ less than the shower cutoff $Q_0$.

\mcsection{Evaluation of matrix elements}
\label{sec:app_mcs}
In this section we review in more detail the generation and evaluation
of hard subprocess matrix elements and the related 
methods to integrate over the phase space of outgoing particles.
%%%%%%%%%%%%%%%%%%%%%%%%%%%%%%%%%%%%%%%%%%%%%%%%%%%%%%%%%%%%%%%%%%%%%%
%%%%%%%%%%%%%%%%%%%%%%%%%%%%%%%%%%%%%%%%%%%%%%%%%%%%%%%%%%%%%%%%%%%%%%
\mcsubsection{Matrix element calculation}
\label{Sec:TL_ME}
%%%%%%%%%%%%%%%%%%%%%%%%%%%%%%%%%%%%%%%%%%%%%%%%%%%%%%%%%%%%%%%%%%%%%%
%%%%%%%%%%%%%%%%%%%%%%%%%%%%%%%%%%%%%%%%%%%%%%%%%%%%%%%%%%%%%%%%%%%%%%
%%%%%%%%%%%%%%%%%%%%%%%%%%%%%%%%%%%%%%%%%%%%%%%%%%%%%%%%%%%%%%%%%%%%%%
\paragraph{Summation versus sampling}
%%%%%%%%%%%%%%%%%%%%%%%%%%%%%%%%%%%%%%%%%%%%%%%%%%%%%%%%%%%%%%%%%%%%%%

When calculating the cross section for a given process, an integral
over the relevant phase space has to be performed, typically through
Monte Carlo methods.  Then, for each phase-space point (\ie for each
set of incoming and outgoing momenta) the matrix element squared has
to be evaluated.  This involves a summation and averaging over the
unobserved quantum numbers of the outgoing and incoming particles,
respectively.  On the level of squared amplitudes the summation can be
performed analytically, involving algebraic relations such as
completeness relations or the Dirac trace algebra, which yields an
analytical result for the squared matrix elements, that can in turn be
expressed by Lorentz-invariant combinations of the four-momenta of the
involved particles.  This method is typically applied for matrix
elements of low final-state multiplicity, which are pre-computed and
implemented explicitly in the event generators.  On the level of
numerically evaluated amplitudes, on the other hand, one may choose
between a summation over all quantum states such as helicities and
colours for a given phase-space configuration and a Monte Carlo
sampling over these states, together with the phase-space integration.
The computational complexities for summation and sampling for matrix
elements with $n$ external lines naively differ by the
$n^{\mathrm{th}}$
power of the number of possible states, \ie usually ${\cal
  O}(2^n)$ for the possible helicity assignments and ${\cal
  O}(3^{n_3}8^{n_8})$ for the possible colour assignments of $n_3$
external quarks and $n_8$ external gluons.  By suitably eliminating
common subexpressions of the matrix elements, one can, however, often
reduce these naive factors quite considerably.  The decision whether
to sum or to sample over helicity and colour degrees of freedom is
therefore strongly dependent on the process and the particle
multiplicity.  Corresponding comparisons have been presented, \eg
in~\cite{Gleisberg:2008fv}. This issue will not be discussed further
here.  We will, however, discuss different techniques to efficiently
deal with coloured states, as the sum over colours usually poses the
most severe restriction on the ability to evaluate high-multiplicity
matrix elements.
 
%%%%%%%%%%%%%%%%%%%%%%%%%%%%%%%%%%%%%%%%%%%%%%%%%%%%%%%%%%%%%%%%%%%%%%
\paragraph{Pre-computed matrix elements} 
%%%%%%%%%%%%%%%%%%%%%%%%%%%%%%%%%%%%%%%%%%%%%%%%%%%%%%%%%%%%%%%%%%%%%%

Traditionally, the matrix elements acting as seeds for event generation have
been related to processes with low multiplicity of two or maximally three
particles in the final state.  For such processes, analytical results are
usually available, and consequently they are used in all event generators.
With the advent of multijet merging methods and agreements for interface
structures, the importance of incorporating such matrix elements directly
into MC programs has diminished, such that in the new generation of 
event generators only a few pre-computed analytic matrix elements are available.
On the other hand, due to the incorporation of matching methods for NLO 
matrix elements, some explicit next-to-leading order results are now direct
inputs for event generation, just as leading-order results were previously.
This situation is likely to change, as with the advent of general methods
to automate the computation of NLO virtual corrections the need for explicit
calculations may slowly disappear. 

\paragraph{The helicity method}
%%%%%%%%%%%%%%%%%%%%%%%%%%%%%%%%%%%%%%%%%%%%%%%%%%%%%%%%%%%%%%%%%%%%%%

Textbook methods of squaring full matrix elements and summing over helicity
and colour through the application of completeness relations yields a rather 
large number of terms: for $N$ Feynman diagrams in this method $N(N-1)/2$ contributions
must be evaluated.  An obvious way of reducing this number is to 
directly evaluate the amplitudes, yielding complex numbers, before summing 
and squaring them and before sampling over the phase space.  
In order to compute the individual numerical values of the amplitudes, an 
efficient representation in terms of external momenta and helicities is mandatory.  
A first solution to this problem was achieved in~\cite{Kleiss:1985yh}.  
The basic idea is to replace all momentum-dependent terms appearing
in an amplitude through suitably chosen spinor products. This substitution 
can always be achieved, since spinors are the simplest representations of 
the Lorentz group and their products correspondingly yield the minimal 
representation of a Lorentz-invariant complex number. One can, for example, 
identify the numerator of a fermion propagator as
\begin{equation}
\begin{split}
p\!\!\!/+\mu \;=\; &
\frac12\sum\limits_\lambda
        \left[u(\lambda,p)\bar u(\lambda,p)\;
                               \left(1+\frac{\mu}{\sqrt{p^2}}\right)+
             v(\lambda,p)\bar v(\lambda,p)\;
                              \left(1-\frac{\mu}{\sqrt{p^2}}\right)\right]\;,
\end{split}
\end{equation}
and the polarization vector of a spin-1 boson with momentum $p+q$ can be 
written as
\begin{equation}
\begin{split}
\epsilon_\mu(p+q) \;=\; & \frac{1}{\sqrt{4p\cdot q}}
             \bar u(\lambda,q)\gamma_\mu u(\lambda,p)\,.
\end{split}
\end{equation}
In such a way, and employing a Chisholm identity for terms of the form
$\bar u\gamma^\mu u\times\bar u\gamma_\mu u$, every amplitude containing
fermion interactions can be decomposed into spinor products of the form 
$\bar uu$ and $\bar vv$, see~\cite{Kleiss:1985yh,Ballestrero:1992ed,
  Ballestrero:1992dv,Ballestrero:1994ti}.  
In the context of tree-level matrix-element generators, the corresponding 
elementary building blocks are usually referred to as Lorentz functions.
They are implemented in a similar form in any of the automated 
matrix-element generators listed above.

%%%%%%%%%%%%%%%%%%%%%%%%%%%%%%%%%%%%%%%%%%%%%%%%%%%%%%%%%%%%%%%%%%%%%%
\paragraph{Feynman-diagram based methods}
%%%%%%%%%%%%%%%%%%%%%%%%%%%%%%%%%%%%%%%%%%%%%%%%%%%%%%%%%%%%%%%%%%%%%%

Having at hand the basic Lorentz functions to decompose amplitudes into terms that can
be evaluated numerically in a straightforward manner, the remaining problem is
the generation of these amplitudes.  Traditionally this is achieved through
the construction of Feynman diagrams -- an algorithm with improved efficiency using 
recursive relations will be discussed later.  The diagrammatic approach has been followed 
for instance in \Madgraph~\cite{Stelzer:1994ta} and \Amegic~\cite{Krauss:2001iv}.  
In both programs, Feynman-diagram-like topologies, i.e\ trees with binary or tertiary 
vertices, are generated and then filled with the actual interactions given by the physics
model in question.  The resulting objects are translated into so-called helicity amplitudes,
\ie into products of the Lorentz functions discussed in the previous
paragraph.  In so doing, some manipulations may be performed, trying to
identify common subexpressions and either factoring them out or storing them 
such that identical pieces need to be calculated only once.
In both cases, the programs write out the helicity amplitudes in a high-level
programming language to be compiled and linked to the original program. The resulting
libraries are then employed to calculate cross sections, to generate parton-level events
and to pass these events on to a parton-shower simulation, for instance, using Les Houches 
Event Files, see \AppRef{sec:MEinterfaces}.

%%%%%%%%%%%%%%%%%%%%%%%%%%%%%%%%%%%%%%%%%%%%%%%%%%%%%%%%%%%%%%%%%%%%%%
\paragraph{Skeletons}
%%%%%%%%%%%%%%%%%%%%%%%%%%%%%%%%%%%%%%%%%%%%%%%%%%%%%%%%%%%%%%%%%%%%%%

In \Herwigpp only a few pre-computed squared matrix elements are 
available.  The authors of this code have, however, compensated for this by 
a low-level matrix-element generator, which is capable of constructing
helicity amplitudes for processes with up to four external particles (\ie
$2\to2$ scattering and $1\to3$ decay processes).  Depending on the spin 
of those particles, the algorithm identifies all possible topologies
($s$, $t$, and $u$-channel as well as four-point interactions) for the 
process in question, with the corresponding propagators being specified by
the Feynman rules given in an internal format.  These topologies are then
directly mapped onto the respective prefabricated helicity amplitudes.
This algorithm greatly alleviates the task of integrating the cross sections 
efficiently: the knowledge of topologies and propagators allows for a direct 
translation into prefabricated integration channels, forming a multi-channel
integrator, see \AppRef{Sec:PS_ME}. For further details of the 
implementation of this algorithm we refer to~\cite{Gigg:2007cr}.

%%%%%%%%%%%%%%%%%%%%%%%%%%%%%%%%%%%%%%%%%%%%%%%%%%%%%%%%%%%%%%%%%%%%%%
\paragraph{Recursive techniques} 
%%%%%%%%%%%%%%%%%%%%%%%%%%%%%%%%%%%%%%%%%%%%%%%%%%%%%%%%%%%%%%%%%%%%%%
There are several techniques for computing tree-level matrix elements that employ 
different versions of recursive relations. With increasing number of particles 
involved in the scattering they are superior to diagram-based methods,
as they naturally implement an optimal common subexpression elimination.
One such method, which we shall consider as an example of a recursive 
technique in this context, is the Berends-Giele algorithm~\cite{Berends:1987me,
  Berends:1987cv,Kleiss:1988ne,Berends:1988yn,Berends:1989ie}.
It has recently been improved to incorporate an efficient way
to deal with colour~\cite{Duhr:2006iq}, rendering it essentially equivalent 
to the Dyson-Schwinger methods employed for instance in 
\Helac~\cite{Draggiotis:2002hm}, and comparable in efficiency with the 
\ALPHA\ algorithm of~\cite{Caravaglios:1995cd}, implemented in 
\Alpgen~\cite{Mangano:2002ea} and \OmegaCode~\cite{Moretti:2001zz}. 
 
\begin{figure}[t]\begin{center}
  \includegraphics[width=\textwidth]{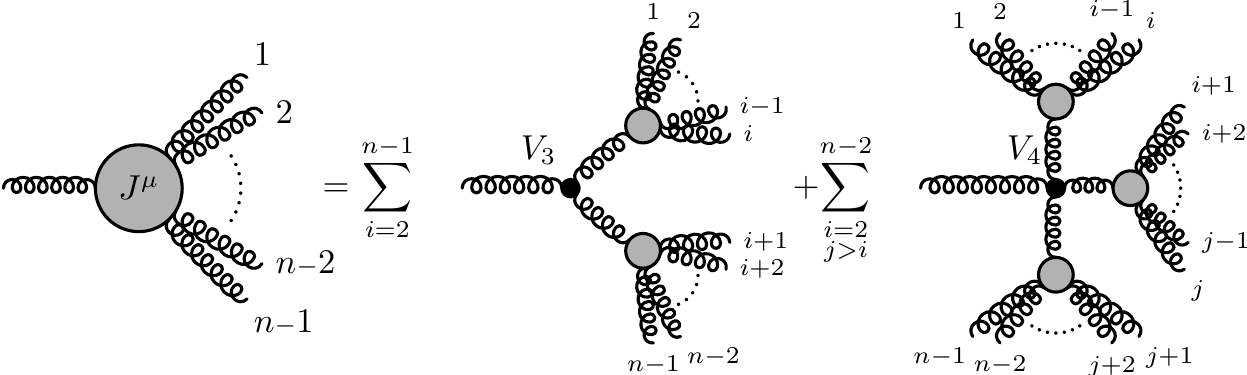}
  \end{center}
  \caption{Pictorial representation of the Berends-Giele 
  recursive relations.\label{fig:BG}}
\end{figure}
The basic idea of the Berends-Giele recursion algorithm can be summarized as 
follows. At first, an $n-1$-point gluon 
off-shell current, $ J^\mu$, is defined, which represents the sum 
of all colour-ordered Feynman diagrams with $n-1$ external on-shell legs 
and a single off-shell leg with polarization $\mu$.  This off-shell current 
can then be decomposed into lower-point off-shell currents, which are joined 
by the elementary gluon interaction vertices, thus forming a bigger part of
the full scattering amplitude. This algorithm is schematically depicted in 
\FigRef{fig:BG}.  A full $n$-gluon amplitude is finally obtained by 
amputating the off-shell propagator and contracting the remaining quantity 
with the external polarization of gluon $n$.  

Similar recursions exist for the off-shell quark currents~\cite{Berends:1987me} 
and for the full Standard Model~\cite{Gleisberg:2008fv}.  They can, in fact, 
be defined for any theory allowing the construction of Feynman diagrams.
A further improvement of this method was recently obtained through a 
decomposition of all four-particle vertices into three-particle ones.
Such a decomposition reduces the computational complexity for
many-particle final states, as the numerical effort grows approximately like 
$N^n$, with $N$ the average number of legs in elementary vertices of the theory.

%%%%%%%%%%%%%%%%%%%%%%%%%%%%%%%%%%%%%%%%%%%%%%%%%%%%%%%%%%%%%%%%%%%%%%
\paragraph{Treatment of colour} 
%%%%%%%%%%%%%%%%%%%%%%%%%%%%%%%%%%%%%%%%%%%%%%%%%%%%%%%%%%%%%%%%%%%%%%

Several methods have been suggested over the past decades to optimize the
computation of amplitudes including QCD particles with respect to the 
colour degrees of freedom.  There are two essentially different approaches:
the textbook method would be to compute colour-ordered quantities, \ie 
sets of Feynman diagrams or off-shell currents with all colour information 
combined into kinematics-independent prefactors.  When assembling the full 
matrix element, colour-factors and kinematics-dependent functions are then
treated separately.  An alternative approach is to directly include colour 
in the diagrams or off-shell currents and to devise, for example, recursive 
relations which depend on the colour quantum numbers.

The latter approach has been very successful in the past, leading to the 
construction of advanced tree-level matrix-element generators, capable of 
dealing with very large final-state multiplicities~\cite{Kanaki:2000ey, 
  Draggiotis:2002hm,Mangano:2002ea,Gleisberg:2008fv}.
The textbook approach, on the other hand, is often much more convenient to use,
especially when insight into the analytical structure of the computation is 
necessary.  It also usually leads to a significant acceleration of 
matrix-element computations for low-multiplicity final states.

Although a number of possible colour bases exist~\cite{Mangano:1987xk,
  DelDuca:1999ha,DelDuca:1999rs}, which have been used for
several numerical comparisons in the past~\cite{Duhr:2006iq}, the 
one that is widely adopted today is the colour-flow 
basis~\cite{Kanaki:2000ey,Maltoni:2002mq}. The reason for its superior 
speed in the computation of large-multiplicity QCD amplitudes lies not only 
in the milder growth in the number of possible colour-ordered amplitudes, but 
also in the fact that every colour coefficient multiplying the 
kinematics-dependent functions in squared matrix elements consists only 
of delta functions, which are trivial to evaluate in a Monte Carlo program.

%%%%%%%%%%%%%%%%%%%%%%%%%%%%%%%%%%%%%%%%%%%%%%%%%%%%%%%%%%%%%%%%%%%%%%
%%%%%%%%%%%%%%%%%%%%%%%%%%%%%%%%%%%%%%%%%%%%%%%%%%%%%%%%%%%%%%%%%%%%%%
\mcsubsection{Phase-space integration}
\label{Sec:PS_ME}
%%%%%%%%%%%%%%%%%%%%%%%%%%%%%%%%%%%%%%%%%%%%%%%%%%%%%%%%%%%%%%%%%%%%%%
%%%%%%%%%%%%%%%%%%%%%%%%%%%%%%%%%%%%%%%%%%%%%%%%%%%%%%%%%%%%%%%%%%%%%%
%%%%%%%%%%%%%%%%%%%%%%%%%%%%%%%%%%%%%%%%%%%%%%%%%%%%%%%%%%%%%%%%%%%%%%
\paragraph{MC integration and sampling methods}
%%%%%%%%%%%%%%%%%%%%%%%%%%%%%%%%%%%%%%%%%%%%%%%%%%%%%%%%%%%%%%%%%%%%%%
Once the matrix element for a given process has been constructed, one is left with 
the task of performing the related integral over the phase space of the 
initial- and final-state particles. Taking account of the unknown
momentum fractions of the initial-state partons, the dimension of this phase space 
is $(3n-4)+2$ at hadron colliders, with the integrand (the matrix element 
squared) typically exhibiting a challenging structure with pronounced peaks.  
While the large dimensionality renders traditional quadrature methods useless 
and enforces the usage of MC techniques, the difficult structures 
pose a serious threat to the convergence of the integration.  
This means that highly advanced sampling algorithms need to be introduced,
which are dubbed phase-space integrators or ``integration channels''.
From the formal point of view, all integration channels dealing with the same 
final state are equal, as they must finally yield exactly the same value 
of the MC integral. However, they usually differ greatly in the
rate of convergence and in practice one would obviously prefer to use 
the channel leading to the smallest error in the shortest time.

%%%%%%%%%%%%%%%%%%%%%%%%%%%%%%%%%%%%%%%%%%%%%%%%%%%%%%%%%%%%%%%%%%%%%%
\paragraph{Multi-channel integration}
%%%%%%%%%%%%%%%%%%%%%%%%%%%%%%%%%%%%%%%%%%%%%%%%%%%%%%%%%%%%%%%%%%%%%%
To generate an adequate phase-space integrator for realistic 
$2\to n$-particle processes, several existing channels can be combined 
using the multi-channel method~\cite{Kleiss:1994qy}. 
Symbolically one can write a single channel as a map $X$ from uniformly 
distributed random numbers $\vec x\in[0,1]^{3n-4}$
to the four-momenta $\vec p=(p_1,\ldots,p_n)$ of final-state particles,
The corresponding MC weight $g$ is then given by
\begin{equation}
  \frac{1}{g}=\frac{\done\Phi_n(X(\vec x))}{\done\vec x}
\end{equation}
where $\Phi_n$ represents the $n$-particle phase space.
The multi-channel method now combines several maps $X_i$ into a new map
$\rm X$ as follows:
\begin{equation}\label{eq:gen_point_mc}
  {\rm X}(\vec x,\tilde\alpha)=X_k(\vec x)\;,\quad\text{for}\quad
  \sum_{l=1}^{k-1}\alpha_l<\tilde\alpha<\sum_{l=1}^{k}\alpha_l\;,
\end{equation}
requiring an additional random number $\tilde\alpha$ and arbitrary
coefficients $\alpha_k$ with $\alpha_k>0$ and $\sum_k \alpha_k=1$. 
The corresponding MC weight is given by
\begin{equation}\label{eq:gen_weight_mc}
  G=\sum_k \alpha_k\; g_k\;.
\end{equation}
The coefficients $\alpha_k$ can be adapted to minimize the variance of the 
phase-space integral.

%%%%%%%%%%%%%%%%%%%%%%%%%%%%%%%%%%%%%%%%%%%%%%%%%%%%%%%%%%%%%%%%%%%%%%
\paragraph{Brief review of phase-space factorization}
%%%%%%%%%%%%%%%%%%%%%%%%%%%%%%%%%%%%%%%%%%%%%%%%%%%%%%%%%%%%%%%%%%%%%%
Consider the $2\to n$ scattering process in \EqRef{Eq::Master_For_XSec}, 
where we label incoming particles by $a$ and $b$ and outgoing particles by 
$1\ldots n$.  The corresponding partonic $n$-particle differential phase 
space element reads
\begin{equation}
  \begin{split}
  {\rm d}\Phi_n(a,b;1,\ldots,n)=&
    \left[\,\prod\limits_{i=1}^n\frac{{\rm d}^4 p_i}{(2\pi)^3}\,
    \delta(p_i^2-m_i^2)\Theta(p_{i0})\,\right]\,\\
    &\qquad\times(2\pi)^4\delta^{(4)}\left(p_a+p_b-\sum_{i=1}^n p_i\right)\;,
  \end{split}
\end{equation}
where $m_i$ are the on-shell masses of the outgoing particles.  Following 
Ref.~\cite{James:1968gu}, the $n$-particle phase space can be factorized as
\begin{equation}\label{eq:split_ps}
  \begin{split}
  {\rm d}\Phi_n(a,b;1,\ldots,n)=&\;
    {\rm d}\Phi_{n-m+1}(a,b;\pi,m+1,\ldots,n)\,\\
    &\qquad\times \frac{{\rm d} s_\pi}{2\pi}\,
    {\rm d}\Phi_m(\pi;1,\ldots,m)\;,
  \end{split}
\end{equation}
where $\pi=\{1\ldots m\}$ indicates an $s$-channel virtual particle.
\EqRef{eq:split_ps} allows one to decompose 
the complete phase space into building blocks corresponding to
$s$- and $t$-channel two-body decay processes of the form 
${\rm d}\Phi_{2}(\{12\};1,2)$ and ${\rm d}\Phi_{2}(a,b;1,2)$.
We refer to these objects as phase-space 
vertices, while the integral ${\rm d} s_\pi/2\pi$, introduced in 
\EqRef{eq:split_ps}, is called a phase-space propagator. 
There is a close correspondence between matrix element computation and 
phase-space generation, justifying this notation.
Even though $s$- and $t$-channel decay seem identical, since 
both represent a solid angle integration, in practice one would use 
different sampling strategies~\cite{Byckling:1969sx}. 

%%%%%%%%%%%%%%%%%%%%%%%%%%%%%%%%%%%%%%%%%%%%%%%%%%%%%%%%%%%%%%%%%%%%%%
\paragraph{Sequential algorithm for phase-space integration}
%%%%%%%%%%%%%%%%%%%%%%%%%%%%%%%%%%%%%%%%%%%%%%%%%%%%%%%%%%%%%%%%%%%%%%
\label{sec:sequential_phasespace}
One of the most efficient general approaches to sampling the phase space 
of multi-particle processes is to employ a sequential algorithm, constructing
the full phase space based on the pole structure of one of the Feynman diagrams
that contribute to the matrix element. This technique was suggested 
very early on in the history of MC programs~\cite{Byckling:1969sx}. 
It is then also possible to construct a separate integrator for each possible 
graph and employ multi-channel methods to optimize the integration~\cite{Kleiss:1994qy}. 
The method provides a general way to adapt to the assumed pole structure of 
arbitrarily complicated matrix elements. It is nowadays widely used 
by the most advanced general-purpose phase-space generators~\cite{
  Kanaki:2000ey,Papadopoulos:2000tt,Cafarella:2007pc,
  Krauss:2001iv,Maltoni:2002qb,Alwall:2007st,Gleisberg:2008fv}. 
The core algorithm can be formulated as recursive relations in terms 
of the phase-space propagators and vertices.

The difference between the various phase-space generators available
today is usually only {\it how} these recursive equations are employed.
If the basic building blocks are 
used to build ``phase-space diagrams'', we obtain an integrator which is
suitable for combination with a diagram-based matrix-element generator.
If the recursion is implemented as is, the resulting phase-space generator
is best combined with a recursive method to compute the matrix elements.
\if{false}
In the latter case it is important to note that while momenta 
labelled by $\alpha$ in \EqRef{eq:ps_building_blocks} 
may correspond to an off-shell internal particle, $b$ always indicates 
a fixed external incoming particle. This eventually allows the reuse of 
MC weights, just as currents are reused in recursive matrix-element 
computations. 
\fi

%%%%%%%%%%%%%%%%%%%%%%%%%%%%%%%%%%%%%%%%%%%%%%%%%%%%%%%%%%%%%%%%%%%%%%
\paragraph{Other algorithms}
%%%%%%%%%%%%%%%%%%%%%%%%%%%%%%%%%%%%%%%%%%%%%%%%%%%%%%%%%%%%%%%%%%%%%%
Other algorithms for phase-space integration exist, which are 
often less general, but potentially more efficient for their purpose.
One of them is the HAAG method~\cite{vanHameren:2002tc},
which is designed to produce momenta distributed approximately according to 
a QCD antenna function for an $n$-particle process, which reads
\begin{equation}\label{eq:antenna}
  A_{n}(p_0,p_1,...,p_{n-1})=\frac{1}{
    (p_0 p_1)(p_1 p_2)...(p_{n-2} p_{n-1})(p_{n-1} p_0)}.
\end{equation}
Different antennae can be obtained from permutations of the momenta $\{p_i\}$.
Generally, like the sequential phase-space integrator described above,
HAAG relies on phase-space factorization over time-like intermediate 
momenta.  The main difference lies in the sequence of factorization 
and in the sampling technique for the basic vertices,
which resembles the phase-space sampling in a dipole shower.

An important, simple but universally applicable phase-space integrator is 
Rambo~\cite{Kleiss:1985gy}. It is widely used because the underlying
algorithm requires no information about the integrand. This makes
Rambo the preferred default choice if no time is to be spent on the 
construction of a dedicated integration channel for the process in question.
Rambo assumes an unconstrained phase space, \ie a phase space where 
four-momentum conservation does not hold, to generate initial particle
momenta. These momenta are in turn boosted and rescaled to arrive at 
a physically meaningful phase-space point. The conformal transformation 
thus applied is associated with an additional weight. Rambo can be used
for massless and massive particles alike, where massive particles simply
require an additional step in the algorithm.

%%%%%%%%%%%%%%%%%%%%%%%%%%%%%%%%%%%%%%%%%%%%%%%%%%%%%%%%%%%%%%%%%%%%%%
%%%%%%%%%%%%%%%%%%%%%%%%%%%%%%%%%%%%%%%%%%%%%%%%%%%%%%%%%%%%%%%%%%%%%%
\mcsubsection{Interface structures}
\label{Sec:Inter_ME}
%%%%%%%%%%%%%%%%%%%%%%%%%%%%%%%%%%%%%%%%%%%%%%%%%%%%%%%%%%%%%%%%%%%%%%
%%%%%%%%%%%%%%%%%%%%%%%%%%%%%%%%%%%%%%%%%%%%%%%%%%%%%%%%%%%%%%%%%%%%%%
\label{sec:MEinterfaces}
%%%%%%%%%%%%%%%%%%%%%%%%%%%%%%%%%%%%%%%%%%%%%%%%%%%%%%%%%%%%%%%%%%%%%%
\paragraph{Les Houches Event Files}
%%%%%%%%%%%%%%%%%%%%%%%%%%%%%%%%%%%%%%%%%%%%%%%%%%%%%%%%%%%%%%%%%%%%%%
The Les Houches Event File (LHEF) format offers a simple structure for 
transferring parton-level events to general purpose event generators that
subsequently accomplish parton showering and hadronization. While the 
original version proposed in \cite{Boos:2001cv} was based on exchanging two
Fortran common blocks, in the recent version the information is 
embedded in a minimal XML-style file structure \cite{Alwall:2006yp}. 

All information specifying the actual run that 
produced the events, \eg the incoming beams, their 
energies and the PDF set used, is collected in a header structure.
This is supplemented by information on the 
cross sections and the event weighting strategy.

For each parton-level event all necessary information is stored in a 
separate structure, listing, amongst other things, the incoming and 
outgoing particle momenta, their flavour and potential mother--daughter 
relations as well as the event's weight. Most importantly for subsequent 
showering, each event carries a definite colour flow, 
determined according to some algorithm by the matrix-element generator 
code.

The LHEF format to output parton-level events is supported by all the 
major matrix-element generator programs and has proved to be a robust tool
for interfacing them with general-purpose event generators,
greatly boosting the set of available processes for the latter. 

%%%%%%%%%%%%%%%%%%%%%%%%%%%%%%%%%%%%%%%%%%%%%%%%%%%%%%%%%%%%%%%%%%%%%%
\paragraph{Binoth Les Houches Accord} 
%%%%%%%%%%%%%%%%%%%%%%%%%%%%%%%%%%%%%%%%%%%%%%%%%%%%%%%%%%%%%%%%%%%%%%
It is apparent from \EqRef{Eq::NLO_XSec_Subtracted}
that calculating a cross section at NLO is a very modular 
task. This is exploited by the Binoth Les Houches Accord, see 
Ref.~\cite{Binoth:2010xt}. It defines
a standard for passing the virtual times Born contribution of an one-loop 
calculation to a tree-level MC program that deals with the 
generation of the corresponding Born and real-emission processes, as 
well as the differential and integrated subtraction terms. 

In an initialization stage the one-loop provider (OLP) and the MC 
program exchange information on the calculational scheme. Then, for a given 
set of Born level momenta, the OLP returns the coefficients of the 
$1/\epsilon^2$ and $1/\epsilon$ poles and the finite term. This is sufficient 
to compose the full cross section calculation within a tree-level generator 
that implements the necessary subtraction terms \cite{Gleisberg:2007md,Frederix:2008hu,Frederix:2009yq,Czakon:2009ss}. This interface 
structure was used recently to calculate the NLO corrections to $W+3$ jets 
\cite{Berger:2009ep} and $t\bar{t}+2$ jets \cite{Bevilacqua:2010ve}.

%%%%%%%%%%%%%%%%%%%%%%%%%%%%%%%%%%%%%%%%%%%%%%%%%%%%%%%%%%%%%%%%%%%%%%
\paragraph{Implementing your own ME into MCs}
%%%%%%%%%%%%%%%%%%%%%%%%%%%%%%%%%%%%%%%%%%%%%%%%%%%%%%%%%%%%%%%%%%%%%%

Of course, the various event generators also support, to varying degrees,
implementations of matrix elements by their users.  This option is particularly
interesting for models with unorthodox particle content or to study small
fragments of larger models.

\mcsection{Top quark mass definitions}
\label{sec:top-quark-masses}
One of the important applications of Monte Carlo event generators is in
the experimental measurement of Standard Model parameters.  An example
where they are particularly heavily used is the top quark mass
determination.  As we will discuss in this Appendix, this application
warrants a deeper investigation of precisely how the top quark mass is
defined.  Our aim is not to review this entire field, but rather to give
just enough background information to set the scene for a discussion
of the mass definition used in event generators.  For more technical
details we refer to the literature and in particular
\cite{Hoang:2008yj,Fleming:2007qr,Fleming:2007xt,Jain:2008gb},
whose approach we largely follow.

Most of our discussion applies equally well to any coloured massive
object (\ie with mass in the perturbative regime, $m\gg\Lambda_{QCD}$):
the bottom quark and more marginally the charm quark, but also any new
coloured particles
that are discovered at the LHC, such as squarks, excited quarks or other
quark partners.  However, we will see that the width of the particle
plays an important role in our discussion and it seems likely that the
top quark is unique in this regard: its decay width (1.5~GeV) is above
the typical scale of confinement so the top decays before it can
hadronize and its production and decay should, in principle, be fully
calculable in perturbation theory.  At the same time, its width is not
so far above the confinement scale and is certainly a lot smaller than
its mass, so events containing top quarks are able to evolve
significantly between its production and decay and parton showers and
high-order perturbative effects are very important.  The bottom quark's
lifetime is much longer and in many, but not all, BSM scenarios those of
new coloured particles are much shorter.

We begin our discussion by recalling that in renormalized quantum field
theory, parameters that appear in the Lagrangian do not have a unique
physical interpretation, but rather are theoretical constructs that
serve as stepping stones to making physical predictions.  In particular,
for each parameter that we renormalize, we have to choose what quantity
to keep fixed, corresponding to the choice of
renormalization scheme.  In the case of particle masses, at one loop
order, we have to consider self-energy corrections that are divergent in
the ultraviolet,
\begin{equation}
  m_0 \to m_0 + \Sigma(m_0),
\end{equation}
where
\begin{equation}
  \Sigma(m) = \frac34C_F\,\frac{\alphaS}{\pi}\,m\left(\frac1\epsilon
  +\mbox{finite}\right)+\mathcal{O}(\alphaS^2)
\end{equation}
is the on-shell quark self-energy in $d=4-2\epsilon$ dimensions.  The
choice of scheme corresponds to a choice of mass parameter
$m^{\mathrm{scheme}}$,
\begin{equation}
  \label{eq:deltam}
  m^{\mathrm{scheme}}=m_0+\delta m
\end{equation}
and a reexpression of $\Sigma$ as a function of $m^{\mathrm{scheme}}$,
such that in
\begin{equation}
  m_0\to
  m^{\mathrm{scheme}}+\Sigma'(m^{\mathrm{scheme}}),
  \qquad \Sigma'(m)=\Sigma(m)-\delta m,
\end{equation}
$\Sigma'$ is finite.  The text-book wisdom is that the choice of scheme
is a purely technical issue, because at a given order of perturbative
theory the corresponding ambiguity is one order higher and therefore, if
calculated to sufficiently high order, the scheme-dependence becomes
irrelevant.  However, this means firstly that it remains a very
important practical issue, because one scheme may result in a
perturbative expansion that converges much more rapidly than another.
If we use the systematic rate of convergence as a criterion for our
preferred choice of scheme, and find that this rate is different for
different physical observables, we will conclude that the `best' choice
of scheme is an observable-dependent statement.  And secondly, the fact
that QCD perturbation theory is at best an asymptotic series means that
one is not able to calculate to infinite orders of perturbation theory
and one must seek a scheme that is well-defined also at the
non-perturbative level.

Before proceeding to discuss specific schemes that are in use, we
briefly mention that if one includes electroweak corrections in the
self-energy, then one obtains an imaginary part from the fact that the
top quark can decay to a quasi-on-shell W boson.  Including this in the
all-orders quark propagator, one obtains the imaginary part that gives
rise to the width term in the Breit-Wigner distribution.  Thus, from a
technical point of view, one can view the renormalized top quark mass as
a complex parameter whose imaginary part gives the top width.

Of the several top quark mass definitions on the market, we can divide
them into two categories: long-distance, which practically means the
pole mass scheme, and short-distance, for example the
$\overline{\mathrm{MS}}$ mass or jet mass schemes, which we define
briefly below.

The pole mass is defined by analogy with the mass definition used in
most QED calculations.  Conceptually, one imagines taking the particle
to infinity and measuring its classical mass in isolation.  Even though
this cannot be physically done for a quark in QCD, one can make it an
operational definition at any finite order of perturbation theory, with
the mass parameter defined to be the real part of the position of the
pole in the complex momentum space.  At the one-loop level, this amounts
to defining $\delta m=\Sigma(m)$ in \EqRef{eq:deltam}.

The archetypal short-distance scheme is the $\overline{\mathrm{MS}}$
one\footnote{Note that the choice of renormalization scheme used for
  particle masses is totally independent of the choice of
  renormalization scheme used for coupling constants. In particular,
  using $\overline{\mathrm{MS}}$ for $\alphaS$ does not require us to
  use the $\overline{\mathrm{MS}}$ scheme also for the top mass. In
  fact, the two schemes are unrelated to each other, except
  operationally: in both one subtracts only the epsilon pole and
  associated universal constants.}.  There, one defines
\begin{equation}
  \delta m(\mu) = \frac34C_F\,\frac{\alphaS}{\pi}\,m
  \,\frac{(4\pi)^{\epsilon}}{\Gamma(1-\epsilon)}\,\frac1\epsilon\,.
\end{equation}
That is, one subtracts only the divergent term itself and associated
universal $\epsilon$-dependent constants.

The difference between the masses in any two schemes can be calculated
as a perturbative series in $\alphaS$.  In particular, the difference
between the pole and $\overline{\mathrm{MS}}$ masses is simply the
ultraviolet-regular part of the self-energy.  Crucial
information about the mass schemes can be obtained by examining the
infrared behaviour of this difference.  At one-loop level, it contains
the integral over gluon loop momenta, weighted by the running coupling
evaluated at the scale of the loop momentum,
\begin{equation}
  \label{eq:C6}
  m^{\mathrm{pole}}-m^{\overline{\mathrm{MS}}}
  \;\stackrel{q\ll m}\sim\;
  C_F\int \frac{\dthree q}{2(2\pi)^3}\,\frac{\alphaS(q)}{q^2}
  \;\sim\; C_F\int \done q\,\alphaS(q).
\end{equation}
This integral is ill-defined in all-order perturbation theory, since it
involves an integral over the region where $\alphaS$ becomes large.
In a perturbative
expansion in powers of $\alphaS(\mu)$, this shows up as a set of
factorially-growing terms, such that perturbation theory does not
converge.  Technically, this gives rise to an ambiguity in the all-order
result, known as the renormalon ambiguity, of
order $\Lambda_{QCD}$: the bottom line is that one cannot,
perturbatively, relate $m^{\mathrm{pole}}$ and
$m^{\overline{\mathrm{MS}}}$ to each other with an accuracy of better
than $\Lambda_{QCD}$.  This indicates that one (or possibly both) of
these definitions is unsuitable for making perturbative calculations
with an accuracy better than this.

Further insight can be gained by calculating simple physical quantities
in the two schemes.  For example, one can calculate the static
interquark potential and show that it has exactly the same renormalon
ambiguity as the self-energy correction.  Therefore a prediction of the
total energy of a static quark-antiquark system in the pole mass scheme,
which absorbs all of the self-energy into the mass definition, leaves a
renormalon ambiguity in the prediction of this physical quantity.  On
the other hand, short-distance mass schemes do not subtract it,
allowing it to cancel between the self-energy and the potential, leaving
a perturbatively-calculable physical prediction.  This argument shows
that, for this observable, a short-distance mass is preferable.  In
fact, for every
observable that has been analysed in sufficient detail to make this
comparison, the same conclusion has been reached.  The practical results
also bear it out: the perturbative series for total top production
cross sections, the top quark decay width and electroweak corrections
such as the $\rho$ parameter, all converge significantly faster if
expressed in terms of the $\overline{\mathrm{MS}}$ mass rather than the
pole mass.  But is this what is measured experimentally?

It is possible to extract a value of the top mass from a measurement of
the $t\bar{t}$ cross section \cite{Abazov:2009ae,Langenfeld:2009wd},
which is unambiguously the $\overline{\mathrm{MS}}$ mass, but this is
considerably less precise than direct measurements from the final-state
properties.
These direct measurements are highly non-trivial conceptually,
precisely because the top quark is not isolated, but rather is produced
as part of a system, evolves by the emission of gluons, decays to a
$b$~quark, which evolves further and then hadronizes to form a jet.  While
there are many refinements in the experimental techniques, they are all
based in one way or another on the measurement of this jet momentum, and
of the decay products of the W that accompanies it (either a lepton and
neutrino or two jets).  Our goal is therefore to understand the
connection between the properties of this jet and the mass of the top
quark that contributed to it.  However, it also contains hadrons
produced by radiation from other partons in the event, including the
initial-state partons, and by the underlying event.  In the absence of a
first-principles understanding of these effects, the experiments model
them with event generators, so that the experimental measurement can
effectively be thought of as a measurement of the top mass parameter of
the particular event generator used.  We assume that this measurement
itself is well understood, and concentrate on the final step of the
analysis: the relation of this parameter to some quantity that can be
defined perturbatively and related to other mass schemes, for example
the $\overline{\mathrm{MS}}$ one.

As a point of principle, it is not possible to make this connection.
Parton shower algorithms are based on leading logarithmic perturbation
theory and as such are not accurate enough to fix the scheme~--
different schemes will only differ by next-to-leading logarithmic
corrections.  Nevertheless, by speculating about how an ideal all-order
algorithm would work, we can obtain an order-of-magnitude result for the
mass parameter that appears in parton shower algorithms.

This argument is facilitated by the approach developed
in~\cite{Hoang:2008yj}.  This showed that all short-distance mass
schemes in use can be defined perturbatively with reference to the pole
mass, an auxiliary mass scale, $R$, and the scale used to renormalize
$\alphaS$, $\mu$,
\begin{equation}
  \label{eq:Revolution}
  m^{\mathrm{pole}} = m(R,\mu)+R\left[\sum_{n=1}^\infty
    \alphaS^n(\mu)\,C_n\!\left(\frac{\mu}R\right)\right],
\end{equation}
where the series in square brackets does not depend explicitly on $m$,
only implicitly through $R$. Renormalization group arguments can then be
used to derive the joint dependence of $m$ on $R$ and $\mu$, which has a
leading logarithm at the $n^{\rm th}$ order $\ln^n\!R/\mu$.  Moreover,
since the perturbation theory in which $m$ is used must also be $\mu$
dependent, large logarithms may arise at all orders of the perturbative
expansion of the observable being calculated, or the expression for $m$,
or both, unless $\mu$ and $R$ are chosen to be of order the physical
scale for the observable being calculated.  One also observes in
calculating the terms in \EqRef{eq:Revolution} that the renormalon
ambiguity arising from the square brackets is equal to that in
$m^{\mathrm{pole}}$, showing that the short-distance mass $m$ does not
contain a renormalon ambiguity.

Different
schemes fall into different classes, with the $\overline{\mathrm{MS}}$
scheme having $R\sim m$, threshold mass schemes such as the 1S, PS and
kinetic mass schemes typically used in B~physics having $R\sim\alphaS m$
and the jet mass scheme discussed below having $R\sim\Gamma_t$, the top
quark decay width.  Using the insight from this renormalization group
approach, one can view $R$ as a new factorization scale above which
physics is integrated out into the mass definition.
Since the infrared contribution to the self-energy (the right-hand side
of \EqRef{eq:C6}) is positive definite, one expects in this picture that
the series in brackets will be positive in general, which it is for all
of the mass definitions in practical use.

With this physical picture in mind, we can describe the action of an
idealized parton shower event generator.  It would describe the
production and evolution of the system containing a top quark using a
properly-matched combination of fixed-order matrix elements and parton
showers to sufficient
accuracy (at least next-to-leading order and next-to-leading logarithmic
respectively).  This evolution would describe the state of the system
down to scales of order the top decay width, whereupon the top quark
would decay.  It would also describe the evolution of the partons
involved in the decay from the scale of the energy release ($\sim m$)
down to the top width.  Finally, for the evolution of the system at
scales from $\Gamma_t$ down to the parton shower's infrared cutoff, the
system of partons produced by the previous steps should be considered
the external partons that emit, including the $b$ quark but not the $t$
quark.  Consideration of this evolution shows that the jet distributions
are affected by physics at all scales, but that only the physics at
scales above $\Gamma_t$ is sensitive to the value of the top quark mass.
Therefore the mass that is reconstructed from such a measurement has, in
principle, all logarithmic physics at scales above $\Gamma_t$ integrated
into it.  We can conclude that perturbation theory will converge
quickest with a mass definition defined at a reference scale
$R\sim\Gamma_t$.  This was illustrated in
\cite{Fleming:2007xt,Jain:2008gb} for the
simpler case of $e^+e^-\to t\bar{t}$, where it was explicitly shown that
a suitably-defined jet mass scheme indeed gave quicker convergence, with
much smaller order-to-order changes in the shape of the top hemisphere
mass distribution than in the $\overline{\mathrm{MS}}$ scheme, for
example.  The final step of the argument is to state that, if the
idealized all-orders calculation converges quickest with such a scheme,
then the scheme-independent leading-order leading-log results will be
most similar to them if their mass parameter is chosen to be of order
the jet mass.

In practice, current parton shower algorithms do not interrupt the
evolution at scale $\Gamma_t\sim1.5$~GeV, although an implementation was
attempted in \pythiasix and some effects of it studied for $e^+e^-\to
t\bar{t}$ \cite{Khoze:1994fu}.  They rather continue it down to
their infrared cutoffs $Q_0\sim1$~GeV.  That is, they shower the events
as if $\Gamma_t<Q_0$.  Despite the small difference between these two
scales, it means that in principle one can repeat the argument above and
state that the parton shower results are most similar to an all-orders
calculation in a scheme in which $R\sim Q_0$.  That is, we can state as
the final result for the likely relation between the top quark mass
measured using a given Monte Carlo event generator (``MC'') and the pole
mass as~\cite{Hoang:2008xm}
\begin{equation}
  m^{\mathrm{pole}} = m^{\mathrm{MC}}+Q_0\Bigl[\alphaS(Q_0)\,c_1+\ldots\Bigr],
\end{equation}
where $Q_0\sim1$~GeV and $c_1$ is unknown, but presumed to be of
order~1 and, according to the argument above, presumed to be positive.
Given that $\alphaS(\mbox{1~GeV})$ is also of order~1, this
states that $m^{\mathrm{pole}}$ could be of order 1~GeV higher than the
value measured by the Tevatron experiments (and hence that
$m^{\overline{\mathrm{MS}}}$ could be of order 1~GeV higher than the
value obtained by assuming that the measured value is actually
$m^{\mathrm{pole}}$).  Since the current experimental uncertainty is
$\pm1.1$~GeV~\cite{:1900yx},
clarifying this relation clearly demands more attention.

% Local Variables: 
% mode: LaTeX
% TeX-master: "../mcreview"
% End: 

%% References

\clearpage
\addcontentsline{toc}{section}{References}
\bibliographystyle{elsarticle-num}
\bibliography{mcreview}






% \end{thebibliography}

\end{document}